%% file: RCW79_langu_corr.tex
\documentclass{aa} %[structabstract]
\pdfoutput=1
\usepackage{xspace}
\usepackage{graphicx}
\usepackage{natbib}
\usepackage{amssymb}
\usepackage{caption}
\usepackage{subfig}

\usepackage{color}
\usepackage{breakurl}
\usepackage[breaklinks]{hyperref}
\bibpunct{(}{)}{;}{a}{}{,}

\def\um{$\mu$m}

\def\msun{$M_{\odot}$}
\def\lsun{$L_{\odot}$}
\def\cm2{cm$^{-2}$}

\def\vel{km\,s$^{-1}$}
\newcommand{\thco}{$^{13}$CO}
\newcommand{\twco}{$^{12}$CO}
\newcommand{\eighco}{C$^{18}$O}

\newcommand{\jybeam}{Jy $\mathrm{beam^{-1}}$}

\newcommand{\nhh}{$N_{\mathrm{H_2}}$}
\def\hii{H\,{\sc{ii}}}
\newcommand{\imcoor}{$\alpha_{2000}=13^{\mathrm{h}}40^{\mathrm{m}}20\fs928,\ \delta_{2000}=-61\degr42\arcmin20\farcs16$}
\newcommand{\herschel}{$Herschel$}

\graphicspath{{Figures/}}
\begin{document}

   \title{\herschel\thanks{\herschel\ is an ESA space observatory with science
   instruments provided by European-led Principal Investigator consortia and with
   important participation from NASA.} observations of the Galactic \hii~region RCW\,79 }

   \author{Hong-Li Liu\inst{1,2,3}, Miguel Figueira\inst{1}, Annie Zavagno\inst{1}, Tracey Hill\inst{4}, Nicola Schneider\inst{5,6,7},
   Alexander Men'shchikov\inst{8}, Delphine Russeil\inst{1},
    Fr$\acute{\rm e}$d$\acute{\rm e}$rique Motte\inst{9,10},
     J$\acute{\rm e}$r$\acute{\rm e}$my Tig$\acute{\rm e}$\inst{1},
    Lise Deharveng\inst{1}, L. D. Anderson\inst{11,12}, Jin-Zeng Li\inst{2},
   Yuefang Wu\inst{13}, Jing-Hua Yuan\inst{2}, \and  Maohai Huang\inst{2}}

   \titlerunning{{\it Herschel} observations of RCW\,79}
    \authorrunning{Liu et al.}
    \institute{Aix Marseille Univ, CNRS, LAM, Laboratoire d'Astrophysique de Marseille, Marseille, France \\
        \email{hong-li.liu@lam.fr}%, annie.zavagno@lam.fr}
        \and National Astronomical Observatories, Chinese Academy of Sciences, 20A Datun Road, Chaoyang District, 100012, Beijing, China
         \and
          University of Chinese Academy of Sciences, 100049, Beijing,  China
                \and Joint ALMA Observatory, 3107 Alonso de Cordova, Vitacura, Santiago, Chile
                \and Univ. Bordeaux, LAB, CNRS, UMR 5804, 33270, Floirac, France
                \and  CNRS, LAB, UMR 5804, 33270, Floirac, France
                \and I. Physik. Institut, University of Cologne, 50937 Cologne, Germany
                \and  Laboratoire AIM Paris–Saclay, CEA/DSM–CNRS–Universit$\acute{\rm e}$ Paris Diderot, IRFU, Service d'Astrophysique, Centre d'Etudes de Saclay, Orme des Merisiers, 91191 Gif-sur-Yvette, France
                \and   Institut de Planétologie et d'Astrophysique de Grenoble (IPAG), Univ. Grenoble Alpes/CNRS-INSU, BP 53, 38041 Grenoble Cedex 9, France
                \and Laboratoire AIM Paris-Saclay, CEA/IRFU - CNRS/INSU - Universit$\acute{\rm e}$ Paris Diderot, Service d'Astrophysique, B$\hat{\rm a}$t. 709, CEA-Saclay, 91191, Gif-sur-Yvette Cedex, France
                \and Department of Physics and Astronomy, West Virginia University,
Morgantown, WV 26506, USA ; Also Adjunct Astronomer at the National Radio Astronomy Observatory,
P.O. Box 2, Green Bank, WV 24944, USA
                \and Adjunct Astronomer at the Green Bank Observatory
                \and Department of Astronomy, Peking University, 100871 Beijing, China
          \\}

%\date{Received on 17/10/2016, and accepted on 30/01/2017}

  \abstract
    {Triggered star formation around \hii~regions could be an important process. The Galactic \hii~region RCW\,79 is
    a prototypical object for triggered high-mass star formation. }
    {We aim to obtain a census of the young stellar population observed at the edges of the \hii~region and to determine
    the properties of the young sources in order to characterize the star formation processes that take place at the
    edges of this ionized region. }
    {We take advantage of \emph{Herschel} data from the surveys HOBYS, ``Evolution of Interstellar Dust", and Hi-Gal
to extract compact sources. We use the algorithm {\it getsources}. We complement the \emph{Herschel} data
    with archival 2MASS, {\it Spitzer}, and WISE data to determine
    the physical parameters of the sources (e.g., envelope mass, dust temperature, and luminosity)
    by fitting the spectral energy distribution.}
    {We created the dust temperature and column density maps along with the column density
    probability distribution function (PDF) for the entire RCW\,79 region. We obtained a sample of 50 compact
     sources in this region,
    $96\%$ of which are situated in the ionization-compressed layer of cold and dense gas
    that is characterized by the column density PDF with a double-peaked lognormal distribution.
    The 50 sources have sizes of $\sim 0.1-0.4$\,pc with a typical value of $\sim 0.2$\,pc, temperatures of $\sim 11-26$\,K, envelope masses of $\sim 6-760$\,\msun, densities
    of $\sim 0.1-44\times 10^5$\,cm$^{-3}$, and luminosities of $\sim 19-12712$\,\lsun.
         The sources are classified into 16 class\,0, 19 intermediate,
     and 15 class\,I objects. Their distribution follows the evolutionary
     tracks in the diagram of bolometric luminosity versus envelope mass ($L_{\rm bol}-M_{\rm env}$) well.
     A mass threshold of 140\,\msun, determined from the $L_{\rm bol}-M_{\rm env}$ diagram, yields 12 candidate massive
      dense cores that may form high-mass stars.
      The core formation efficiency (CFE) for
      the 8 massive condensations shows an increasing trend of the CFE with density. This suggests that the denser the
      condensation, the higher the fraction of its mass transformation into dense cores, as previously observed in other high-mass star-forming regions.
      }
    {}
    \keywords{ISM: H\,{\sc{ii}} region-stars: formation-stars: massive-ISM: individual objects: RCW\,79}
    \maketitle

\section{Introduction}\label{sect:intro}
\hii~regions or bubbles are ubiquitous in the Milky Way. Taking advantage of  the {\it Spitzer}-GLIMPSE \citep{ben03}
and MIPSGAL \citep{car05} surveys, \citet{chu06,chu07} cataloged the first largest sample of
about 600 infrared (IR) dust bubbles in longitudes $|l|\leq 65\degr$.
The comparison of these bubbles with the \hii~region catalog of \citet{pal03} indicates that about $12-25\%$ of the bubbles are overlapping with
\hii~regions \citep{chu06,chu07}. This fraction is probably a lower limit because the \hii~region catalog is incomplete, especially for
the \hii~regions with small diameters \citep{chu07}. Indeed, the fraction can reach $\sim~86\%$ \citep{deh10} and
even more \citep{ban10,and11}.
These results
imply a significant correlation between bubbles and Galactic \hii~regions. Based on the same surveys, a larger sample of more than 5000 IR bubbles
has been visually identified by citizen scientists recruited online \citep{sim12}.
Moreover, using data from the all-sky Wide-Field Infrared Survey Explorer (WISE) satellite, \citet{and14} have made a catalog of over
8000 Galactic \hii~regions and \hii~region candidates by searching for their characteristic mid-infrared (MIR) bubble
morphology.

It is suggested that triggered star formation might occur around \hii~regions or bubbles.
For instance, \citet{deh10} studied the association of 102 Churwell's bubbles with the dense condensations
revealed by the ATLASGAL 870\,\um\ continuum survey data \citep{sch09}. Their study suggested that more than $25\%$ of bubbles
may have triggered the formation of high-mass stars. In addition, analyzing the association
of Red Mid-course Space Experiment \citep[MSX,][]{pri01} massive young stellar objects\footnote{A massive
young stellar object (MYSO) is an embedded infrared source that is luminous enough to be a young O- or B-type star, but
has not yet formed an \hii~region \citep[e.g.,][]{urq07,mottr07}} (MYSOs) with 322 Churwell's bubbles,
\citet{tho12} suggested that about $14-30\%$ of high-mass star formation in the Milky Way might
have been triggered by the expanding \hii~regions or bubbles. Similarly, the study of the association
of MYSOs with 1018 bubbles from the \citet{sim12} catalog indicated that
around $22\pm2\%$ of MYSO formation might have been induced by the expansion of the \hii~regions or bubbles \citep{ken12}.
These results suggest that triggered star formation around \hii~regions or bubbles
may be an important process, especially for high-mass star formation \citep[e.g.,][]{deh10,ken16}.

Triggered star formation may cause the increase of the clump and/or core formation efficiency (CFE),
which is analogous to the star formation efficiency \citep[SFE, e.g.,][]{mot07,bon10b,ede12}. For example,
in the W3 giant molecular cloud, which is a
Galactic high-mass star-forming region, \citet{moo07} found that the CFE is around $5-13\%$ in the undisturbed cloud, but
about $25-37\%$ in the feedback-affected region, which is indicative of an increase in CFE.
Furthermore, \citet{ede12} reported a local increase in CFE in the W43 \hii~region,
which may be associated with the triggering of star formation in its vicinity \citep{bal10}. In addition,
the increases in CFE have been predicted by the simulation of \citet{dal07}. Their simulation with
feedback from an \hii~region results in an SFE approximately one-third higher than in the control run without feedback.

In spite of its importance, triggered star formation remains difficult to be clearly identified \citep[e.g.,][]{elm11,dal13,liu15,dal15}.
It is difficult to distinguish stars formed by triggering from those forming spontaneously \citep{elm11,dal15}. The
high surface density of YSOs observed at the edge of \hii~regions or bubbles is often assumed to
be a result of triggered star formation \citep{zav06,deh09,tho12,ken12,liu15,liu16,yad16,nan16}. However,
these YSOs might either be redistributed by the expansion of \hii~regions or bubbles,  or they might form in situ \citep{elm11,liu15,dal15}.
Moreover, the incompleteness of YSOs, especially for the youngest protostars that are deeply embedded in dense clouds,
prevents us from restoring their true spatial distribution \citep{liu15}. These facts make it difficult to
draw a convincing conclusion on triggered star formation around ionized regions.

\emph{Herschel} observations \citep{pil10}, with their unprecedented angular resolutions and sensitivities
in the far-IR regime, allow us to study the young stellar population
in detail and to assess the importance of triggered
star formation. The  large wavelength coverage of \emph{Herschel} (70$-$500\,\um) and the high sensitivity allow the detection of highly embedded YSOs (e.g., class\,0 objects) that cannot be detected at shorter wavelengths because of their low luminosity and high extinction \citep{zav10}. Therefore,
 \emph{Herschel} observations allow us to obtain a census of the YSO population in all evolutionary stages around ionized region for a better study of the impact of high-mass stars on their surrounding.
Indeed, more class\,0 candidates have been detected with \textit{Herschel \textit{\textup{observations}}}
in different star-forming regions \citep[e.g.,][]{mot10,zav10,hen10,deh12,gia12,sam14,deh15}. Furthermore, the spectral
energy distribution (SED) of YSOs can be better constrained by the large FIR wavelength coverage,
leading to more accurate estimates of their physical parameters such as the envelope mass, dust temperature, and
bolometric luminosity. The luminosity-mass diagram is a  useful tool to infer
the evolutionary properties of the YSOs \citep{mol08,mol16}.  Moreover,
{\it Herschel} images allow us to obtain reliable catalogs of compact
sources (core/clumps, \citet{fig16},
and Tig\'{e} et al. 2016, submitted). This allows us to estimate the CFEs on smaller scales toward \hii~regions. This is crucial for studying the influence of
 ionized regions on local star formation.

RCW\,79 is a textbook example of an \hii~region where triggered star formation might have taken place
\citep[][hereafter ZA06]{zav06}.
  In this paper, we analyze the star formation observed around this region,
using  \emph{Herschel} data from the surveys HOBYS\footnote{The Herschel imaging survey of OB Young Stellar objects (HOBYS) is a Herschel key programme. See \url{http://hobys-herschel.cea.fr}} \citep{mot10} and  ``Evolution of Interstellar Dust" \citep{Abe10}, complemented with 2MASS, {\it Spitzer,} and WISE data.
Our purposes are to search for clumps
that may form high-mass stars, to explore the star formation evolutionary scenarios, and to
investigate the CFEs in RCW\,79.  This paper is organized as follows: we present
 RCW\,79 in Sect.\,\ref{sect:present}, the \emph{Herschel} observations together with other archival data sets are described
in Sect.\,\ref{sect:obs}, the results are presented in Sect.\,\ref{sect:result}, and the discussion is given in Sect.\,\ref{sect:discuss}, followed by
our conclusions in Sect.\,\ref{sect:conclude}.

\section{Presentation of RCW\,79}\label{sect:present}
%%%%%%%%%%%%%fig:tcolor%%%%%%%%%%%%%%%%
\begin{figure}[!ht]
\centering
\includegraphics[width=3.4 in]{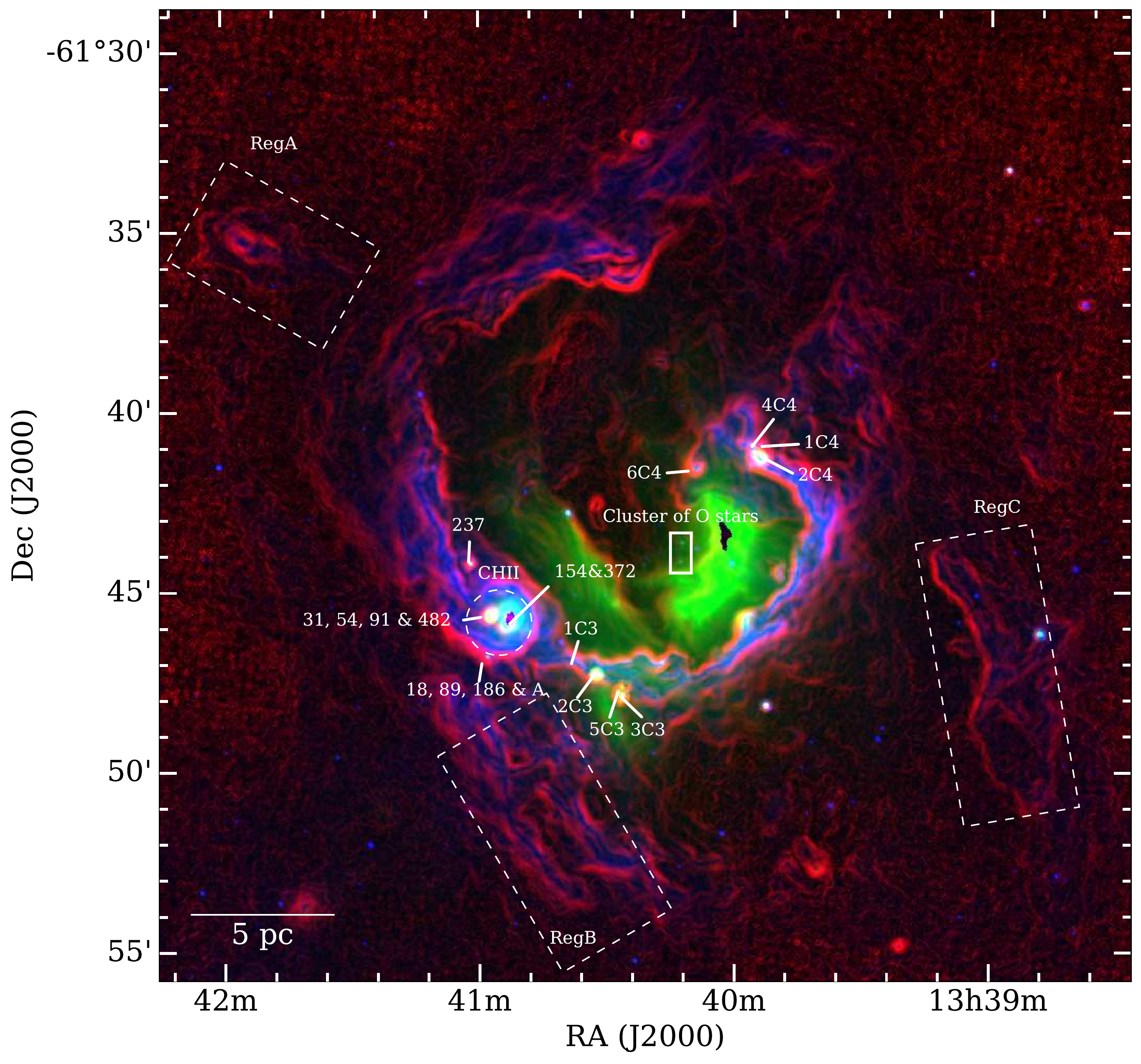}
\caption{Three-color image of RCW\,79: {\it Spitzer} GLIMPSE 8\,\um\ image (blue), {\it Spitzer} MIPSGAL 24\,\um\ image (green),
and {\it Herschel} 70\,\um\ image (red). The black knot seen at 24\,\um\ in the direction of the \hii~region is a result of saturation.
The 70\,\um\  image is unsharped to better emphasize the YSOs associated with the region.
 Nineteen YSOs are pinpointed with the identity number given by ZA06. The compact \hii~region (C\hii) is delineated with the dashed circle.
The full box locates the exciting stars composed of a cluster of O stars \citep{mar10}.
The dashed boxes mark the other three regions (RegA, RegB, and RegC) associated with RCW\,79. North is up and east is left.}
\label{fig:tcolor}
\end{figure}
%%%%%%%%%%%%%fig:end%%%%%%%%%%%%%%%%
RCW\,79 \citep{rod60} is a bright optical \hii~region ionized by a cluster of a dozen O stars,
the two most massive of which have a spectral type O4-6V/III \citep{mar10}. The ionizing luminosity of
the ionizing stars was estimated to be $10^3$ times higher than the mechanical luminosity of their stellar winds \citep{mar10}, indicating
 a radiation-driven \hii~region. This region
is spatially encompassed by an almost complete dust ring (see Fig.\,\ref{fig:tcolor}), with
a diameter of $\sim12\arcmin$ , which corresponds to 12.8\,pc at a distance of 4.3\,kpc (ZA06).
The ring structure was revealed  with a velocity range of
 $-50$ to $-44$\,\vel\ in the observations of \twco, \thco, and \eighco\ ({\it J=1-0}) \citep{sai01}.
It is in good agreement with the velocity range of $-51$ to $-40$\,\vel\ measured for ionized gas (ZA06),
indicative of a good association of the dust ring with the \hii~region.

Figure\,\ref{fig:tcolor} shows the composite three-color image of RCW\,79
 where blue, green, and red code the {\it Spitzer} 8.0 and 24\,\um\
       and the {\it Herschel} 70\,\um, respectively.
An unsharp masking\footnote{The module {\it scipy.ndimage} for multidimensional gradient magnitude using Gaussian
derivatives is available at \burl{http://docs.scipy.org/doc/scipy/reference/py-modindex.html}.} was applied to the  70\,\um\ image to
filter out diffuse emission and enhance the contrast of intense emission in the images.
As shown in Fig.\,\ref{fig:tcolor}, the dust ring is seen at both 8.0 and 70\,\um.
Emission at 8.0\,\um\ mainly comes from  polycyclic aromatic hydrocarbons (PAHs) at 7.7 and 8.6\,\um,
indicative of photoionization regions \citep[PDRs, e.g.,][]{pav13} which indicate the interplay between
ionized gas and the adjacent neutral cloud.
The 70\,\um\ emission mainly traces hot components such as very small grains (VSGs) or
warm material heated by protostars. Therefore, the appearance of the dust ring at both 8.0 and 70\,\um\
demonstrates that the enclosed \hii~region is  interacting with and heating its vicinity.
The 24\,\um\ emission is predominantly distributed in the
direction of the \hii~region. This spatial distribution is in good agreement with the fact that
24\,\um\ emission mainly arises from hot dust, which can reach rather high temperatures after absorbing
high-energy photons \citep[e.g.][]{deh10,liu16}.

SEST-SIMBA 1.2 mm continuum observations with an angular resolution of $24\arcsec$
 revealed three highest mass fragments in
  the dust ring (see ZA06).
  For this, ESO-NTT SOFI near-IR and {\it Spitzer} GLIMPSE mid-IR data were combined to
  study the young stellar population observed toward this region.
Nineteen class\,I YSO candidates (see Fig.\,\ref{fig:tcolor}) were found to be associated
with the three fragments. In addition, one compact \hii~region (C\hii) is embedded in
the most massive fragment in the southeast region of the ring.
\citet{mar10} furthermore observed 8 out of the 19 YSOs with the near-IR integral field spectragraph SINFONI
mounted on the VLT telescope.  All present spectral features typical of YSOs.
All lines have velocities similar to that of the ionized gas, confirming the association of these YSOs with the region.
The dust ring is opened in the northwest.
The H$\alpha$ velocity field of ionized gas shows a flow through the hole with a few \vel\
(see Fig.\,12 of ZA06). This flow was interpreted as a champagne phenomenon, indicating a strong interaction
of RCW\,79 with its surrounding material (ZA06). Moreover, combining the model of \citet{whi94},
ZA06 found that the ring of collected gas had enough time to fragment during the lifetime of RCW\,79, and that
the radius and mass of the fragments basically agree with the values predicted by the model.
Therefore, ZA06 concluded that the
YSOs at the edge of RCW\,79 might have been triggered by the expanding \hii~region.

An elongated clump (i.e., RegA) is located about 6\,pc away from the northeast edge of RCW\,79 (see Fig.\,\ref{fig:tcolor}).
This clump is associated with 8\,\um\ emission, suggesting
influences from ionized gas. The northeastern ring orthogonal to the
clump appears diffuse relative to its neighbors. This diffuse characteristic implies that
the clump could be photoionization-shaped by the leaking photons from the \hii~region through the
diffuse ring, as shown in Fig.\,\ref{fig:TdNH2}(c).
Additionally, there are two other extended filamentary features (e.g., RegB and RegC) situated to the south and  southwest of RCW\,79, respectively.
Based on the observations of \twco, and \thco\ ({\it J=1-0}) \citep{sai01}, these two features have the same
velocity as RCW\,79, which means that they may be associated. As shown in Fig.\,\ref{fig:tcolor},
these two features are associated with PDRs, as seen in 8\,\um\ emission. Likewise, the two features
 could be a consequence of photoionization by the leaking photons from RCW\,79, as discussed in Sect.\,\ref{sect:dustproperty}.

\section{Observations and data reduction}\label{sect:obs}
\subsection{Herschel observations}
%%%%%%%%%%%%%%%%%%%%%%%%%%%%%%%%%%%%%%%%%%%%%%%%%%%%%%%%%%%%%%%%%%%%%%%%%%%
\begin{table}
\caption{{\it Herschel} observational parameters}
\label{tbl:obsHerschel}
\resizebox{9cm}{!}{
\centering
\begin{tabular}{lcccc}
\hline\hline
Instrument & Size    & Time & ObsIDs    & Date        \\
           & arcmin  & s    &           & yyyy-mm-dd  \\
\hline

{\it PACS}  &  $30\times30$ &           \phantom{0}2768 & 1342188880, 1342188881 & 2010-01-03 \\
{\it SPIRE} &  $30\times30$ &           \phantom{0}837  & 1342192054             & 2010-03-10 \\

\hline
\end{tabular}
}
\end{table}
%%%%%%%%%%%%%%%%%%%%%%%%%%%%%%%%%%%%%%%%%%%%%%%%%%%%%%%%%%%%%%%%%%%%%%%%%%
RCW\,79 was observed as part of the HOBYS \citep{mot10} and ``Evolution of Interstellar Dust" \citep{Abe10} guaranteed
time key programs. The Photodetector Array Camera \&~Spectrometer \citep[PACS,][]{pog10} at 100 and 160\,\um\ and the Spectral
and Photometric Imaging Receiver \citep[SPIRE,][]{gri10} at 250, 350, and 500\,\um\ were equipped to carry out both
surveys with  scan speeds of
$20\arcsec$ per second for PACS and  $30\arcsec$ per second for SPIRE. The angular resolutions of these
five bands in order of increasing wavelength are 6$\farcs7$, 11$\farcs4$, 18$\farcs2$, 25$\arcsec$, and 36$\arcsec$. Table~\ref{tbl:obsHerschel} lists the observation parameters including
the mapping size, the total integration time, the observation
identification number, and the observation date.

The {\it Herschel} data were processed using slightly modified versions of the default PACS and SPIRE pipelines
built into the {\it Herschel} interactive processing environment (HIPE) software v.\,10. The pipelines produced level\,2 data,
which to some extent suffer from striping artifacts in the in-scan directions and flux decrements around bright zones of emission
resulting from the median-filtering baseline removal. To remove both artifacts, the {\it Scanamorphos}
software \citep{rou12}, version 9, was used to create the final level\,2 maps without the ``Galactic" option.
Additionally, the astrometry of all the maps
was adjusted to be consistent with each other and with higher
resolution {\it Spitzer} data.

The 70\,\um\ image from the Hi-Gal survey \citep{mol10} was also retrieved to complement our observations. Its measured
angular resolution is 10$\farcs7$ \citep{tra11}. The detailed descriptions of the preprocessing of the data up to usable
high-quality image can be found in \citet{tra11}.

The absolute calibration uncertainty for PACS is estimated to be $10\%$ at 70 and 100\,\um\ and $20\%$  at 160\,\um\
(see PACS observers' manual\footnote{\burl{http://herschel.esac.esa.int/Docs/PACS/pdf/pacs\_om.pdf}}), while
for SPIRE it is within $10\%$ for all bands (see SPIRE observers'
manual\footnote{\burl{http://herschel.esac.esa.int/Docs/SPIRE/spire\_handbook.pdf}}).
\subsection{Archival data}
To carry out a multiwavelength analysis of this region,
ancillary infrared data were taken from the IRSA Archive.\footnote{\burl{http://irsa.ipac.caltech.edu/frontpage/}}
The {\it J, H, Ks} images at 1.25, 1.65, and 2.17\,\um\ with a
resolution of $4\arcsec$ were retrieved from the Two Micron All Sky Survey \citep[2MASS,][]{skr06}. In addition,
the images of the {\it Spitzer} Infrared Array Camera (IRAC) at 3.6, 4.5, 5.8, and 8.0\,\um, together with
the Multiband imaging photometer for {\it Spitzer} (MIPS) at 24\,\um, were obtained from the GLIMPSE \citep{ben03}
and MIPSGAL \citep{car05} surveys, respectively. The resolutions in the IRAC bands are better than $2\arcsec,$ and the resolution
is $6\arcsec$ in the MIPS 24\,\um\ band. Moreover, the 12\,\um\ image with a resolution of
$6\farcs5$  was used from the WISE survey \citep{wri10}.
This survey provides the images in four wavelength bands, but the 3.4 and 4.6\,\um\ bands were not
 taken into account because their resolutions are lower than those of the IRAC  3.6 and 4.5\,\um\ bands.
 We did not make use of the 22\,\um\ data either because negative values appear in the
majority of pixels of the image covering RCW\,79.

\section{Results}\label{sect:result}
\subsection{Dust temperature and column density maps}\label{sect:dustproperty}
%%%%%%%%%%%%%fig:TdNH2%%%%%%%%%%%%%%%%
\begin{figure*}[!ht]
\centering
\includegraphics[width=3.4 in]{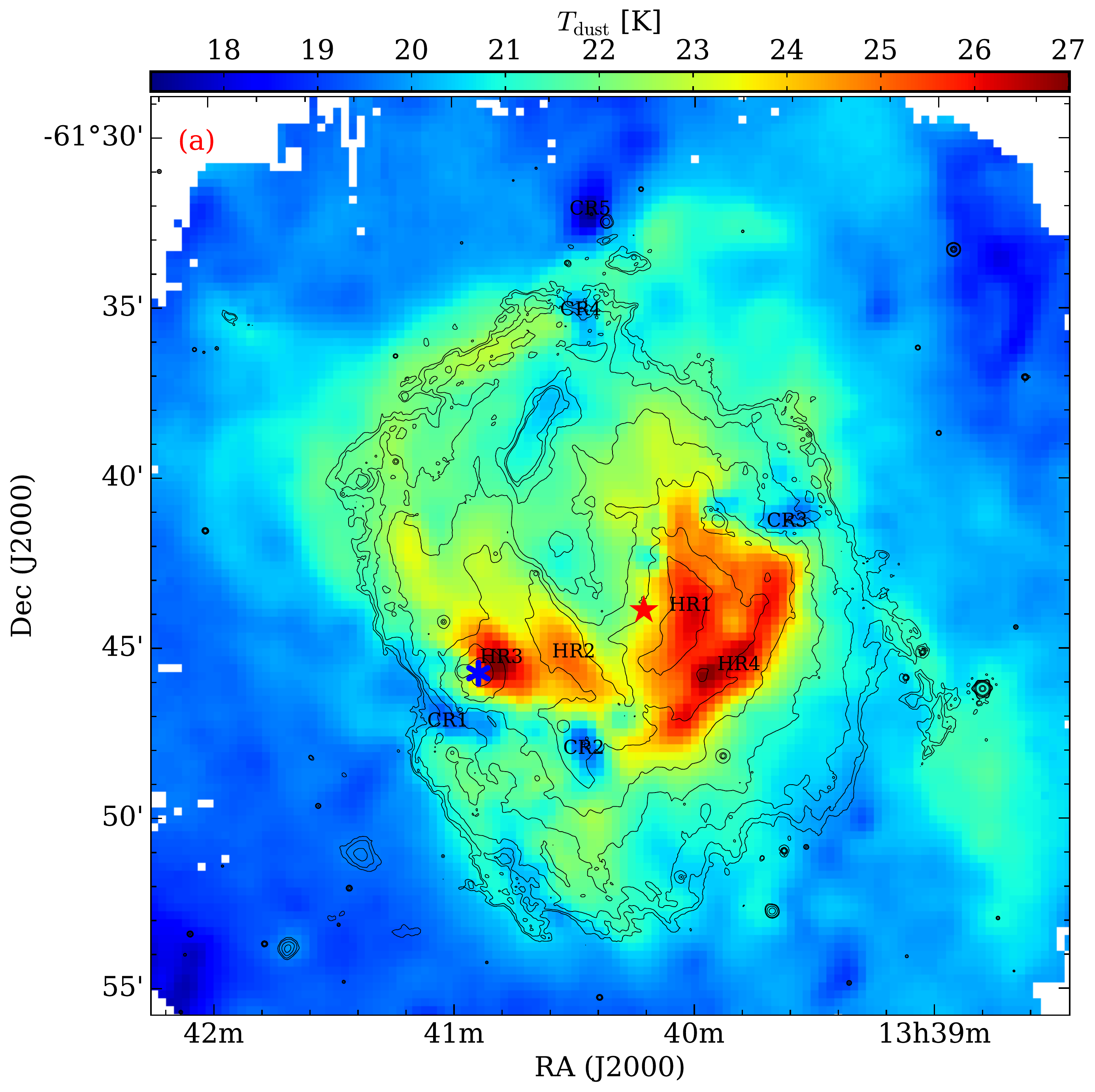}
\hskip 0.00000001cm
\includegraphics[width=3.4 in]{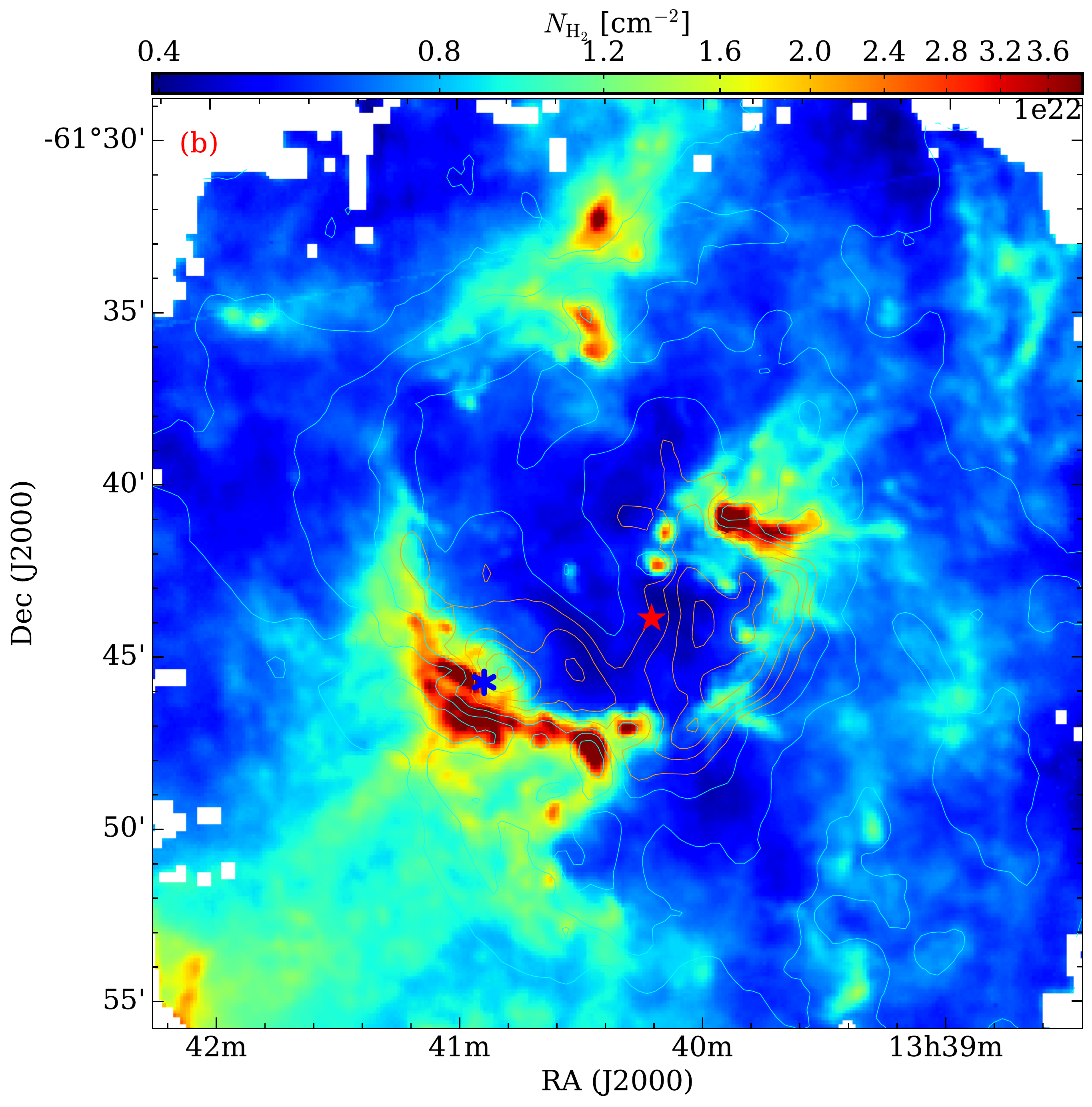}
\includegraphics[width=3.4 in]{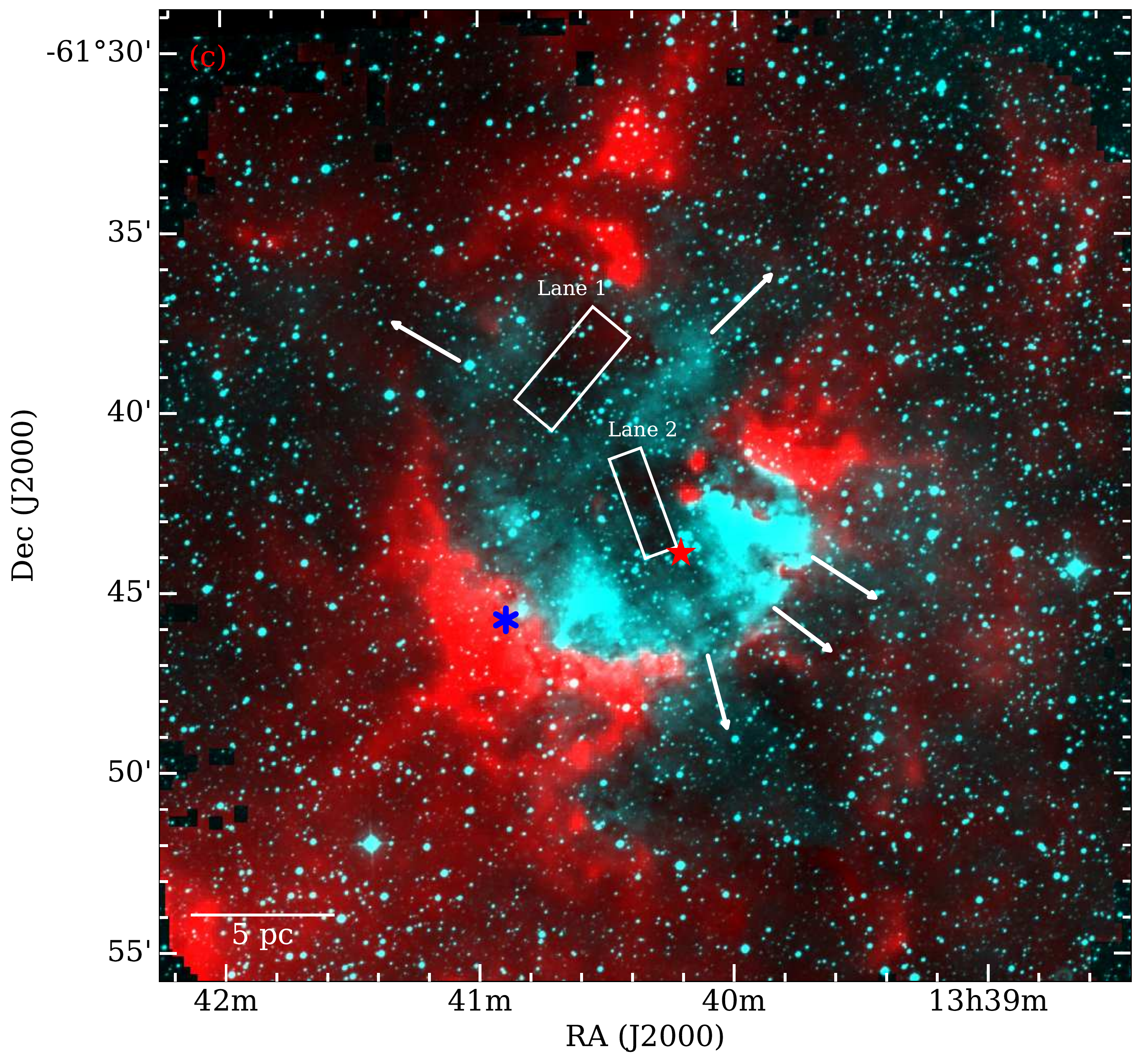}
\caption{(a) 36$\arcsec$ resolution dust temperature  map (color scale) overlaid with {\it Spitzer}-MIPSGAL 24\,\um\ emission (black contours).
The  contour levels are 40, 44, 55, 84, 163, 374, 940, and 2459 MJy\,sr$^{-1}$. HRs\,1-4 show the four hottest regions and CRs\,1-5 the five
coldest regions in RCW\,79.
(b) $18\farcs2$ resolution column density map with dust temperature contours.  The cyan and orange temperature contours start from 20, and 23\,K,
respectively,  with a step of 1\,K. (c): two-color composite image consisting of the $18\farcs2$ resolution column density map (in red) and
the SuperCOSMOS H$\alpha$ image (in turquoise). Arrows denote directions of leaking ionizing photons and rectangles outline dark
H$\alpha$ emission. A scale bar of 5\,pc is shown on the bottom left. In all panels, the star and asterisk symbolize the center position
of a cluster of exciting stars for the classical and compact \hii~regions, respectively.}
\label{fig:TdNH2}
\end{figure*}
%%%%%%%%%%%%%fig:end%%%%%%%%%%%%%%%%
The dust temperature ($T_{\mathrm{dust}}$) and column density (\nhh) maps of RCW\,79 were created using
a modified blackbody model to fit the SEDs pixel by pixel, as described by \citet{hil12a,hil12b}.
Before the SED fitting, all {\it Herschel} images except for the 70\,\um\ image were convolved
to the resolution of the 500\,\um\ band  and then regridded to the same pixel size as
that of the 500\,\um\ image. Emission at 70\,\um\  was excluded in the SED fitting because it
can be contaminated by emission from small grains in hot PDRs.
In the SED fitting, a dust opacity law of  $\kappa_{\nu} = 0.1 \times (\nu/1 \mathrm{THz})^{\beta}$
 was adopted with a gas-to-dust mass ratio of 100 \citep{bec90}.  $\beta=2$ was fixed to be
 consistent with other papers of the HOBYS consortium.
Additionally, to reveal more small structures, a high-resolution ($18\farcs2$) column density map
was made based on the method of \citet{hil12a}. The 36$\arcsec$
resolution $T_{\mathrm{dust}}$
and $18\farcs2$ resolution \nhh\ maps
are presented in Fig.\,\ref{fig:TdNH2}.

 In Fig.\,\ref{fig:TdNH2}\,(a) the dust
temperature distribution on the large scale  almost agrees with
the 24\,\um\ emission.
As mentioned in Sect.\,\ref{sect:present}, this 24\,\um\  emission traces hot dust
heated by high-energy ionizing photons, which can be demonstrated by
 the good spatial coincidence of 24\,\um\ emission
with ionized gas seen by the H$\alpha$ emission (see Fig.\,\ref{fig:TdNH2}\,(c)).
On the small scale, we see the four main hottest regions (HRs\,1-4) with a range of 23.5 to 27\,K (see Fig.\,\ref{fig:TdNH2}\,a-b). The first two (HR1 and HR2)
are located in the direction of the \hii~region, spatially overlapping with dense ionized gas
(see Fig.\,\ref{fig:TdNH2}\,(c)). The other two (HR3 and HR4) are located on the southwestern and southeastern edges of RCW\,79, respectively,
 where they are exposed to ionized gas. These four hottest regions could in large part be a consequence of their exposure to the heating of ionized gas.
Additionally, the hottest of the four regions (HR3) is situated on the southeastern edge (see Fig.\,\ref{fig:TdNH2}\,a-b), but it is
the farthest from the cluster of ionizing stars of RCW\,79. Given the association of  this region with
the C\hii~region, the highest temperature can be attributed to an additional heating from the C\hii~region.
In contrast, there are five cold regions (CRs\,1-5)  with lower temperatures
($\sim 17.5-20$\,K, see Fig.\,\ref{fig:TdNH2}\,(a-b)).
These five coldest regions are all spatially coincident with the column density peaks.
The first two regions (CR1 and CR2) lie on the southern edge, the third region (CR3) lies on the western edge, and the remaining two regions (CR4 and CR5) lie in
the northern area of RCW\,79. For these regions, the anticorrelation between their temperatures and column densities
suggests that their low temperatures could  arise, in part, from a lower
penetration of the external heating from the \hii~region into dense regions \citep{liu16}.

Figure\,\ref{fig:TdNH2}\,(c) shows the column density map (red) superimposed on H$\alpha$ emission (turquoise) from the SuperCOSMOS survey \citep{par98}.
The shell seen in the column density distribution encompasses ionized gas traced by H$\alpha$ emission,
suggesting the strong impact that the enclosed \hii~region has on its surroundings.
The influence on the column density structure is discussed in more detail in Sect.\,\ref{sect:pdf}.
Of interest are two
dark lanes of H$\alpha$ emission in the direction of the \hii~region. Such characteristics
have also been observed in RCW\,120 \citep{and15}. The dark lanes of H$\alpha$ emission
have been attributed to optical absorption by foreground material. As shown in Fig.\,\ref{fig:TdNH2}\,(c), lane\,1 is indeed
associated with column density enhancement with respect to its surrounding. Therefore, the dark lane\,1 of ionized gas
could be a result of optical absorption by foreground material. In contrast, lane\,2 is not clearly related with enhanced column densities, indicating that lane\,2 of ionized gas may
not be caused predominantly by optical absorption of foreground clouds. In fact, lane\,2 of
the ionized gas is spatially well coincident with the central cavity,  as inscribed  at 24\,\um\
in Fig.\,\ref{fig:TdNH2}\,(a).  On the basis of hydrodynamical simulations,
 it is suggested that a young \hii~region should
not be strongly affected by stellar winds at the beginning of its evolution, but the winds eventually become stronger,
giving rise to a very hot dust cavity \citep{Cap01,Fre03,Fre06}. Such a cavity is probably shown in the
24\,\um\ emission tracing hot dust \citep{wat08,liu15}. However, the dust cavity  could also be caused by either
the radiation pressure of ionizing stars
or dust destruction by their intense radiation \citep{Ino02,Kru09,mar10}.
In lane\,2, ionized gas dark emission indicates no intense radiation in this cavity.
Additionally, \citet{mar10} concluded that the effect of the stellar winds on the dynamics of RCW\,79
is rather limited. Therefore, we suggest that   lane\,2 may arise from the dispersion of ionized gas
by the strong radiation pressure from the cluster of ionizing stars.
Diffuse ionized gas emission is observed outside the ionized region, indicated by arrows in Fig.\,\ref{fig:TdNH2}\,(c).
These emissions can be due to the leaking of ionizing photons through density holes in the PDR, as has been suggested
in \citet{zav07} (see their Fig.\,2).

\begin{figure*}[htb]
\begin{centering}
\includegraphics [width=8.8cm, angle={0}]{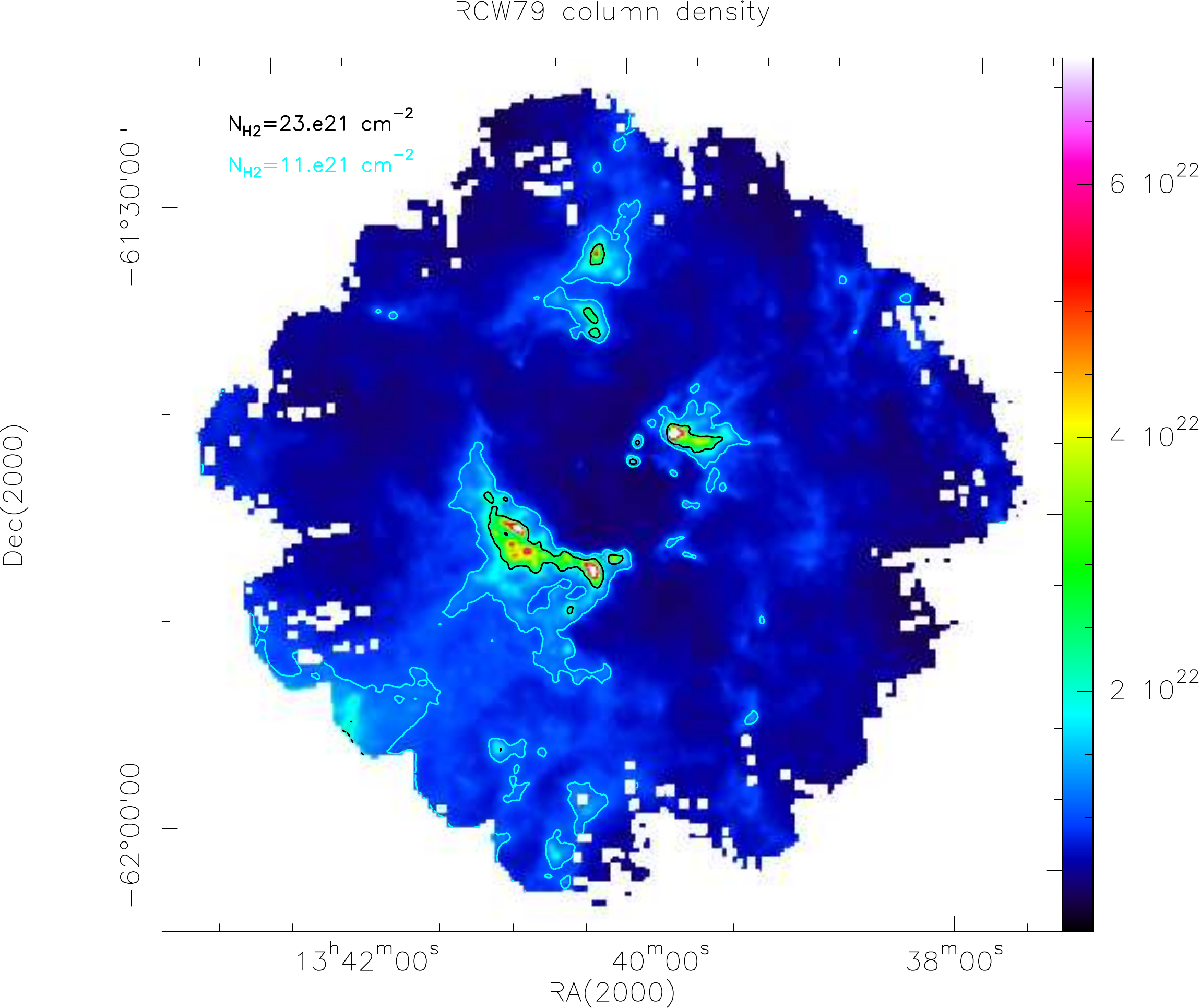}
\includegraphics [width=6.8cm, angle={90}]{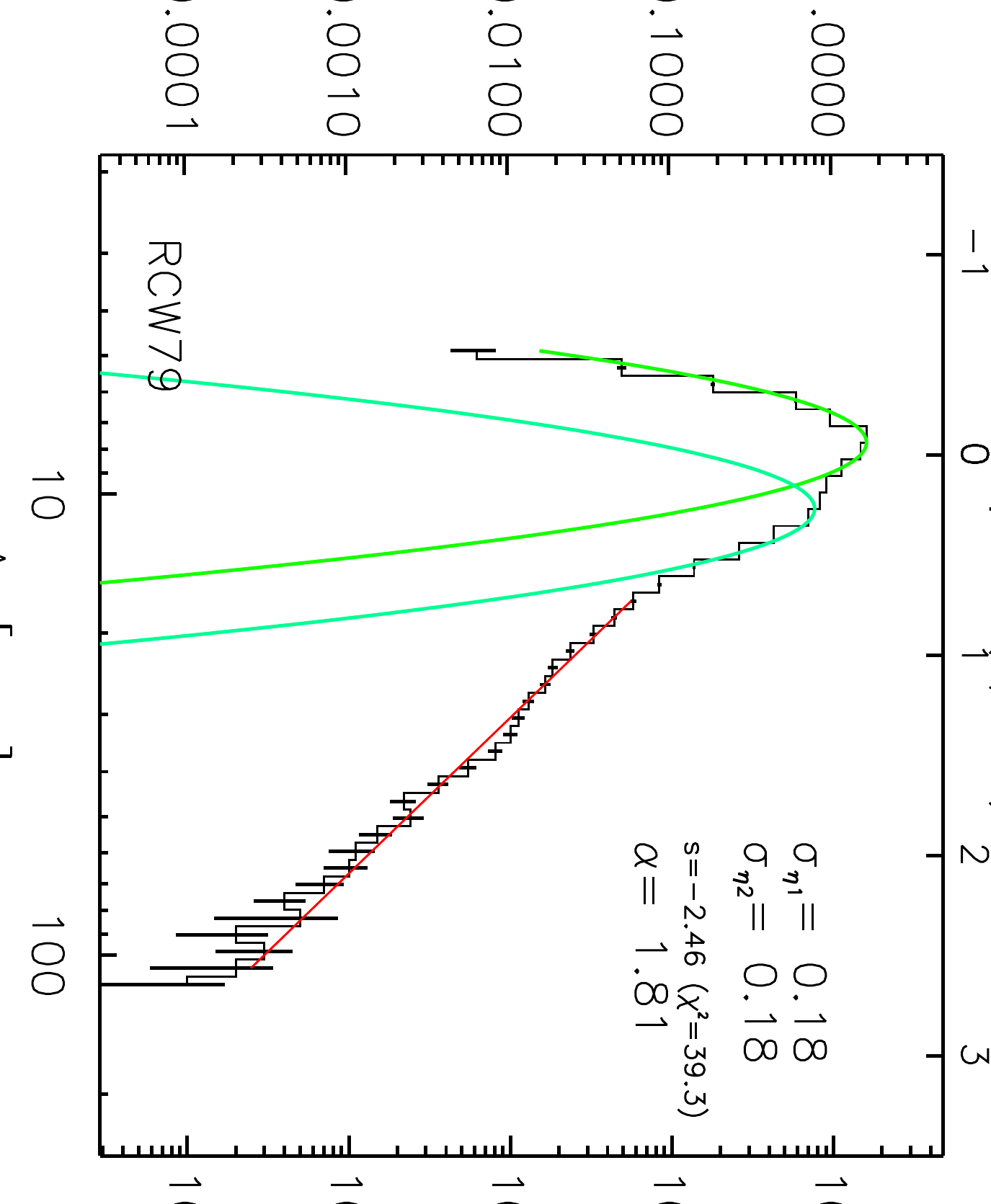}
\caption{{\bf Left:} High-angular (18$''$.2) resolution column density
  of RCW\,79.  The contour levels characterize the gas in the
  compressed layer (above A$_{\rm v} \sim$10--11) and the gravitationally
  dominated gas (above A$_{\rm v} \sim$20). {\bf Right:} PDF of the column
  density map. The distribution is characterized by two lognormals
  (dark and light green) with similar widths of $\sigma$=0.18), and a
  power-law tail with a least-squares fit in red. The slope $s$ of the
  power-law tail is --2.46. }
\label{pdf-figure}
\end{centering}
\end{figure*}

\subsection{Probability distribution function}\label{sect:pdf}

The impact of different physical processes (e.g., turbulence, gravity,
or pressure) on the column density structure of a whole molecular cloud
or parts of it can be studied using probability distribution functions
(PDFs) of the column density.  PDFs are frequently used in
observations and theory \citep[e.g.,][for an overview]{kai09,sch15a,fed12}. Using {\sl
  Herschel} dust maps, various studies showed that a lognormal shape
of the PDF is consistent with low-density gas, dominated by turbulence
\citep{sch13}, and a single or double power-law tail appears
for dense star-forming clouds \citep[e.g.,][]{hil11,rus13,sch15b,kon15}, which is attributed
to the effects of gravity \citep[e.g.,][for an overview]{gir14}.
 Molecular clouds surrounding \hii~regions \citep{sch12,tre14a}
  showed PDFs with two lognormal
distributions (``double-peak" PDFs) or PDFs with a larger width, followed
by a power-law tail. These observations were interpreted as an
expansion of the ionized gas into the turbulent molecular cloud
(representing the first lognormal form of the PDF), leading to a compression zone
with higher densities that in turn cause the second peak, but are still dominated by
turbulence. The widths of the two lognormal distributions of the PDFs and the distance between the peaks
depend on the relative importance of ionization pressure and turbulent
ram pressure \citep{tre14a}.

For this paper, we constructed a PDF\footnote{To be consistent with
  other {\sl Herschel} studies, we used the visual extinction $A_{\rm v}$
  instead of the hydrogen column density, linked by $N_{\rm H_2} =A_{\rm v} \times 0.94 \times 10^{21}$\,cm$^{-2}$\,mag$^{-1}$ \citep{boh78}.}
  of RCW\,79 from the
whole area observed with {\sl Herschel}, using the high- and
low-angular resolution maps (18$''$.2 and 36$''$, respectively). The
PDFs do not differ much \citep[see][for resolution effects on PDFs]{sch15a,oss16}, therefore we
only present the PDF of the high-resolution map here. We did not perform a
background subtraction, as recommended by \citet{sch15a},
because the map of RCW\,79 is too small to clearly define a background
level. The column density at the map borders is on the order of a few
A$_{\rm v}$, but may still contain parts of the associated molecular cloud.

Figure~\ref{pdf-figure} (left) shows the column density map and the
corresponding PDF (right).  The PDF has a complex structure, and the best
results (we performed KS-tests for which we fit different distributions)
were obtained with two lognormal distributions in the lower column density
range and one power-law tail for higher column densities. The widths
of the lognormal forms are $\sigma$=0.18 for both, and the peaks are around
A$_{\rm v}$ = 7 and A$_{\rm v}$ = 11. Starting at A$_{\rm v}$ = 20, the distribution is
better described by a power-law tail with a slope of s=--2.46, which
corresponds to $\alpha$=1.8 for a spherical density distribution with
$\rho \propto r^{-\alpha}$. Our interpretation of these results is
that the first lognormal form shows the turbulent gas of the associated
molecular cloud (in dark blue scale in the column density map), followed by the
compressed shell component (indicated by the gas component between the
light blue and black contours in Fig.\,\ref{pdf-figure}). This turbulent
gas layer starts to fragment, and gravity takes over in the densest
parts of the compressed shell, forming clumps and finally cores. The
gravitational collapse of the embedded cores then leads to the
power-law tail. The exponent $\alpha$=1.8 assuming a spherical density
distribution is consistent with the exponent $\alpha$=1.5--2 predicted
from theory \citep{shu77,whi85}.  These results are fully consistent
with what is found in \citet{tre14a} for RCW\,120, which is also
an \hii~region bubble. The only difference is that the peaks of the
two lognormal distributions of the unperturbed lower density gas and the compressed shell are
closer together, implying that the density contrast in RCW\,79 is lower.
The second lognormal form indicates the compression from
ionized gas that might have created the necessary condition for triggered star formation in RCW\,79.

\subsection{Compact sources}\label{sect:compactsources}
\subsubsection{Source extraction}\label{sect:sourExtract}
The algorithm \emph{getsources} \citep{men10,men12,men13} was used to extract compact sources from the images at all {\it Herschel}
wavelengths from 70 to 500\,\um. Full details on the source extraction can be found in Appendix\,\ref{sect:app-Sectraction}.
The resulting catalog returned by \emph{getsources} contains the identity number of sources,
unique coordinates, peak and integrated fluxes with respective errors, and FWHM major and minor
sizes with a position angle at each wavelength, as presented in Table\,\ref{tbl:herschel_flux_ori}.
In our work, 317 sources were initially extracted as candidate compact sources
(Sig$_{\rm mono}$ $ \geq 7$)\footnote{Sig$_{\rm mono}$
is the detection significance given by \emph{getsources}. A source with Sig$_{\rm mono}$ $ \geq 7$ is
regarded as reliable \citep{men12}.} within a region of
25$\arcmin \times$ 25$\arcmin$ centered at \imcoor.

To pick out the most reliable compact sources, two selection criteria were applied to  the 317 sources:
\begin{enumerate}
\item A minimum of three measured integrated fluxes with good qualities for a  better constraint on
the SED fitting in Sect.\,\ref{sect:sedfit}. They must include the flux at a reference wavelength (160 or 250\,\um,
see Sect.\ref{sect:app-SEDfit}). The signal-to-noise (S/N) ratios of the integrated and corresponding peak fluxes
at each wavelength have to be greater than 2 to guarantee the good qualities of the flux measurements.
\item An axis ratio of $<2$ and a deconvolved size (see Eq.\,\ref{eq-sizecorrect}) of $<0.4$ pc for
preliminarily selected compact sources. The latter is arbitrarily determined and is based on the fact that  most of sources in the catalog have sizes of $<0.4$\,pc.  Objects that do not fulfill
the above criteria may be cloud fragments or filament pieces.
\end{enumerate}
Applying these criteria, we end up with a sample of 100 candidate compact sources.

\subsubsection{Graybody SED fitting}\label{sect:sedfit}
Assuming optically thin dust emission,
we adopted a single-temperature graybody function to fit the SED of sources
between 100 and 500 \um\ (see Appendix\,\ref{sect:app-SEDfit}).
The SED fitting was performed for the 100 candidate compact
sources to derive their physical parameters including the dust temperature $T_{\rm d}$ and envelope mass $M_{\rm env}$.
As mentioned in Sect\,\ref{sect:dustproperty}, 70\,\um\ emission was not included in the SED fitting.
Before the SED fitting, we scaled fluxes to the same aperture at the reference wavelength (see Sect.\,\ref{sect:app-SEDfit})
 and made the corresponding color
corrections. Detailed descriptions of the flux scaling and color correction methods can be found
in Appendix\,\ref{sect:app-SEDfit}. In the SED fitting, the dust emissivity spectral index,
$\beta$, was set at  2 and was not left as a free parameter. This is the value adopted for the
analysis of the HOBYS survey. Moreover, \citet{men16} suggested that
variable $\beta$ during the SED fitting for mass derivation leads to huge biases and should never be
used \citep{men16}. We must keep in mind that absolute values in masses are at least a factor
of 2--3 \citep{men16}.
After the SED fitting, we kept 89 sources selected by the goodness
of their fit ($\chi^2/N_{data}<10$).

\subsubsection{Infrared counterparts}\label{sect:IRfeatures}

%%%%%%%%%%%%%fig:NH2HR compact sources%%%%%%%%%%%%%%%%
\begin{figure*}[!ht]
\centering
\includegraphics[width=6.0 in]{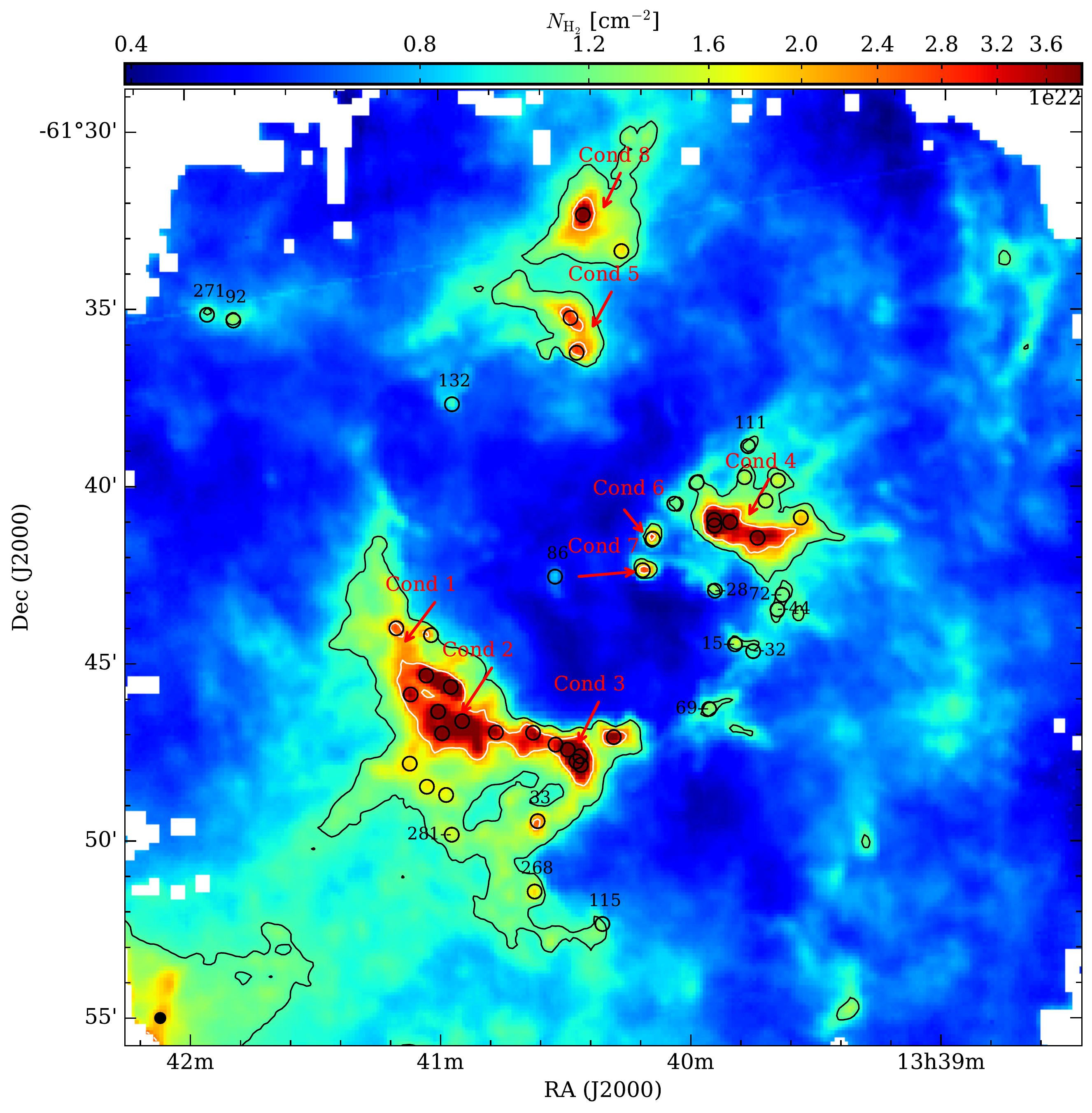}
\caption{Distribution of 50 compact sources overlaid on the high-resolution ($18\farcs2$) column density map.
The black and white contours delineate the gas in the compressed layer (above \nhh~$=11\times10^{21}$\,\cm2 marked with the black contour) and the gravitationally
dominated gas (above \nhh~$=23\times10^{21}$\,\cm2 marked with the white contour, see Sect.\,\ref{sect:pdf}), respectively.
The arrows pinpoint the eight condensations as named by ZA06. A beam size of
 $18\farcs2$ is shown with the black full circle on the bottom left.}
\label{fig:NH2HRCP}
\end{figure*}
%%%%%%%%%%%%%fig:end%%%%%%%%%%%%%%%%

We searched for the possible IR features
of the 89 sources in the 1.25 to 70\,\um\ range, including point-like objects,
absorption against local bright background emission, and filament- or extended-structure emission.
A point-like object, with a size of $<10 \arcsec$ depending on the resolution of the IR data (see Sect.\,\ref{sect:obs}),
may be an indicator
of ongoing star formation, absorption is indicative of cold gas lying in front of the hot dust, and
the IR filament or extended-structure implies an accumulation of either hot dust
or PDRs in which there may be
embedded point-like objects that are not easily separated from bright background emission.

Taking advantage of {\it Herschel} images complemented with the $18\farcs2$ resolution
column density map and other IR data, we made a  plot consisting of
16 $1\farcm5 \times 1\farcm5$ images for each source (see Fig. \ref{fig:appmultimage}) to pinpoint  the IR
counterparts.
The IR sources were searched for within 5$\arcsec$, which corresponds to 0.1\,pc at the distance of RCW\,79.

Following the appearance of their infrared emission, we classified the sources into three groups:
\begin{enumerate}
\item $Group\,0$ compact sources with IR absorption or without any detectable IR counterparts,
\item $Group\,IM$ (i.e., intermediate) compact sources with filament or extended-structure IR emission,
\item $Group\,I$  compact sources with point-like IR counterparts.
\end{enumerate}
By visually inspecting the plots in Fig. \ref{fig:appmultimage}, we identified 14 $Group\,0$  and 11 $Group\,I$ sources. The remaining 64 out of 89 sources are found to be only spatially coincident with filament (or extended) emission.
These sources can be either compact sources or pieces of clouds.
To distinguish the compact sources  from the pieces of clouds, we requested that the compact sources
have at least three good flux measurements at wavelengths $\geq100$\,\um.  The good photometry at each wavelength is defined by
the photometry ellipse centered on the corresponding density peak. For example, the photometry ellipses of source\,1 at wavelengths 100 to 500\,\um\
are centered on the respective density peaks (see Fig.\,\ref{fig:appmultimage}).
As a result, we picked out 25 $Group\,IM$ sources.
 In all, we obtained 14 $Group\,0$, 25 $Group\,IM$, and 11 $Group\,I$ compact sources.
The physical parameters of these 50 compact sources are summarized
in Cols. 1-6 of Table\,\ref{tbl:compactSour_para_proto}, including the ID, the coordinates ($R.A.$ and $Dec.$), the deconvolved size ($D_{\rm dec}$),
the dust temperature ($T_{\rm d}$), the envelope mass ($M_{\rm env}$), and the number density ($n_{_{\rm H_2}}$).
The errors of $T_{\rm d}$ and $M_{\rm env}$, given by the MPFITFUN\footnote{\url{http://cow.physics.wisc.edu/~craigm/idl/mpfittut.html}} procedure, mainly arise from the photometric flux error at each wavelength
(see Table\,\ref{tbl:herschel_flux_corr}). The 50 compact sources have sizes ranging from 0.1 to 0.4\,pc with an average value of 0.2\,pc,
temperatures ranging from 11 to 26\,K, envelope masses ranging from 6 to 760\,\msun, and number density ranging from 0.1
to $44\times 10^{5}$\,cm$^{-3}$. In the following, we should keep in mind that
due to their large typical size of 0.2\,pc, these compact sources may be made of
different sources in different evolutionary stages.

Figure\,\ref{fig:NH2HRCP} displays the spatial distribution of the 50 compact sources overlaid on the high-resolution column density map.
They are predominantly  concentrated on the local density peaks, which has also been observed in other star-forming regions like the Rosette molecular cloud \citep{sch12}.
Interestingly, $96\%$ of the sources are observed toward the compressed layer (within the black contour in Fig.\,\ref{fig:NH2HRCP}, see Sect.\,\ref{sect:pdf}). This distribution
not only indicates that these compact sources are exposed to the influence of the \hii~region, but it is also
suggestive of more efficient formation of compact sources in the layer of compressed gas than in other regions away from the \hii~region.  Moreover,
26 out of 50 compact sources are found to be associated with the gravitationally bound gas (within the white contour). Coupled with their associated IR counterparts, the 26 compact sources have a high probability of forming stars through gravitational collapse.

\subsubsection{Luminosity}\label{sect:luminosity}
The bolometric luminosities $L_{\rm bol}$ for the 50 compact sources are calculated as
\begin{equation}\label{eq:luminosity}
L_{\rm bol} = 4\pi\,D^2 \int{F_\nu \,{\rm d}\nu}.
\end{equation}
 The bolometric luminosity of the 39 $Group\,0$ and $Group\,IM$ sources that have no IR point-like objects was obtained by integrating the graybody SED fits over frequencies.
For the 11 $Group\,I$ sources, IR fluxes were added to the calculation.
To obtain the IR fluxes, we cross matched these 11 sources with those
in the archival GLIMPSE\,I \citep{ben03}, MIPSGAL \citep{gut15}, and ALLWISE \citep{wri10} catalogs  within a search radius of
$5\arcsec$. These catalogs all have been cross matched with the 2MASS survey. Therefore,
if there is any, the near IR fluxes can be simultaneously retrieved.
All $Group\,I$ sources except for source\,28 can be well matched in the catalogs. For source\,28 we used the software DS9 to perform aperture photometries
at the wavelengths where the IR counterpart exists.
Table\,\ref{tbl:NIR_MIR_phot} gives a summary of the resulting IR fluxes for the 11 compact sources.
In addition, the submillimeter luminosities ($L_{_{\rm submm}}^{_{\lambda >\rm 350}}$) for all 50 compact sources were
derived from the integrated luminosity of the resulting SED fit over wavelengths $\geq 350$\,\um.  The derived
$L_{\rm bol}$ and $L_{_{\rm submm}}^{_{\lambda >\rm 350}}$ are given in Table\,\ref{tbl:compactSour_para_proto}.
An uncertainty of $10\%$ for both $L_{\rm bol}$ and $L_{_{\rm submm}}^{_{\lambda >\rm 350}}$ is estimated from the flux errors.
The bolometric luminosities of the 50 compact sources range from 19 to 12712\,\lsun.

\begin{table*}
\centering
\caption{Derived parameters for the 50 compact sources}
\label{tbl:compactSour_para_proto}
\resizebox{15cm}{!}{
\begin{tabular}{crrrrrrrr}
\hline\hline
ID & $R.A.$ & $Dec.$ & $D_{\mathrm{dec}}^{\rm eff}$ & $T_\mathrm{d}$ & $M_\mathrm{env}$ & $n_{_{\rm H_2}}^a$ & $L_\mathrm{bol}$ & $L_\mathrm{smm}/L_\mathrm{bol}$  \\
   & J2000 & J2000 & pc & K & \msun\ & $10^5$\,cm$^{-3}$ & \lsun\ &  \\
\hline
\input ./tbl/compactSour_para_proto.tbl
\hline
\end{tabular}
}
\tablefoot{\\
$^a$ The number density is approximated as
$n_{_{\rm H_2}}=M_\mathrm{env}/(4\pi/3 \times (D_{\mathrm{dec}}^{\rm eff}/2)^3 \times \mu \times m_{\rm H})$ by
 assuming a sphere entity for compact sources, where the mean molecular weight $\mu$ is assumed to be 2.8 and
 $m_{\rm H}$ is the mass of a hydrogen atom.}
\end{table*}

\subsection{Condensations in RCW\,79}\label{sect:appCond}
%%%%%%%%%%%%%fig:NH2HR%%%%%%%%%%%%%%%%
\begin{figure*}[!ht]
\centering
\includegraphics[width=2.52 in]{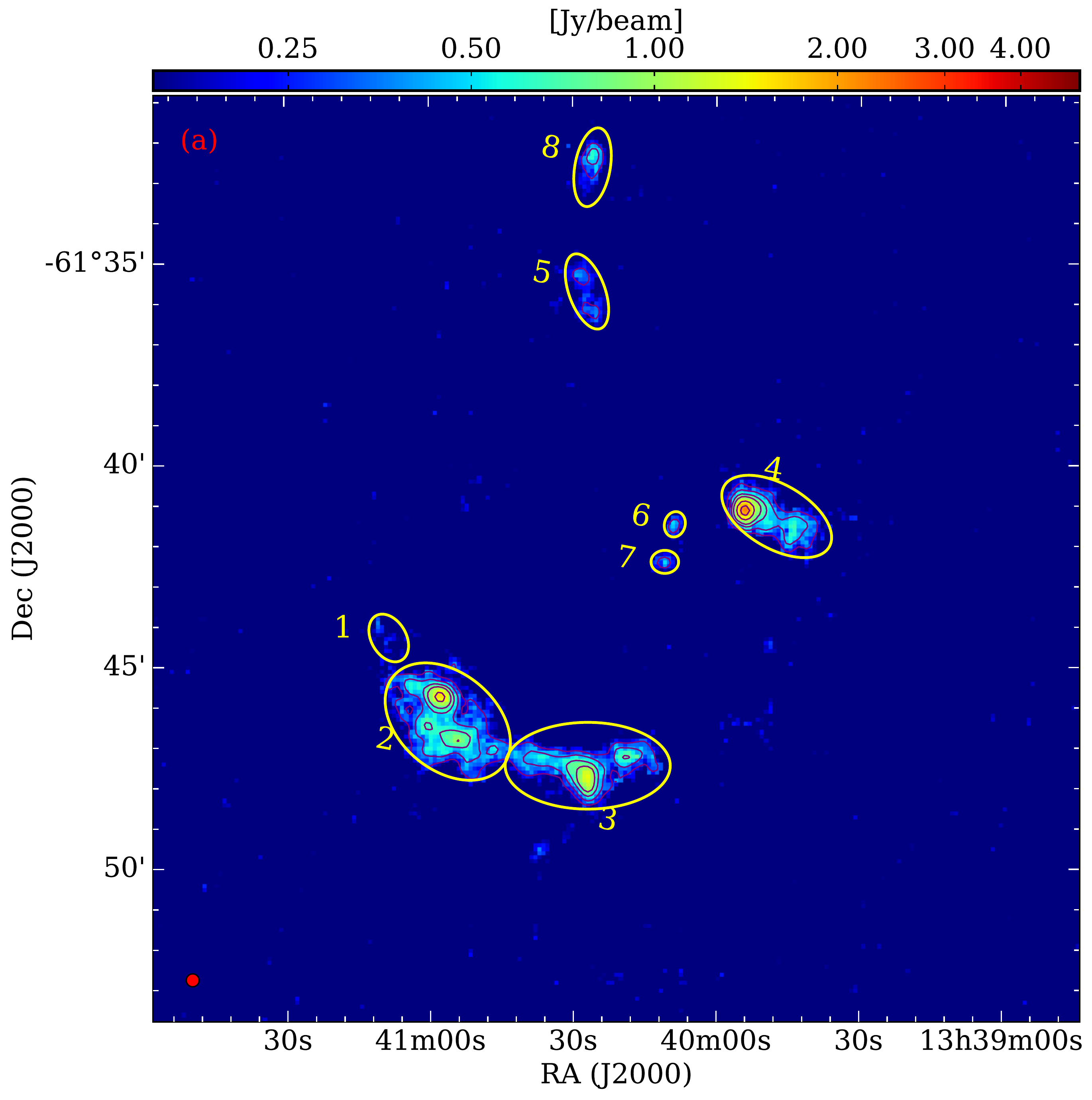}
\hskip 0.000000000001cm
\includegraphics[width=2.2 in]{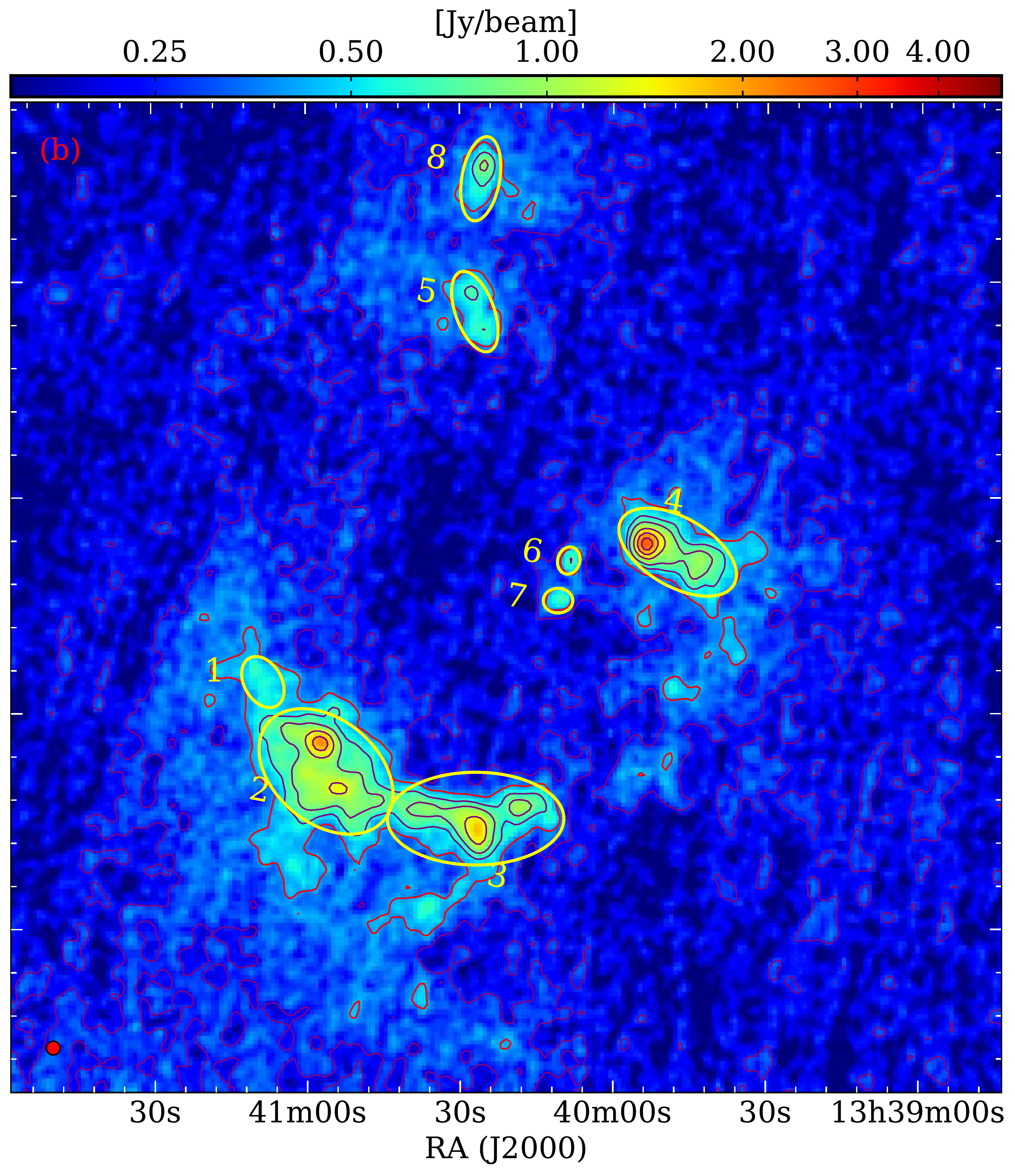}
\hskip 0.000000000001cm
\includegraphics[width=2.2 in]{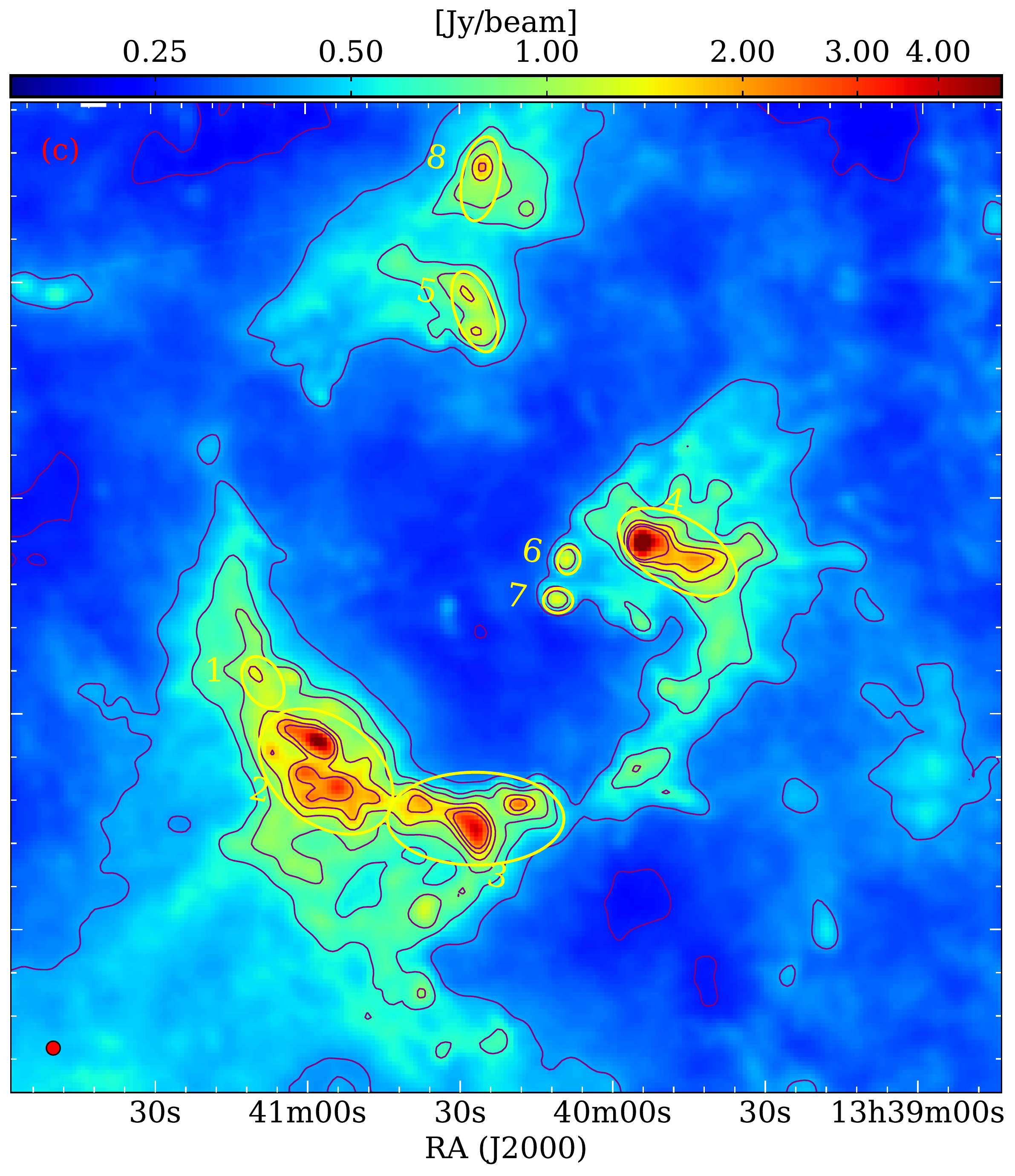}
\caption{(a) Observed APEX 870\,\um\ map, (b) APEX+Planck combined map, (c) modeled 870\,\um\ map
by inverting from the high-resolution column density map. In all panels, the color bars are
identical, going from 0.15 to 5.0\,Jy\,beam$^{-1}$ in logarithmic scale. The purple contours show levels
of 0.25, 0.40, 0.60, 0.80, 1.20, 1.60, and 2.0\,Jy\,beam$^{-1}$. It
is noteworthy that the sensitivity of the observed 870\,\um\ map is $0.05-0.07$\,Jy\,beam$^{-1}$.  A beam size
of $19\farcs2$ is shown on the bottom left (full red circle). Eight condensations are roughly
depicted in ellipses and labeled as given in ZA06.}
\label{fig:dust870}
\end{figure*}
%%%%%%%%%%%%%fig:end%%%%%%%%%%%%%%%%

\begin{table*}
\caption{Observed condensation parameters}
\label{tbl:Condpara_obs}
\resizebox{18cm}{!}{
\centering
\begin{tabular}{crrrrrrrrrr}
\hline\hline
Name & $R.A.$ & $Dec.$ & $R_{\rm maj}$ & $R_{\rm min}$ & $F_{\rm 870 \mu m}$  & $F_{\rm 870 \mu m}^{Planck}$
& $F_{\rm 1.2mm}^{\rm a}$ & $\overline{T_{\rm d}}$ & $\overline{N_{\rm H_2}}$ & $\overline{N_{\rm H_2}^{\rm bg}}$ \\
     &  J2000 & J2000  & $\arcsec$     & $\arcsec$      & mJy                  & mJy
&  mJy            & K                      & $10^{22}$\,\cm2\             & $10^{22}$\,\cm2\                      \\

\hline
\input ./tbl/Condpara_obs.tbl
\hline
\end{tabular}
}
\tablefoot{\\
$^{\rm a}$ $F_{\rm 1.2mm}$ is the 1.2\,mm flux retrieved from Table\,2 of ZA06.}
\end{table*}

\begin{table*}
\centering
\caption{Derived condensation parameters}
\label{tbl:Condpara_der}
\resizebox{13cm}{!}{
\begin{tabular}{crrrrrrr}
\hline\hline
Name & $R_{\rm maj}\times R_{\rm min}$ & $M_{\rm 870 \mu m}$ & $M_{\rm 870 \mu m}^{Planck}$ & $M_{\rm 1.2 mm}$ & $M_{\rm HOBYS}$ & $CFE$ & $n_{\rm Cond}$ \\
     & pc$\times$ pc                   &  $M_{\odot}$        & $M_{\odot}$                  & $M_{\odot}$      & $M_{\odot}$     &       & $10^3$\,cm$^{3}$  \\

\hline
\input ./tbl/Condpara_der.tbl
\hline
\end{tabular}
}
\end{table*}

Figure\,\ref{fig:NH2HRCP} shows several cold, dense, and gravitationally bound regions marked in the white contour that were detected in the
SEST\,1.2\,mm observations (see Fig.\,3 of ZA06) and called condensations. They are also
revealed in the APEX\,870\,\um\ and APEX$+$Planck\footnote{
APEX$+$Planck map is a combination of the 870\,\um\ map from the ATLASTAL survey \citep{sch09} with the 850\,\um\ map
with large angular scale ($8\arcmin$) detection from the Planck/HFI instrument. The combined map
can, to some extent, recover the lost information on the distribution of diffuse gas emission resulting from
the data reduction used for ground-based observations.} maps (see Fig.\,\ref{fig:dust870}\,(a)$-$(b)).
To better define these condensations in our data, we take advantage of the APEX$+$Planck map because of its recovery of
 diffuse gas emission on large angular scales \citep{cse16}.
This recovery is rather obvious in RCW\,79 (see Fig.\,\ref{fig:dust870}\,(a)$-$(b)). In the map,
eight condensations are separated with the ellipses that roughly contain
the majority of the fluxes within a contour level of 0.4\,Jy\,beam$^{-1}$ ($\sigma=0.05-0.07$\,Jy\,beam$^{-1}$).

Four groups of mass estimates were calculated  for each of the eight condensations, using the APEX\,870\,\um, APEX$+$Planck,
SEST\,1.2\,mm fluxes, and the high-resolution column density map. First of all, the total fluxes
$F_{\rm 870 \mu m}$  and $F_{\rm 870 \mu m}^{Planck}$ were integrated over the ellipse of
each condensation from the APEX\,870\,\um\  and APEX$+$Planck maps, respectively. The integrated 1.2\,mm flux
$F_{\rm 1.2 mm}$ was retrieved from Table\,2 of ZA06. With the measured fluxes,
the respective mass $M_{\rm 870 \mu m}$, $M_{\rm 870 \mu m}^{Planck}$,
 and $M_{\rm 1.2 mm}$ can be calculated according to Eq.\,\ref{eq:graybody} with the same dust opacity law
 as the one adopted in Sect.\,\ref{sect:dustproperty}. In the calculations, the average
 temperature ($\overline{T_{\rm d}}$) over each condensation from the dust temperature map was taken into account.
 It is noteworthy that we recalculated the mass $M_{\rm 1.2 mm}$ even though it had previously been calculated in ZA06.
 The reason is that ZA06 adopted a different dust opacity law and the dust temperature for
 each condensation was kept in the range of $20-30$\,K due to the lack of the dust temperature map.
 Additionally, we made use of the high-resolution column density map with resolution similar to the 870\,\um\ data.
 After measuring the mean column density ($\overline{N_{\rm H_2}}$) and background column density
 ($\overline{N_{\rm H_2}^{\rm bg}}$) over the ellipse of each condensation, we approximated the mass
 $M_{\rm HOBYS}$ by $(\overline{N_{\rm H_2}}-\overline{N_{\rm H_2}^{\rm bg}})\times A$,
 where A equals $\pi R_{\rm maj}\times R_{\rm min}$, the area of each condensation.
 The background emission ($\overline{N_{\rm H_2}^{\rm bg}}$) for each condensation was
 considered by subtracting its nearby diffuse emission,
 with the aim of minimizing the contribution of contamination from the line of sight to the mass estimates.
 All measured parameters for the condensations are listed in
 Table\,\ref{tbl:Condpara_obs}. The derived parameters are summarized in Table\,\ref{tbl:Condpara_der}.

 The four mass groups are compared to illustrate the contribution of the flux recovery of ground-based
 observations to the mass estimate. As indicated in Table\,\ref{tbl:Condpara_der}, the mass $M_{\rm 870 \mu m}^{Planck}$
 is on average $2.6\pm0.5$ times higher than $M_{\rm 870 \mu m}$. This demonstrates the loss of large-scale emission
  in the APEX\,870\,\um\ map that is due to the drawback of ground-based continuum observations. This drawback
 can be caused in the data processing in which emission at angular scales larger than a fraction of the field of view of the
 telescope is filtered out in the subtraction of the sky noise \citep{cse16}. Particularly,
 the mass $M_{\rm 870 \mu m}$ of the condensation\,1 (cond.1) is most severely affected by the
 loss of large-scale diffuse gas emission, by 3.8 times less than $M_{\rm 870 \mu m}^{Planck}$.
 This may be related to the least compact property of cond.1 of the eight condensations,
   as indicated by the contrast of the peak to mean intensities (see Fig.\,\ref{fig:dust870}\,a).
  The comparison of the mass $M_{\rm 870 \mu m}^{Planck}$ with $M_{\rm 1.2 mm}$ shows that $M_{\rm 1.2 mm}$ is on
   average a factor $1.9\pm0.5$ lower than $M_{\rm 870 \mu m}^{Planck}$,
  indicating that the 1.2\,mm ground-based observations might have suffered from the loss of large-scale emission as well.
  Furthermore, space-based {\it Herschel} observations could be regarded as almost no loss of large-scale emission with respect
  to the ground-based observations. In comparison, $M_{\rm HOBYS}$ is found to be on average a factor $1.6\pm0.1$
  as massive as $M_{\rm 870 \mu m}^{Planck}$, suggesting that the mass $M_{\rm 870 \mu m}^{Planck}$ is not yet
  completely recovered. This incomplete recovery may be attributed to the poor resolution of the Planck/HFI instrument, which cannot
  be demonstrated in this work because of the lack of higher resolution space observations at 870\,\um\ or the wavelengths close to it.
  However, we can try to confirm the result of this incomplete recovery. Taking advantage of
  the high-resolution column density map, we modeled 870\,\um\ emission according to
  Eq.\,1 of \citet{liu16}. The resulting map is displayed in Fig.\,\ref{fig:dust870}\,(c). Compared with the APEX$+$Planck combined map,
  the modeled 870\,\um\ emission can recover more extended diffuse emission, as shown in Fig.\,\ref{fig:dust870}\,(b)$-$(c), which is
  in support of the incomplete recovery result in the APEX$+$Planck map.

  The comparisons of the four groups of masses led to a relation of $M_{\rm HOBYS} >
  M_{\rm 870 \mu m}^{Planck} > M_{\rm 1.2 mm} > M_{\rm 870 \mu m}$. $M_{\rm 870 \mu m}^{Planck} > M_{\rm 1.2 mm} > M_{\rm 870 \mu m}$
  indicates that there are mass losses in ground-based observations due to the drawback in the data reduction.
  $M_{\rm HOBYS} >  M_{\rm 870 \mu m}^{Planck}$  shows that the APEX$+$Planck combined map has not fully recovered the mass losses.
  It is worth noting that $M_{\rm HOBYS}$ may be uncertain since the background subtraction may not be estimated accurately  due to the uncertain background emission level. Therefore,
  $M_{\rm 870 \mu m}^{Planck}$ and $M_{\rm HOBYS}$ are conservatively treated as lower and upper limits for the eight condensations.

\section{Discussion}\label{sect:discuss}
\subsection{Evolutionary state of compact sources}
\subsubsection{Luminosity-mass diagram}\label{sect:LM}
\begin{figure*}[!ht]
\centering
\includegraphics[width=7.0 in]{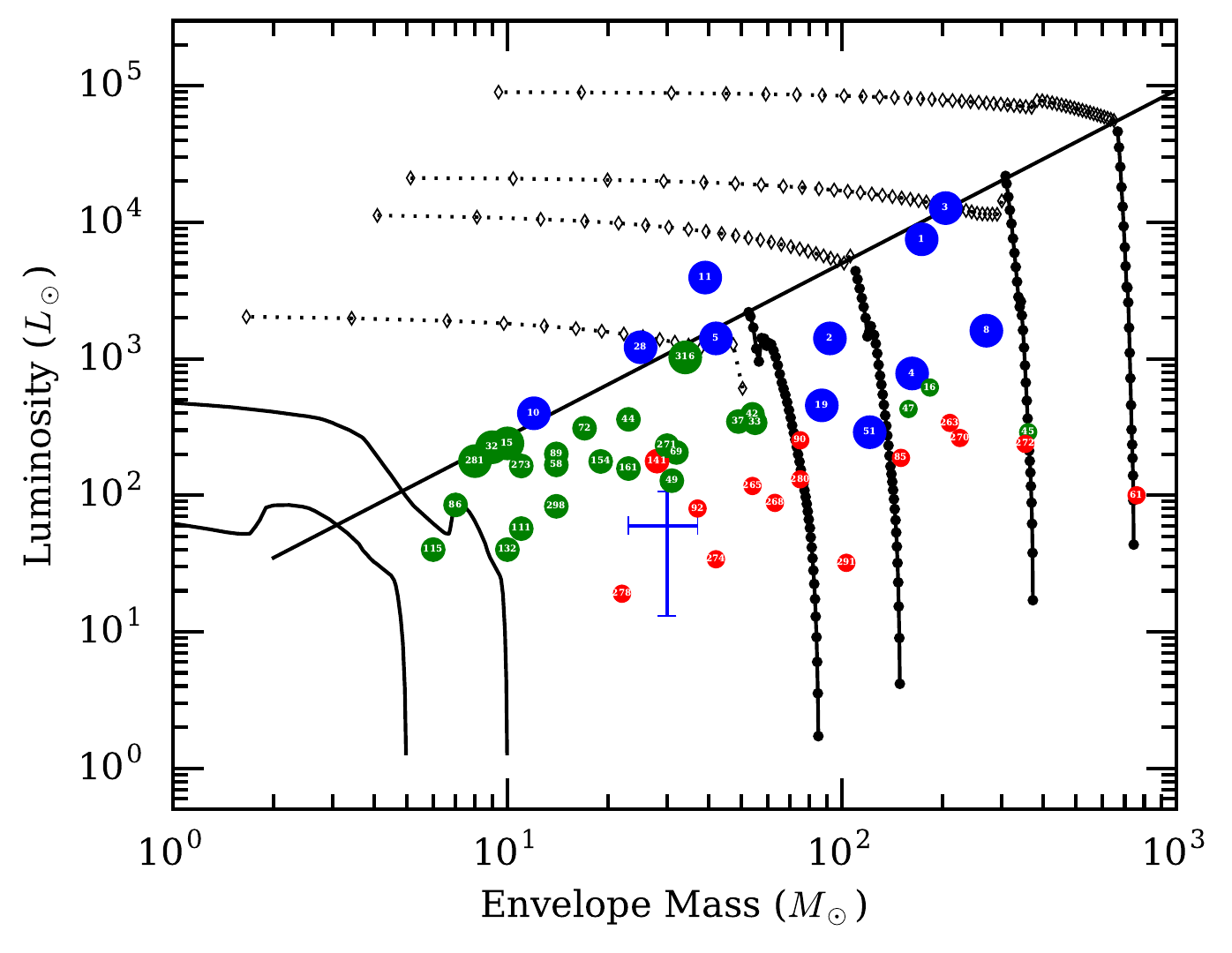}
\caption{ Bolometric luminosity vs. envelope mass ($L_{\mathrm{Bol}}$-$M_{\mathrm{Env}})$ diagram.
Red circles show the compact sources associated with 8\,\um\ absorption, green circles indicate
sources associated with IR extended emission, and blue circles symbolize sources associated with IR point-like objects.
 As described in Sect\,\ref{sect:EvolTrend}, these compact sources are classified as class\,0
 (small circles), intermediate objects (medium circles),
 and class\,I (large circles). Typical error bars are shown.}
\label{fig:LM}
\end{figure*}

A promising evolutionary tool for young protostellar cores is the diagram that shows luminosity versus envelope mass ($L_{\rm bol}-M_{\rm env}$) \citep[e.g.,][]{sar96,mol08,mol16}. It has first been used for low-mass objects \citep{sar96} and was then extended
to high-mass objects \citep{mol08}. The diagram of \citet{mol08} is a result
of simulations based on the assumption of a scaled-up version of classical inside-out collapse scenario
\citep{shu77}. The simulations can be summarized as two phases of accelerating accretion and envelope
clean-up. During the accelerating accretion, its rate increases with time since the inside-out collapse
wave propagates outward, reaching out to material dominated by supersonic speeds and where higher accretion
rates are possible \citep{mck03}. Despite the fast accretion onto a central protostar, its envelope mass
 decreases slowly with time due to the main mass loss by powerful molecular outflows. In contrast, the bolometric
luminosity of the protostellar core increases drastically due to the contributions from accretion, deuterium burning,
and core contraction. In the $L_{\rm bol}-M_{\rm env}$ diagram (see Fig.\,\ref{fig:LM}) the tracks
(the dot-marked full thick lines) all start at very
low luminosity and proceed almost vertically toward the zero-age main sequence (ZAMS), which marks the end
of the accelerating accretion phase. In the early ZAMS phase, the bolometric luminosity of the core remains constant,
while its envelope mass continues to be expelled through radiation and molecular outflows. The evolution of the prostellar core then enters
the second envelope clean-up phase in which a large amount of material is expelled predominantly by energetic radiation
and powerful outflows, and then the luminosity mainly arises from the protostar and the persisting residual accretion.
The envelope clean-up phase shown in the $L_{\rm bol}-M_{\rm env}$ diagram is that the protostellar core evolves along an almost
horizontal path (the dashed diamond-marked lines), which ends when the envelope is all almost expelled and the protostar becomes optically visible.
% According to the distribution of massive objects in the diagram, they can be grouped into
%Class\,0, I and II objects analogue to low-mass YSOs.
Although this diagram has been widely adopted for the
investigation of young massive and dense cluster-progenitor clumps, it should be treated with caution due
to several intrinsic caveats \citep{mol08,tra15}. For instance, the evolutionary tracks were initially
modeled for single cores, not for clumps, while our compact sources with a typical size of 0.2\,pc can consist of multiple sources in different evolutionary stages. Therefore, the evolutionary stages of the 50 compact sources derived from this diagram have to be taken with caution in the following analysis.

%%%%%%%%%%%%%fig:LM%%%%%%%%%%%%%%%%
%%%%%%%%%%%%%fig:end%%%%%%%%%%%%%%%%

\subsubsection{Evolutionary trend}\label{sect:EvolTrend}

\begin{table*}
\centering
\caption{Statistics of IR features and  evolutionary diagnostics of 50 compact sources}
\label{tbl:cms_statistics}
\resizebox{13cm}{!}{
\begin{tabular}{cccc}
\hline\hline
Sources associated with IR features & Class\,0 & Intermediate object & Class\,I  \\
\hline
\input ./tbl/compact_source_statistic_tex.tbl

\hline
\end{tabular}
}

\end{table*}

Figure\,\ref{fig:LM} shows the distribution of the 50 compact sources in the $L_{\rm bol}-M_{\rm env}$ diagram.
{\it Group\,0} sources are coded in red, {\it Group\,IM } in green, and {\it Group\,I} in blue.
In Fig.\,\ref{fig:LM} the {\it Group\,0} compact sources are distributed above
the starting points from the bottom of all the evolutionary tracks, which could be regarded as
 the beginning of the accretion phase.  Their distribution suggests that
they may be protostellar sources although they are associated with 8\,\um\ absorption,
which is indicative of no active ongoing star formation.
This may be attributed to the fact that the low-luminosity protostars deeply embedded in the dense clumps
are not observable in the near- and mid-IR continuum observations.
Given the above consideration, {\it Group\,0} will be
treated as protostellar sources in the following analysis. However, it is worthwhile
to determine the nature of these {\it Group\,0} sources using millimeter interferometric observations of molecular tracers of
 ongoing star-forming activities (e.g., outflows and infall). Furthermore,
the majority of {\it Group\,0} sources can be distinguished from the {\it Group\,IM } and {\it Group\,I}
sources. Most of {\it Group\,0} sources are located in the lower part of the $L_{\rm bol}-M_{\rm env}$ diagram
than the other two groups. This fact suggests an evolutionary trend of {\it Group\,0} sources
being the least evolved of the three groups.

The $L_{_{\rm submm}}^{_{\lambda >\rm 350}}/L_{_{\rm bol}}$ ratio between the submillimeter ($\lambda >\rm 350$\,\um) and bolometric luminosities
is thought to be useful for exploring the nature of protostellar cores and their evolutionary trend \citep[e.g.,][]{and00,bon10a}.
According to \citet{bon10a}, sources with $L_{_{\rm submm}}^{_{\lambda >\rm 350}}/L_{_{\rm bol}}\geq 0.03$
are classified as class\,0 objects, sources with $0.01<L_{_{\rm submm}}^{_{\lambda >\rm 350}}/L_{_{\rm bol}}<0.03$
as intermediate (IM) objects, and sources with $L_{_{\rm submm}}^{_{\lambda >\rm 350}}/L_{_{\rm bol}}\leq 0.01$ as class\,I objects.
The so-called IM objects are regarded as those in the transition regime where Class\,0 and I objects are not readily distinguishable.
This classification scheme sorts the 50 compact sources into  17 class\,0, 23 IM, and 10 class\,I protostellar cores.
We observed that 1  out of 17 class\,0 (number 51), and 5 out of 23 IM (numbers 2, 4, 7, 8, and 19) prostellar cores are associated
with point-like IR counterparts. These sources, except for source
4, were identified as class\,I objects by
the mid-IR color-color diagrams in ZA06. However, source 4 was not mentioned in ZA06. Likewise, we make use of the color-color scheme of
\citet{all04}, adopted in ZA06, to investigate the nature of this source. As a result, the colors of $[3.6]-[4.5]=\sim 1.5$ and
$[5.8]-[8.0]=\sim 1.2$ for source 4 suggest that it is a class\,I object.  Finally, we obtain
16 class\,0, 19 IM, and 15 class\,I prostellar cores, which are marked in small, medium, and large circles in Fig.\,\ref{fig:LM}, respectively.
Table\,\ref{tbl:cms_statistics} summarizes the association of the three classes of objects with the three groups of
sources as defined in Sect.\,\ref{sect:IRfeatures}. By comparison,  $81\%$ of class\,0 objects are found to be associated with
 {\it Group\,0} sources with 8\,\um\ absorption, $95\%$ of IM objects
 associated with {\it Group\,IM}  sources with extended IR features, and
 $100\%$ of class\,I objects associated with {\it Group\,IM/I} sources. These results demonstrate that the {\it Group\,0} sources
are indeed in the earliest evolutionary stages of the three groups since the majority of the {\it Group\,0} sources can be categorized into
class\,0 objects. Hereafter, the definition of {\it Group} will be replaced with that of class since the latter is more unbiased and physically meaningful.

As shown in Fig.\,\ref{fig:LM}, all the class\,0, IM, class\,I protostellar cores  follow the evolutionary tracks well.
The class\,I protostellar cores are the most luminous, approaching the ZAMS phase as indicated by the solid line
given by \citet{mol08}. The IM and class\,0 objects
are less luminous, situated below the  class\,I objects in the $L_{\rm bol}-M_{\rm env}$ diagram. These results are
well consistent with the known evolutionary trend that class\,0 objects are less evolved and have
lower luminosities than class\,I objects \citep[e.g.,][]{and93,and00,dun08,eva09,eno09}.

\subsubsection{Properties of compact sources in different phases}
%%%%%%%%%%%%%fig:histrogram%%%%%%%%%%%%%%%%
\begin{figure*}[!ht]
\centering
\includegraphics[width=2.3 in]{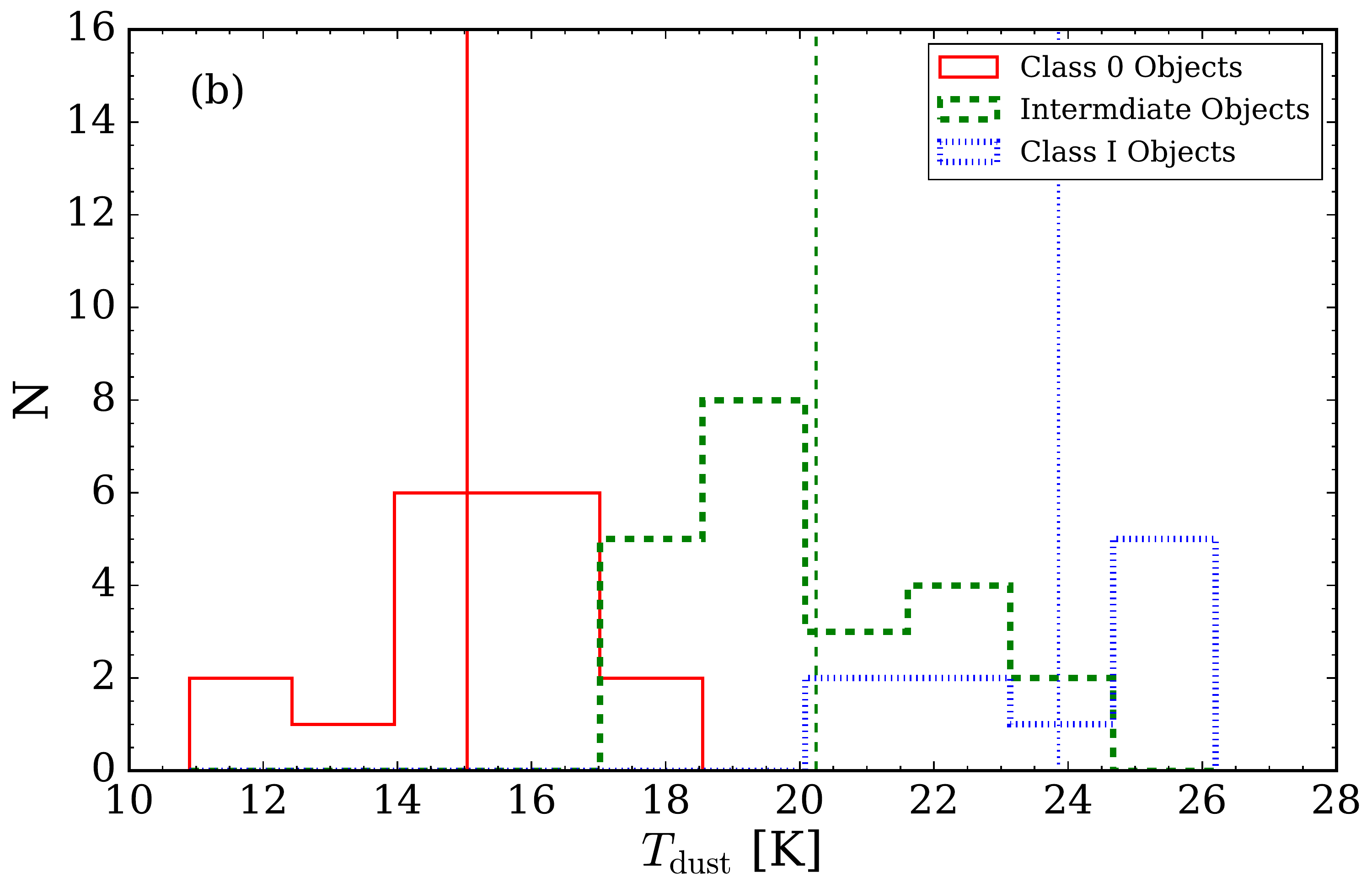}
\hskip 0.00000001cm
\includegraphics[width=2.3 in]{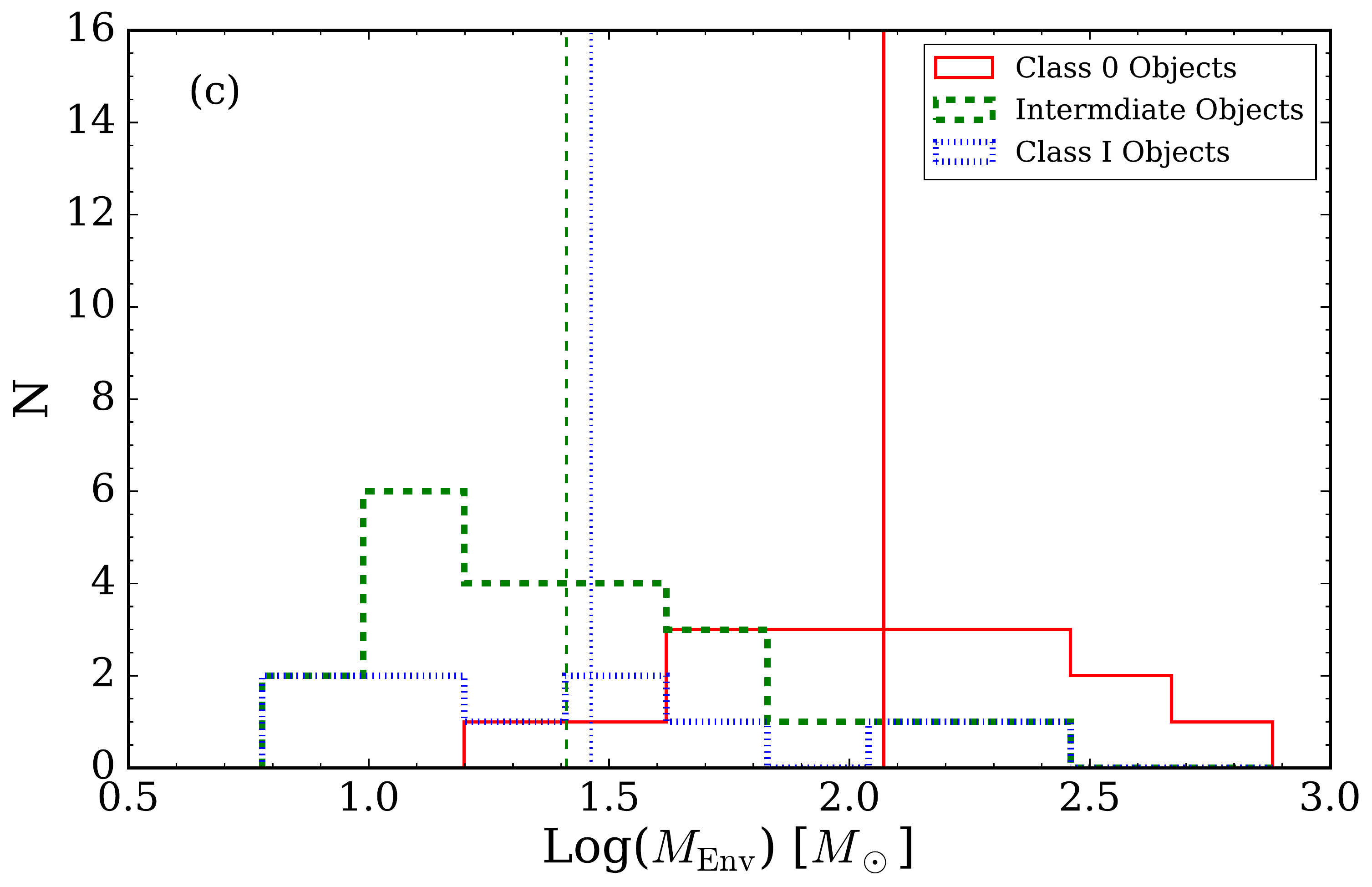}
\hskip 0.00000001cm
\includegraphics[width=2.3 in]{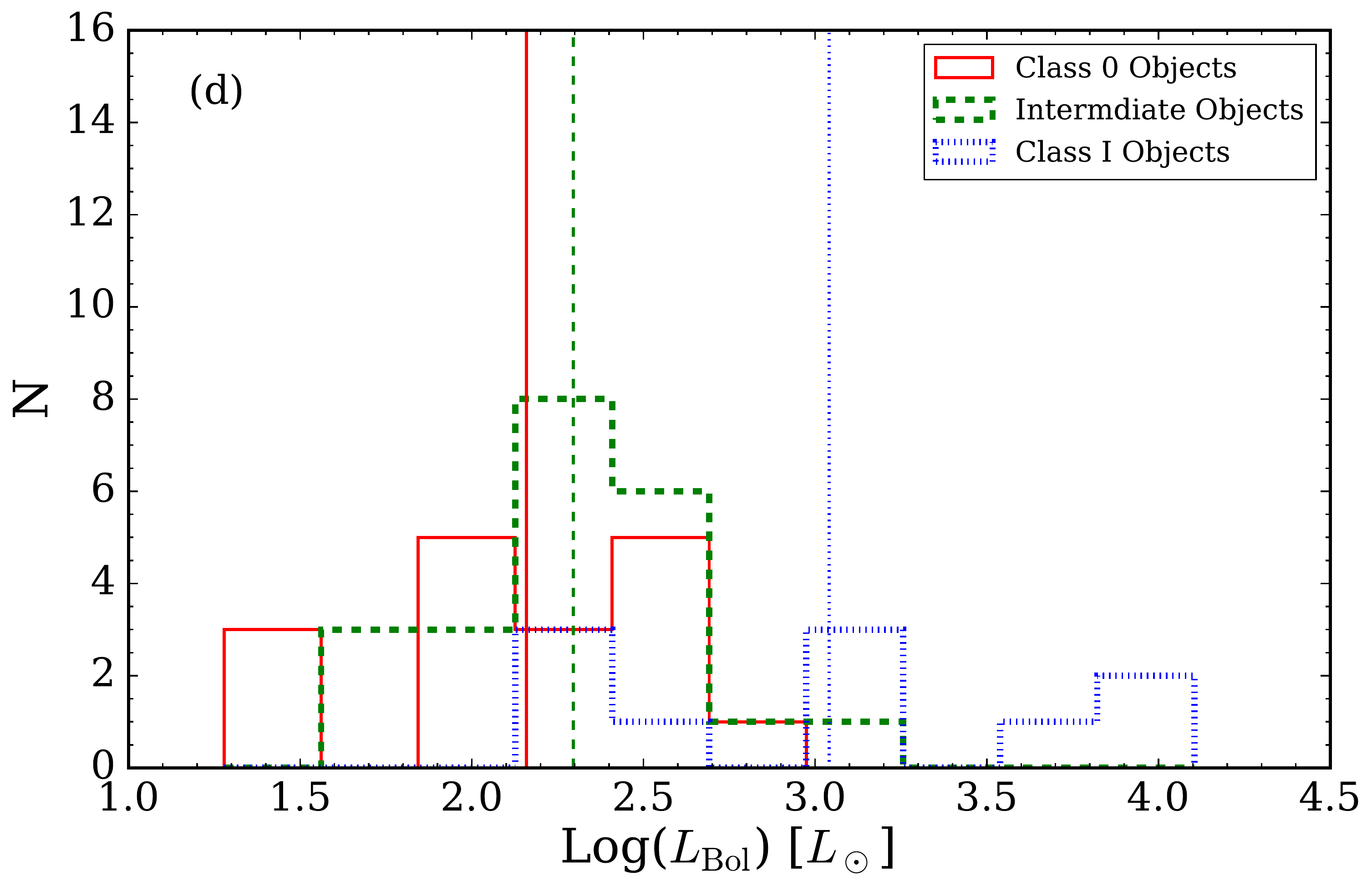}
\caption{Histograms of the  dust temperature (a),  envelope mass (b), and
bolometric luminosity (c) for the 50 compact sources. Average values are marked with the solid, dashed, and dash-dotted  lines for
class\,0, IM, and class\,I objects, respectively.}
\label{fig:hist}
\end{figure*}
%%%%%%%%%%%%%fig:end%%%%%%%%%%%%%%%%

Figure\,\ref{fig:hist} shows the distributions of three physical parameters ($T_{\rm d}$, $M_{\rm env}$, and $L_{\rm bol}$)
for the three classes of objects. In Fig.\,\ref{fig:hist}(a), the source temperature increases as a function of
evolutionary stage, with average values of 15.0$\pm$1.7, 20.2$\pm$2.0, and 23.4$\pm$1.9\,K for the class\,0, IM, and class\,I protostellar
cores, respectively. This trend with increasing temperature can be understood as the central protostar(s) being
the dominant heating engine of a core. The envelope mass distributions of the sources are shown in Fig.\,\ref{fig:hist}(b). The mean
masses are 118$\pm$2, 26$\pm$3, and 30$\pm$3\,\msun\ for the class\,0, IM, and class\,I protostellar cores, respectively.
We note that the lower mean mass of the IM protostellar cores than that of the
class\,I protostellar cores may not be true. This may arise
from the exclusion of the 70\,\um\ flux in the graybody SED fitting. Such exclusion underestimates the dust temperature, especially for the class\,I protostellar cores, leading to a corresponding overestimated envelope mass \citep[see more discussions in][]{tra15}. Despite this possible overestimation of class I envelope mass, a decreasing mass trend from class\,0 to IM and class\,I protostellar cores is present.
This suggests that class\,0 protostellar cores should have more mass for entering the
accelerating accretion phase and class\,I protostellar cores have to decrease in mass due to the envelope
clean-up process (see Sect.\,\ref{sect:LM}, and Fig.\,\ref{fig:LM}). The luminosity distributions of the sources
are shown in Fig.\,\ref{fig:hist}(c), with mean values of 144$\pm$3, 197$\pm$2, and 1097$\pm$4\,\lsun\ for the class\,0, IM, and
class\,I protostellar cores, respectively. The increasing luminosity  trend with evolutionary stage is
also in agreement with the evolutionary tracks shown in Fig.\,\ref{fig:LM} (see Sect.\,\ref{sect:LM}).

From the mass and luminosity distributions, their consistences with the evolutionary tracks in Fig.\,\ref{fig:LM}
demonstrate that  the 50 compact sources in RCW\,79 are indeed in the early stages of star formation.
The nature of these compact sources needs to be explored in more detail with
observations of higher angular resolution since their typical size  ($\sim 0.2$\,pc) is not the minimum scale of the final fragmentation (i.e.,
$\sim 0.01-0.04$\,pc) leading to
individual collapsing protostars \citep[e.g.,][Motte et al. in prep.]{bon10b,pal13b,wan14}. A possible way of addressing this is
either to look at the number of IR protostars within the protostellar cores ($\sim 0.2$\,pc),
or to identify cold dense substructures within them using millimeter interferometric observations.

\subsection{High-mass star-forming candidates}\label{sect:HMSF}
%%%%%%%%%%%%%fig:HM%%%%%%%%%%%%%%%%
\begin{figure}[!ht]
\centering
\includegraphics[width=3.4 in]{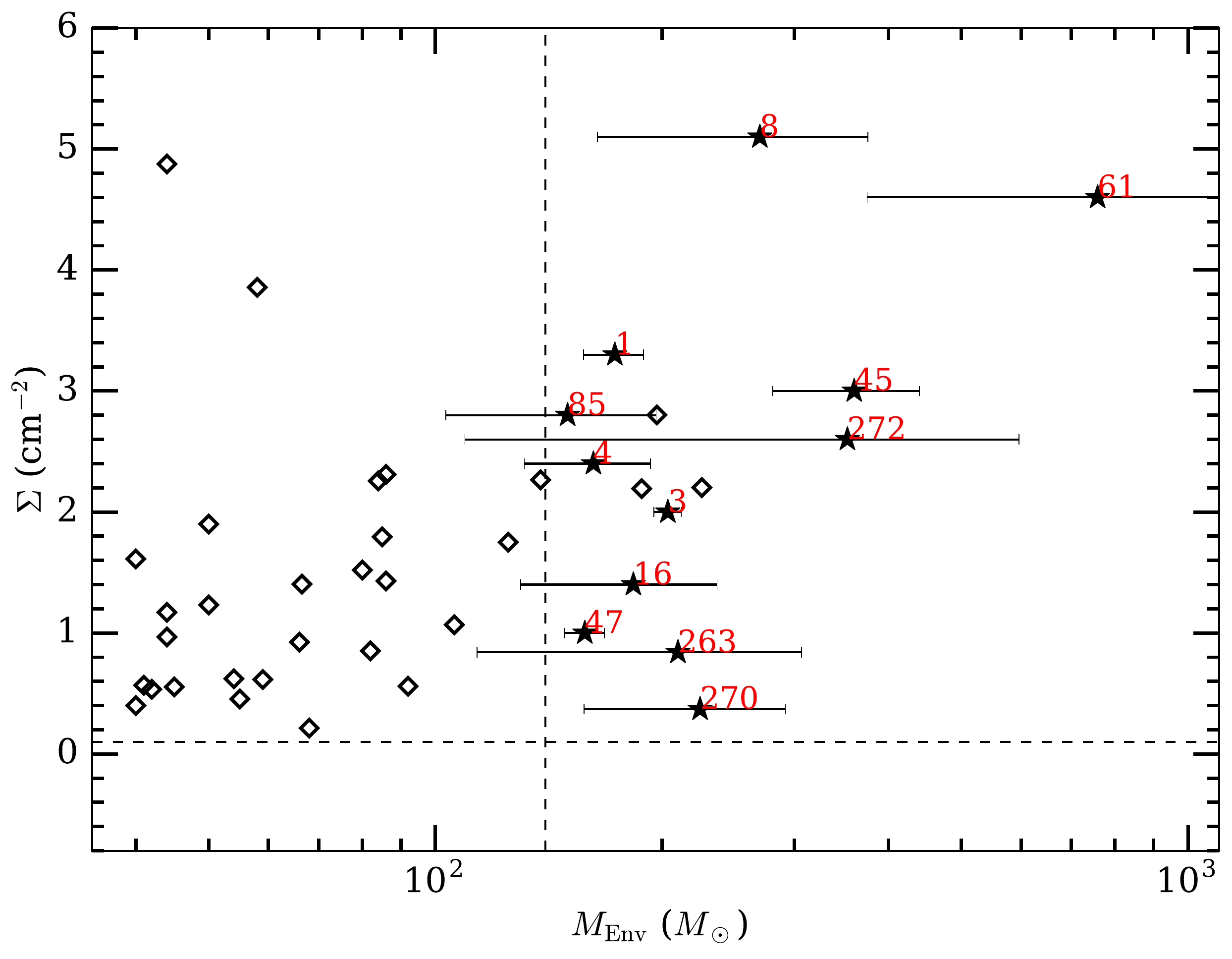}
\caption{Surface density versus envelope mass for the massive dense cores (MDCs) in RCW\,79 (filled stars)
and those in Cygnus\,X \citep[open diamonds,][]{mot07}. The vertical dashed line is the mass threshold
 $>140$\,\msun\ for MDCs in this work. The horizontal dashed line represents a surface density threshold
of  $\Sigma = 0.1$\,g\,cm$^{-2}$ as suggested in the literature \citep[e.g.,][]{but12,tan13,tan14,tra15}.
}
\label{fig:HM}
\end{figure}
%%%%%%%%%%%%fig:end%%%%%%%%%%%%%%%%
\begin{table*}
\centering
\caption{Statistics of massive dense cores}
\label{tbl:MDC_statitical}
\resizebox{15cm}{!}{
\begin{tabular}{crrrrrrr}
\hline\hline
Region & Dist & $M_{\rm th}$  & Size & \multicolumn{3}{c}{$n_{\rm MDC}$\,[cm$^{-3}$]}  &  Ref. \\
\cline{5-7}
      & [kpc]  & [$M_{\odot}$] & [pc] & Min$^a$  & Mean$^a$    & Max$^a$ &   \\

\hline
\input ./tbl/MDC_statitical.tbl
\hline
\end{tabular}
}
\tablefoot{$^{a}$ The range of the volume-averaged density: minimum (Min), mean (Mean), and maximum (Max) densities.  }
\end{table*}

Triggered star formation around \hii~regions may be an important process, especially for high-mass
star formation \citep[e.g.,][Palmeirim et al. submitted]{deh10,tho12,ken12,ken16}. It is therefore worthwhile to search for massive dense
cores (MDCs) that may form high-mass stars in RCW\,79.

We use the L-M$_{\rm env}$ diagram to determine the mass threshold of MDCs. Based on Eq.\,10 of \citet{mol08}, a dense core
with $M_{\rm env}>140$\,\msun\ may form a massive star of $>8$\,\msun. This corresponds to the last three tracks from left to right in
Fig.\,\ref{fig:LM}. An alternative way of determining the mass threshold of MDCs is to
match the compact sources with the known signposts of ongoing high-mass star formation,
such as class\,II methanol maser, bright SiO emission, and ultracompact
\hii~regions \citep[e.g.,][Tig\'{e} et al. 2016, submitted]{mot07,rus10}, leading to
 a lower mass limit for the sources associated with the signposts, which
 can be adopted as the minimum mass for MDCs. Table\,\ref{tbl:MDC_statitical}
gives a statistics of massive  dense cores classified by searching for signs of high-mass star formation
in two nearby high-mass star-forming regions.
The mass threshold of the MDCs in NGC\,6334 is about twice higher than that in Cygnus\,X.
This difference could be caused by different environments (e.g., different number densities in Table\,\ref{tbl:MDC_statitical}, see below),
given the same typical scale of the MDCs and similar mass sensitivities between observations
by \citet{mot07} and Tig\'{e} et al., (2016).
In RCW\,79, we do not have enough signposts of ongoing star formation (only source\,1 with maser emission) to do a statistic, therefore
we adopted the $140$\,\msun\ determined from the L-M$_{\rm env}$ diagram as the mass threshold for candidate MDCs in RCW\,79.
When we assume that the compact sources have a density gradient
of $\rho \propto r^{-2}$ , the minimum
mass of MDCs with a typical size of $\sim~0.2$\,pc (see Table\,\ref{tbl:MDC_statitical})
needs to be  twice higher than that in NGC\,6334. The mass threshold
of $140$\,\msun\ is about 2 and 3.5 times higher
than those in NGC\,6334 and Cygnus\,X, respectively, therefore $140$\,\msun\ may not be the actual minimum mass for the MDCs
in RCW\,79, but is a conservative value.
%Therefore,  the mass of $\sim 170$\,\msun\ is
%kind of massive to be regarded as the minimum mass of MDCs. Therefore, we adopt $140$\,\msun,  which is about 2 times the mass threshold in NGC\,6334,
%as the mass threshold for candidate MDCs in RCW\,79.
There are 12 MDC candidates in RCW\,79 according to these criteria. They have densities raging from 0.9
to $4.6\times 10^6$\,cm$^{-3}$ with an average value of $1.6 \times10^6$\,cm$^{-3}$. In comparison, the MDCs in RCW\,79 are found on average to be
not as dense as those in NGC\,6334, but about 5 times denser than those in Cygnus\,X (see Table\,\ref{tbl:MDC_statitical}).
This is suggestive of
the dense properties of the MDCs in RCW\,79.

The mass surface density $\Sigma$ is believed to be a potential measure to investigate
the environment of massive star formation \citep[e.g.,][]{kru08,tan14,tra15}. The models of \citet{kru08}
suggested $\Sigma \geq 1$\,g\,cm$^{-2}$ as a threshold to avoid the fragmentation of clouds
into low-mass cores, thus allowing high-mass star formation. However, this threshold has been found to be variable
in observations \citep[][]{tan14,tra15}. The surface densities of the 12 candidate MDCs are shown in
Fig.\,\ref{fig:HM}, in which the 33 MDCs of $>40$\,\msun\ in Cygnus\,X are included for comparison.
The surface densities of the MDC candidates in RCW\,79 range from $\sim 0.4-5.1$\,g\,cm$^{-2}$. This range
is in agreement with that of the MDCs in Cygnus\,X, indicating that the
12 candidate MDCs are probably the precursors of high-mass star formation. Ten out 12 MDCs in RCW\,79 have  surface densities  $\geq 1$\,g\,cm$^{-2}$. For the other two
MDCs 263 and 270, their large sizes (0.26 pc for MDC 263, and 0.4 pc for MDC 270) among the 12 MDCs may
lead to their lowest surface densities with $\Sigma < 1$\,g\,cm$^{-2}$ as shown in Fig.\,\ref{fig:HM}.
However, their surface densities are still within $0.1 \leq \Sigma \leq 1$\,g\,cm$^{-2}$ observed in several
high-mass cores \citep[e.g.,][]{but12,tan13,tan14}.

Triggering may be an important mechanism for high-mass star formation \citep[e.g.,][]{deh10,tho12,ken12,ken16}.
In the Red MSX MYSOs catalog (see Sect.\,\ref{sect:intro}), only one MYSO object associated with MDC\,1 is
included, showing that 11 out of 12 MDCs  are new additional MYSO candidates with respect to the existing MYSOs catalog.
This can be attributed to the advantages of {\it Herschel} observations, which are more sensitive to embedded YSOs
than those at shorter wavelengths.
If the increase of the MYSOs number in RCW\,79 due to {\it Herschel} observations can be extrapolated to other
\hii~regions or bubbles, then the fraction of high-mass stars observed toward the edges of \hii~regions or bubbles would be higher than
than $14-30\%$ suggested by \citet{deh10,tho12,ken12}. This shows
that it is worthwhile to carry out a more complete statistic
to the MYSOs around \hii~regions or bubbles in the light of {\it Herschel} observations to understand the interplay between HII regions
 and high-mass star formation (Palmeirim et al. 2016, submitted).

\subsection{Star formation in massive condensations}\label{sect:SFinCond}
\subsubsection{Young stellar contents}
Stellar contents of the three most massive condensations have
 partly been discussed in the light of near- and mid-IR data (ZA06), with 19 luminous class\,I YSOs revealed.
 In this section, we continue to discuss the stellar contents of these three condensations and the other five condensations
 to explore the early stages of star formation in them.\\
  \textbf{Condensation\,2}

\begin{figure}[!ht]
\centering
\includegraphics[width=3.4 in]{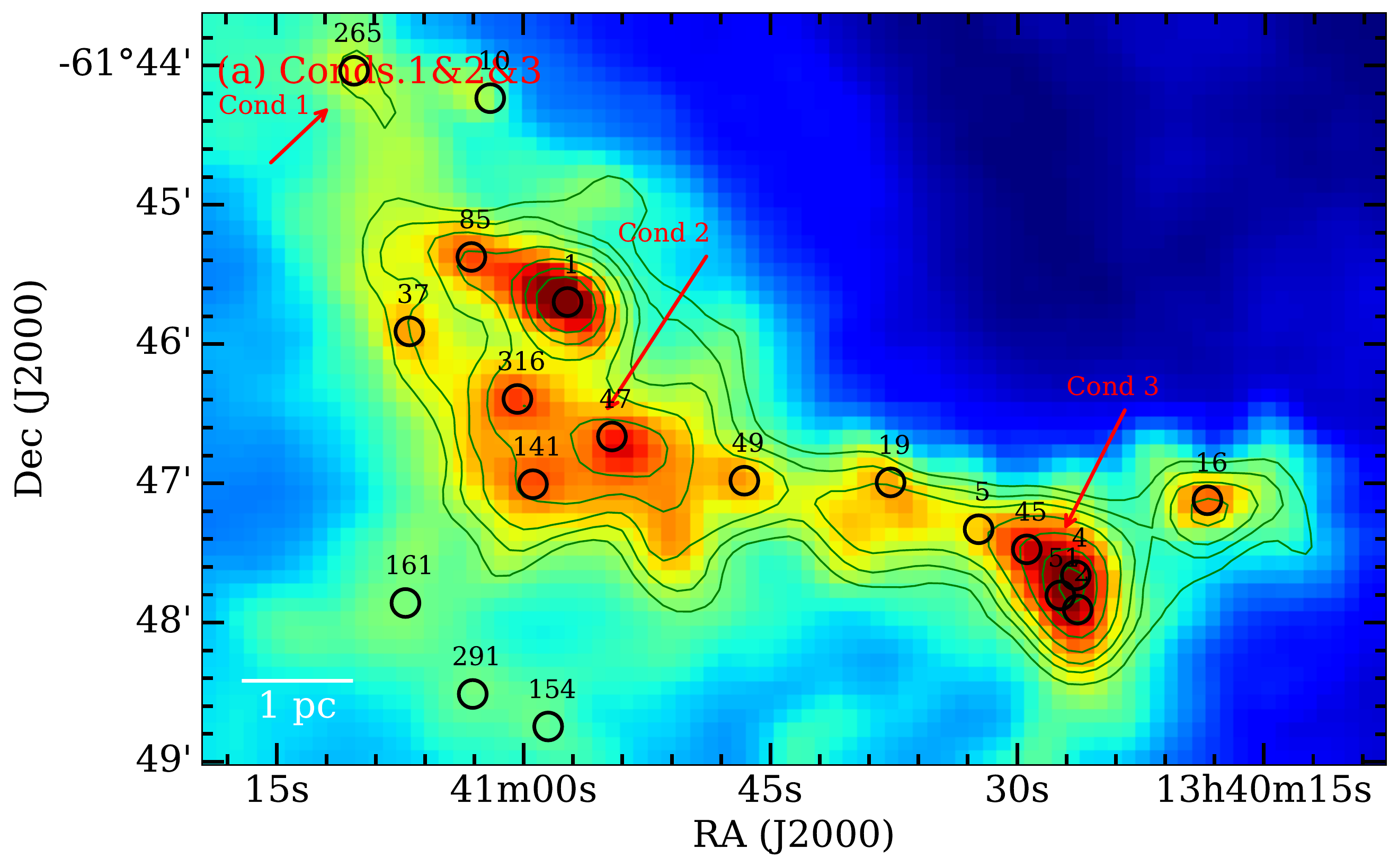}
\hskip 0.00000001cm
\includegraphics[width=3.4 in]{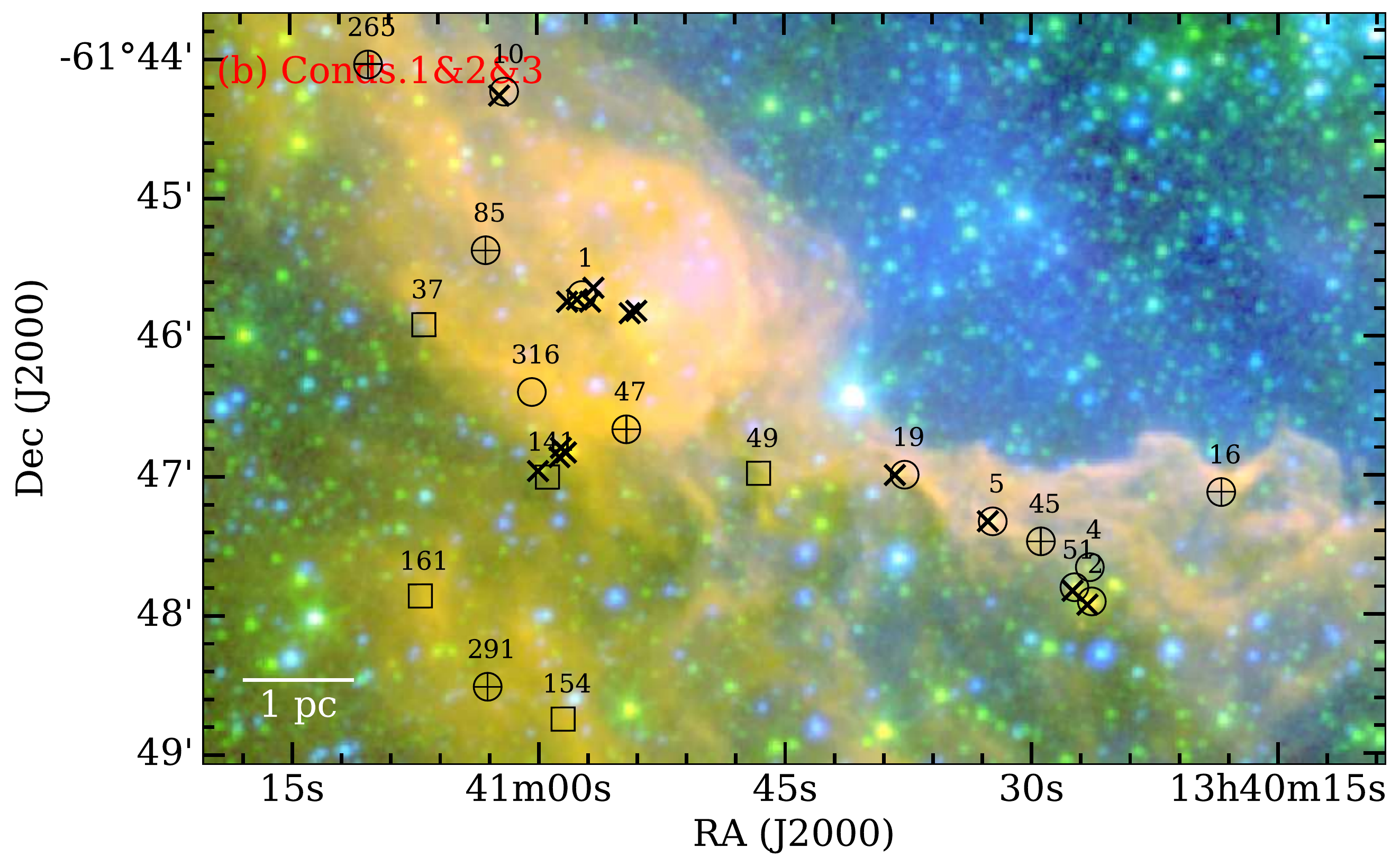}
\caption{(a): Compact sources (circles) overlaid on high-resolution ($18\farcs2$) column density map.
Contours represent 870\,\um\ continuum emission, showing levels
of [0.55, 0.65, 0.75, 0.9, 1.1, 1.3, 1.5]\,Jy\,beam$^{-1}$  ($\sigma = 0.05-0.06$\,Jy\,beam$^{-1}$).
 (b): Compact sources overlaid on a three-color image. The H$\alpha$,
 3.6\,\um, and 8.0\,\um\ emission are coded in blue, green, and red, respectively.
 The compact sources are symbolized in plus circles for class\,0, in squares for IM, and in circles for class\,I objects.
 Crosses denote the class\,I YSOs identified by ZA06. }
\label{fig:Cond123yso}
\end{figure}

 This is the most massive condensation in RCW\,79 with a mass of $\sim 3200-5300$\,\msun.
 It is located in the southeastern edge, facing bright H$\alpha$ emission (see Fig.\,\ref{fig:TdNH2}(c))
 and is associated with the luminous PDRs seen in 8\,\um\ emission (Fig.\,\ref{fig:Cond123yso}(b)), indicating the influence of
 ionized gas on cond.\,2. This suggests that star formation especially in early stages might have been affected by the \hii~region.
 The high-resolution density map reveals several density peaks toward cond.\,2. They are almost spatially coincident with
 the 870\,\um\ continuum emission peaks (see Fig.\,\ref{fig:Cond123yso}(a)),
 which demonstrates the cold and dense characteristics of cond.\,2.
 Additionally, this condensation overlaps most part of the C\hii~region (see Fig.\,\ref{fig:Cond123yso}).
 In the northwest of the C\hii~region is more diffuse gas, which could be attributed either to
  the gas dispersal by the twofold influences  from the C\hii\ region and the \hii~region RCW\,79 or to the
 inhomogeneity of initial medium in which the C\hii~region forms and evolves.

%%%%%%%%%%%%%fig:NH2HR compact sources%%%%%%%%%%%%%%%%
%%%%%%%%%%%%%fig:end%%%%%%%%%%%%%%%%

Seven compact sources are found within cond.\,2, including two class\,0 (sources 47 and 85), three IM (sources 37, 49,
 and 141), and two class\,I (sources1 and 316) objects classified by the ratio of their
 $L_{_{\rm submm}}^{_{\lambda >\rm 350}}/L_{_{\rm bol}}$.  Another nine
 class\,I YSOs (see Fig.\,\ref{fig:tcolor}) are rather luminous in the near IR bands (ZA06). As shown in Fig.\,\ref{fig:Cond123yso},
 three of them are located around source 1, which is  indicative of a small cluster of star formation around source 1. Another two of the nine
 class\,I YSOs are centered on the C\,\hii~region. The remaining four are located on its southern edge, close to source 141.
 Three of the seven compact sources are MDCs 1, 47, and 85 (see Sect.\,\ref{sect:HMSF}).
  Source 1 is situated at the center of the C\hii~region and close to its ionizing stars.
  According to Eq.\,10 of \citet{mol08},  source\,1 may form a high-mass star with a final stellar
  mass of $\sim 9$\,\msun. We note that the final stellar mass is very uncertain at least since the
  compact source with a typical scale of $\sim 0.1$\,pc can probably fragment into smaller fragments with
  smaller scales \citep[e.g.,][]{bon10b}. Moreover, source\,1 is
   surrounded by another three class\,I YSOs, which
  suggests that massive stars form in clusters \citep[e.g.,][]{lad03}. The other two massive sources 47 and 85
  are located at the edge of the C\hii~region, which most likely will lead to a high-mass star of $\sim 8$\,\msun\ for
  each of them. \\
   \textbf{Condensation\,3}

This is a massive condensation with a mass of $\sim 2600-4300$\,\msun, situated in the southern part of RCW\,79.
It is extended along the ionized gas (see Fig.\,\ref{fig:TdNH2}(c)) and is associated with PDR (see Fig.\,\ref{fig:Cond123yso}(b)),
suggesting the interaction between the \hii~region and the condensation.
It is composed of three density peaks as seen in both the
column density map and in 870\,\um\ continuum emission (see Fig.\,\ref{fig:Cond123yso}(a)). The density peaks revealed by the
column density map and 870\,\um\ emission are well consistent in the densest region, but  shifted about 5\arcsec\ in
the two lower-density regions. The other density peaks in the rest of the map seem to be well consistent between the
column density map and 870\,\um\ emission, therefore, the density peaks shifted in cond.\,2 may be caused by physical environments
instead of the astrometry problem.
The temperature distribution of this condensation shows that the two lower density regions are warmer than the densest region
(see Fig.\,\ref{fig:TdNH2}(b)). Therefore, the peak shifts between the column density map and 870\,\um\ emission
could be interpreted as that the temperatures probed by {\it Herschel} observations are higher than by
870\,\um\ observations.

%%%%%%%%%%%%%fig:NH2HR compact sources%%%%%%%%%%%%%%%%
\begin{figure}[!ht]
\centering
\includegraphics[width=3.4 in]{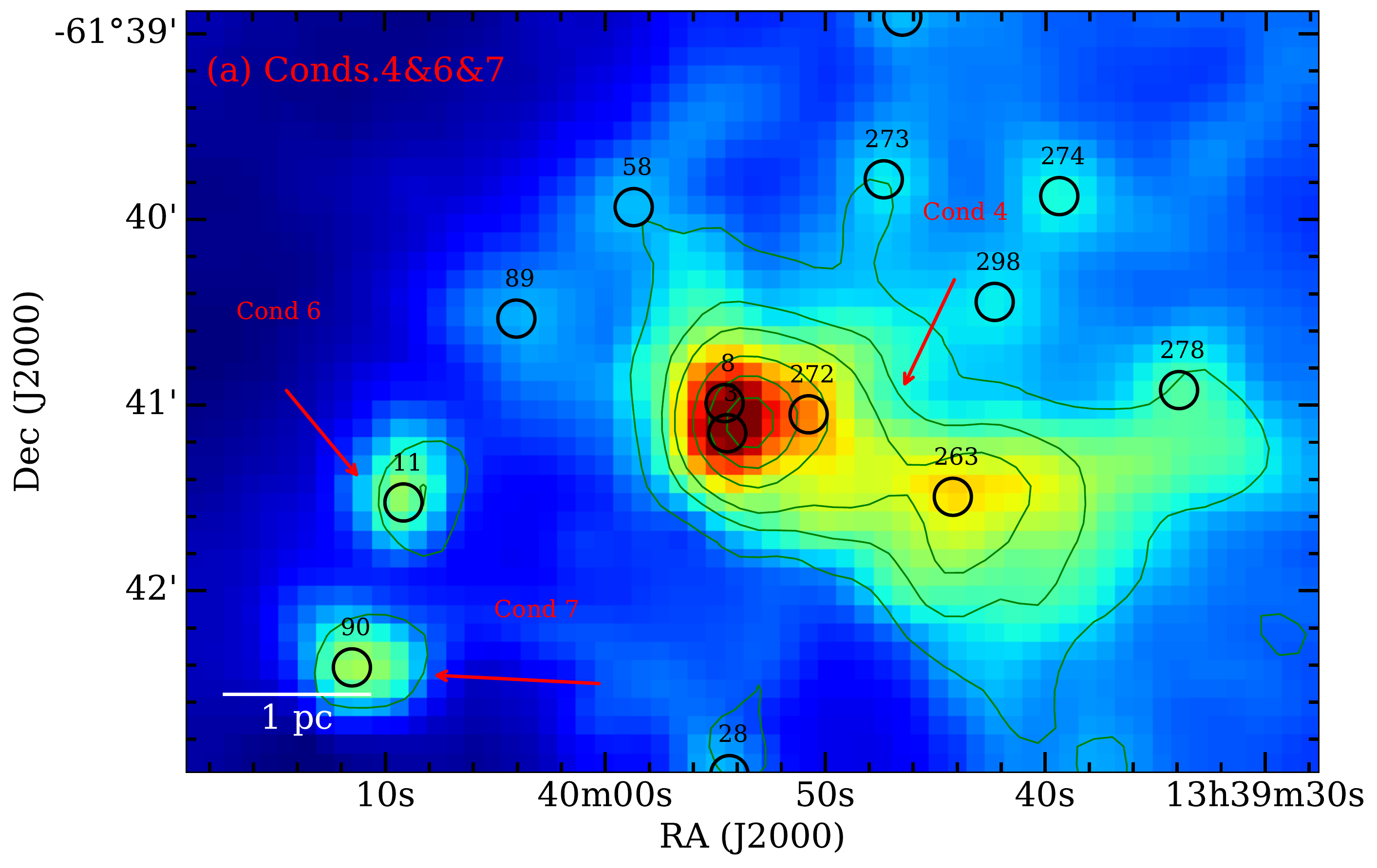}
\hskip 0.00000001cm
\includegraphics[width=3.4 in]{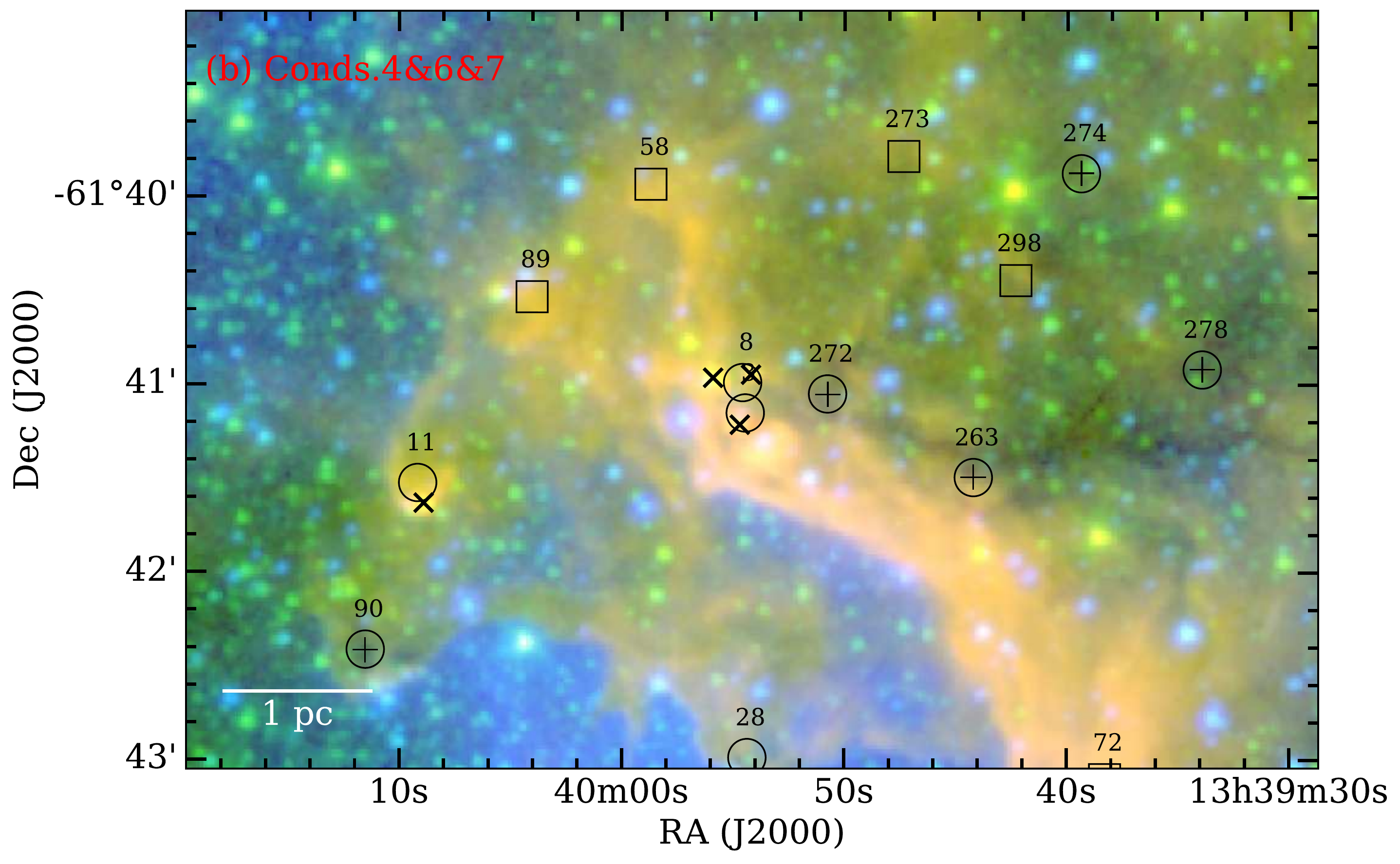}
\caption{As Fig.\,\ref{fig:Cond123yso}, but for condensations\,4, 6, and 7. The contours represent levels of
[0.4, 0.6, 0.8, 1.2, 1.6, 2.0]\,Jy\,beam$^{-1}$.}
\label{fig:Cond467yso}
\end{figure}
%%%%%%%%%%%%%fig:end%%%%%%%%%%%%%%%%

Seven compact sources (sources 2, 4, 5, 16, 19, 45, and 51)
are detected in this condensation. Source 19 is located in one of the two lower-density peaks (see Fig.\,\ref{fig:Cond123yso}(a)).
This source contains a  bright mid-IR point-like object. It is classified as  a class\,I
object (see Fig.\,\ref{fig:LM}), which is in good agreement with the classification
 by ZA06 (see Fig.\,\ref{fig:Cond123yso}(b) and Table\,\ref{tbl:IRfeatures}). As indicated in Fig.\,\ref{fig:LM}, this source may form
an intermediate-mass star of $6-8$\,\msun. Source 16 is associated with the other lower-density peak.
It is a class\,0 candidate and may form a massive star of $\sim~9$\,\msun.
The densest region in cond.\,3 contains five compact sources (sources 2, 4, 5, 45, and 51).
 This region has the highest number of embedded YSOs compared with the other regions in RCW\,79.
 There are two  high-mass YSO candidates (i.e., sources 4 and 45). Source 4 is classified as a class\,I object in Fig.\,\ref{fig:LM},
 associated with a bright star in the near- and mid-IR bands.
 Source\,45 is regarded as class\,0 in Fig.\,\ref{fig:LM}.
 In the densest region, the other three sources 2, 5, and 51 probably form intermediate-mass stars.
 Source 2 is a class\,I object  with IR counterparts in near- and mid-IR bands.
   Source\,5 is luminous at near- and mid-IR
wavelengths and appears as a class\,I object in  Fig.\,\ref{fig:LM}, which agrees with the result of ZA06.
 Source 51 is categorized as class\,0 by the ratio of
$L_{_{\rm submm}}^{_{\lambda >\rm 350}}/L_{_{\rm bol}}$, which
is inconsistent with the classification as a class\,I by the color-color diagram (ZA06).
This inconsistency may be because the class\,I objects detected in the near- and mid-IR bands
 are only overlapping with the compact source in the line of sight, but are not centered inside this compact source, which needs to be
further confirmed. \\
        \textbf{Condensation\,4}

This is a massive condensation, with a mass of $\sim 1800-3100$\,\msun, in the western edge of RCW\,79.
The bright edge of this condensation is adjacent to the ionized gas (see Figs.\,\ref{fig:TdNH2}(c) and \ref{fig:Cond467yso}(b)). This suggests this condensation is under the influence of the ionized gas.
 Condensation\,4 consists of
two density peaks as shown in Fig.\,\ref{fig:Cond467yso}(a). Four compact sources are situated in this
condensation. Sources\,3, 8, 263, and 272  are associated with the densest peaks and
probably form high-mass stars of $8-13.5$\,\msun. Source 3 is detected at mid-IR wavelengths. It is classified
as a class\,I object in Fig.\,\ref{fig:LM}, which is in good agreement with the result of ZA06.
Source 8 is also a class\,I object associated with a point-like source in
mid-IR bands.
Coupled with its high luminosity of $\sim 1600$\,\lsun,  source\,8 is a young massive class\,I object (B2-3V). Sources\, 263 and 272 are classified as massive class\,0 objects
in Fig.\,\ref{fig:LM} and are associated with 8\,\um\ absorption.

Around the condensation are six sources (sources 58, 89, 273, 274, 278, and 298).
They are centered on the local density peaks of more diffuse gas relative to the condensation.
 Sources 58, 89, 273, and 298 are classified as IM candidates according to their luminosities.
They all are associated with IR extended features.
 In addition to the IM candidates, there are two
 class\,0 candidates 274 and 278, which are both characterized with  8\,\um\ absorption. \\
 \textbf{Five other condensations}

 Condensation\,1 is located in the southeastern edge of RCW\,79 and next to cond.\,2, with a mass of $\sim 270-420$\,\msun. In this
 condensation, there is one embedded class\,0 protostar (source
265, see Fig.\,\ref{fig:Cond123yso}).

%%%%%%%%%%%%%fig:NH2HR compact sources%%%%%%%%%%%%%%%%
\begin{figure}[!ht]
\centering
\includegraphics[width=1.7 in]{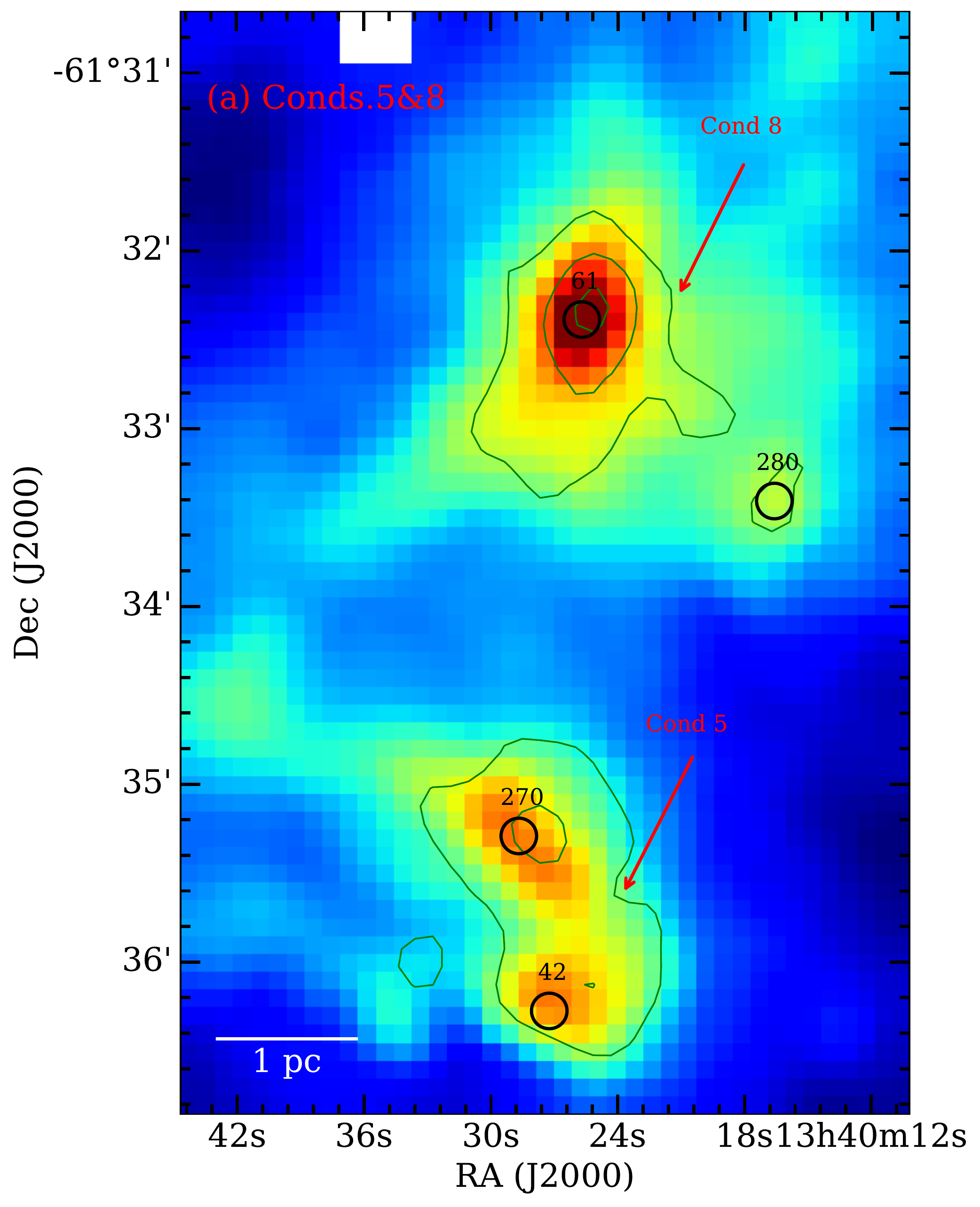}
%\hskip 0.00000001cm
\includegraphics[width=1.7 in]{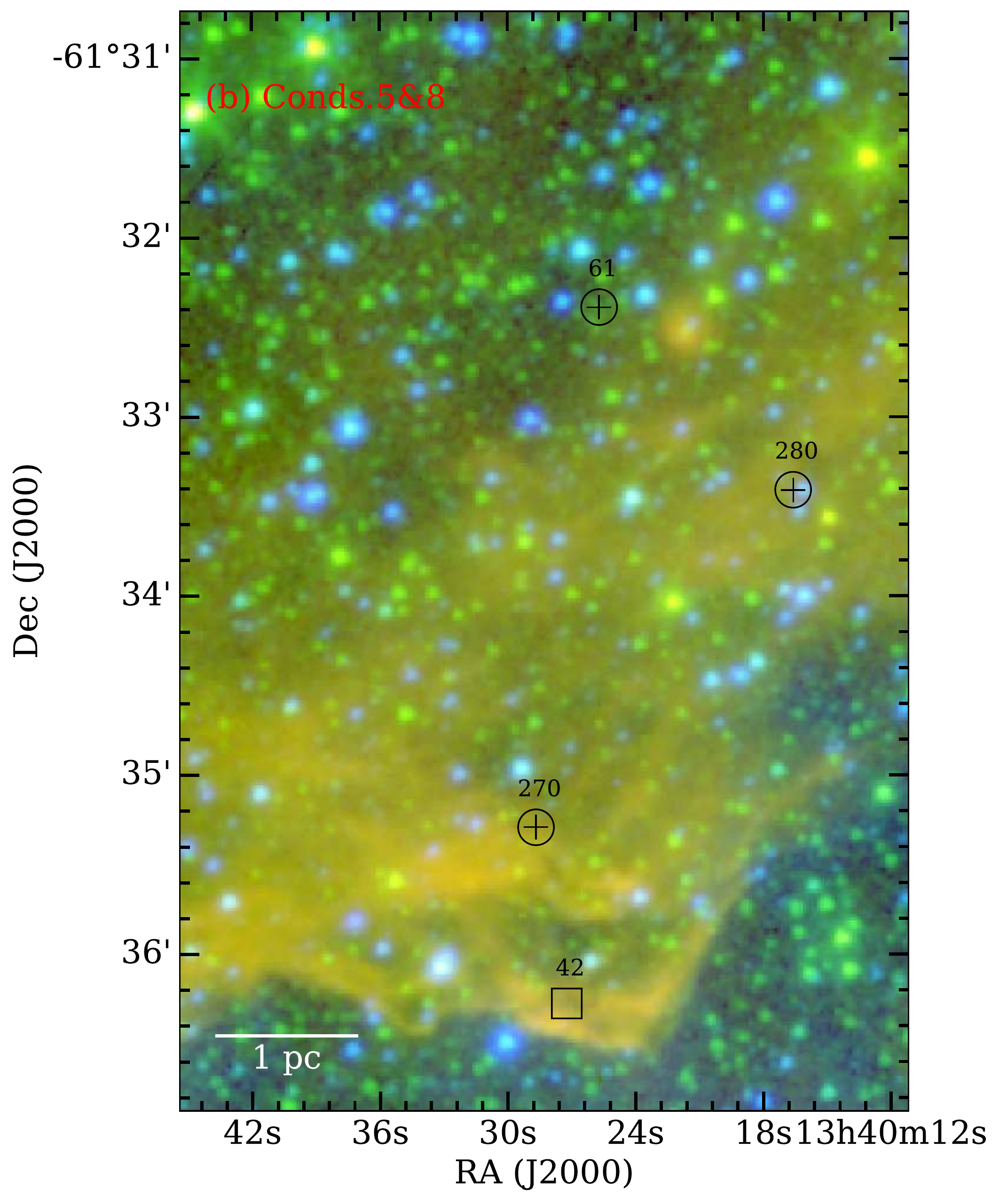}
\caption{As Fig.\,\ref{fig:Cond123yso}, but for condensations\,5 and 8. The contours show levels of
[0.4, 0.6, 0.8]\,Jy\,beam$^{-1}$.}
\label{fig:Cond589yso}
\end{figure}
%%%%%%%%%%%%%fig:end%%%%%%%%%%%%%%%%

 Condensations\,6 and 7 have a mass of $\sim 70-100$\,\msun\
  and $\sim 100-160$\,\msun, respectively. They are located in the direction of the \hii~region,
 indicating that  they may be fully exposed to the ionized gas.
 Of interest are their appearances as  cometary globules (see Fig.\,\ref{fig:Cond467yso}(b)).
 One class\,I object (source 11, see Fig.\,\ref{fig:LM} ) in cond.\,6  and one class\,0 object (source 90) in cond.\,7 are located at the head of
 the two cometary globules, respectively. These two youngest protostars indicate
 that cond.\,6 is more evolved than cond.\,7. Furthermore, the appearance of the cometary globules with
 protostellar cores forming at their heads is reminiscent of the scenario of the radiation-driven  implosion (RDI)
 process \citep[e.g,][]{ber89,lef94}. This  process models the scenario that the pressure exerted by ionized
 gas on the surface of a compact source can lead
 to its implosion, to the formation of a cometary globule with a dense head and a tail extending away from the ionizing source,
 and finally to the collapse of the dense head to form a new generation of stars.

 Figure\,\ref{fig:Cond589yso} displays conds.\,5 and 8 located in the northern part of RCW\,79. They have a mass of $\sim 470-670$\,\msun,
 and $\sim 560-1000$\,\msun, respectively. Condensation\,5 is on the northern edge of RCW\,79 and
 is associated with bright PDRs as seen in 8\,\um\ emission (see Fig.\,\ref{fig:Cond589yso}(b)), but cond.\,8 is about 5\,pc away from the edge and may not be
 very tightly associated with the PDR, as seen in the relatively dark 8\,\um\ emission.
 This implies that cond.\,5 is more affected by ionized gas from the \hii~region than cond.\,8.
  There are one IM (source 42) and one class\,0 (source 270) prototellar core in cond.\,5, and one class\,0
 protostellar core (source 61) in cond.\,8. Of these protostellar cores, two class\,0 (sources 270 and 61) would
 form high-mass stars, as seen in Fig.\,\ref{fig:LM}.
 We note that the mass of source 61 should be treated with caution
because of its large uncertainty of $50\%$. This uncertainty
 may be caused by the flux measurements, which are not good enough at 160\,\um.

Eight massive condensations are observed around RCW\,79. Five of them (conds.\,1, 2, 3, 4, and 5) are distributed on
its edge, associated with bright PDRs, and exposed to ionized gas. These observed characteristics demonstrate that these
condensations must have been inevitably affected by the \hii~region, which is in agreement with the existence of the ionization
compression (see Sect.\ref{sect:pdf}).
According to ZA06 and \citet{wal11}, the
formation of these five condensations may be a result of the ``collect and collapse" (C~\&~C) process\footnote{The `collect and collapse process models a scenario
in which a shell of compressed gas can be accumulated and become denser between the ionization front and the shocked front
during the expansion of an \hii~region, likely leading to a gravitational collapse to form the next generation of stars
in due time \citep{elm77}.}. The two other condensations, conds.\,4, and 5,  may form under the RDI process because they appear as cometary globules with young protostar cores at their heads. Condensation\,8 may be a preexisting compact
source because it is located about 5\,pc away from the edge of RCW\,79. All this suggests that multiple physical processes (e.g., C~\&~C and RDI) may be responsible for the formation of
the condensations around RCW\,79 and even for
the star formation therein, which is consistent with the conclusion of ZA06.

A total of 22 young protostar candidates are detected in the 8 massive condensations.
 Class\,0/I YSOs can be regarded as a new generation of stars since they are
 much younger than the \hii~regions. \citet{eva09} and \citet{eno09} suggested that the entire
 embedded phase for a protostar ($\tau_{\rm Class\,0}+\tau_{\rm Class\,I}$) lasts
 about $5.4\times10^5$ yr. \citet{mar10} estimated a
 photometric age of $2.3\pm0.5$\,Myr for the ionizing cluster of RCW\,79, which is in good agreement with a dynamical age of
 $2.2\pm0.1$\,Myr derived by \citet{tre14}. This means that the age of
 RCW\,79 is about one order of magnitude older than that
 of the class\,0 and/or I YSOs. Given this timescale difference, the youngest protostars in these condensations
 can be affected by the \hii~region in RCW\,79 during its expansion, which is demonstrated by the
 association of the condensations with bright PDR emission (see Figs.\,\ref{fig:Cond123yso} and \ref{fig:Cond467yso}).
  Further to the influences from the  \hii~region, additional influence for cond.\,2 from the associated C\hii~region
  might have affected its nearby youngest protostars. The lifetime of the C\hii~region was estimated to be in the range
 of 0.5 to 2.0\,Myr \citep{mar10}. It is assumed to be 0.5\,Myr in this paper since  the statistical lifetime of C\hii~regions
 has been estimated to be $\sim 6\times10^{5}$\,yr in Cygnus\,X \citep{mot07}. Therefore, the
 C\hii~region may be as young as the class\,0 and/or I YSOs and has had enough time to affect its nearby youngest protostars.

\subsubsection{Core formation efficiency}

%%%%%%%%%%%%%fig:IR and Halpha images with compact sources%%%%%%%%%%%%%%%%
\begin{figure*}[!ht]
\centering
\includegraphics[width=6.0 in]{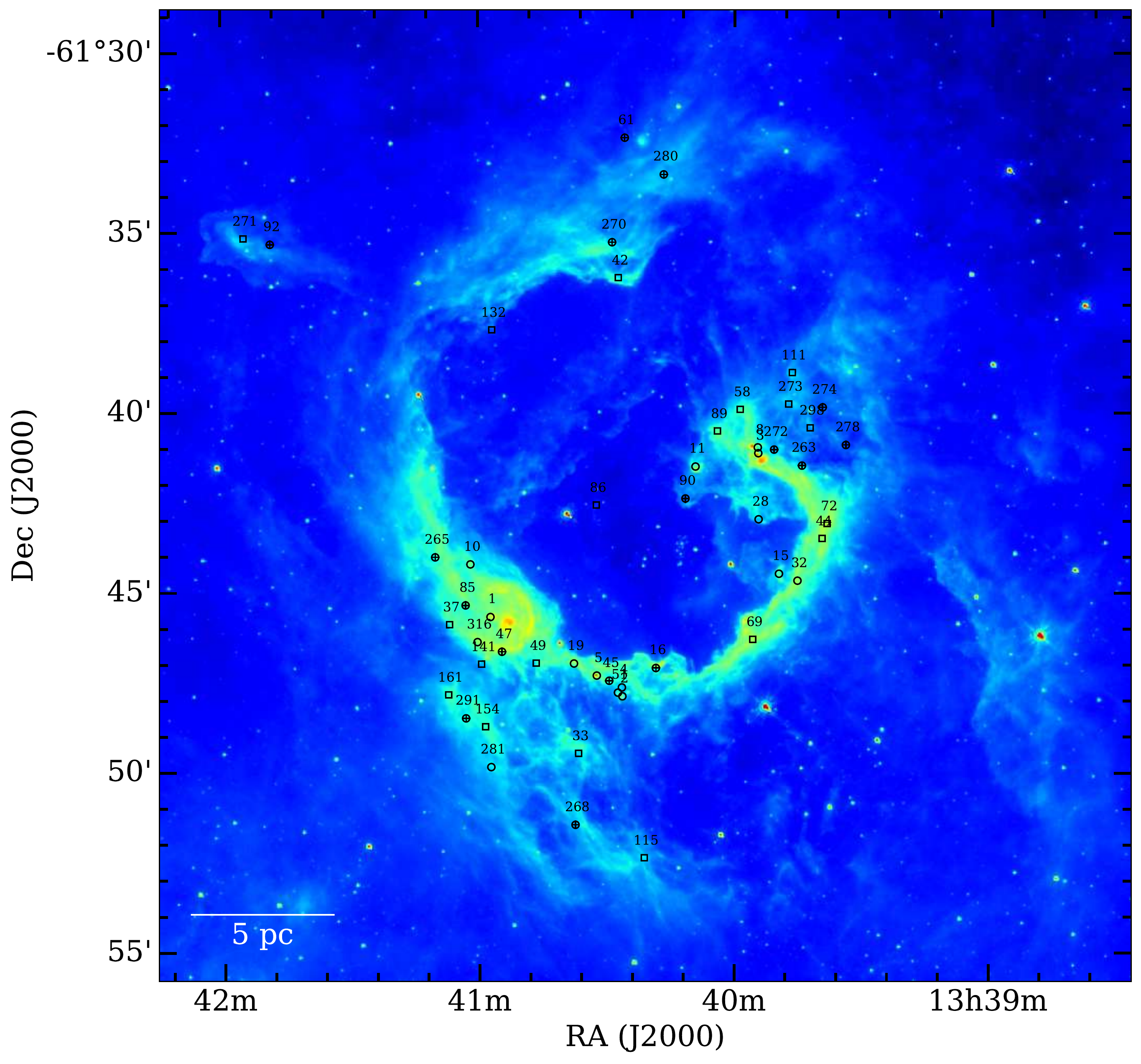}
\caption{Fifty compact sources superimposed on the {\it Spitzer} 8\,\um\ image.
Class\,0, IM, and I objects (see Sect.\,\ref{sect:EvolTrend}) are plotted as plus circles, squares,
and circles, respectively.}
\label{fig:IRhalphaim}
\end{figure*}
%%%%%%%%%%%%%fig:end%%%%%%%%%%%%%%%%

%%%%%%%%%%%%%fig:CFE%%%%%%%%%%%%%%%%
\begin{figure}[!ht]
\centering
\includegraphics[width=3.4 in]{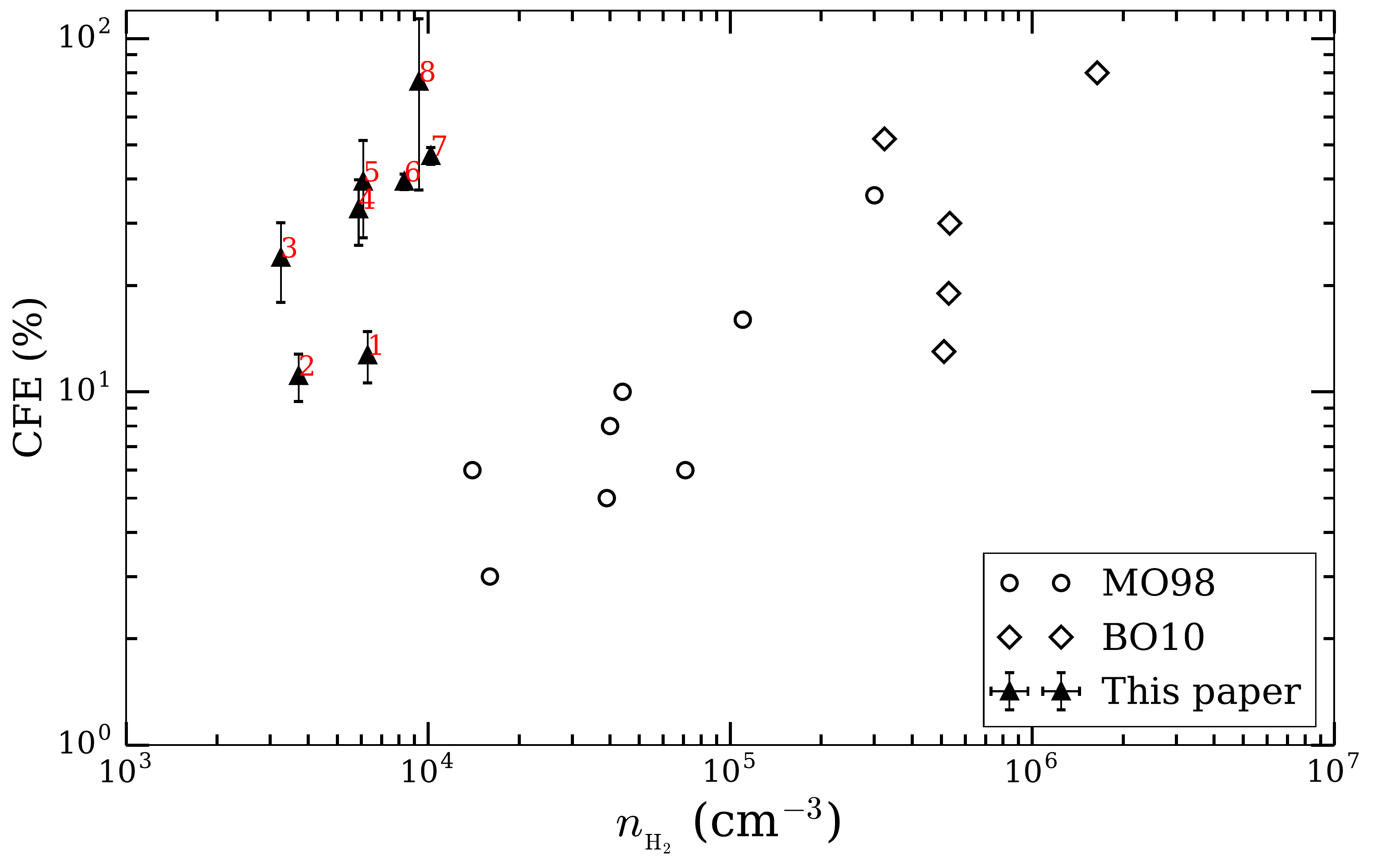} %CFE_density.pdf
\caption{Core formation efficiency as a function of the density. The filled triangles
represent the condensations of pc-scale sizes in RCW\,79. The dense cores of $\sim0.1$ pc
scales in the $\rho$\,Ophiuchus \citep{mot98} and Cygnus X \citep{bon10b} clouds are denoted
as open circles and diamonds, respectively.}
\label{fig:CFE}
\end{figure}
%%%%%%%%%%%%%fig:end%%%%%%%%%%%%%%%%

The association of the 50 compact sources with 8\,\um\ emission (see Fig.\,\ref{fig:IRhalphaim}) demonstrates
the interplay between the \hii~region with star formation.
We observed that 22 out of 50 compact sources
may form a new generation of stars since they are located in the 8 coldest, densest and most massive condensations
that are gravivationally bound (see Fig.\,\ref{fig:NH2HRCP}). Therefore,
we tentatively investigated the core formation efficiency (CFE) for the
8 condensations. The CFE is a measure of which fraction
of molecular gas has been transferred into dense clumps. This quantity is analogous to the SFE and must be treated
as an upper limit to the SFE \citep[e.g.,][]{mot07,bon10b,ede12}.

We estimated the CFE as the ratio of the total mass in compact sources to the mass of their hosting condensation.
To be consistent with the mass of compact sources derived from {\it Herschel} data, $M_{\rm HOBYS}$ was adopted as the mass of the condensations.
The derived CFEs are given in Table\,\ref{tbl:Condpara_der} and range from 11 to $76\%$.
The CFE errors mainly arise from the mass uncertainties of the compact sources.
Compared with the existing results \citep[$5-80\%$, e.g.,][]{mot98,moo07,bon10b,ede12,pal13b,lou14}, the range of 11 to $76\%$
is acceptable.

A trend for an increase in CFE with density has been found in the
 $\rho$\,Ophiuchus \citep{mot98} and  Cygnus\,X clouds \citep{bon10b}
 and has also been confirmed in other star-forming regions \citep{pal13b,lou14}.
To explore if there is a similar trend in RCW\,79, the number densities were approximated as
$n_{_{\rm H_2}} = M_{\rm HOBYS}/(4\pi/3\times R_{\rm min}\times R_{\rm min}\times R_{\rm maj} \times \mu \times m_{\rm H})$
by assuming a 3D geometry for the condensations, where $R_{\rm min}$, and $R_{\rm maj}$ are
their respective minor and major axes (see Table\,\ref{tbl:Condpara_der}),
the mean molecular weight $\mu$ is assumed to be 2.8 \citep{kau08},
and $m_{\rm H}$ is the mass of a hydrogen atom. The derived densities are listed in Table\,\ref{tbl:Condpara_der}. The relation
between the CFE and the density for the condensations is shown in Fig.\,\ref{fig:CFE}, where the
samples from \citet{mot98,bon10b} are included for comparison.
The CFEs of conds.\,1 and 2 are the lowest of all the condensations.
The low CFE of cond.1 may be interpreted in terms of its lowest central concentration.
According to \citet{joh00,kon10},
the degree of the central concentration can be simplified by the internal
column density contrast of a core, which is defined as the peak over mean column densities of
the core. As a result, the internal density contrast of cond.\,1 demonstrates that it is least
centrally concentrated of the eight condensations. For cond.\,2, the lowest CFE could be
in part attributed to strong external pressures imposed by the associated C\hii~region, which
may restrain the mass transformation of the condensation into dense cores by the gas dispersal.

Furthermore, an increasing trend of the CFE with density is present for the other six
condensations (conds.\,3, 4, 5, 6, 7, and 8). This trend is similar to that suggested
by \citet{mot98} and \citet{bon10b}, demonstrating that the denser the condensation,
the higher the fraction of its mass transformation into dense cores.  However, the trend in RCW\,79 is
discontinuous with that found in the $\rho$\,Ophiuchus and  Cygnus\,X clouds.
It is worth noting that \citet{mot98} and \citet{bon10b}
studied the fragmentation of a few 0.1\,pc dense cores into $\sim0.01$\,pc protostellar cores, but
we extend the fragmentation on the scale of a few pc condensations into $\sim~0.2$\,pc dense cores
due to the limited angular resolution in this paper. These
different scales can illustrate the density difference between our condensations and the
dense cores studied in \citet{mot98} and \citet{bon10b} (see Fig.\,\ref{fig:CFE}). Therefore the
above discontinuous appearance could be credited to the different physical scales.
Despite this discontinuity, the slope of the trend is very similar to that demonstrated in other work
\citep[e.g.,][see Fig.\,\ref{fig:CFE}]{mot98,bon10b,pal13b}. Higher-resolution millimeter observations are clearly needed to study the CFE at smaller scales in the condensations observed around RCW 79 to see how the ionization affects the CFE.

\section{Conclusions}\label{sect:conclude}
Using {\it Herschel} observations complemented with {\it Spitzer}, WISE, and 2MASS data,
we have explored the early stages of star formation in the Galactic \hii~region RCW\,79.
The main results can be
summarized as follows:

\begin{enumerate}
        \item The dust temperature is found to be higher in the direction of the \hii~region than in
        the regions away from this region, demonstrating a heating and influence of ionized gas on its vicinity. This
    influence is further demonstrated by the column density probability distribution function (PDF).
    \item The column density PDF consists of two lognormal distributions and a power-law tail.
    The second lognormal distribution reveals the compression from ionized gas. The slope of
    the power-law tail indicates the existence of gravitationally bound regions in RCW\,79.
        \item The most complete sample of 50 reliable compact sources around RCW\,79
        has been assembled for the first time.
        They have sizes from 0.1 to 0.4\,pc with a typical scale of $\sim0.2$\,pc,
        temperatures of $\sim 11 - 26$\,K, masses of $\sim 6 - 760$\,\msun, densities
        of $\sim 0.1 - 44\times 10^{5}$\,cm$^{-3}$ , and bolometric luminosities of $\sim 19 - 12712$\,\lsun.
    The majority ($96\%$) of them are located in the ionization-compressed layer, suggesting that
    they are exposed to the influence of the \hii~region. In addition, $44\%$ of the 50 compact sources that are
    situated in the gravitationally bound regions have a high probability of forming stars.
        \item The 50 compact sources have been classified into three classes of objects:
        16 class\,0, 19 IM, and  15 class\,I,
         by a combination of the ratio of $L_{_{\rm submm}}^{_{\lambda >\rm 350}}/L_{_{\rm bol}}$ \citep{bon10a} with
        the color classification scheme of \citet{all04}. Their distributions in the L-M diagram \citep{mol08} almost follow
        the evolutionary tracks well, indicating
        that the 50 compact sources are indeed in the early stages of star formation.
        \item Twelve massive dense cores (MDCs), candidates of high-mass star precursors,
         have been tentatively identified by a mass threshold of 140\,\msun\ derived
        from the L-M$_{\rm env}$ diagram. Their mass surface densities of $\Sigma =\ \sim 0.4-5.1$\,g\,cm$^{-2}$ agree
        with the existing values observed in several high-mass dense core that may form high-mass stars.
        \item The eight massive condensations contain a total of 22 young protostar cores, which is suggestive of active star formation.
        Their lifetimes, which are much younger than the main \hii~region, the spatial coincidence of the condensations with
        the bright PDR emission and the spatial distribution of compact sources located in the ionized-compression layer
    together indicate that the young protostars are influenced by the \hii~region in the course of its dynamical evolution. For cond.\,2, additional influence from
        its associated C\hii~region might have affected its nearby young protostars.
        \item An increasing trend of the CFE with density has been found in the condensations of RCW\,79, demonstrating that
         a denser condensation has a higher fraction of gas transformation into dense cores.
\end{enumerate}

\begin{acknowledgements}
   This work was carried out under the framework of a joint doctoral promotion program
   between the Chinese Academy of Sciences (CAS) and
   Centre National de la Recherche Scientifique (CNRS), and
   has been supported in part by the Center National d'Etudes Spatiales (CNES) and
   the National Natural Science Foundation of China through the grant 11573036.
   N.S acknowledges support from the DFG (0s 177/2-1 and 177/2-2), and central funds
of the DFG-priority program 1573 (ISM-SPP).
  We thank P. Tremblin for helpful
discussions of the column density probability distribution function, and
P. Palmeirim for the discussion of the dynamical age of the compact \hii~region.
   SPIRE has been developed by a consortium of institutes
led by Cardiff Univ. (UK) with Univ. Lethbridge (Canada); NAOC (China);
CEA, LAM (France); IFSI, Univ. Padua (Italy); IAC (Spain); Stockholm
Observatory (Sweden); Imperial College London, RAL, UCL-MSSL, UKATC,
Univ. Sussex (UK); Caltech, JPL, NHSC, Univ. Colorado (USA). This development
has been supported by national funding agencies: CSA (Canada);
NAOC (China); CEA, CNES, CNRS (France); ASI (Italy); MCINN (Spain);
SNSB (Sweden); STFC (UK); and NASA (USA). PACS has been developed
by a consortium of institutes led by MPE (Germany) with UVIE (Austria); KU
Leuven, CSL, IMEC (Belgium); CEA, LAM(France); MPIA (Germany); INAFIFSI/
OAA/OAP/OAT, LENS, SISSA (Italy); IAC (Spain). This development
has been supported by the funding agencies BMVIT (Austria), ESA-PRODEX
(Belgium), CEA/CNES (France), DLR (Germany), ASI/INAF (Italy), and
CICYT/MCYT (Spain). We have used the NASA/IPAC Infrared Science Archive
to obtain data products from the UKIDSS, \emph{Spitzer}-GLIMPSE, and \emph{Spitzer}-MIPSGAL
surveys.
\end{acknowledgements}

\bibliographystyle{aa}
\bibliography{bibaa}

\clearpage
\begin{appendix}
\section{Compact sources}
\subsection{Source extraction}\label{sect:app-Sectraction}
Compact sources were extracted using the algorithm \emph{getsources} V. 1.140127, a multiscale and multiwavelength
source extraction method \citep{men10,men12,men13}. It works by analyzing decompositions of original images over a broad spectrum
of spatial scales and across all wavebands. The wavelength-dependent single-scale detection images are created after subtracting
 noise and background emission of the corresponding single-scale images. The algorithm then starts to detect
compact sources in the combined detection images by tracking the evolution of their segmentation masks across all spatial
scales. Photometry of the compact sources is measured in the original background-subtracted image at each wavelength, and
the blended sources are deblended in an iterative procedure. The main advantage of this  algorithm is that it improves the detection ability for the compact sources, especially in crowded regions, by means of the fine spatial decompositions of original
images. Another advantage is that the algorithm designed to automatically combine data over all wavebands successfully
averts the difficulty of cross matching independent extraction catalogs at different wavelengths.

In the process of the detection of compact sources, the color-corrected brightness images at 160 and 250\,\um\ were
particularly considered to help \emph{getsources}
focus on real density peaks instead of those only locally illuminated by external heating,
which is an important consideration especially around \hii~regions. More details can be found in
the forthcoming HOBYS consortium paper (Tig\'{e} et al. 2016, submitted).
Additionally, a column density map with high angular resolution was used to validate
the detection of actual density peaks.\
\subsection{SED fittings}\label{sect:app-SEDfit}
To estimate the dust temperature $T_{\rm d}$ and envelope mass $M_{\rm env}$ of our sample of 100 candidate compact
sources, we have performed SED fittings.
Assuming dust emission to be optically thin, we adopted the graybody function for a single temperature
to carry out the SED fitting as below.
\begin{equation}\label{eq:graybody}
F_{\nu} = \frac{\kappa_\nu\,B_\nu(T_{\rm d})\,M_{\rm env}}{D^2},
\end{equation}
where $F_{\nu}$ is the observed flux, $B_\nu(T_{\rm d})$ is the blackbody function and $D$ is the distance.
 With $T_{\mathrm{d}}$ and $M_{\rm env}$ treated as free parameters, the SED fitting was performed by
 invoking the IDL program MPFITFUN \citep{mar09}.

Before the SED fitting, we scaled the fluxes at different wavelengths
as below \citep{ngu11}:
\begin{equation}\label{eq-fluxscaling}
F_{\lambda}^{scaled} = F_{\lambda}^{original}\ \left(\frac{FWHM_{ref}^{dec}}{FWHM_{\lambda}^{dec}}\right),
\end{equation}
where $F_{\lambda}^{scaled}$ and $F_{\lambda}^{original}$ are representatives of respective scaled and
original fluxes, and $FWHM_{ref}^{dec}$ and $FWHM_{\lambda}^{dec}$ represent the deconvolved sizes
(see Equation\,\ref{eq-sizecorrect}) at the reference
wavelength (160 or 250\,\um), and other wavelengths, respectively.
This flux scaling method works
under the assumptions that the source size measured at the reference wavelength is accurate enough, that
emission at all wavelengths is optically thin, and that the flux nearly varies linearly
with the source size.
More descriptions of the principle of this scaling relation are given in \citet{ngu11}.
Regarding the reference wavelength, 160 or 250\,\um\ instead of 100\,\um\ were chosen since emission at 100\,\um\
may be more severely affected by external heating especially in \hii~regions than 160 or 250\,\um.
For the majority of our samples, the 160\,\um\ is regarded as the reference waveband. However, the 250\,\um\ was also considered
for sources whose deconvolved size at 160\,\um\ exceeds that at 250\,\um, which is indicative of more extended emission
at 160\,\um. This extended emission may result from contaminations of small grains heated in PDR regions (Tig\'{e} et al. 2016, submitted).

The deconvolved sizes ($FWHM_\lambda^{dec}$) at different wavelengths were
approximated as follows:
\begin{equation}\label{eq-sizecorrect}
FWHM_{\lambda}^{dec}=\sqrt{{FWHM_{\lambda}^{obs}}^2-{HPBW_{\lambda}}^2},
\end{equation}
where $FWHM_{\lambda}^{obs}$ is the observed size at each wavelength, and $HPBW_{\lambda}$ is the beam size.
This approximation was not applied when the deconvolved size of a source at a special wavelength was
smaller than half the corresponding beam size, which indicates that the source cannot be resolved at this
wavelength. We simply assigned to the unsolved source
a minimum physical deconvolved size of $FWHM_{\lambda}^{dec}=0.5\times HPBW_{\lambda}$.

In the SED fitting, photometric color corrections were applied to all \emph{Herschel}  bands.
The colour correction factors for the PACS wavelengths  vary
with the dust temperature \citep{pog10}. Therefore,
we gradually refined the colour correction factors by an iterative SED fitting procedure
until a variation of a dust temperature of $\leq0.01$ K. For the SPIRE wavelength bands,
their colour correction factors only
depend on the dust emissivity spectral index $\beta$, fixed to 2 in the fitting.
They are $4\%$, $4\%$, $5\%$ at 250, 350, and 500\,\um\ (see SPIRE observers' manual),
respectively. The scaled and color-corrected fluxes are shown in Table\,\ref{tbl:herschel_flux_corr}.

\section{Parameters and figures of compact sources}\label{sect:appcompactsource}

\begin{table*}
\caption{Original fluxes in the {\it Herschel} bands for the 50 compact sources}
\label{tbl:herschel_flux_ori}
\resizebox{18cm}{!}{
\centering
\begin{tabular}{crrrrrrrrrrrr}
\hline\hline
ID & $S_{p,70}$ & $S_{70}$ & $S_{p,100}$ & $S_{100}$ & $S_{p,160}$ & $S_{160}$ & $S_{p,250}$ & $S_{250}$ & $S_{p,350}$ & $S_{350}$ & $S_{p,500}$ & $S_{500}$\\
   & \jybeam\ & Jy & \jybeam\ & Jy & \jybeam\ & Jy & \jybeam\ & Jy & \jybeam\ & Jy & \jybeam\ &Jy \\
\hline
\input ./tbl/Herschel_flux_ori.tbl
\hline
\end{tabular}
}
\end{table*}

\clearpage

\begin{table*}
\centering
\caption{Size-scaled and color-corrected fluxes in the {\it Herschel} bands}
\label{tbl:herschel_flux_corr}
\resizebox{11cm}{!}{
\begin{tabular}{crrrrr}
\hline\hline
ID & $S_{100}$ & $S_{160}$ & $S_{250}$ & $S_{350}$ & $S_{500}$ \\
   & Jy & Jy & Jy & Jy & Jy  \\
\hline
\input ./tbl/Herschel_flux_corr.tbl
\hline
\end{tabular}
}
\end{table*}

\begin{table*}
\centering
\caption{Observed properties of compact sources}
\label{tbl:IRfeatures}
\resizebox{18cm}{!}{
\begin{tabular}{crrrrrr}
\hline\hline
ID & Photometries$^{\rm a}$ & IR\ Counterpart$^{\rm b}$ & Blended\ Source & Hosting\ Condensation$^{\rm c}$ & Class$^{\rm d}$
   & Comments$^{\rm e}$ \\

\hline
\input ./tbl/compact_source_comments_v3_tex.tbl

\hline
\end{tabular}
}
\tablefoot{\\
$^{\rm a}$ Parenthesis and bracket denote the good and poor photometries at the corresponding wavelength, respectively.
The criteria of the good photometry are explained in Sect.\,\ref{sect:IRfeatures}. \\
$^{\rm b}$ Parenthesis and bracket represent the point-like and extended-structure sources, respectively. \\
$^{\rm c}$ Condensations follow the identification number given by ZA06. \\
$^{\rm d}$ Class notes the three classes of sources categorized by
the ratio between submillimeter and bolometric luminosities. \\
$^{\rm e}$ The final classification is determined by a combination of the ratio as mentioned in (b) and their IR colors (see Sect.\,\ref{sect:EvolTrend}.)
}
\end{table*}

\begin{table*}
\centering
\caption{Photometries in the NIR and MIR bands for 11 compact sources}
\label{tbl:NIR_MIR_phot}
\resizebox{18cm}{!}{
\begin{tabular}{crrrrrrrrrrr}
\hline\hline
ID & Designation & $J$ & $H$ & $Ks$ & $I1$ & $I2$ & $I3$ & $I4$ & $12$ \um\ & $24$ \um\  \\
   &             & mJy & mJy & mJy & mJy & mJy & mJy & mJy & mJy & mJy  \\
\hline
\input ./tbl/NIR_MIR_phot_final.tbl

\hline
\end{tabular}
}
\tablefoot{The fluxes for source\,28 are obtained by aperture photometry.
}
\end{table*}

\begin{figure*}[!ht]
\centering
\subfloat{\includegraphics[width=3.4 in]{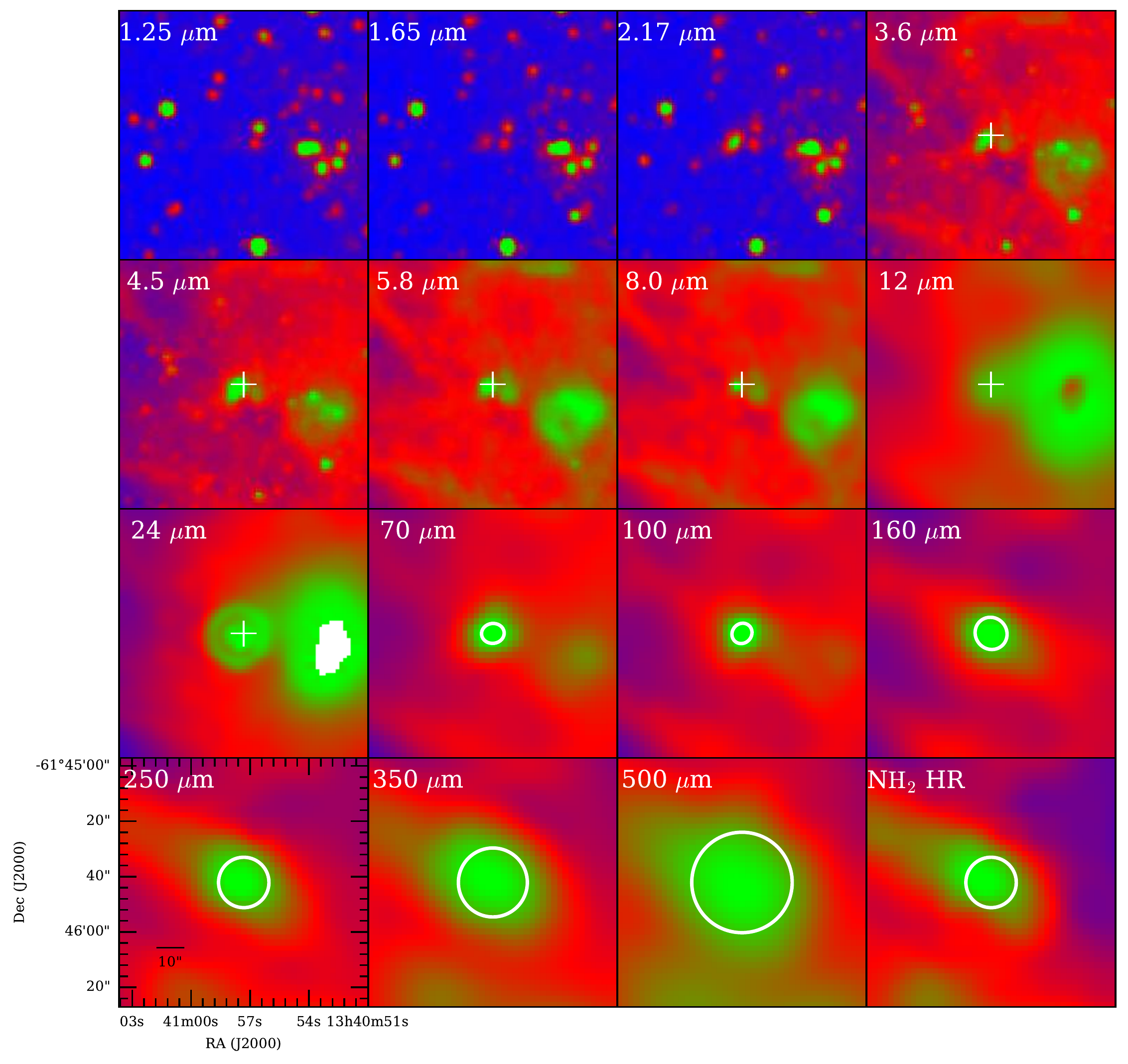}
\hskip 0.00000001cm
\includegraphics[width=3.4 in]{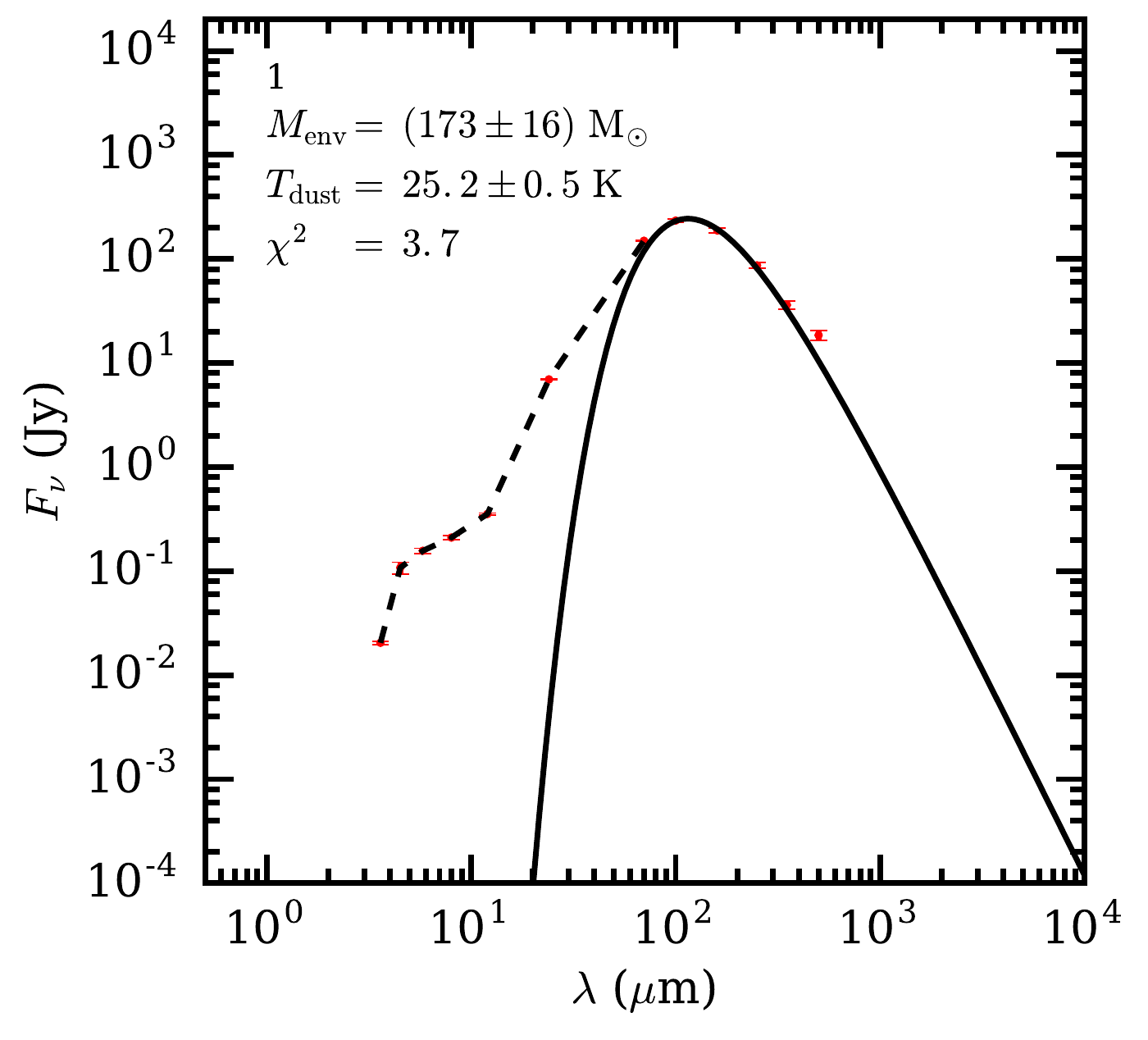}
}
\qquad
\subfloat{\includegraphics[width=3.4 in]{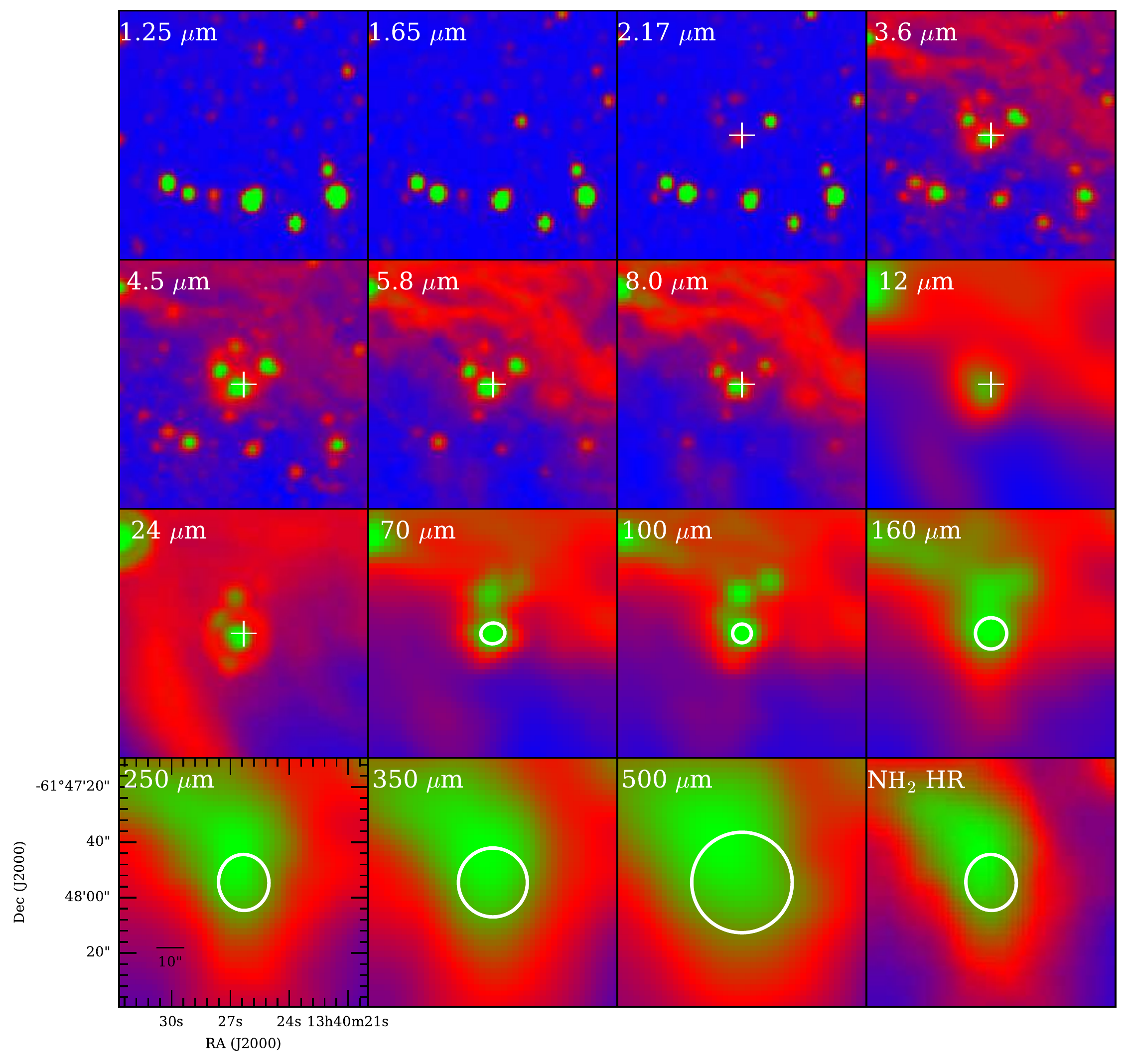}
\hskip 0.00000001cm
\includegraphics[width=3.4 in]{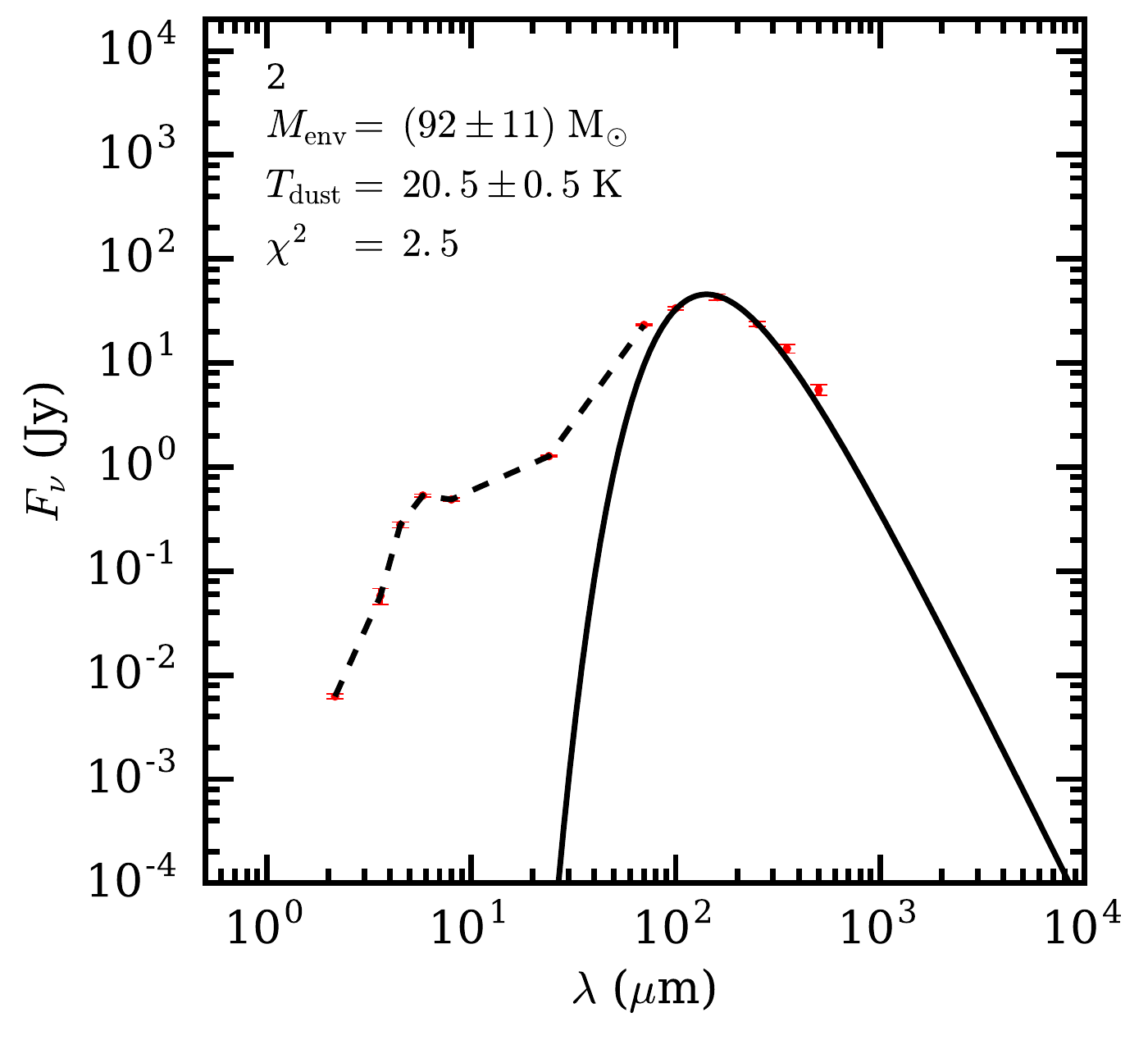}
}
\caption{(Left) plot consisting of multiwavelength images from 1.25 to 500\,\um\ along with the $18\farcs2$ resolution column density map. All images are resampled to the same pixel size ($2\arcsec$) for a better display effect. The coordinates of compact sources with IR point counterparts are shown in the cross and the photometry apertures in the {\it Herschel}  bands
are marked in the ellipse.
(Right) SED fitting plot. The filled circle symbolizes the flux at each wavelength available to each source.
The solid line indicates the graybody fitting to the available fluxes at 100 to 500\,\um\ and
 the dashed line is the connecting line among the available IR fluxes at 1.25 to 70\,\um.}
\label{fig:appmultimage}
\end{figure*}
\begin{figure*}[!ht]
\ContinuedFloat
\centering
\subfloat{\includegraphics[width=3.4 in]{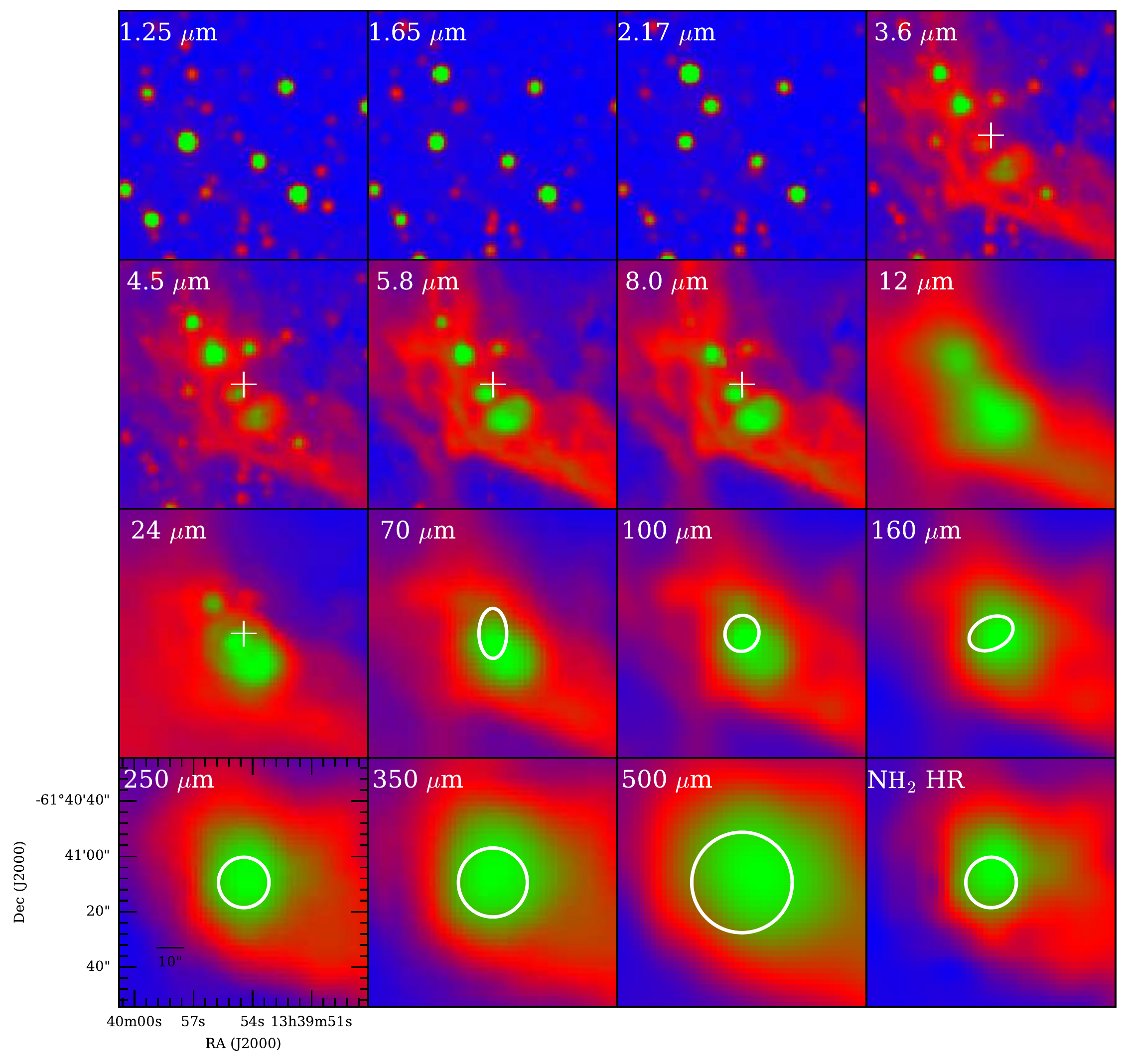}
\hskip 0.00000001cm
\includegraphics[width=3.4 in]{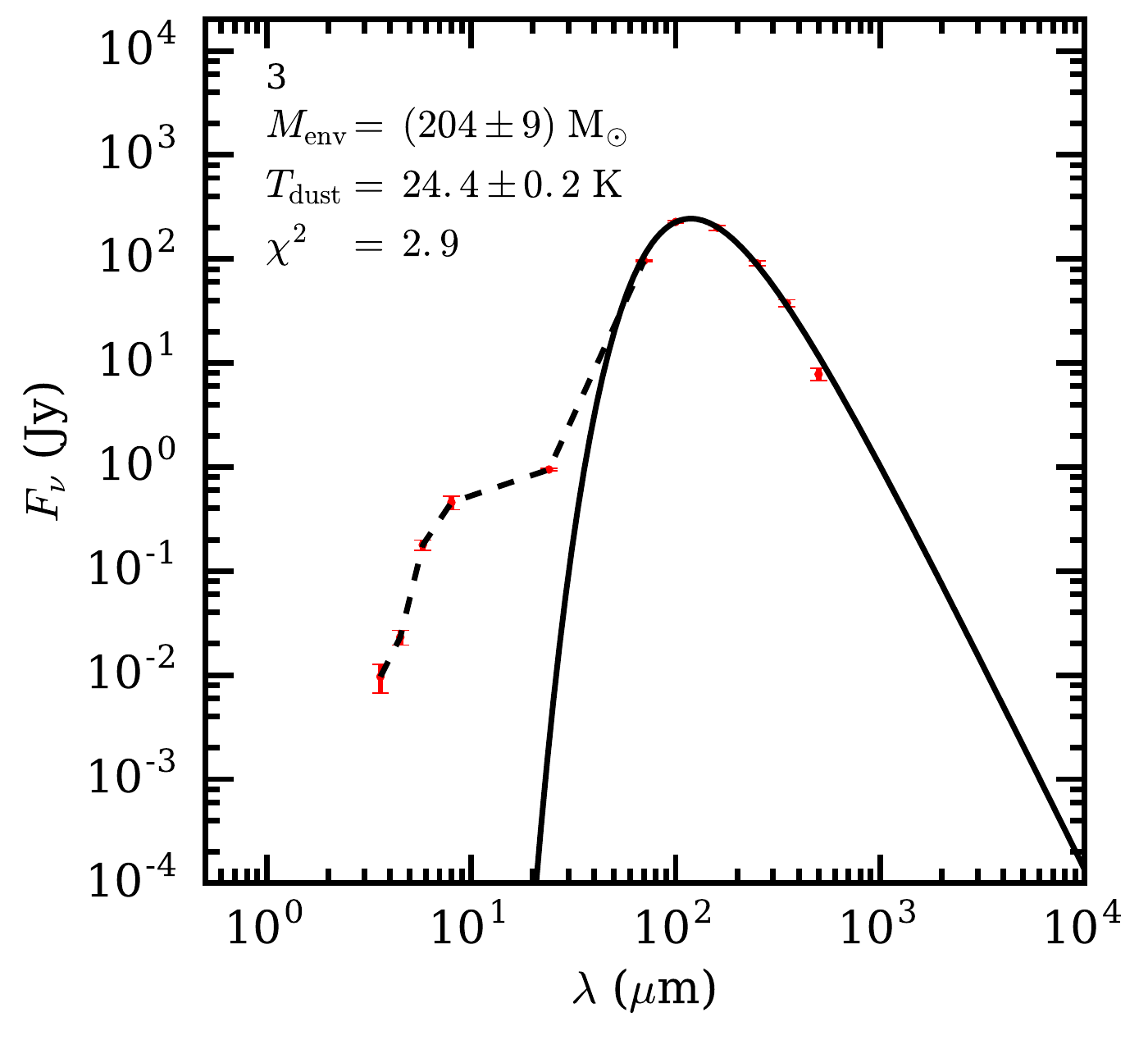}
}
\qquad
\subfloat{\includegraphics[width=3.4 in]{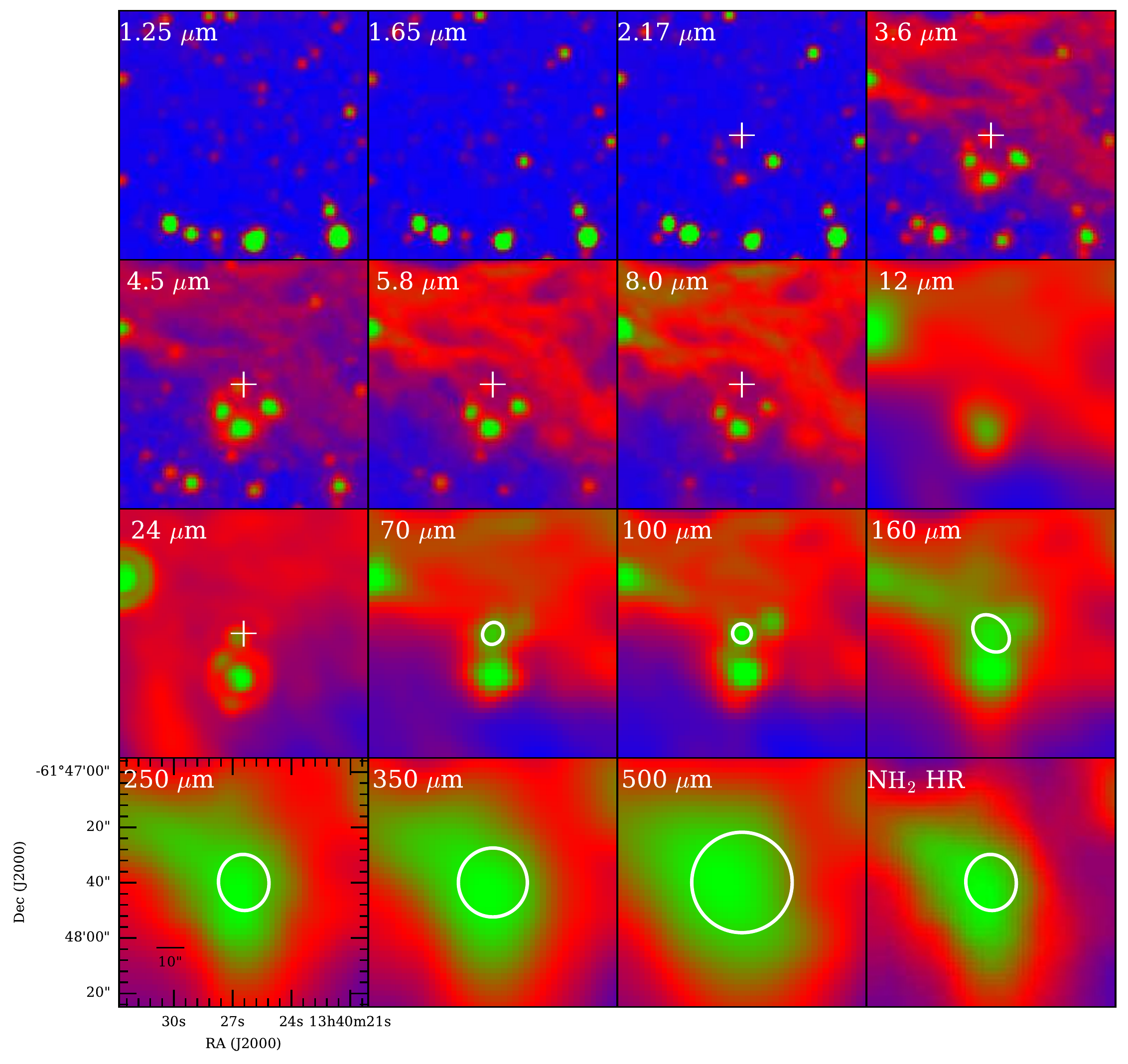}
\hskip 0.00000001cm
\includegraphics[width=3.4 in]{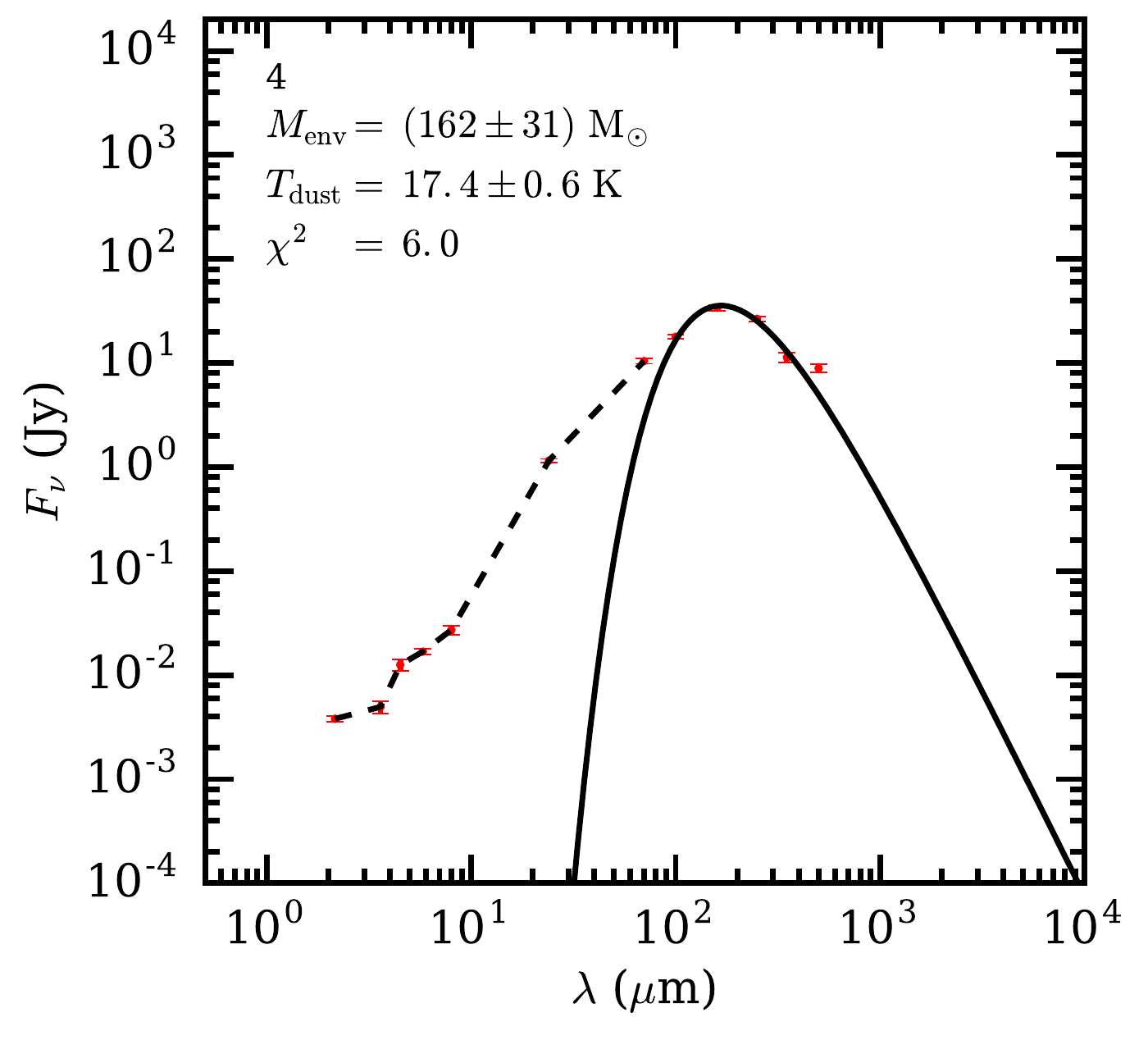}
}
\begin{flushleft}
\caption{ -- continued.}
\end{flushleft}
\end{figure*}
\begin{figure*}[!ht]
\ContinuedFloat
\centering
\subfloat{\includegraphics[width=3.4 in]{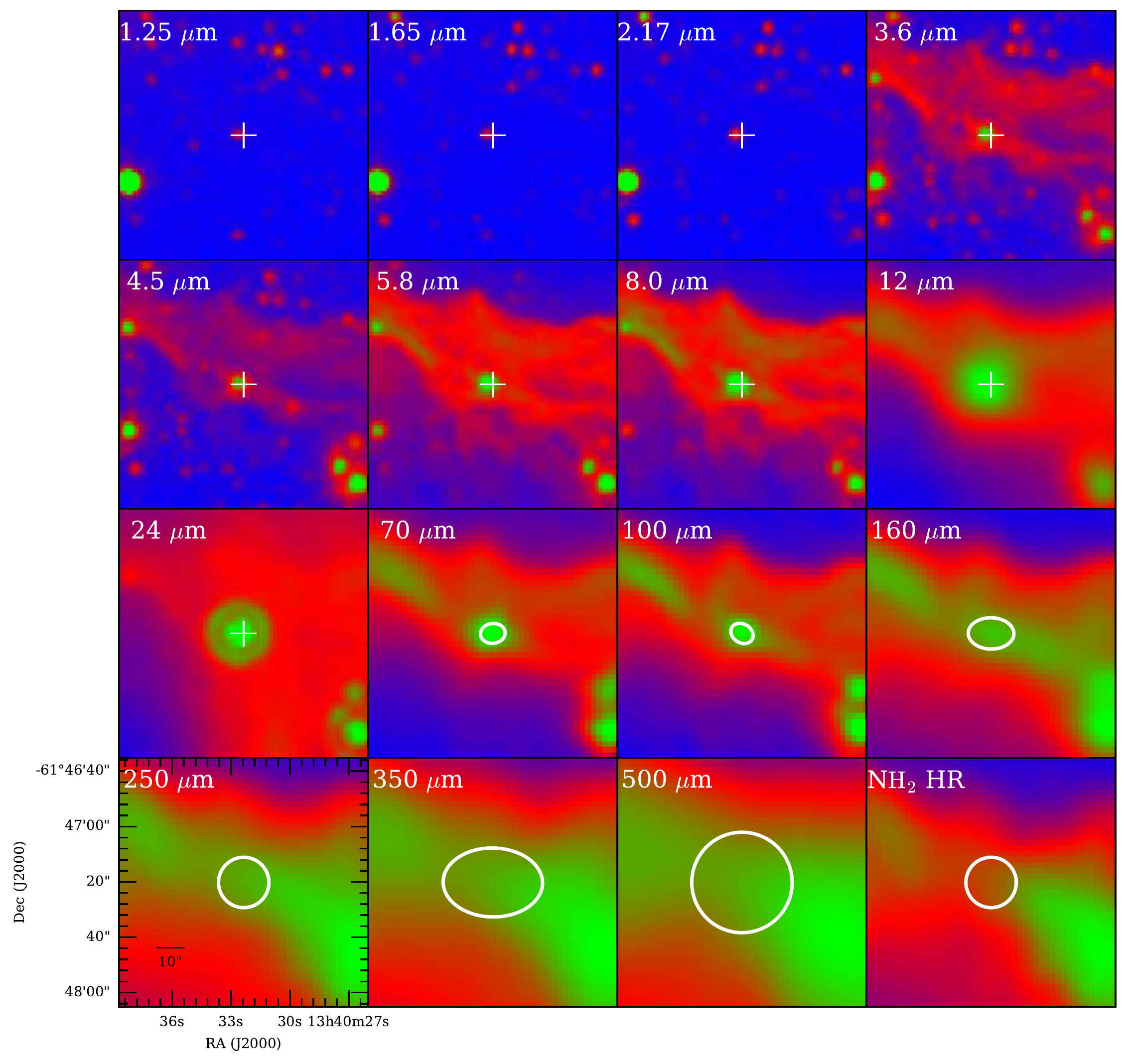}
\hskip 0.00000001cm
\includegraphics[width=3.4 in]{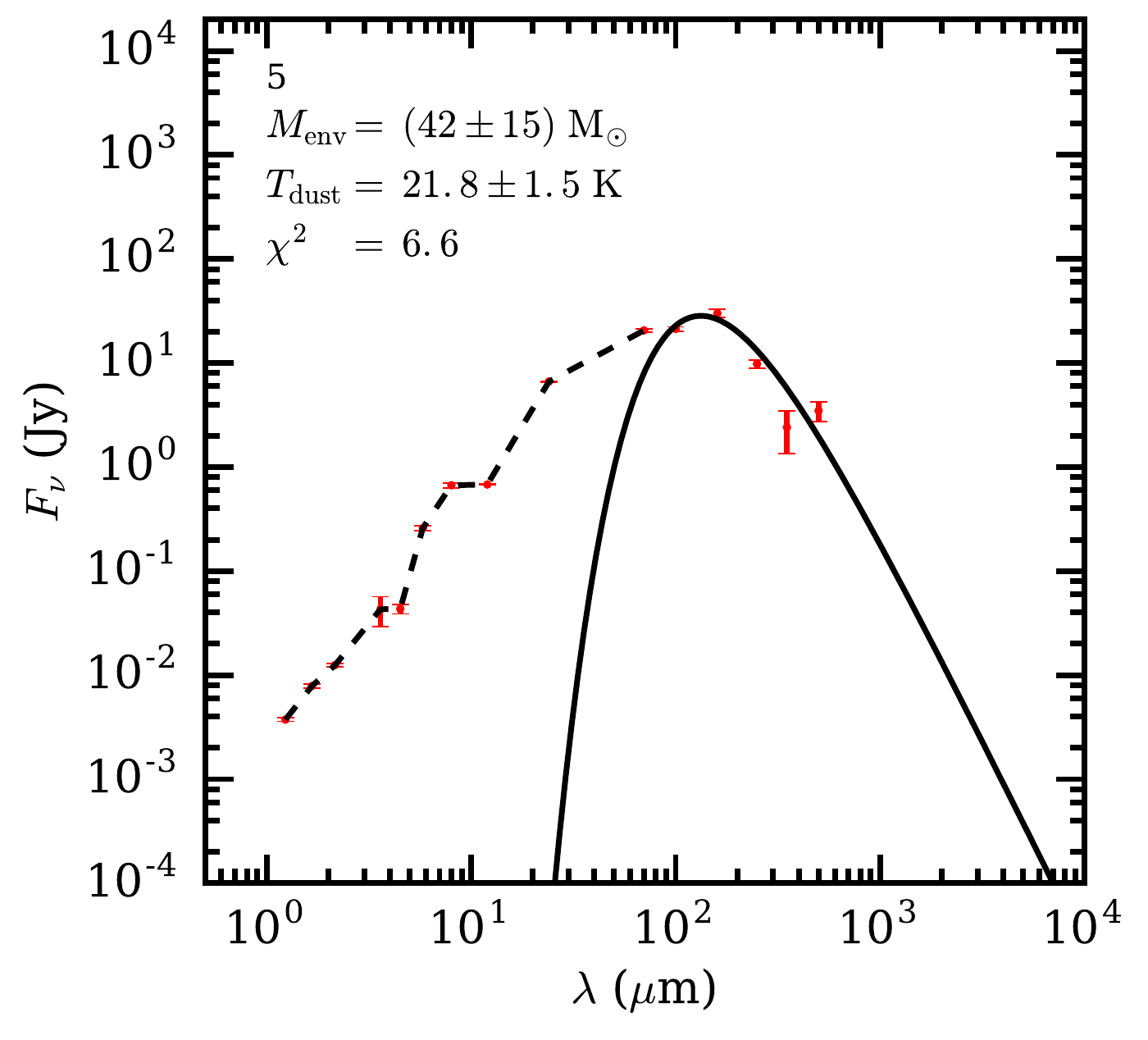}
}
\qquad
\subfloat{\includegraphics[width=3.4 in]{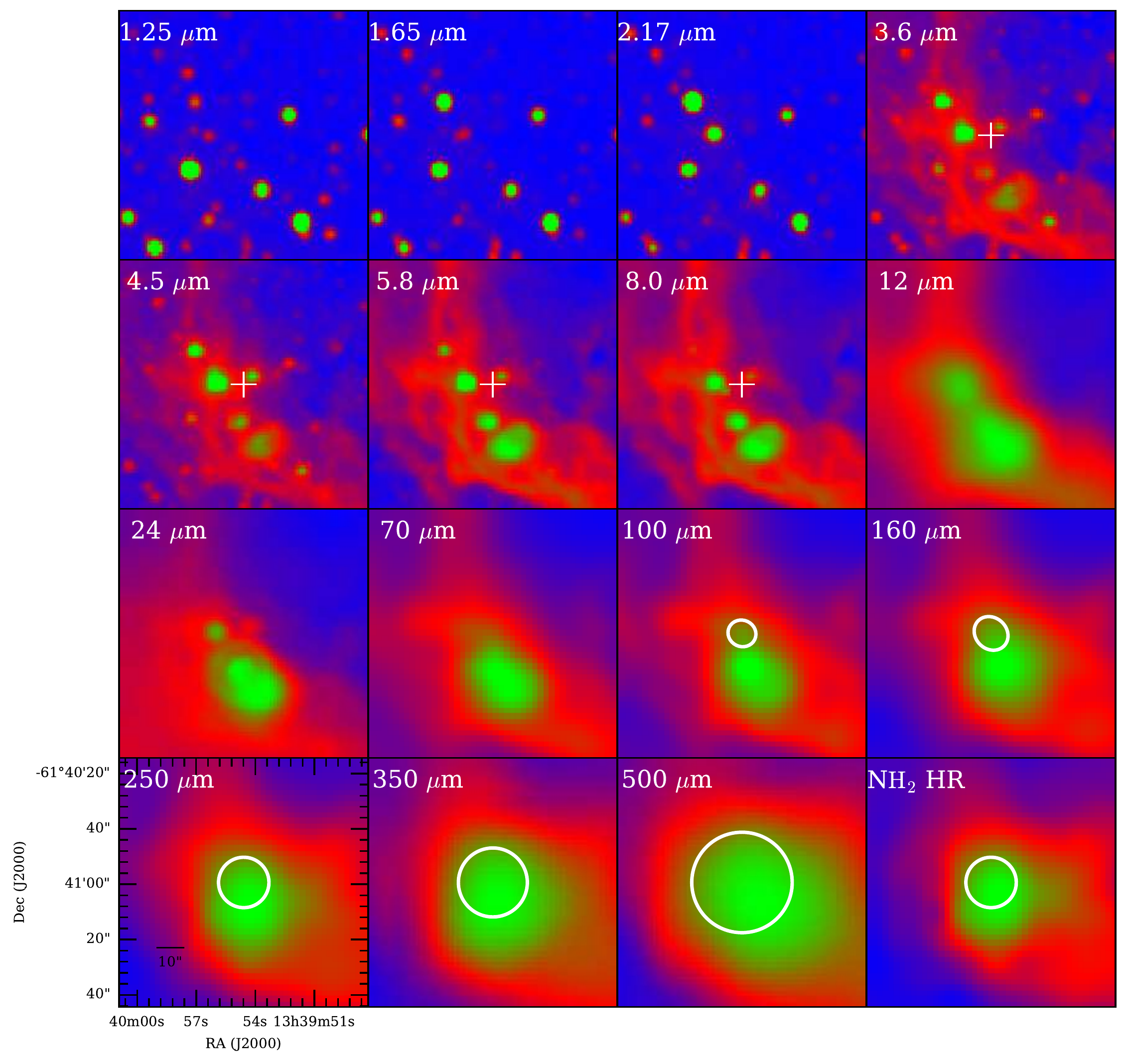}
\hskip 0.00000001cm
\includegraphics[width=3.4 in]{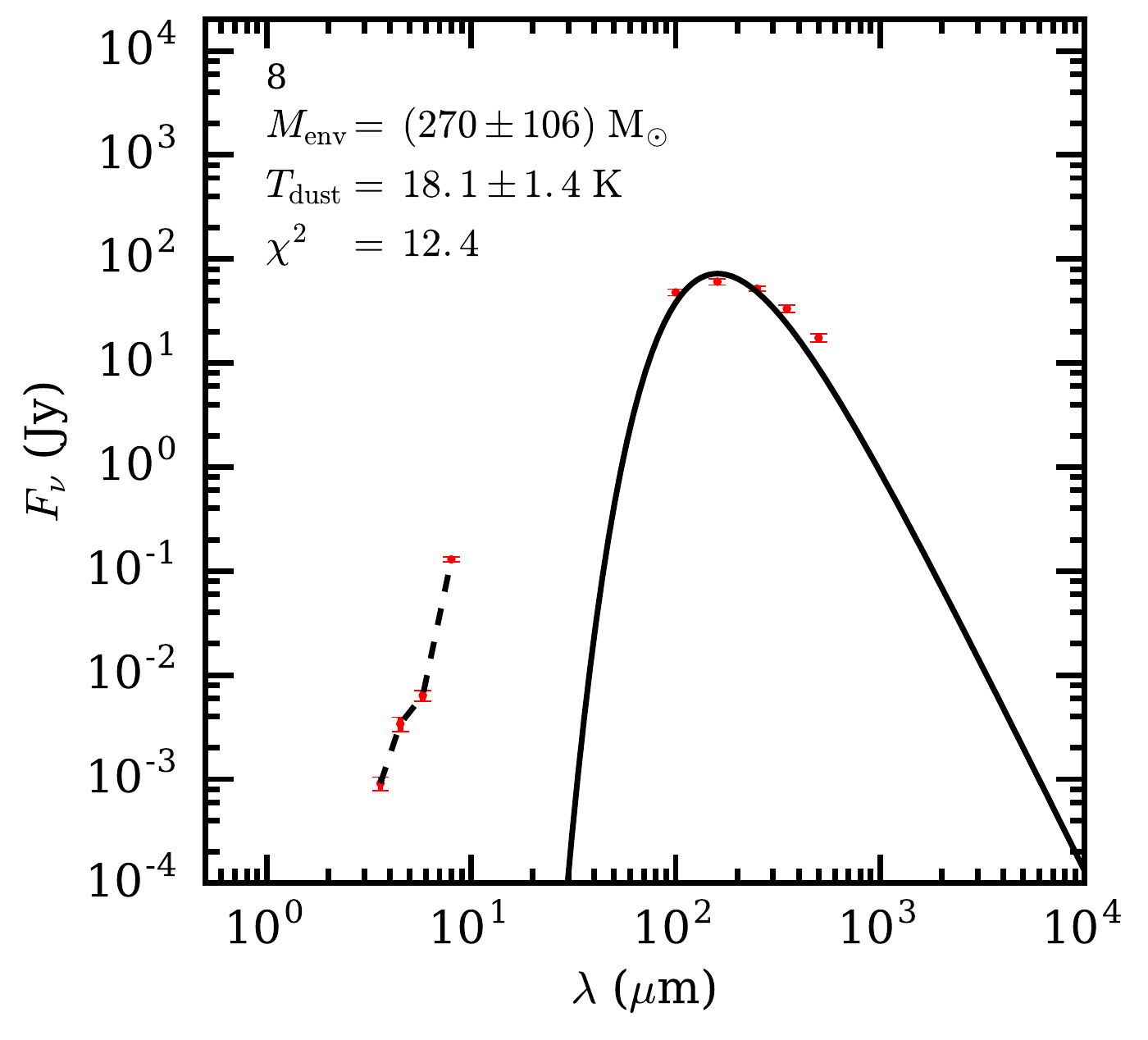}
}
\begin{flushleft}
\caption{ -- continued.}
\end{flushleft}
\end{figure*}
\begin{figure*}[!ht]
\ContinuedFloat
\centering
\subfloat{\includegraphics[width=3.4 in]{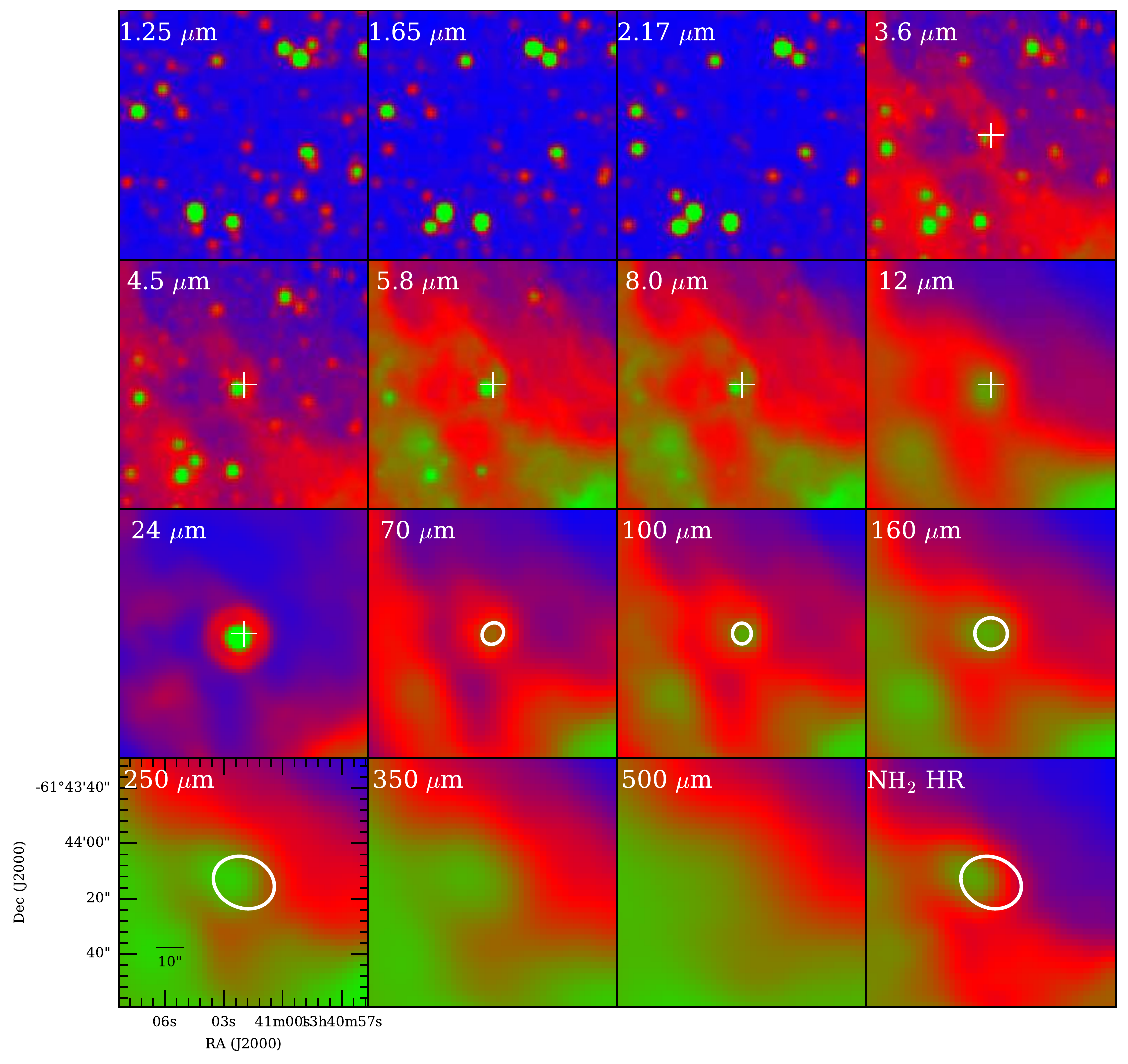}
\hskip 0.00000001cm
\includegraphics[width=3.4 in]{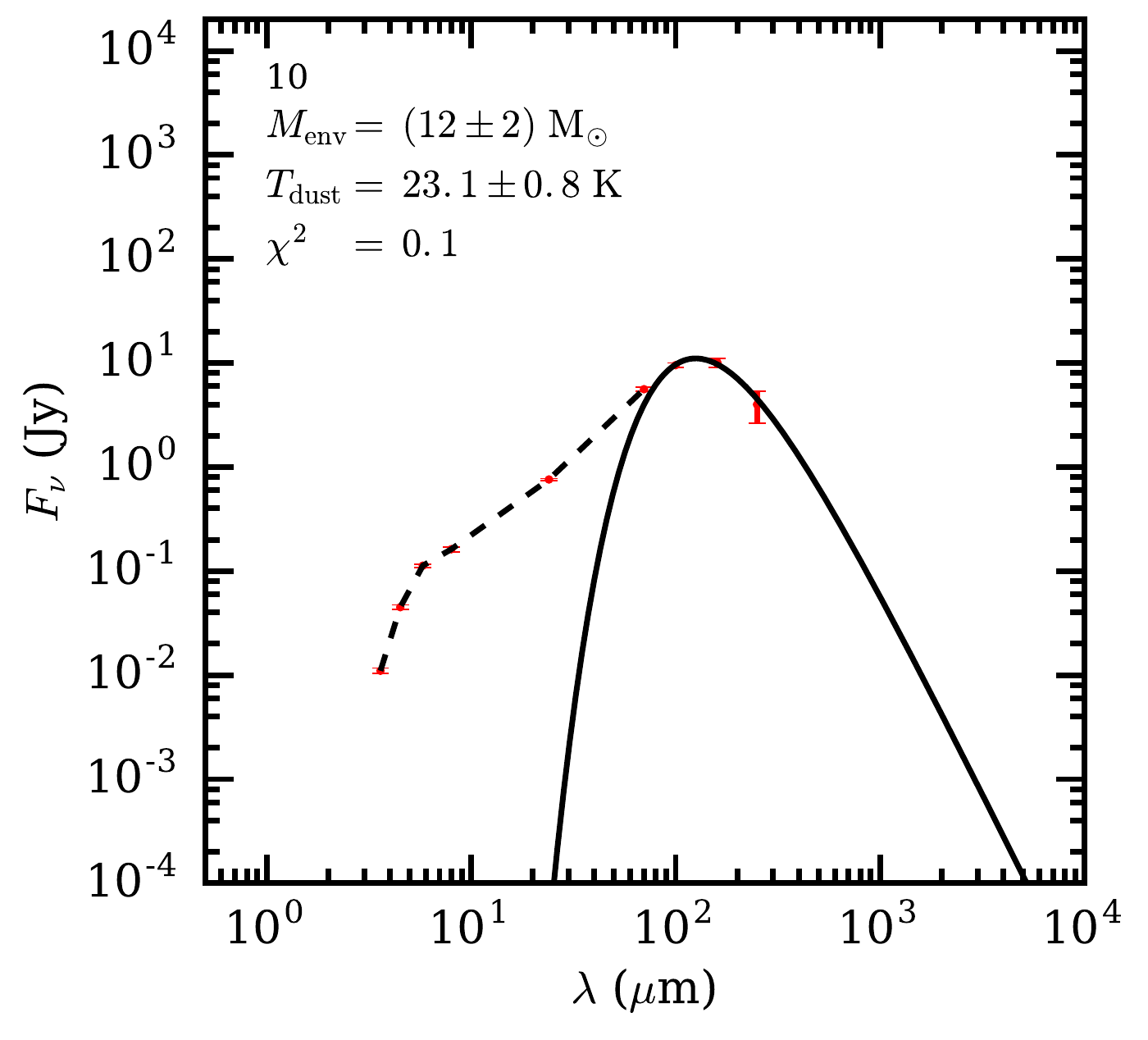}
}
\qquad
\subfloat{\includegraphics[width=3.4 in]{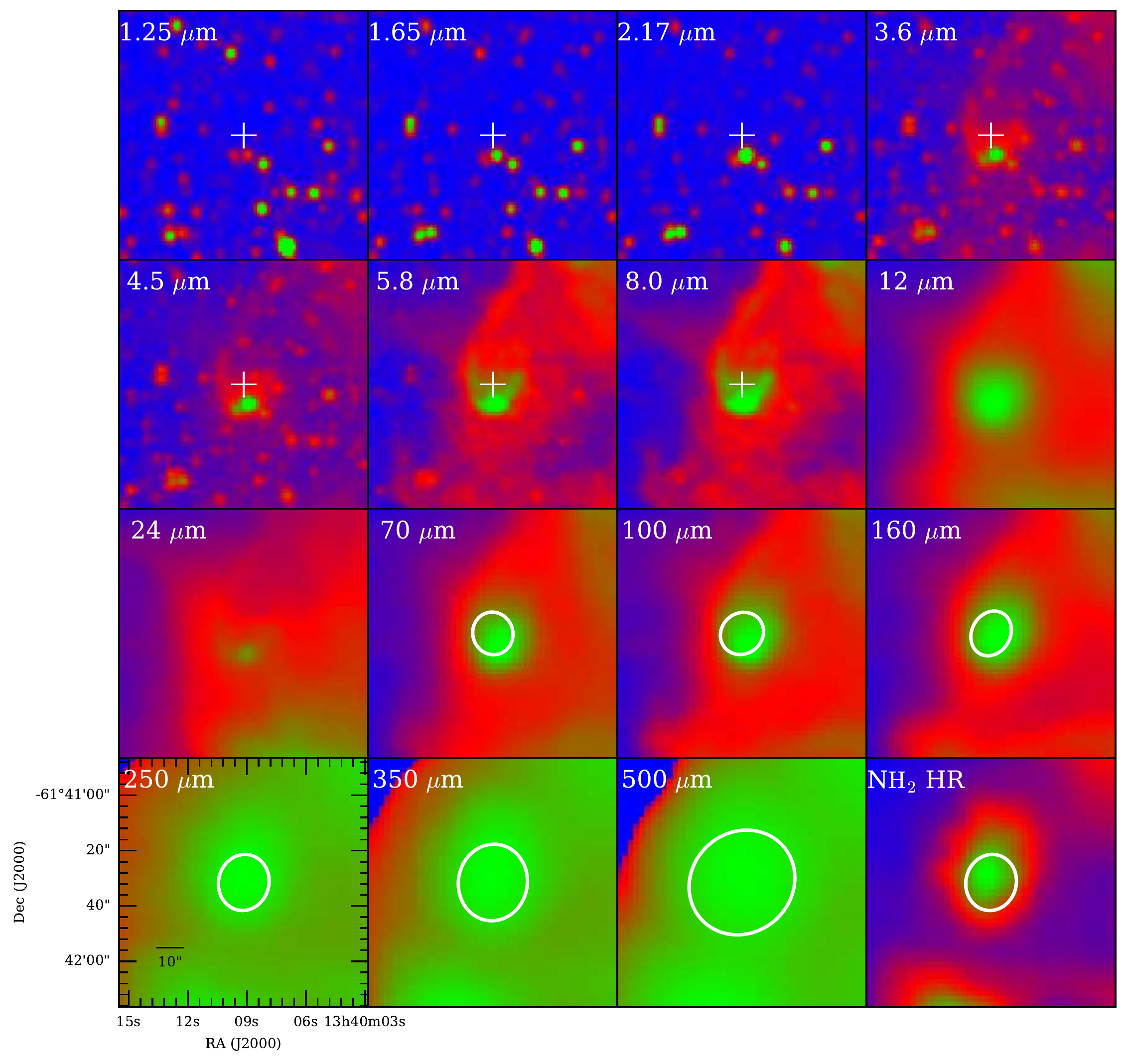}
\hskip 0.00000001cm
\includegraphics[width=3.4 in]{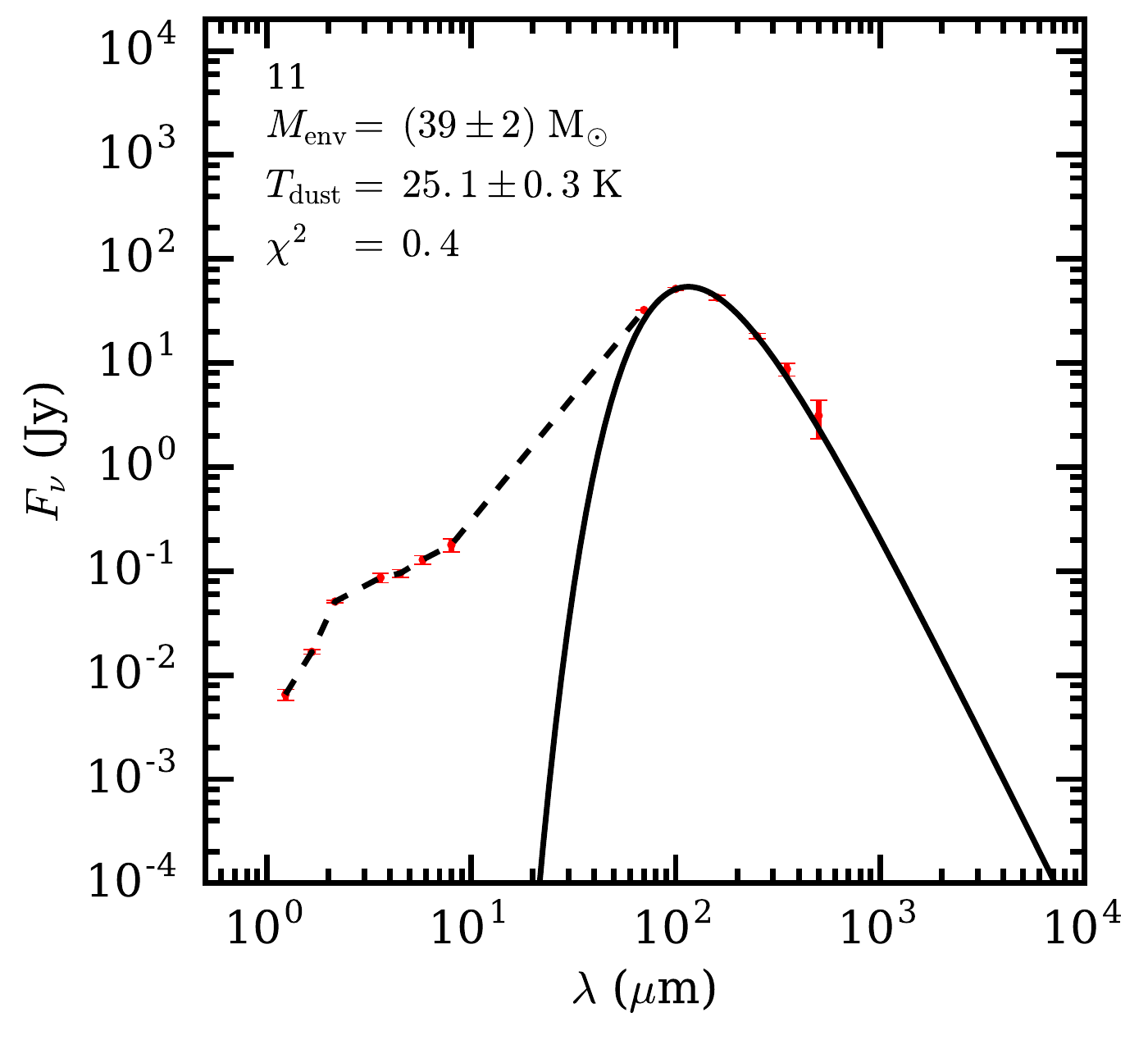}
}
\begin{flushleft}
\caption{ -- continued.}
\end{flushleft}
\end{figure*}
\begin{figure*}[!ht]
\ContinuedFloat
\centering
\subfloat{\includegraphics[width=3.4 in]{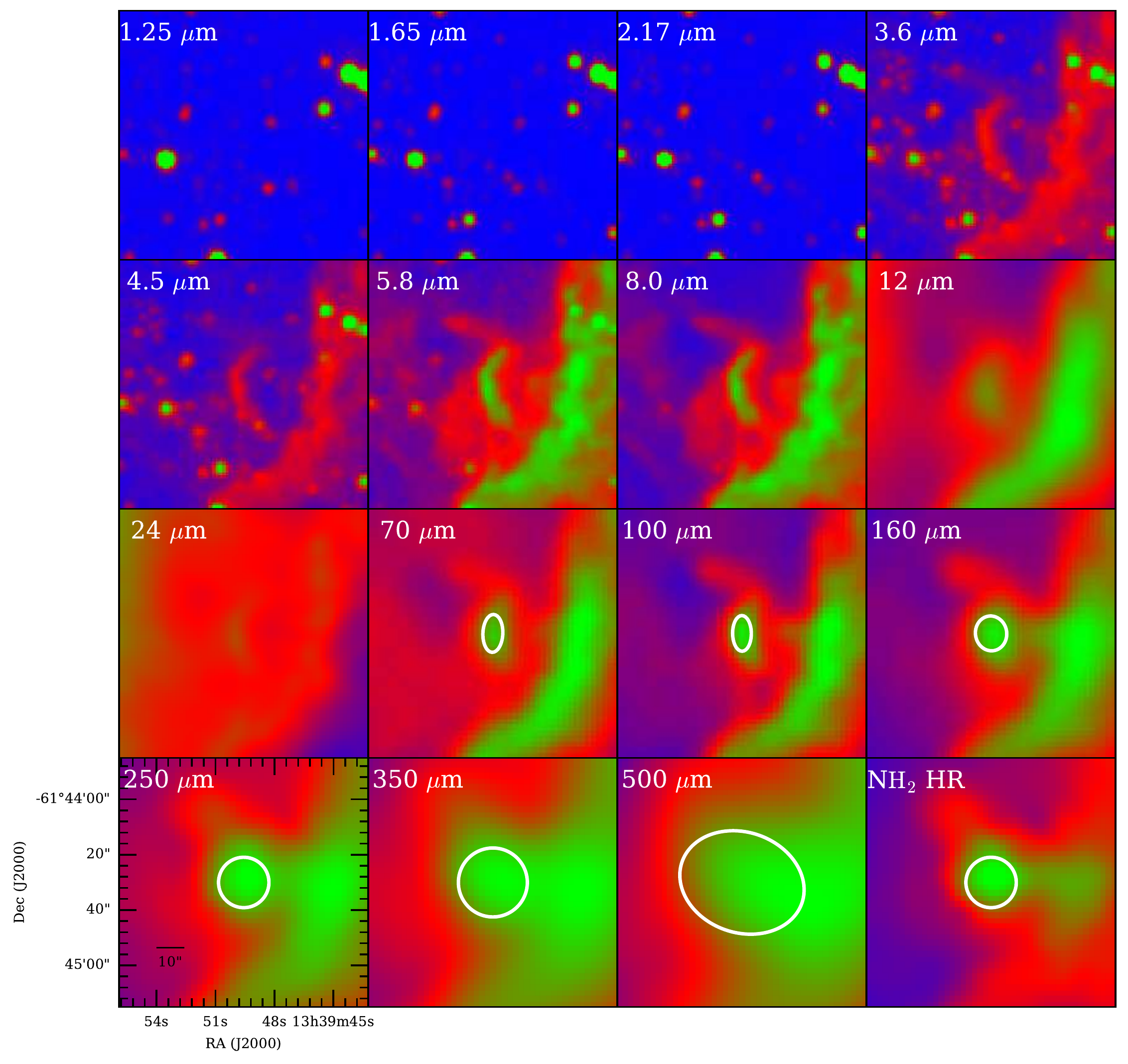}
\hskip 0.00000001cm
\includegraphics[width=3.4 in]{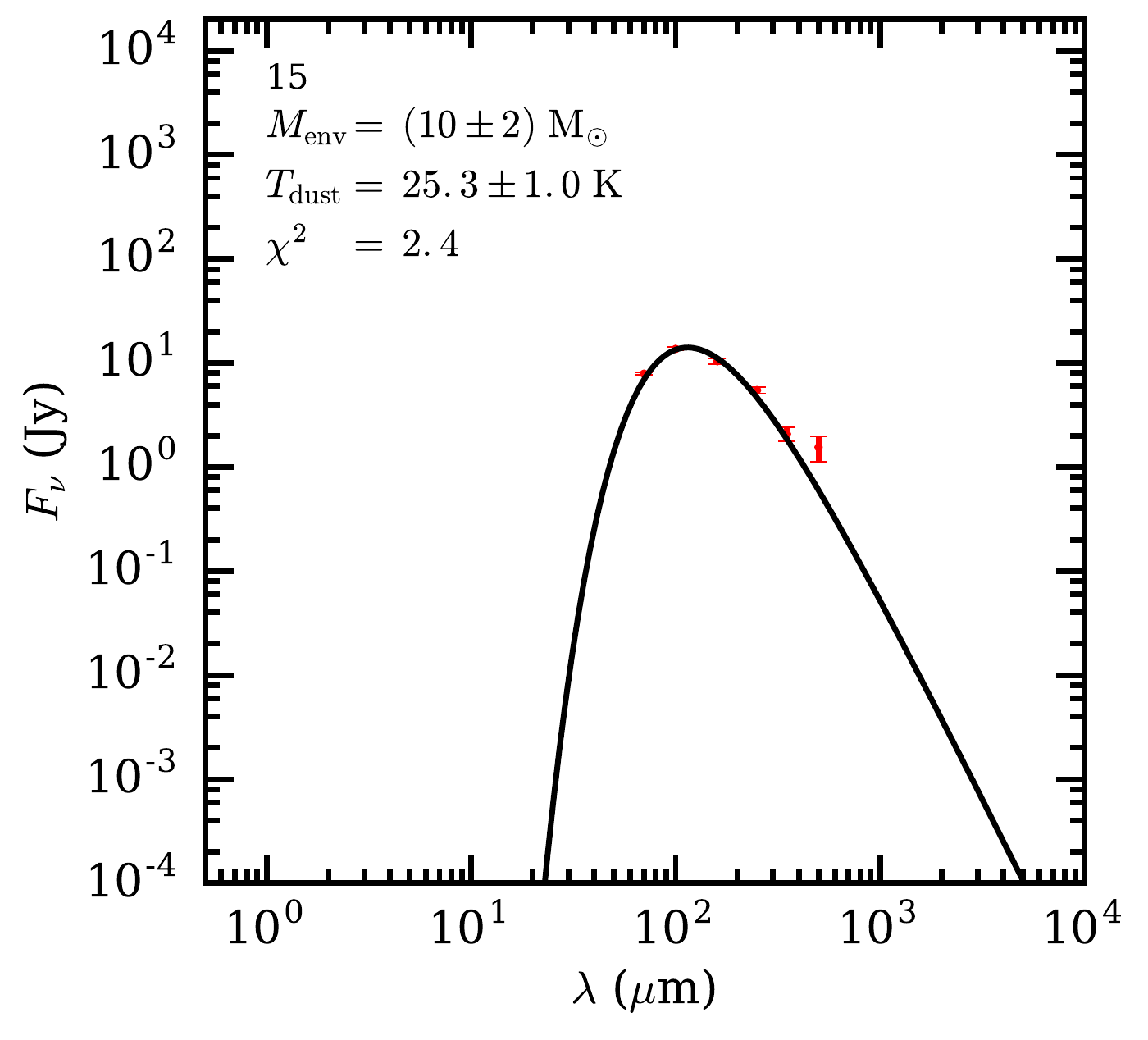}
}
\qquad
\subfloat{\includegraphics[width=3.4 in]{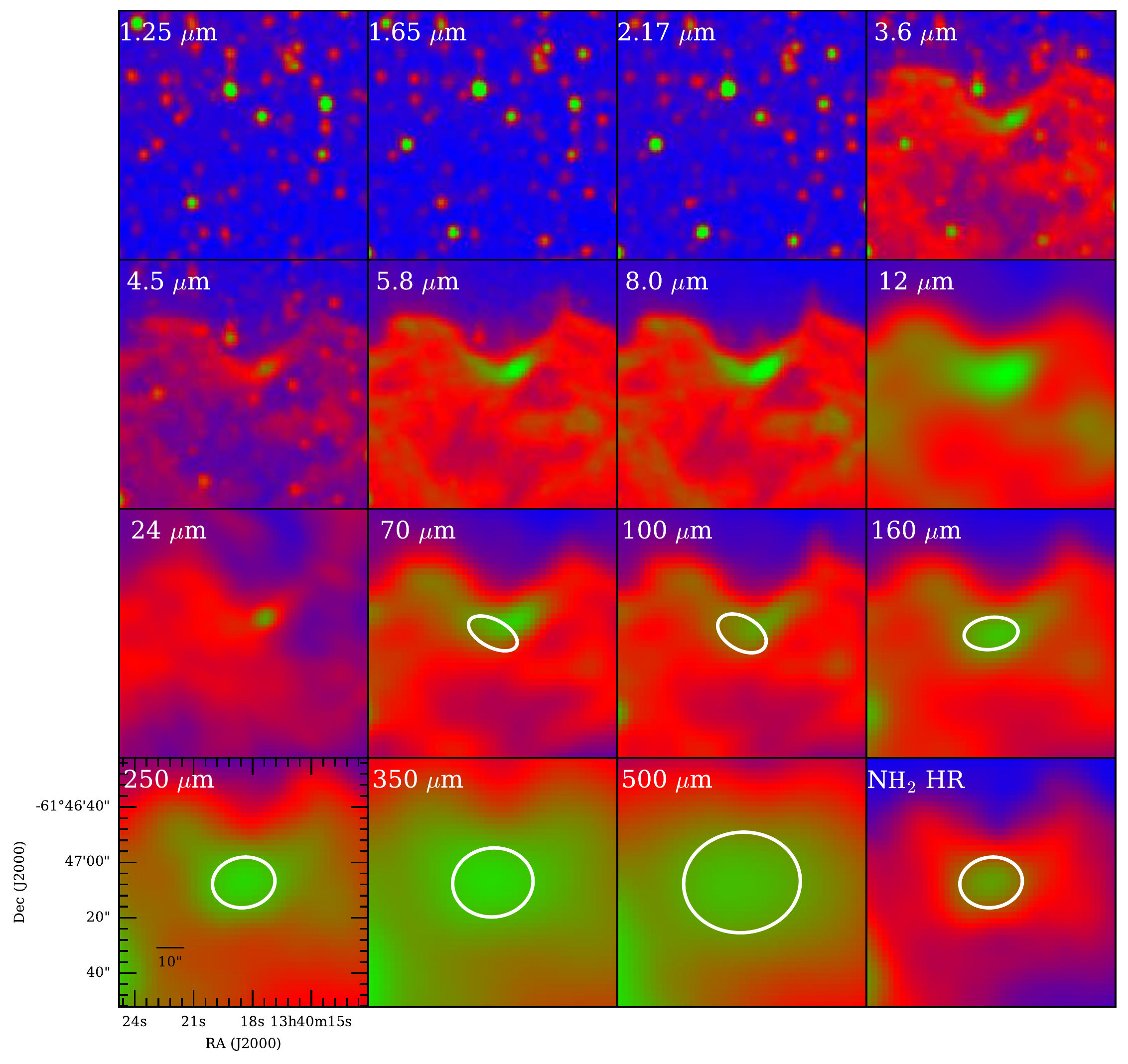}
\hskip 0.00000001cm
\includegraphics[width=3.4 in]{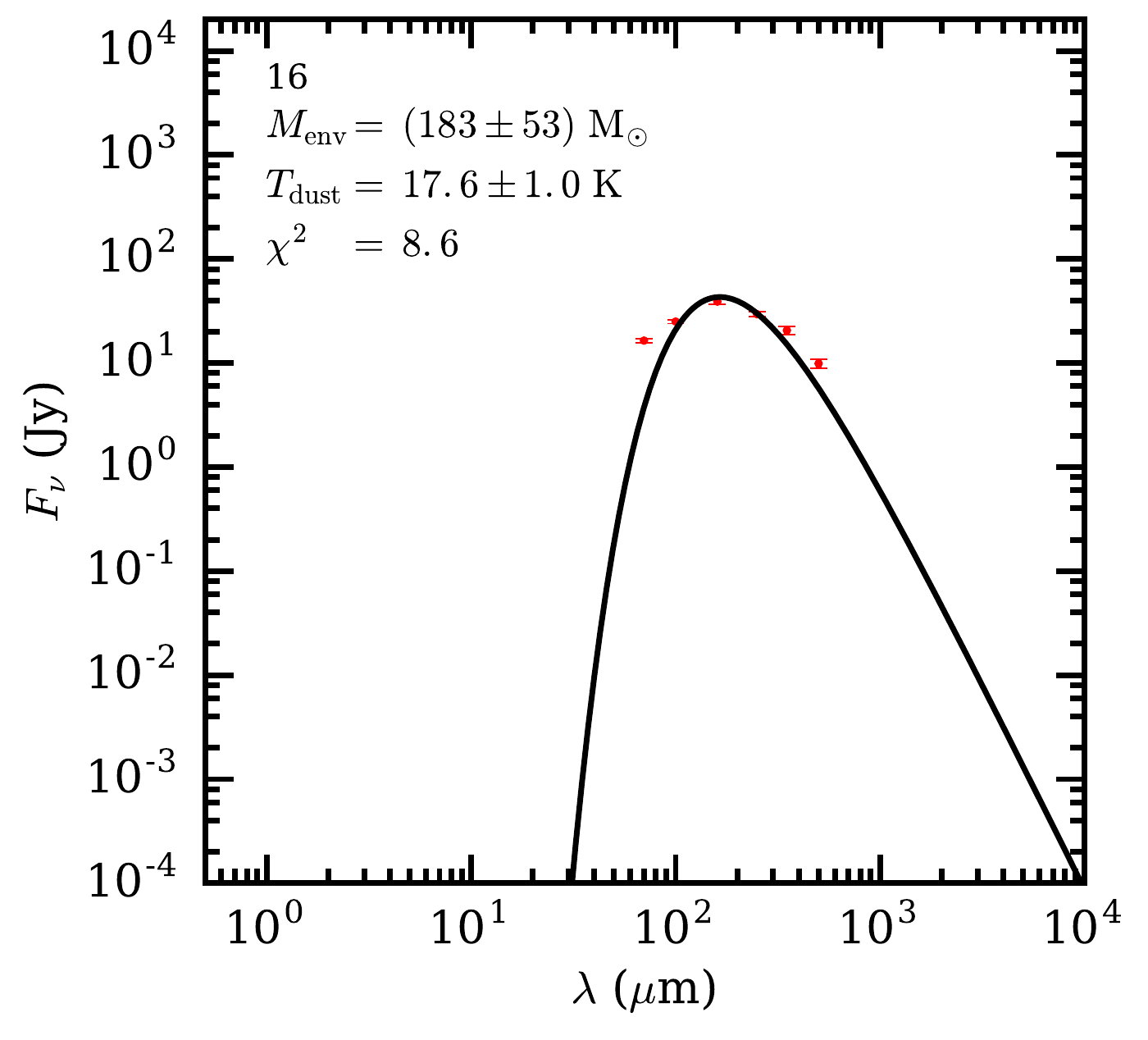}
}
\begin{flushleft}
\caption{ -- continued.}
\end{flushleft}
\end{figure*}
\begin{figure*}[!ht]
\ContinuedFloat
\centering
\subfloat{\includegraphics[width=3.4 in]{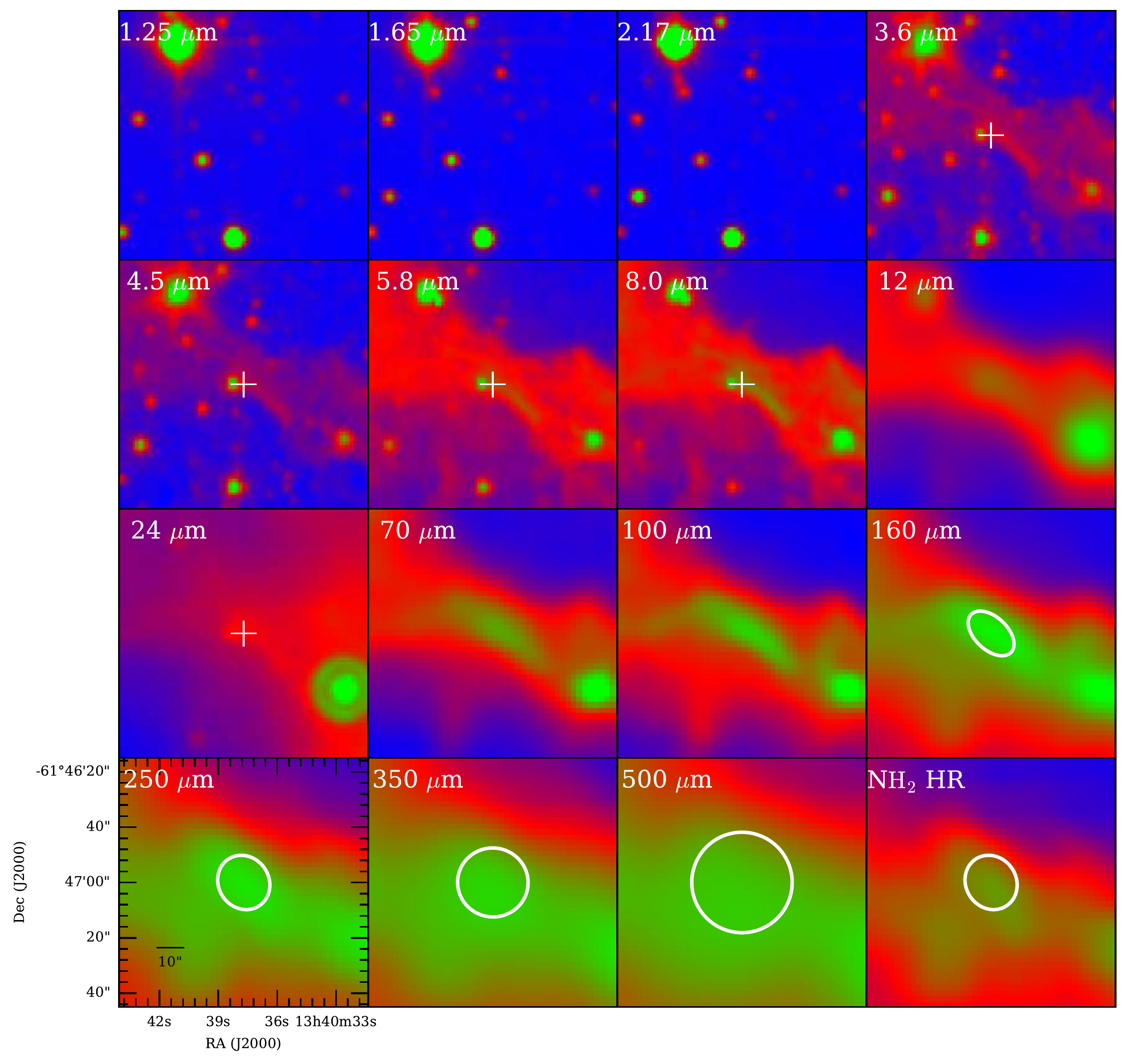}
\hskip 0.00000001cm
\includegraphics[width=3.4 in]{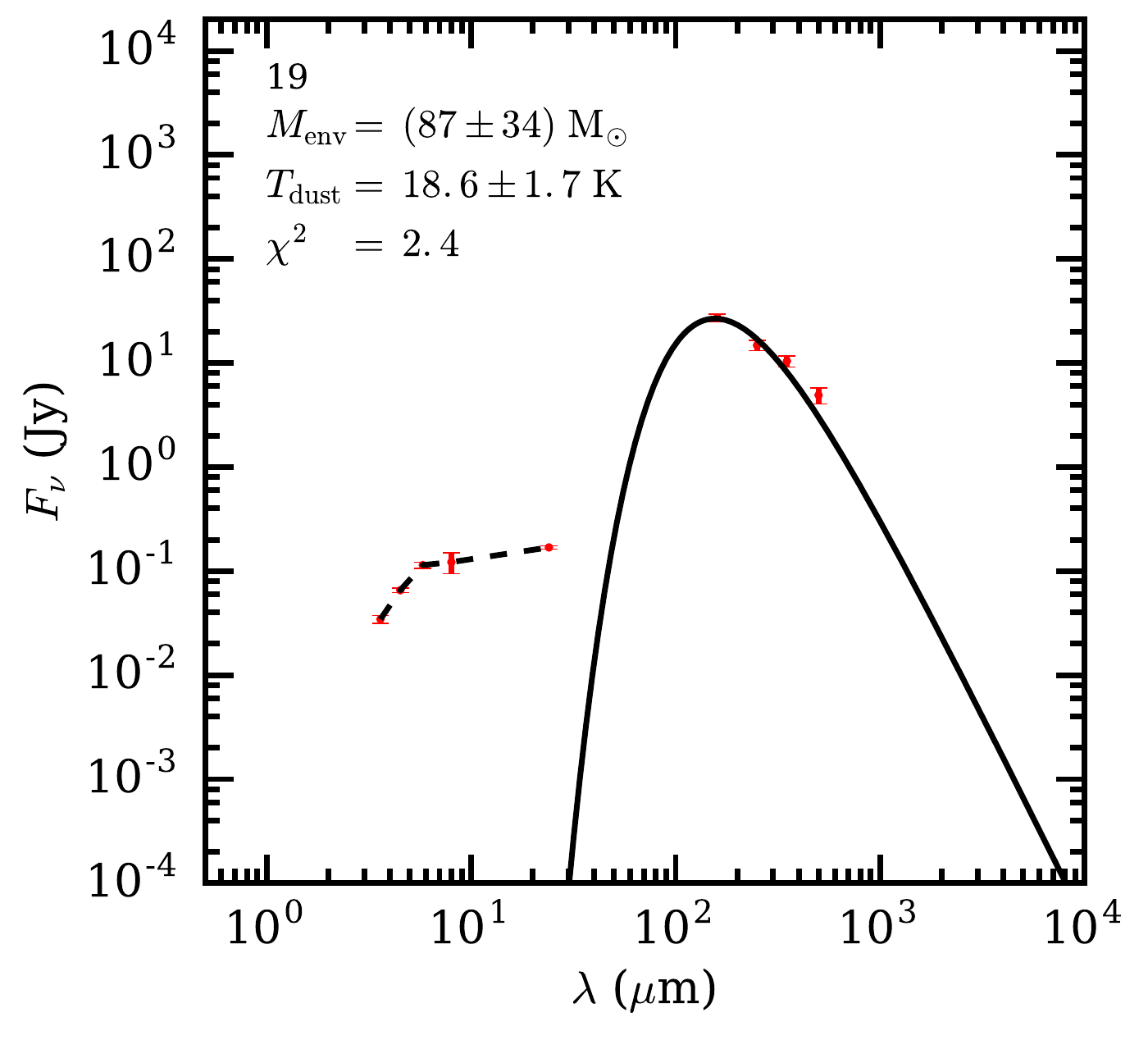}
}
\qquad
\subfloat{\includegraphics[width=3.4 in]{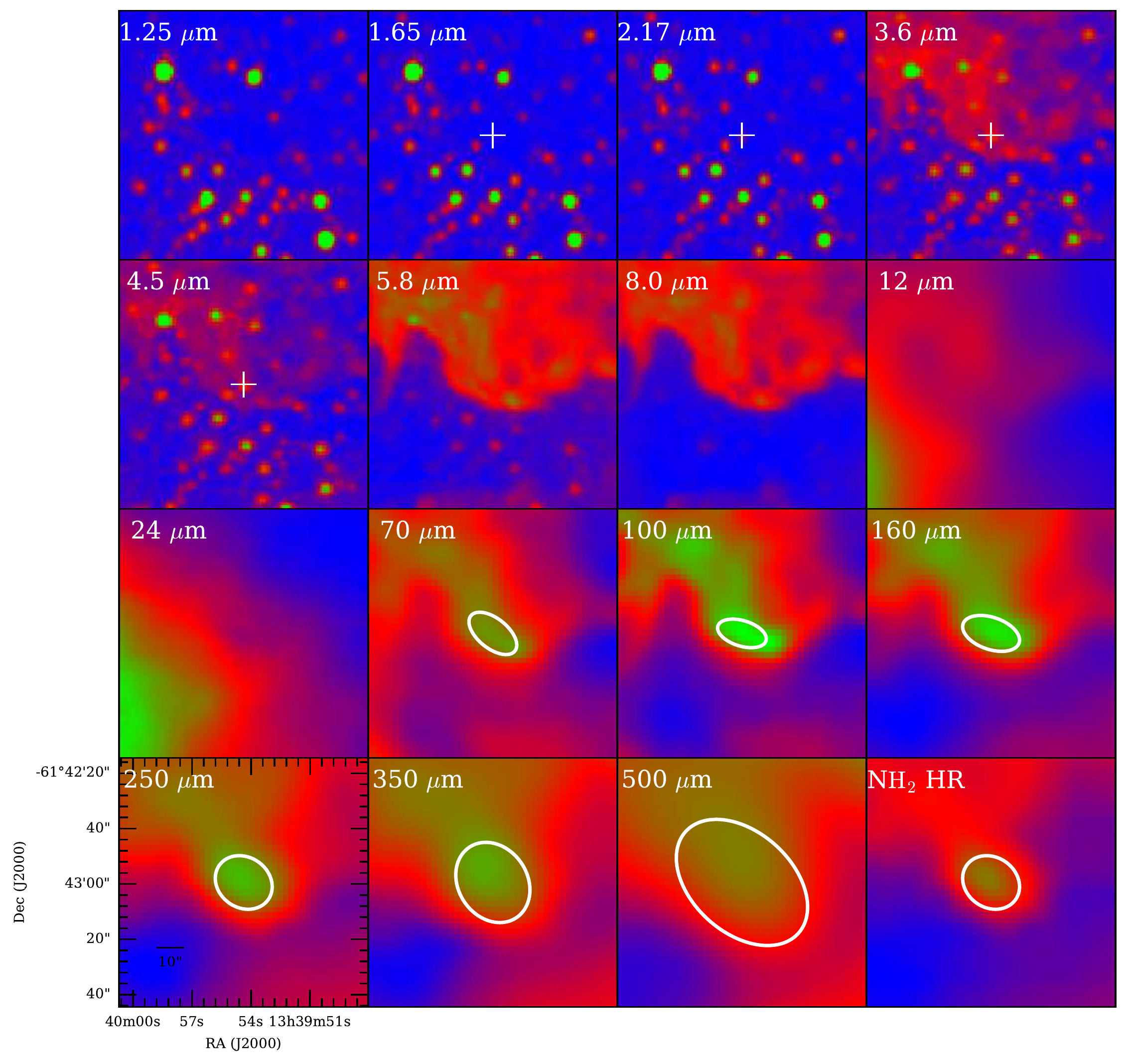}
\hskip 0.00000001cm
\includegraphics[width=3.4 in]{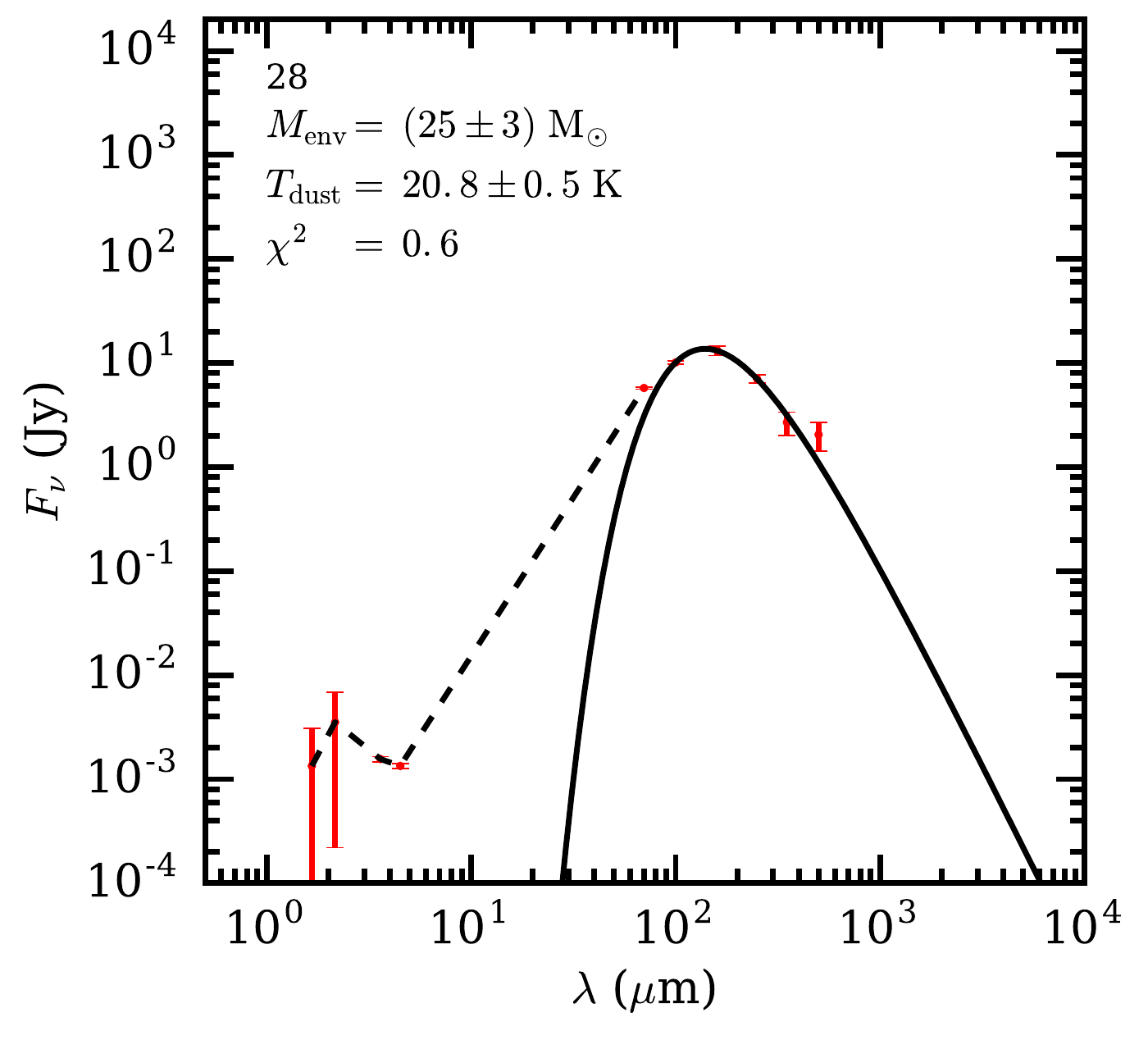}
}
\begin{flushleft}
\caption{ -- continued.}
\end{flushleft}
\end{figure*}
\begin{figure*}[!ht]
\ContinuedFloat
\centering
\subfloat{\includegraphics[width=3.4 in]{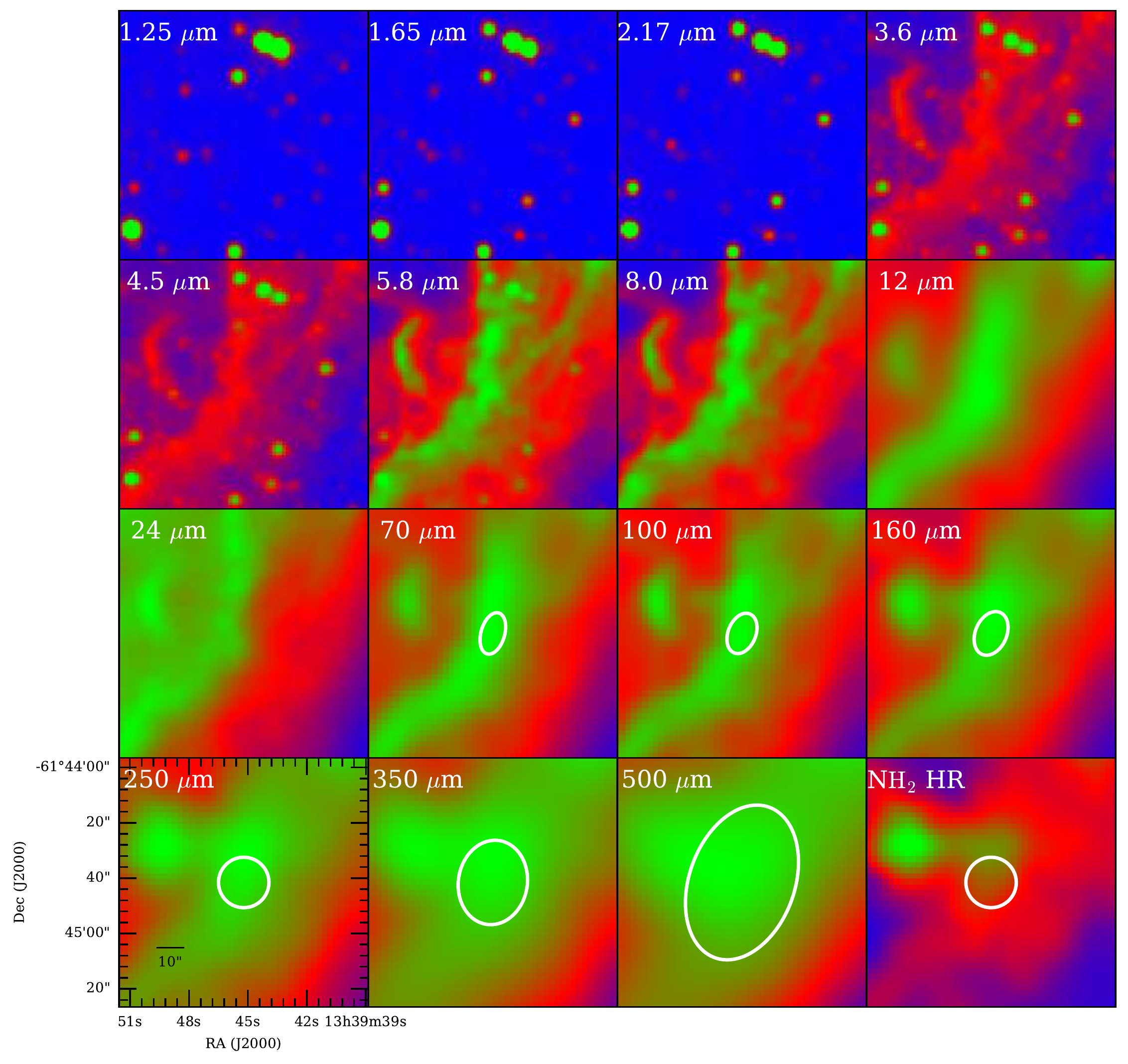}
\hskip 0.00000001cm
\includegraphics[width=3.4 in]{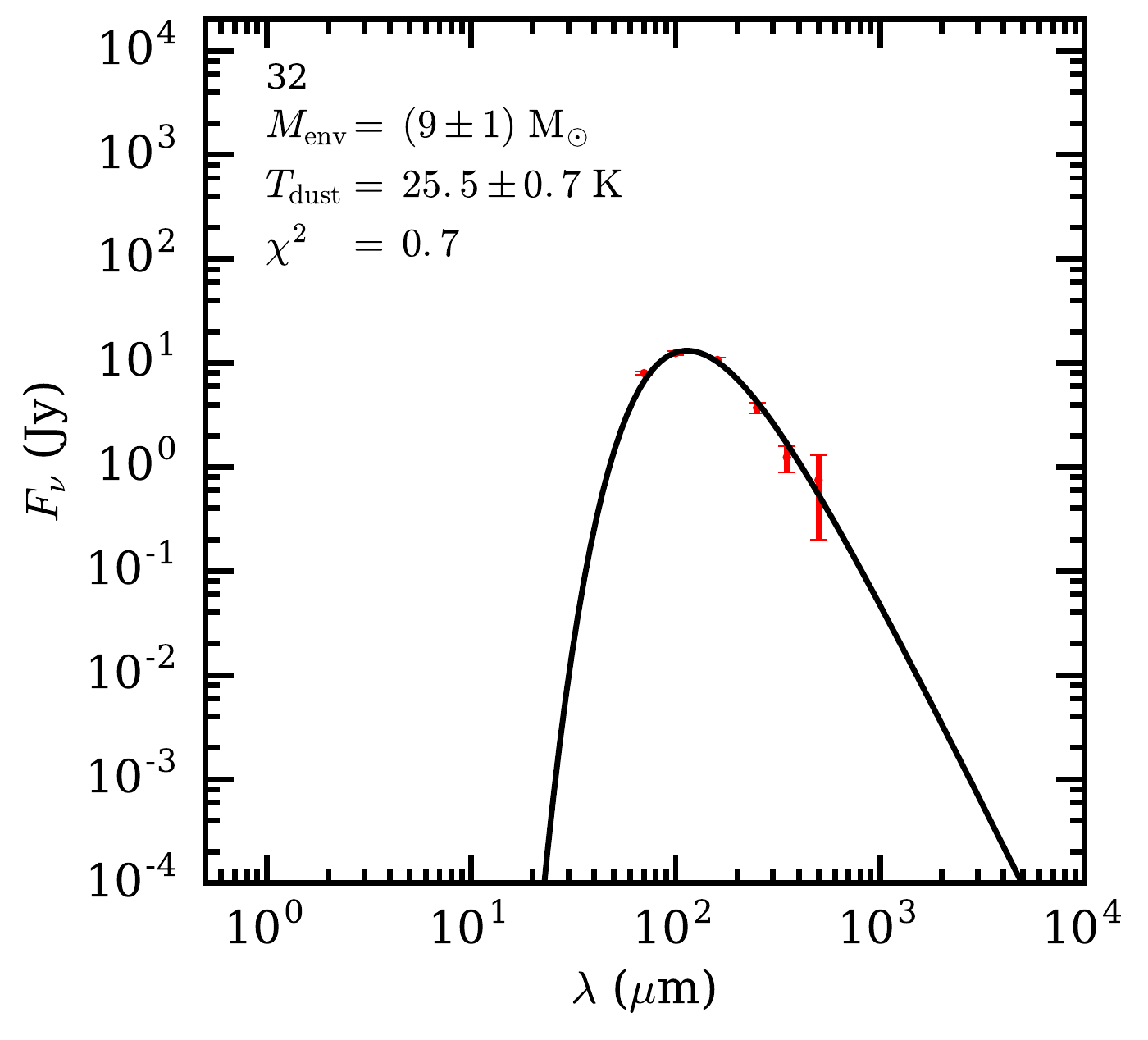}
}
\qquad
\subfloat{\includegraphics[width=3.4 in]{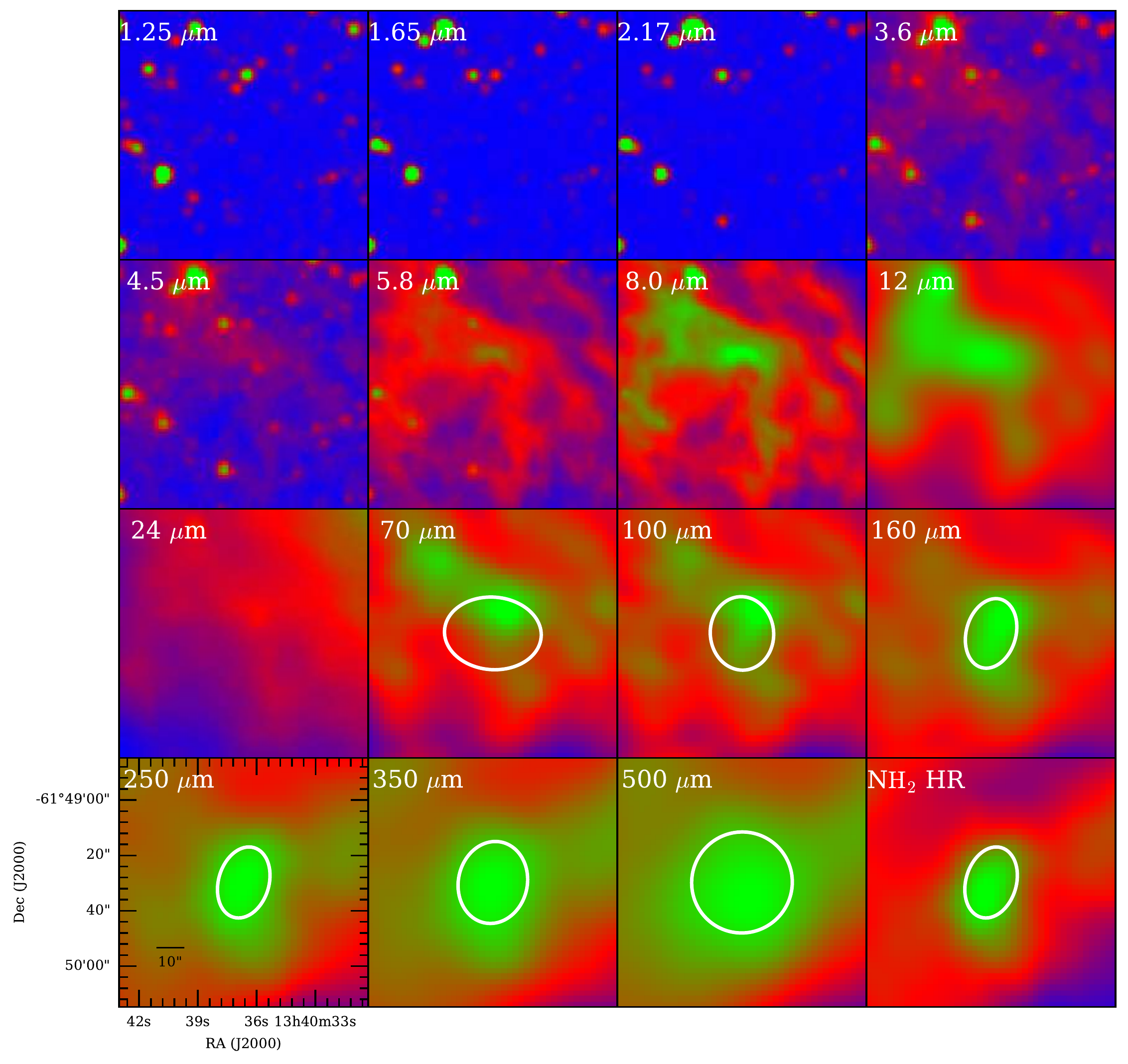}
\hskip 0.00000001cm
\includegraphics[width=3.4 in]{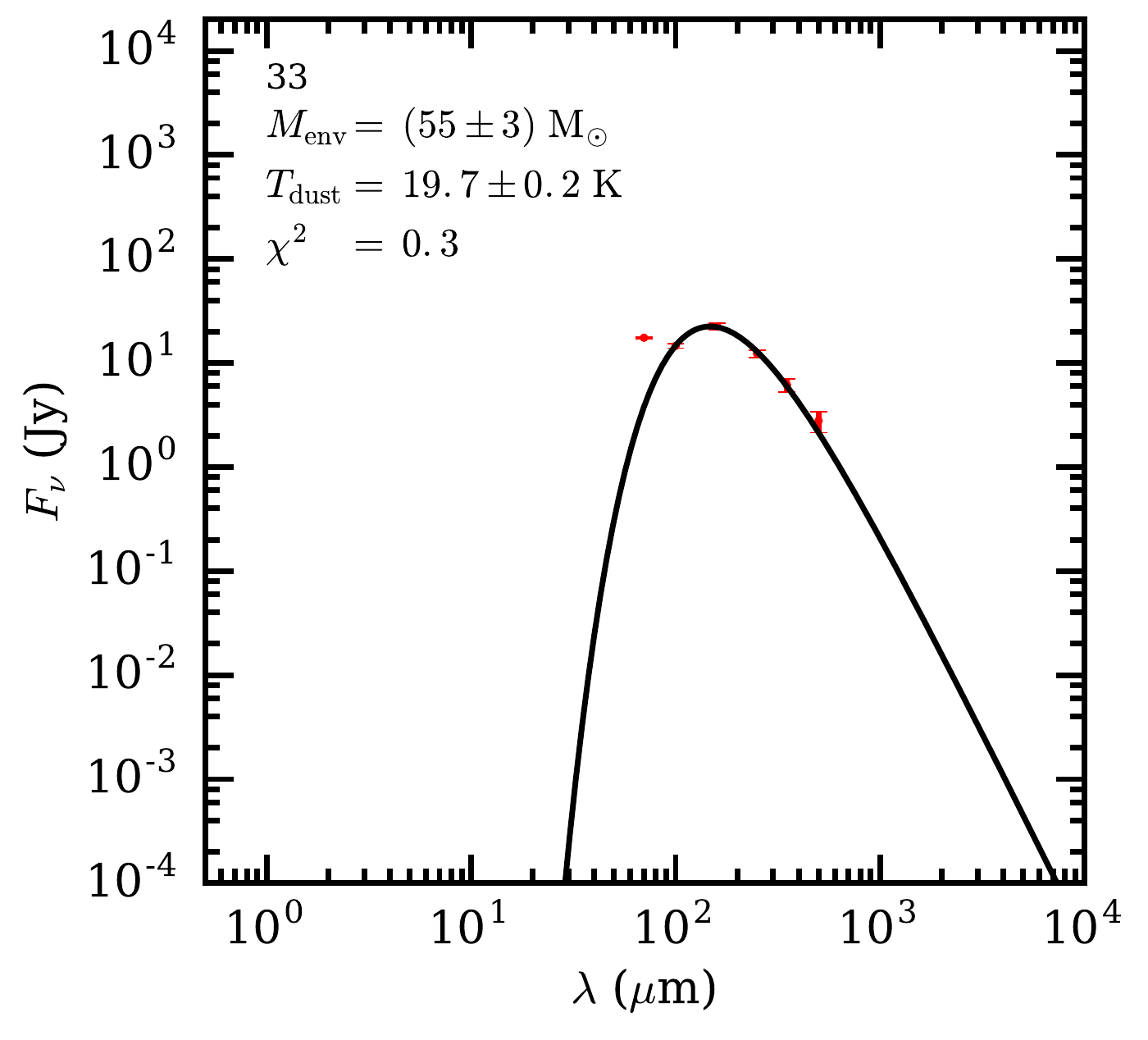}
}
\begin{flushleft}
\caption{ -- continued.}
\end{flushleft}
\end{figure*}
\begin{figure*}[!ht]
\ContinuedFloat
\centering
\subfloat{\includegraphics[width=3.4 in]{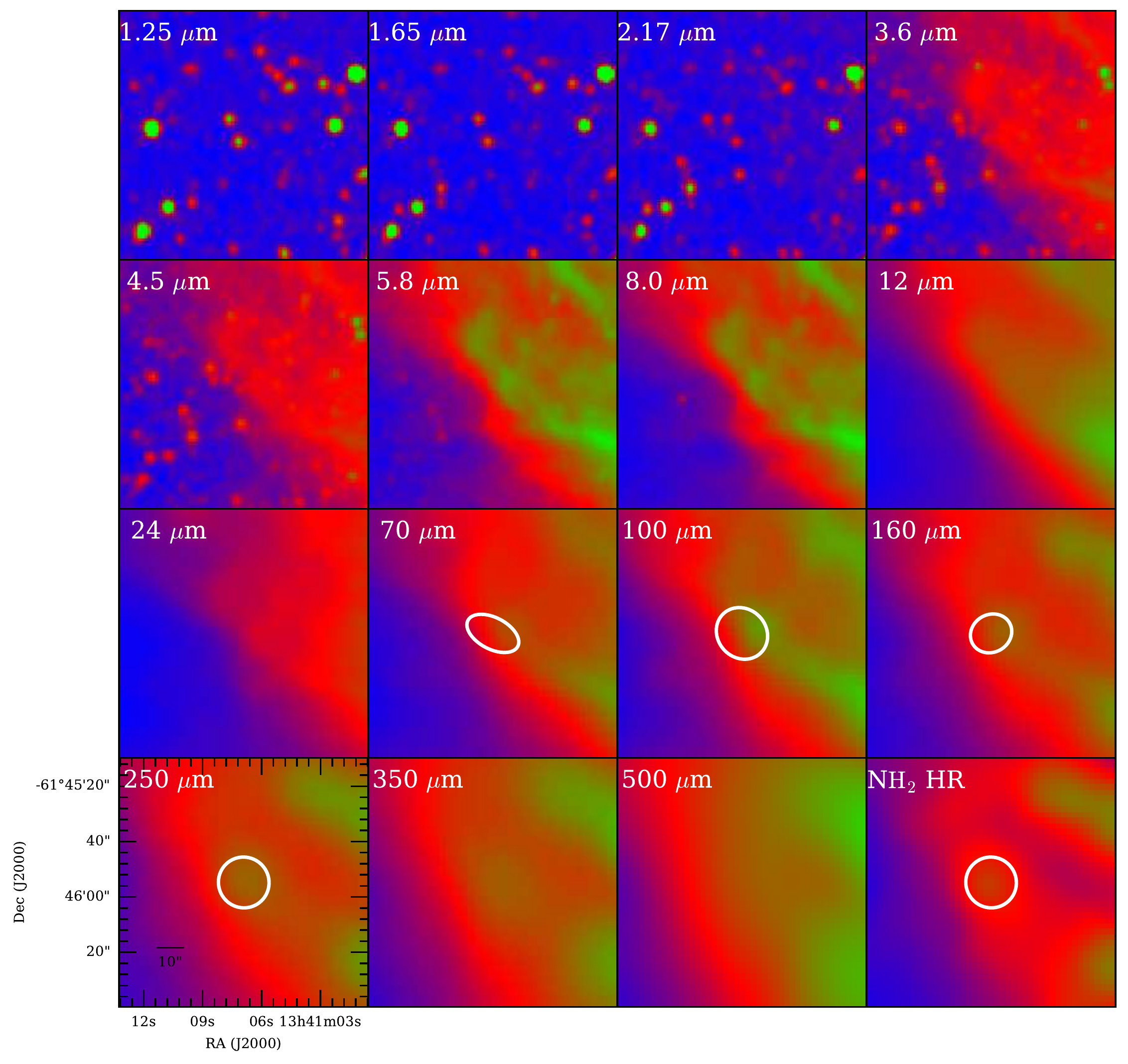}
\hskip 0.00000001cm
\includegraphics[width=3.4 in]{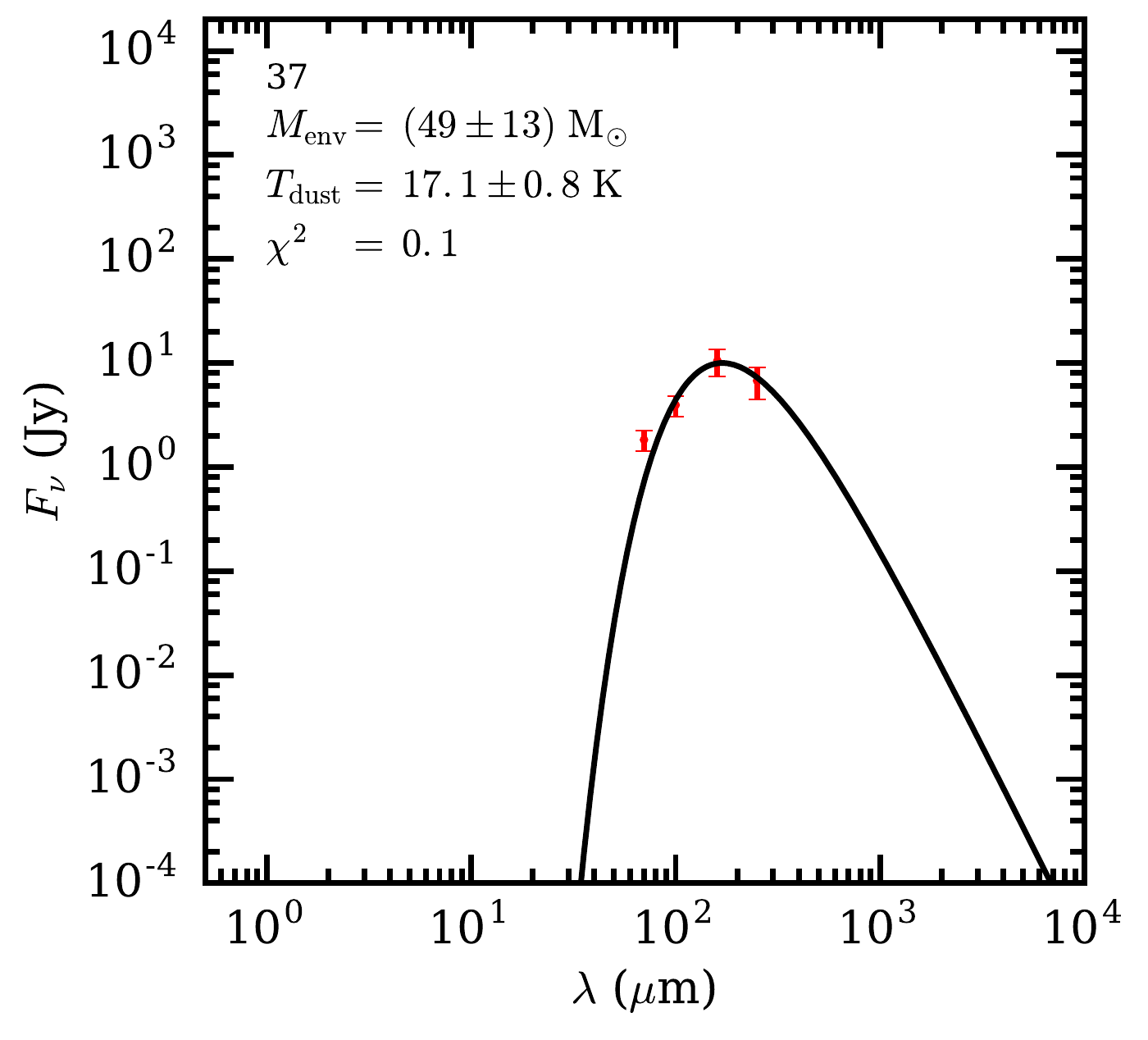}
}
\qquad
\clearpage
\subfloat{\includegraphics[width=3.4 in]{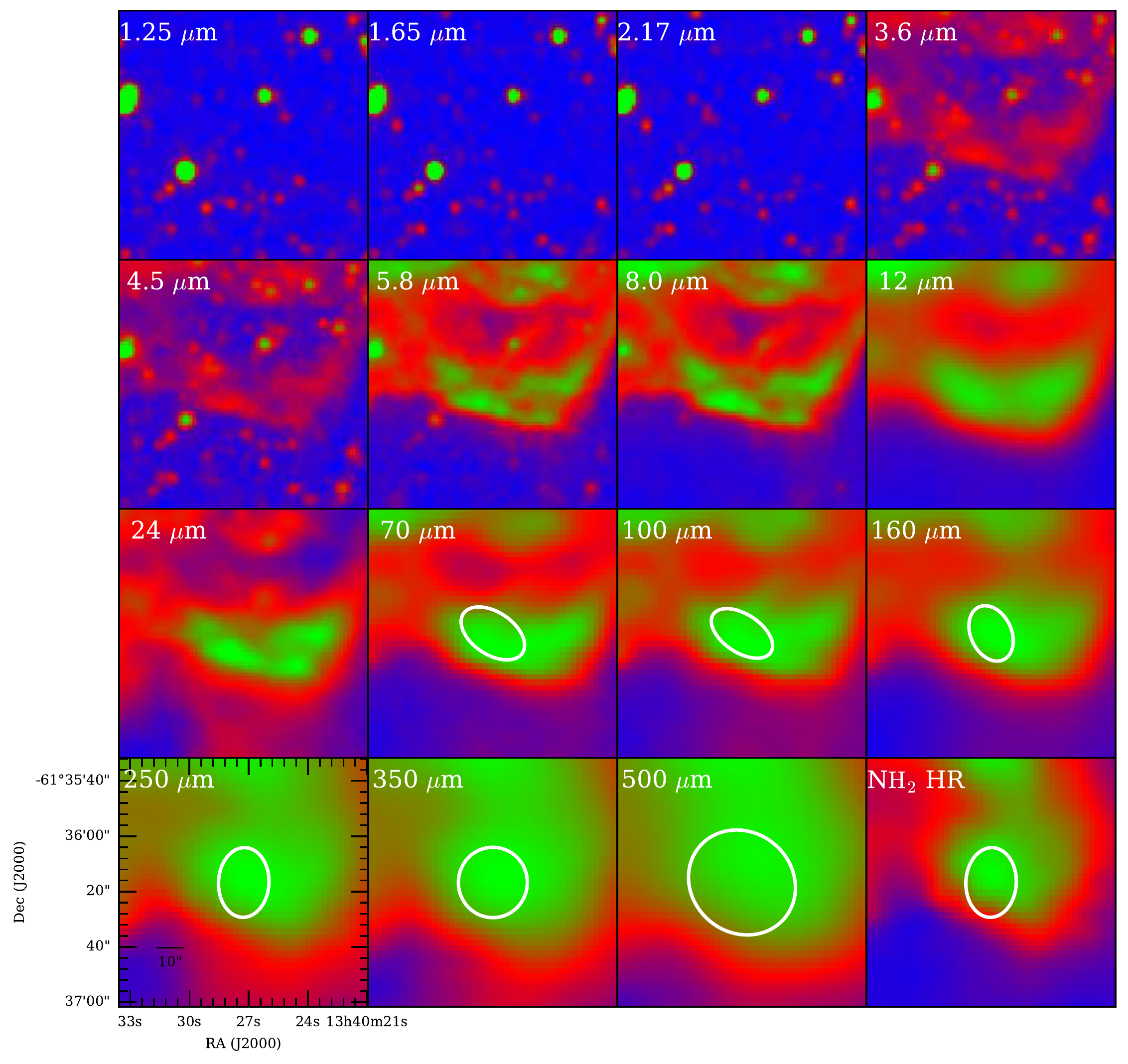}
\hskip 0.00000001cm
\includegraphics[width=3.4 in]{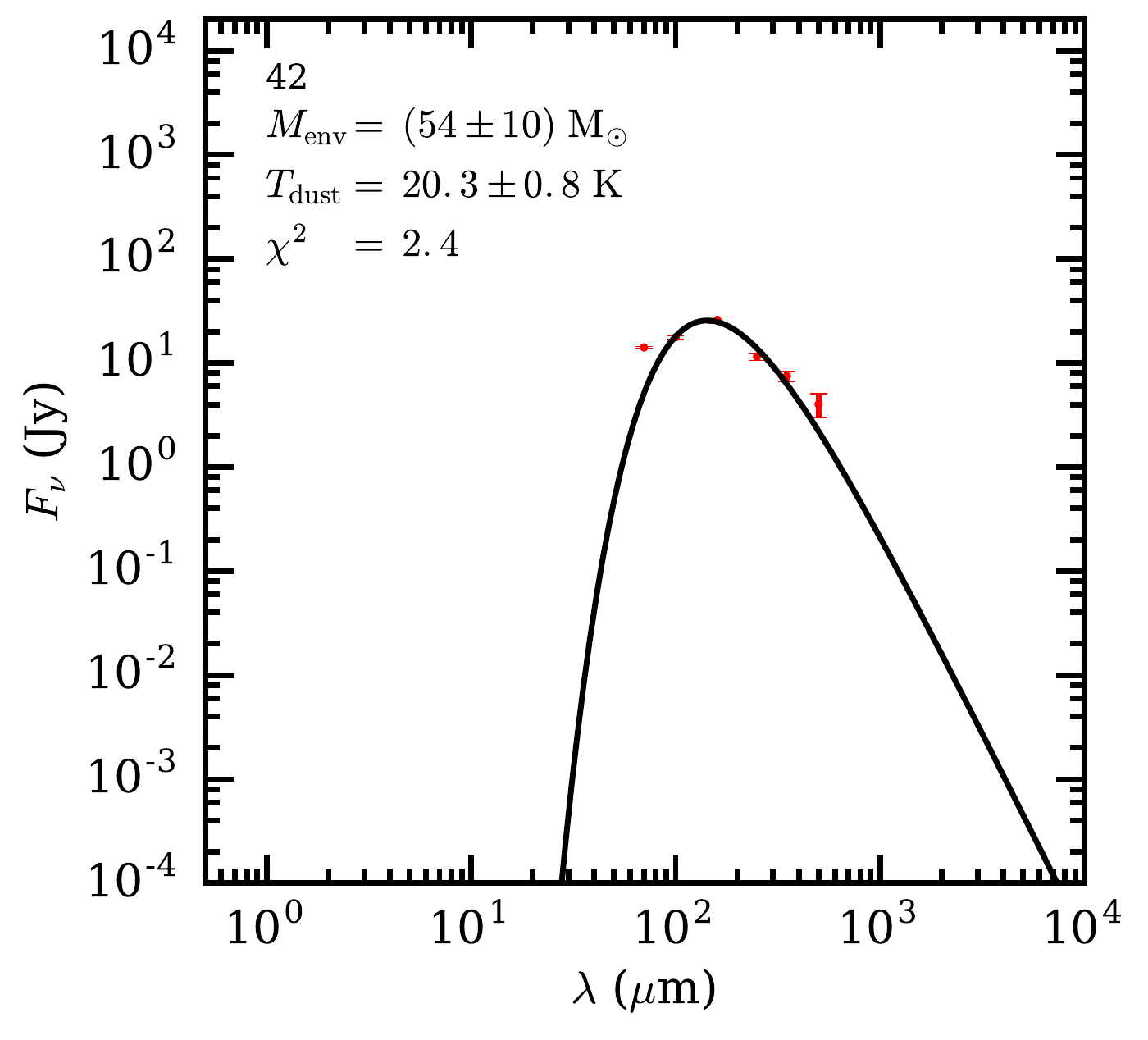}
}
\begin{flushleft}
\caption{ -- continued.}
\end{flushleft}
\end{figure*}
\begin{figure*}[!ht]
\ContinuedFloat
\centering
\subfloat{\includegraphics[width=3.4 in]{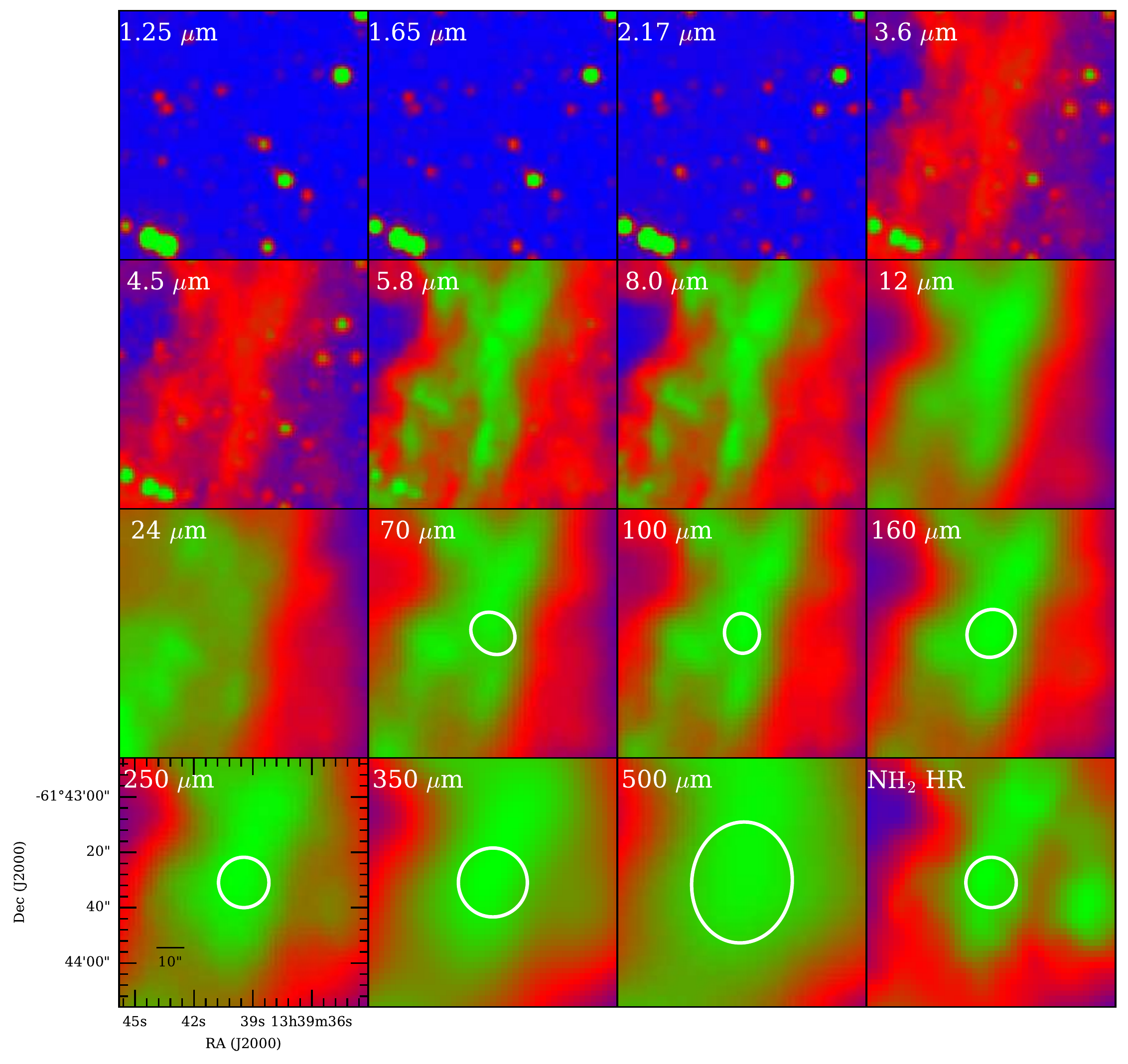}
\hskip 0.00000001cm
\includegraphics[width=3.4 in]{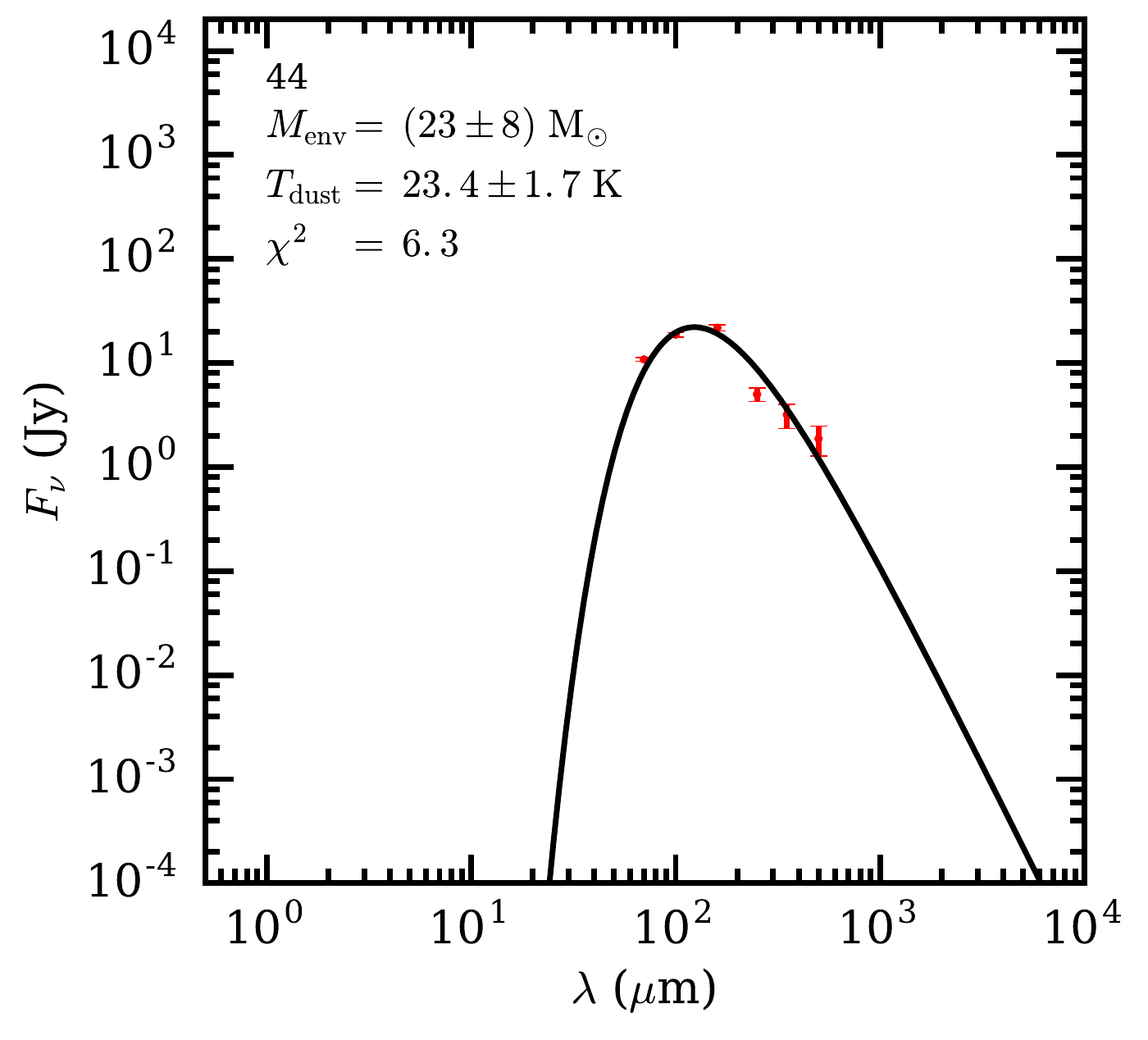}
}
\qquad
\subfloat{\includegraphics[width=3.4 in]{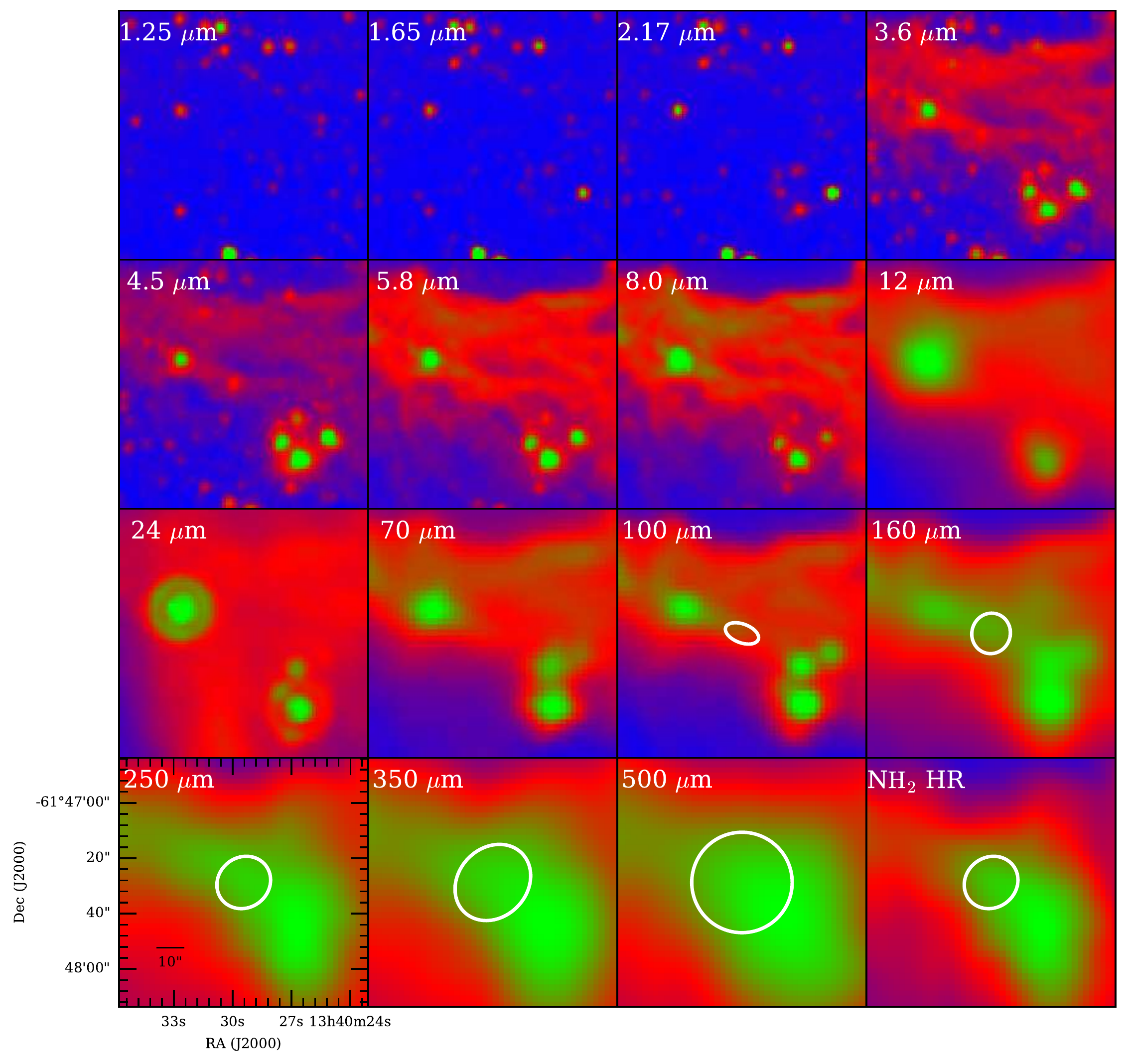}
\hskip 0.00000001cm
\includegraphics[width=3.4 in]{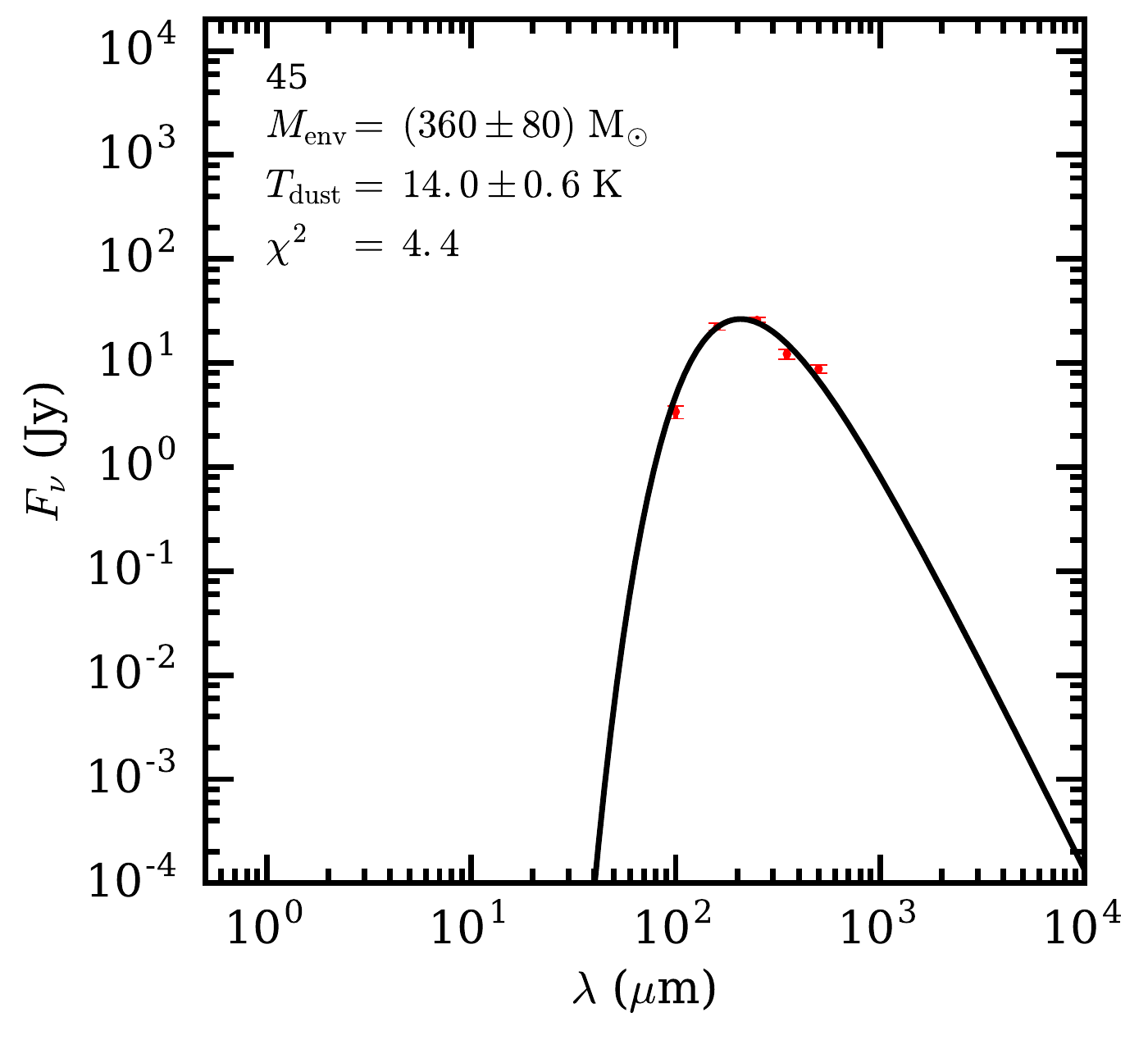}
}
\begin{flushleft}
\caption{ -- continued.}
\end{flushleft}
\end{figure*}
\begin{figure*}[!ht]
\ContinuedFloat
\centering
\subfloat{\includegraphics[width=3.4 in]{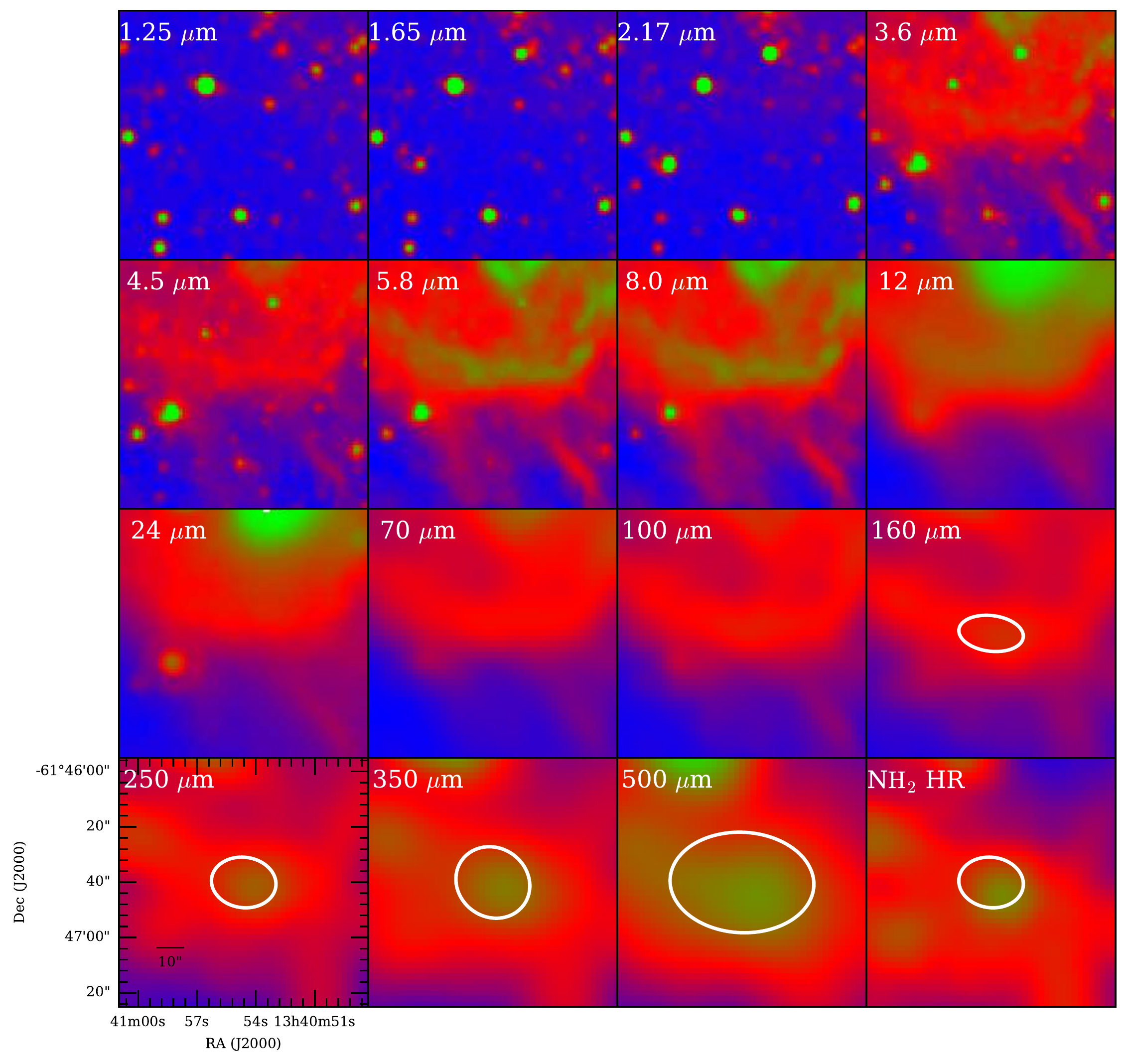}
\hskip 0.00000001cm
\includegraphics[width=3.4 in]{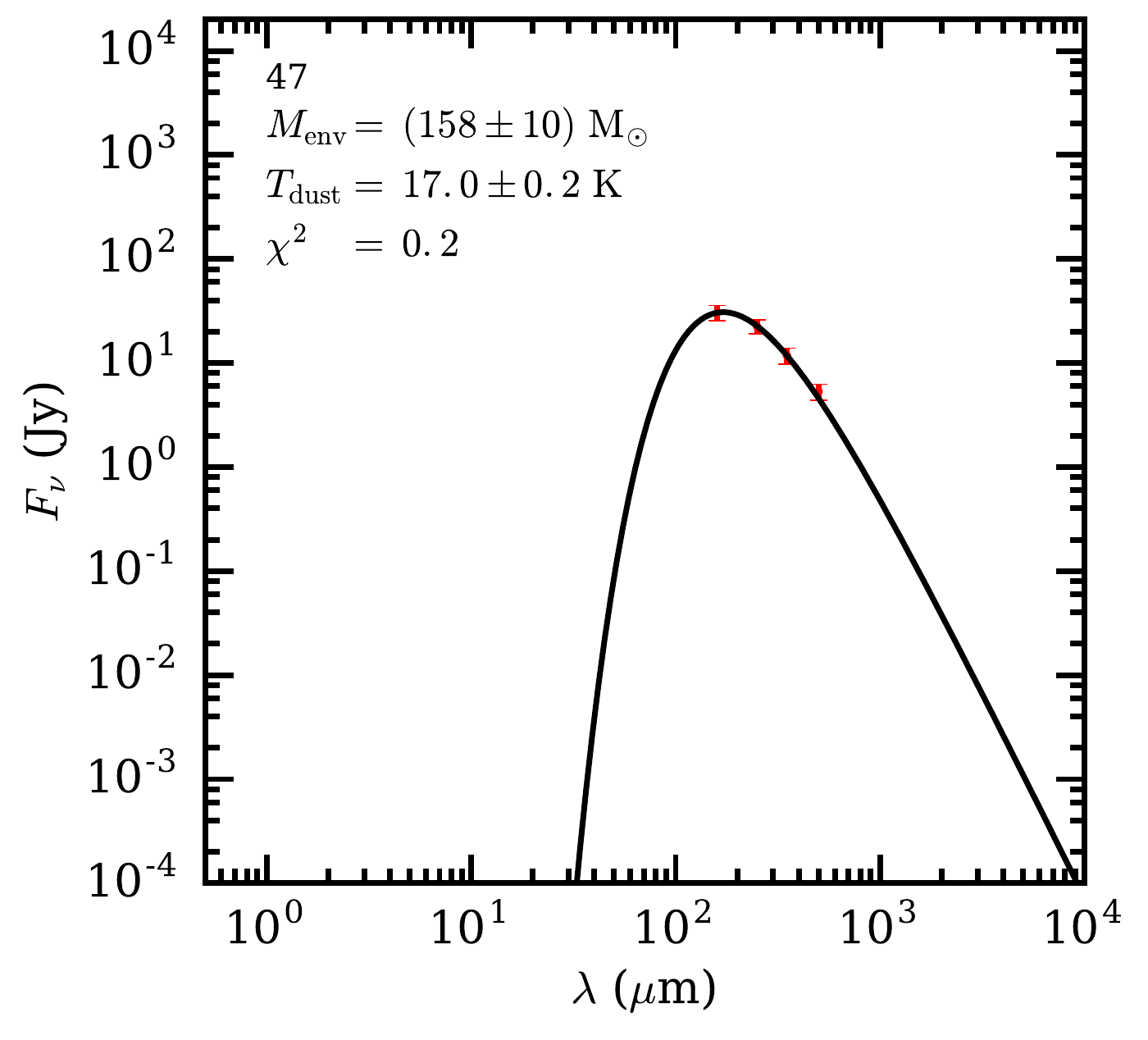}
}
\qquad
\subfloat{\includegraphics[width=3.4 in]{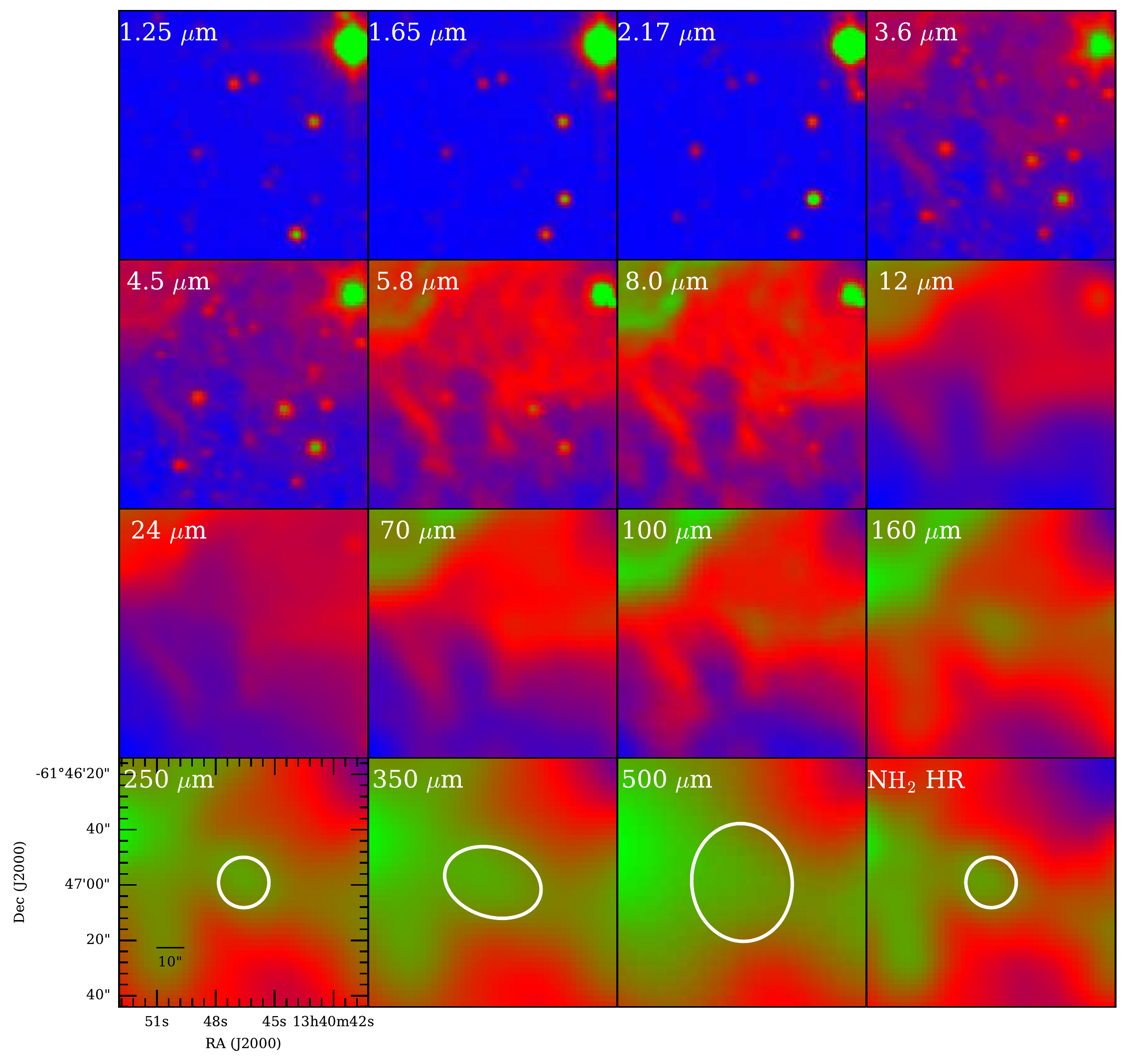}
\hskip 0.00000001cm
\includegraphics[width=3.4 in]{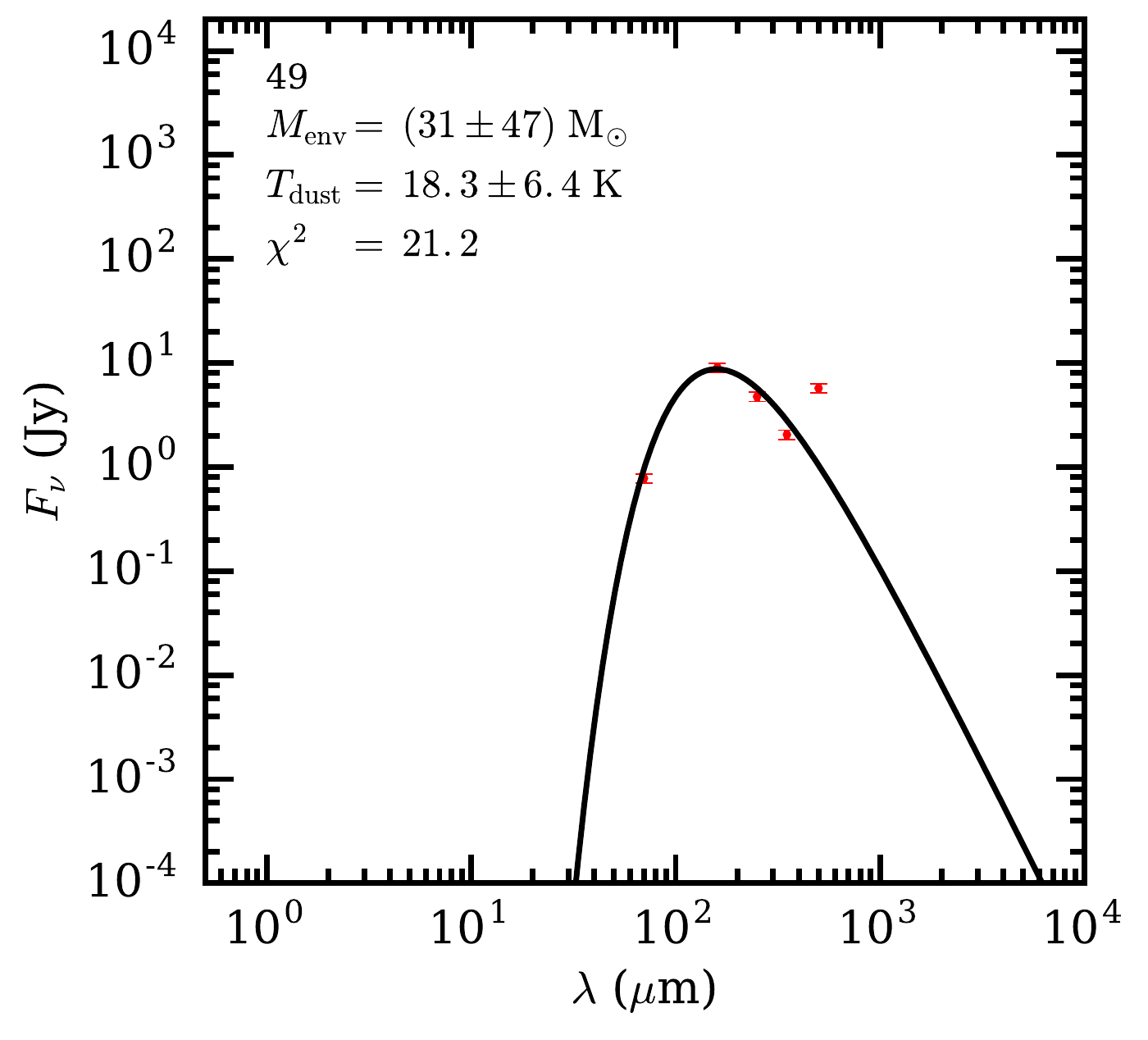}
}
\begin{flushleft}
\caption{ -- continued.}
\end{flushleft}
\end{figure*}
\begin{figure*}[!ht]
\ContinuedFloat
\centering
\subfloat{\includegraphics[width=3.4 in]{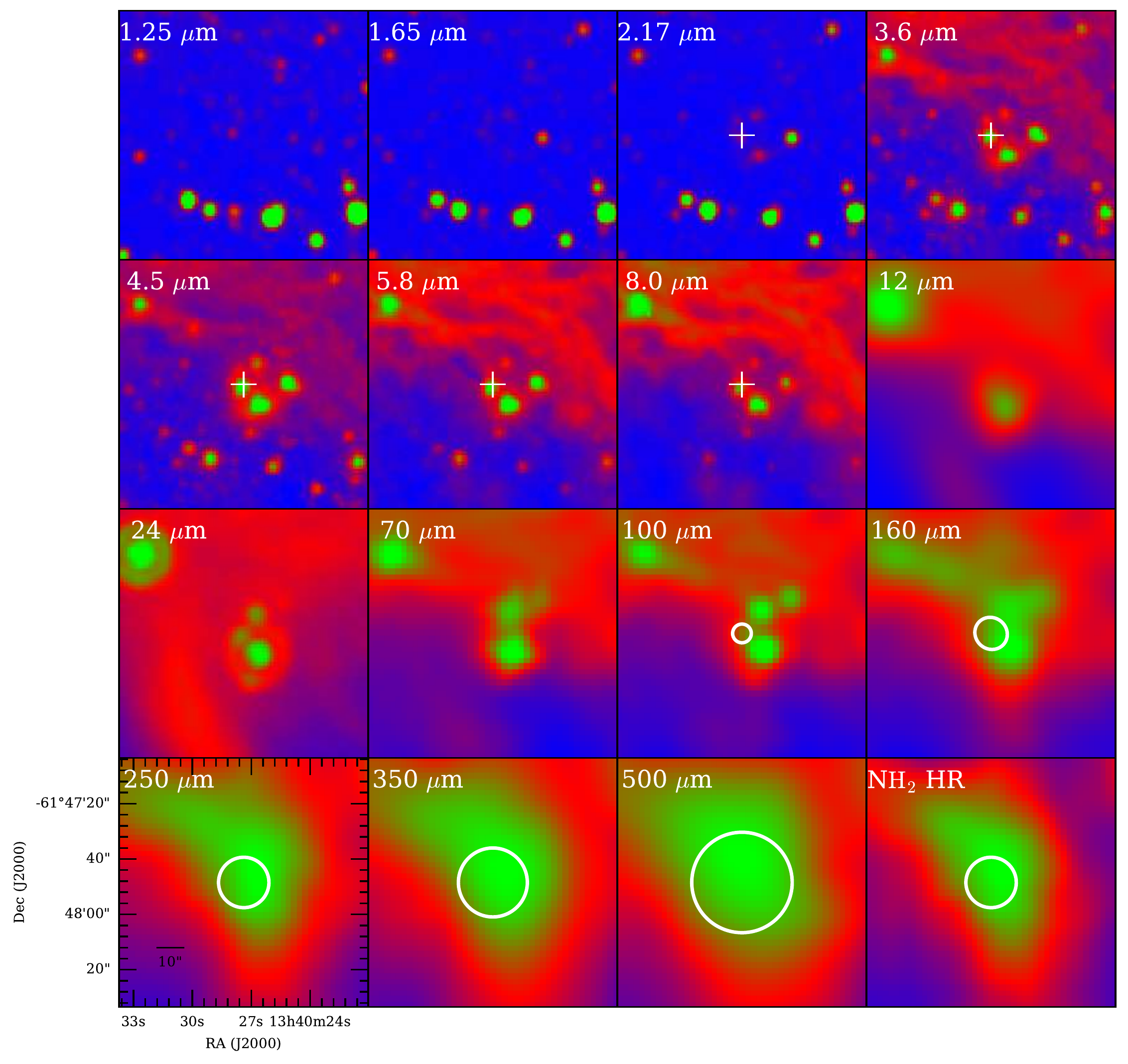}
\hskip 0.00000001cm
\includegraphics[width=3.4 in]{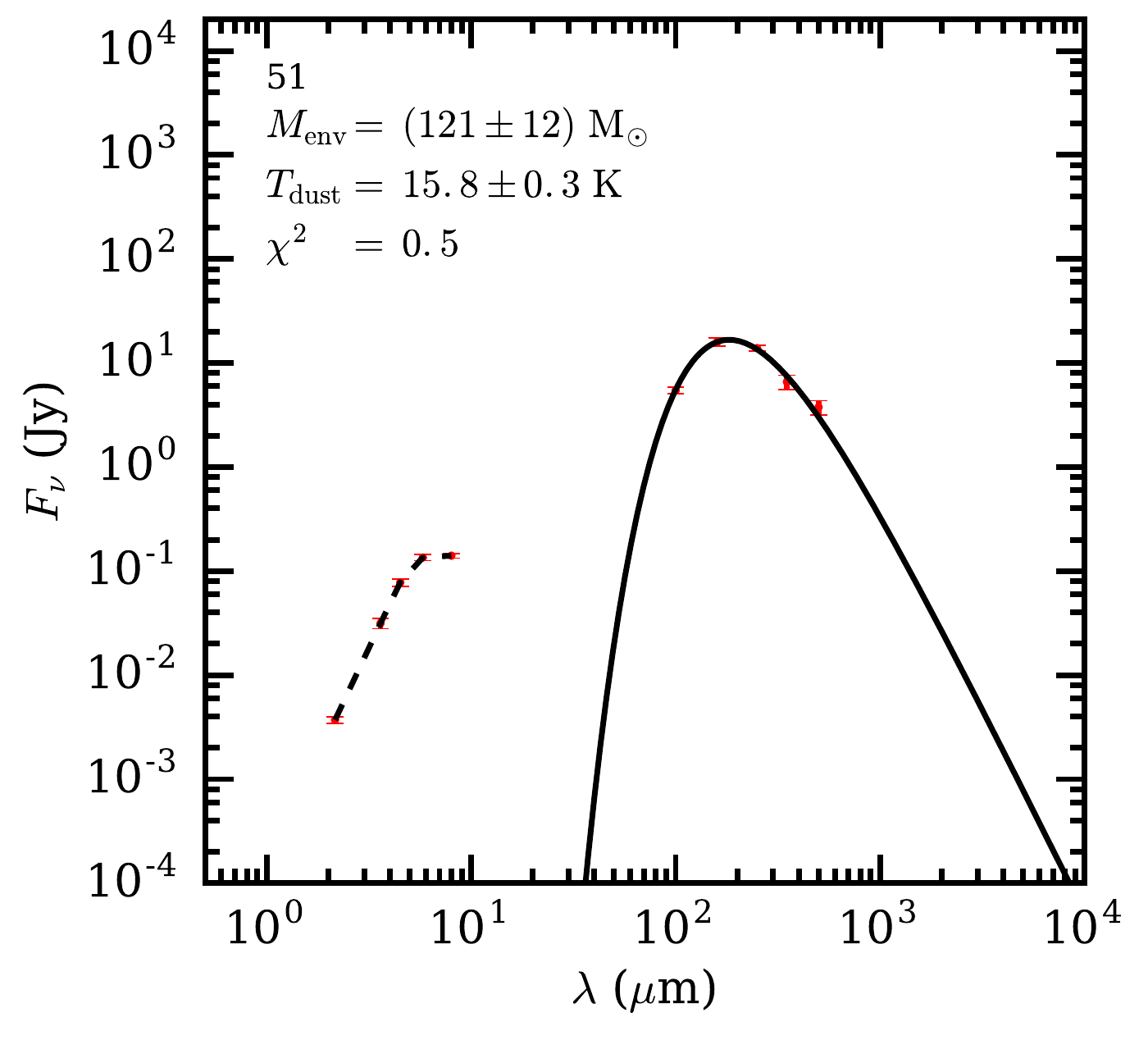}
}
\qquad
\subfloat{\includegraphics[width=3.4 in]{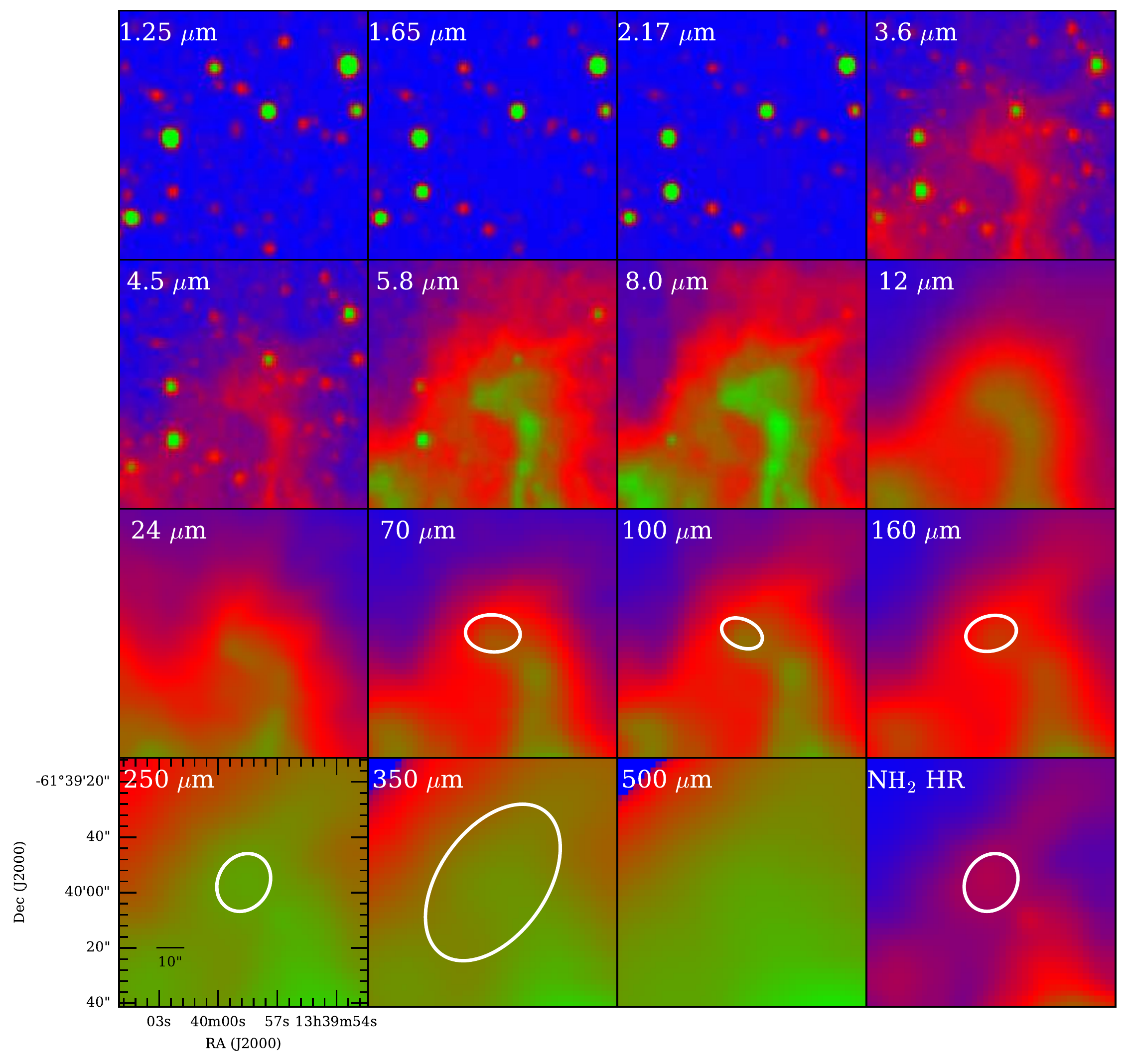}
\hskip 0.00000001cm
\includegraphics[width=3.4 in]{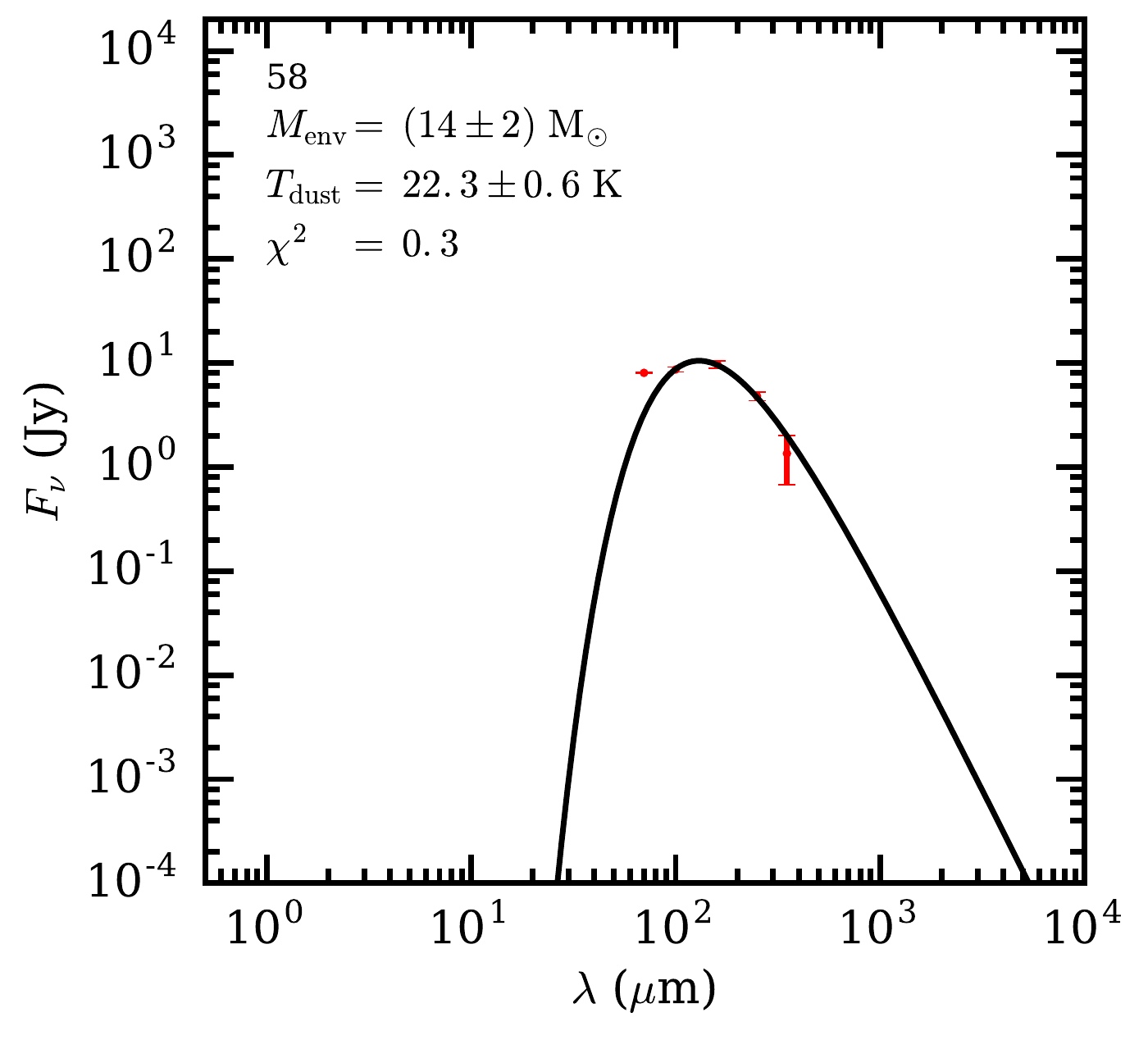}
}
\begin{flushleft}
\caption{ -- continued.}
\end{flushleft}
\end{figure*}
\begin{figure*}[!ht]
\ContinuedFloat
\centering
\subfloat{\includegraphics[width=3.4 in]{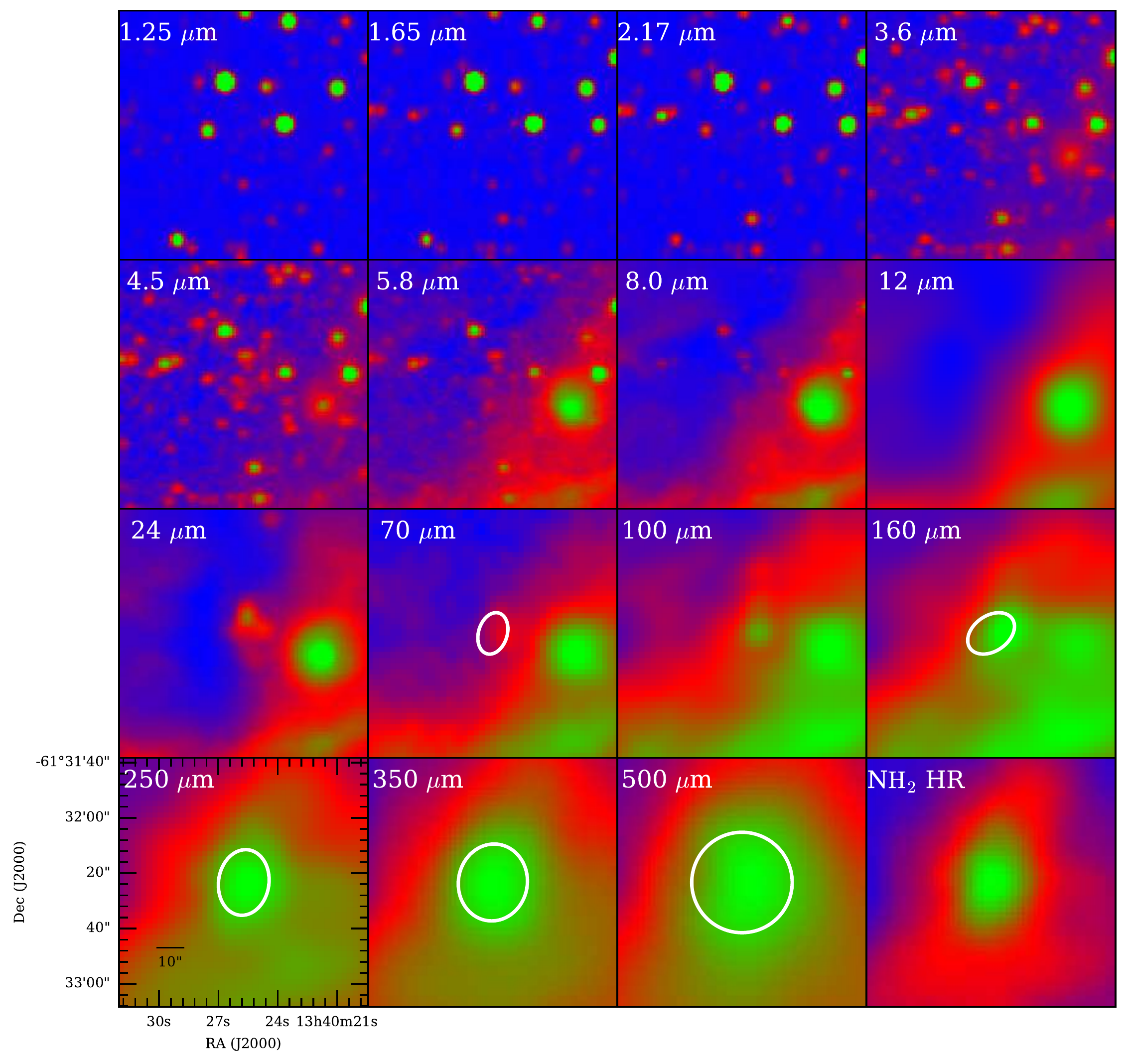}
\hskip 0.00000001cm
\includegraphics[width=3.4 in]{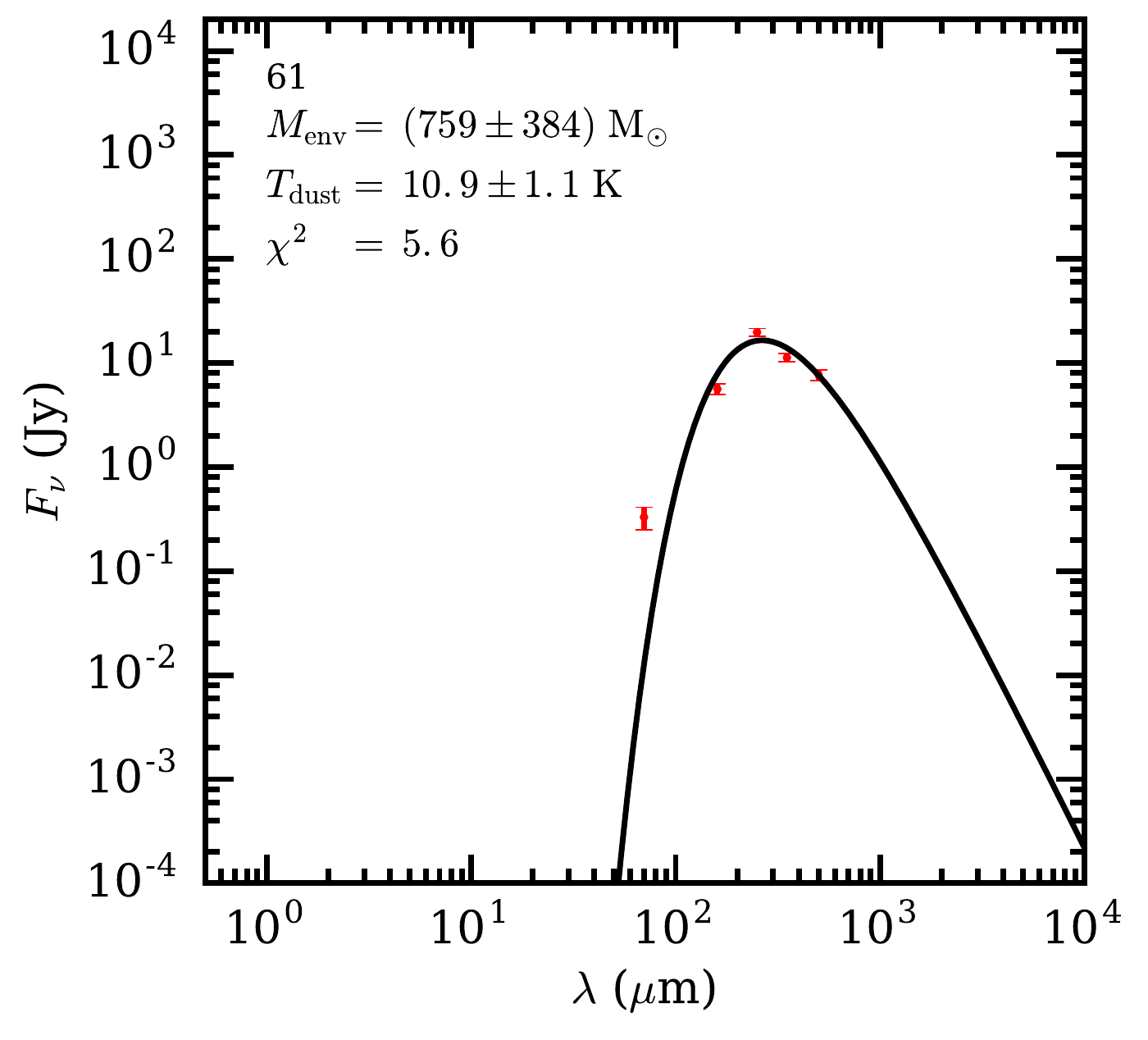}
}
\qquad
\subfloat{\includegraphics[width=3.4 in]{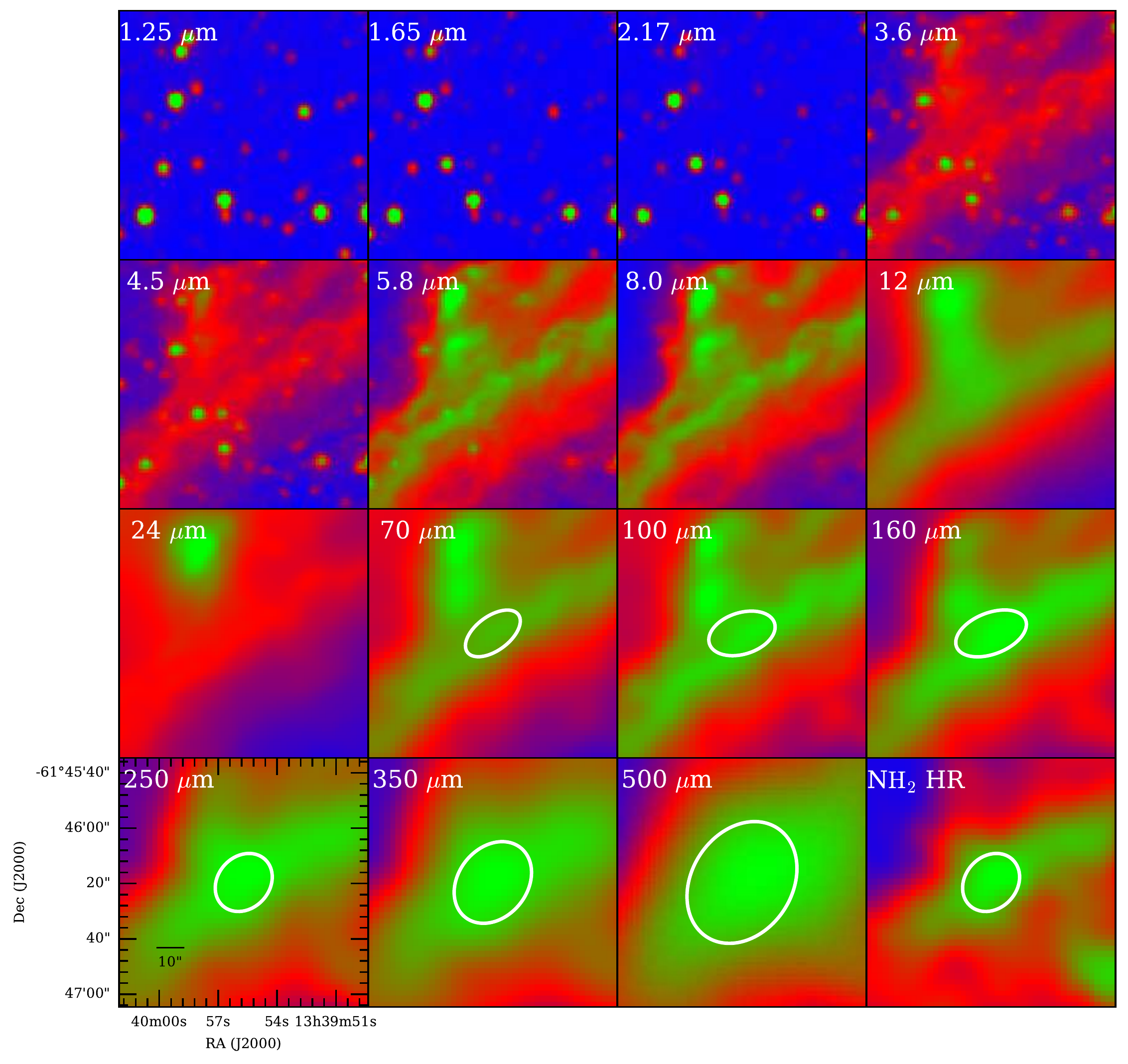}
\hskip 0.00000001cm
\includegraphics[width=3.4 in]{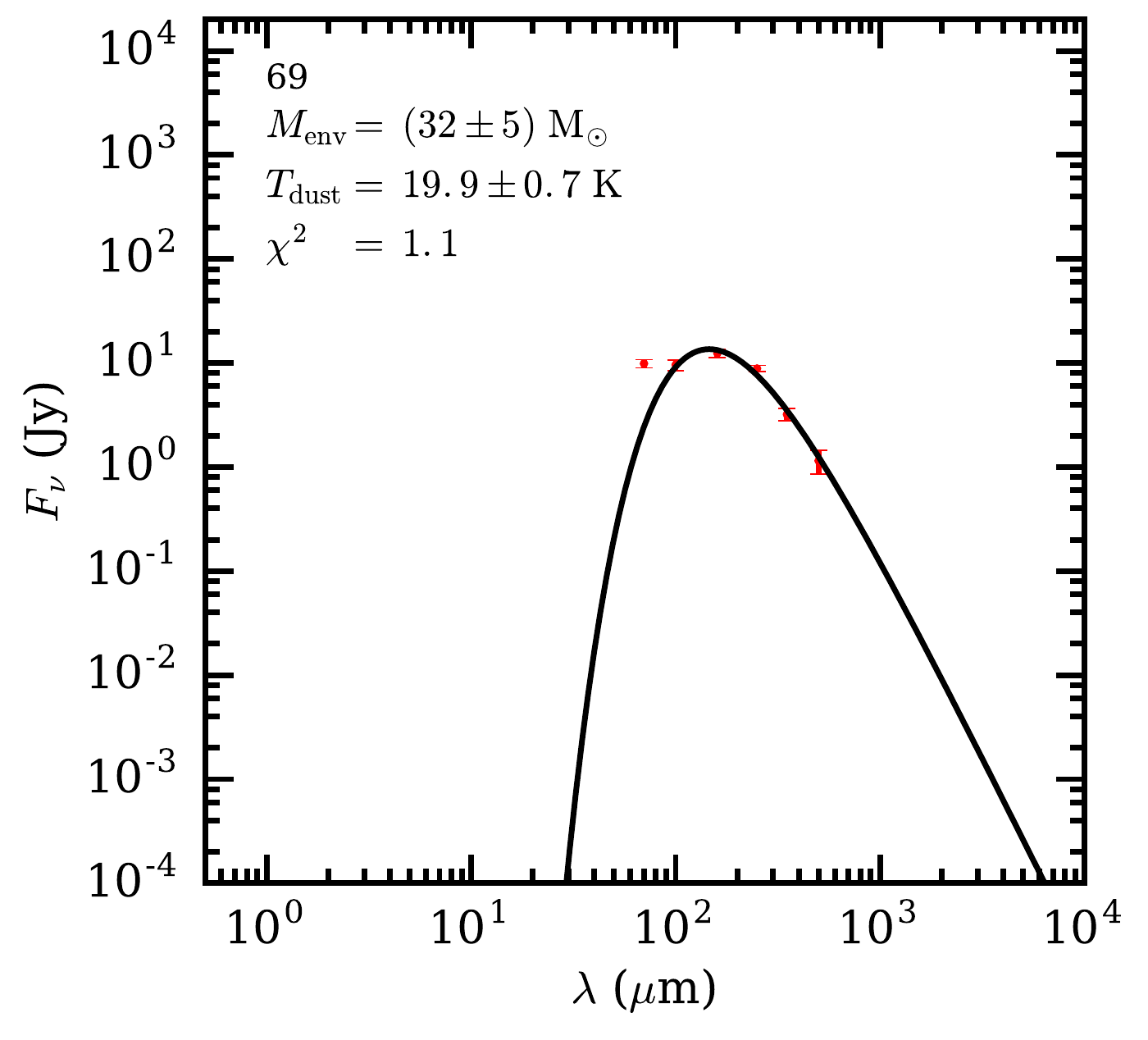}
}
\begin{flushleft}
\caption{ -- continued.}
\end{flushleft}
\end{figure*}
\begin{figure*}[!ht]
\ContinuedFloat
\centering
\subfloat{\includegraphics[width=3.4 in]{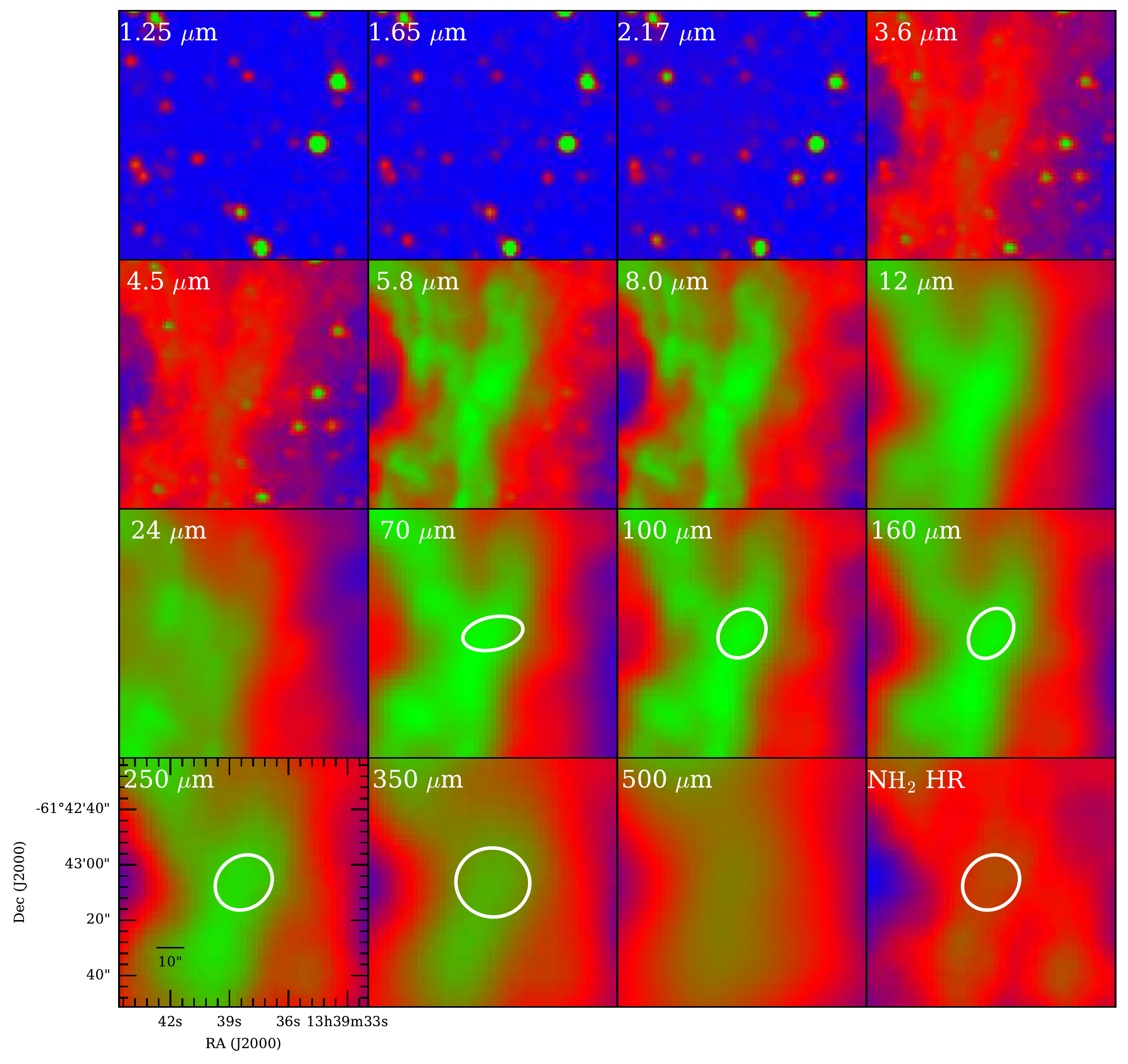}
\hskip 0.00000001cm
\includegraphics[width=3.4 in]{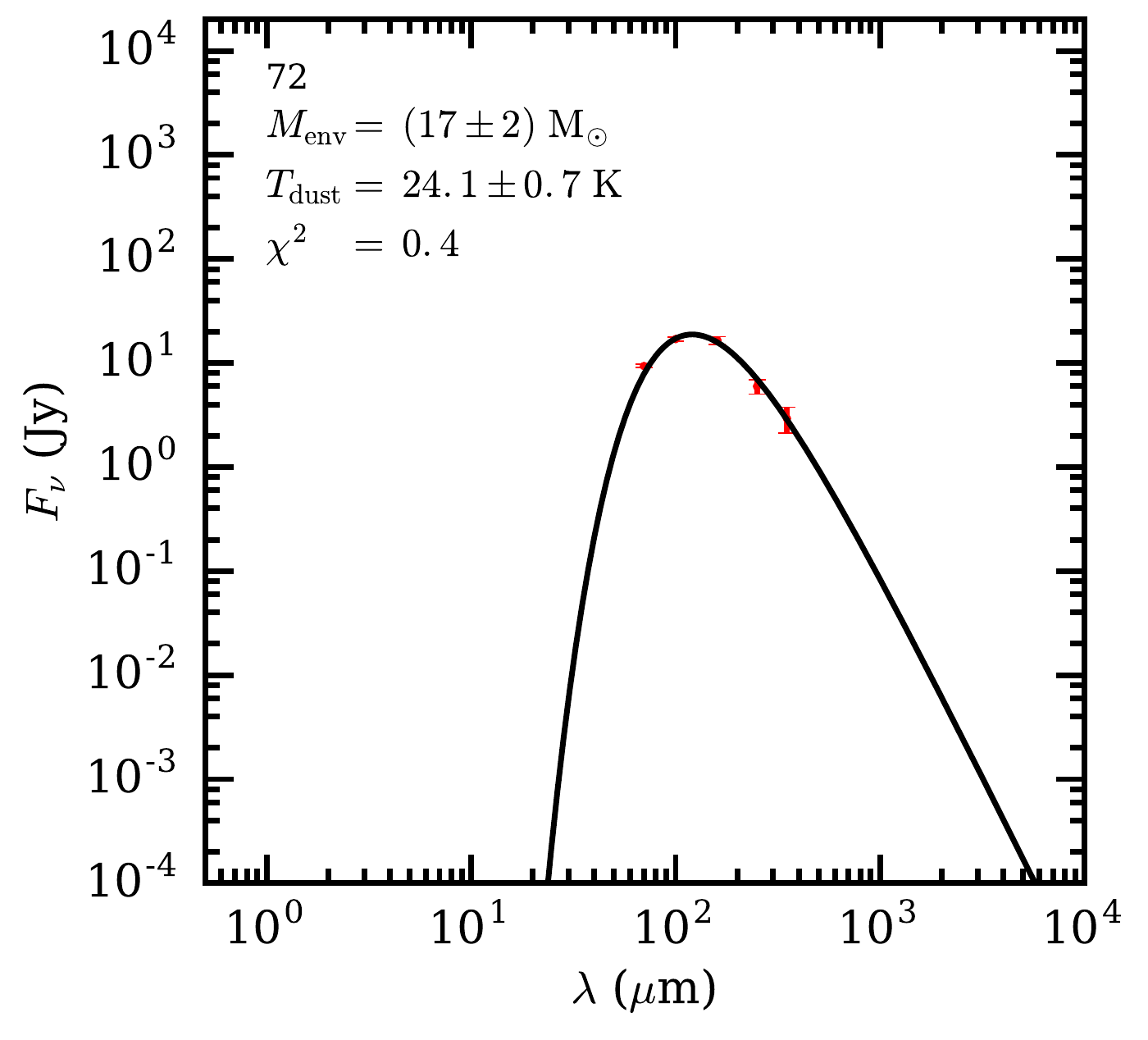}
}
\qquad
\subfloat{\includegraphics[width=3.4 in]{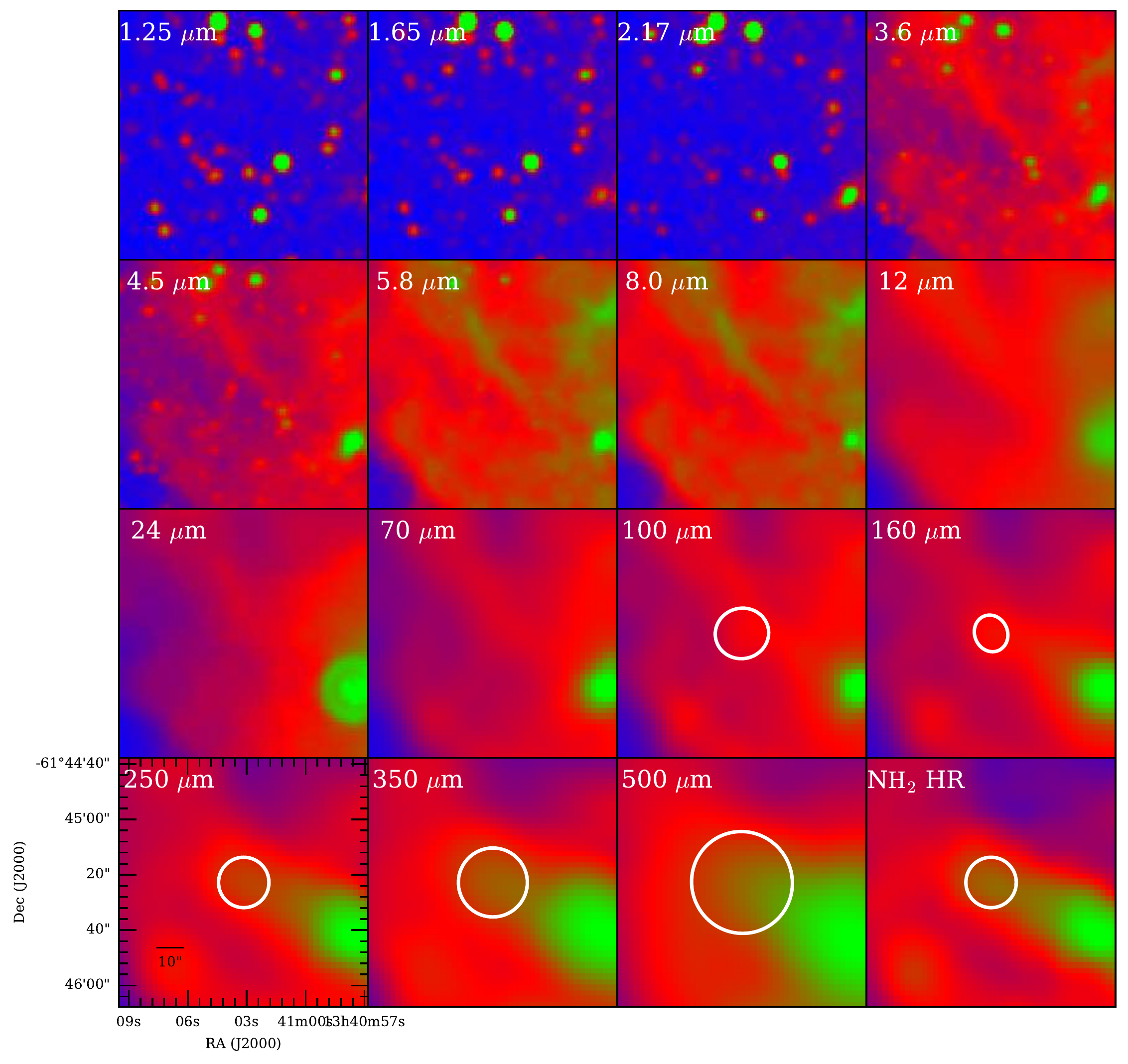}
\hskip 0.00000001cm
\includegraphics[width=3.4 in]{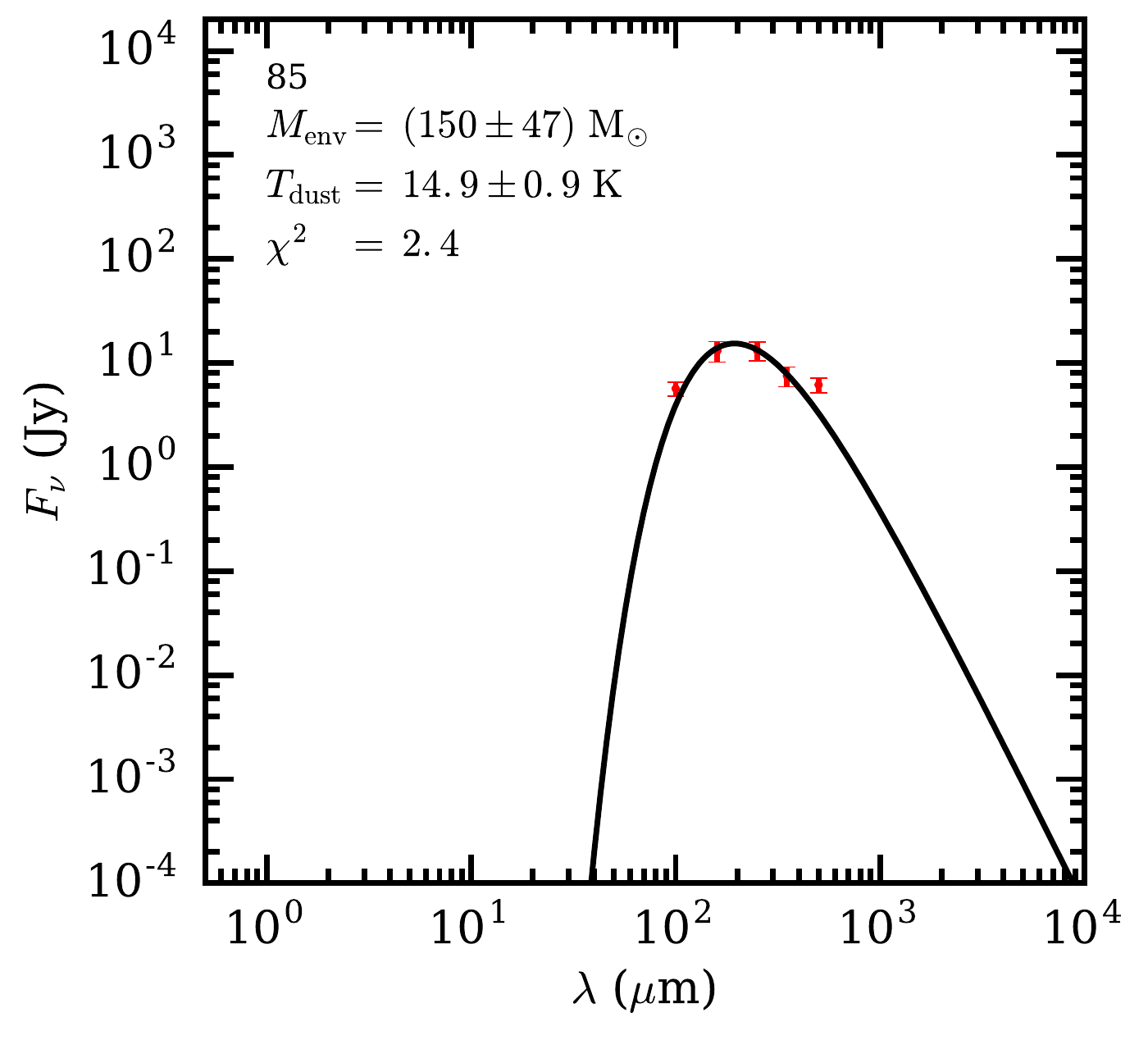}
}
\begin{flushleft}
\caption{ -- continued.}
\end{flushleft}
\end{figure*}
\begin{figure*}[!ht]
\ContinuedFloat
\centering
\subfloat{\includegraphics[width=3.4 in]{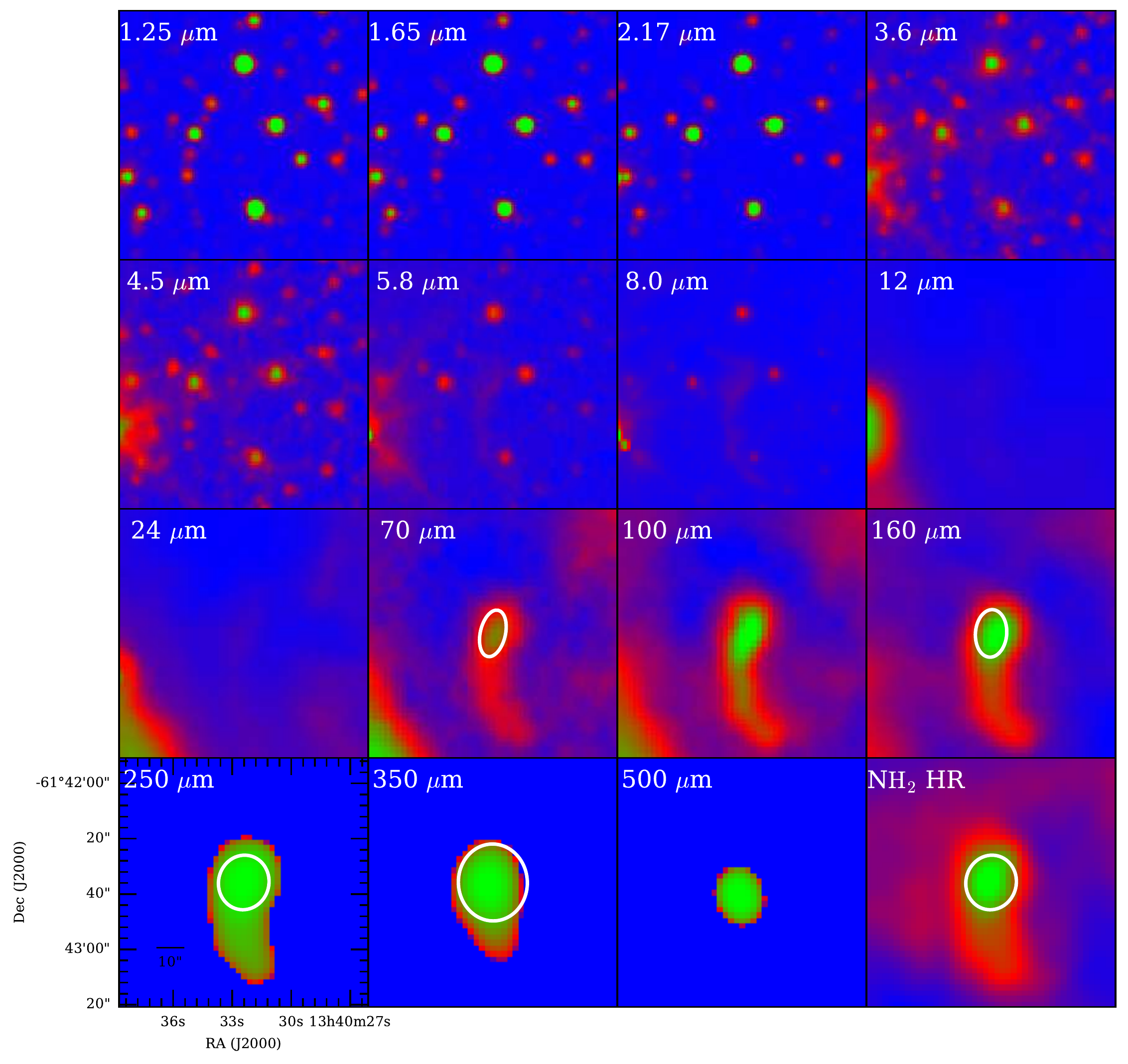}
\hskip 0.00000001cm
\includegraphics[width=3.4 in]{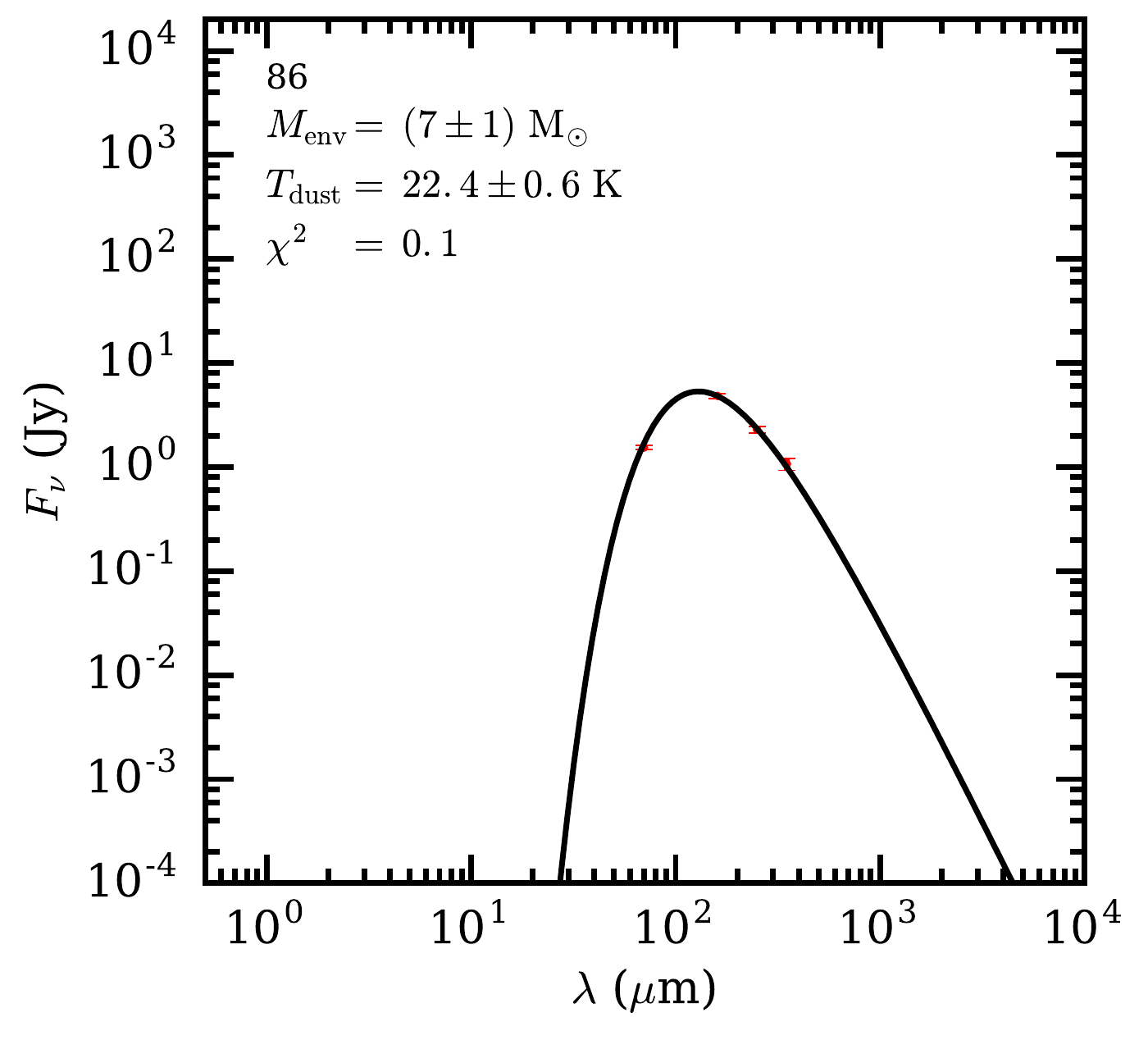}
}
\qquad
\subfloat{\includegraphics[width=3.4 in]{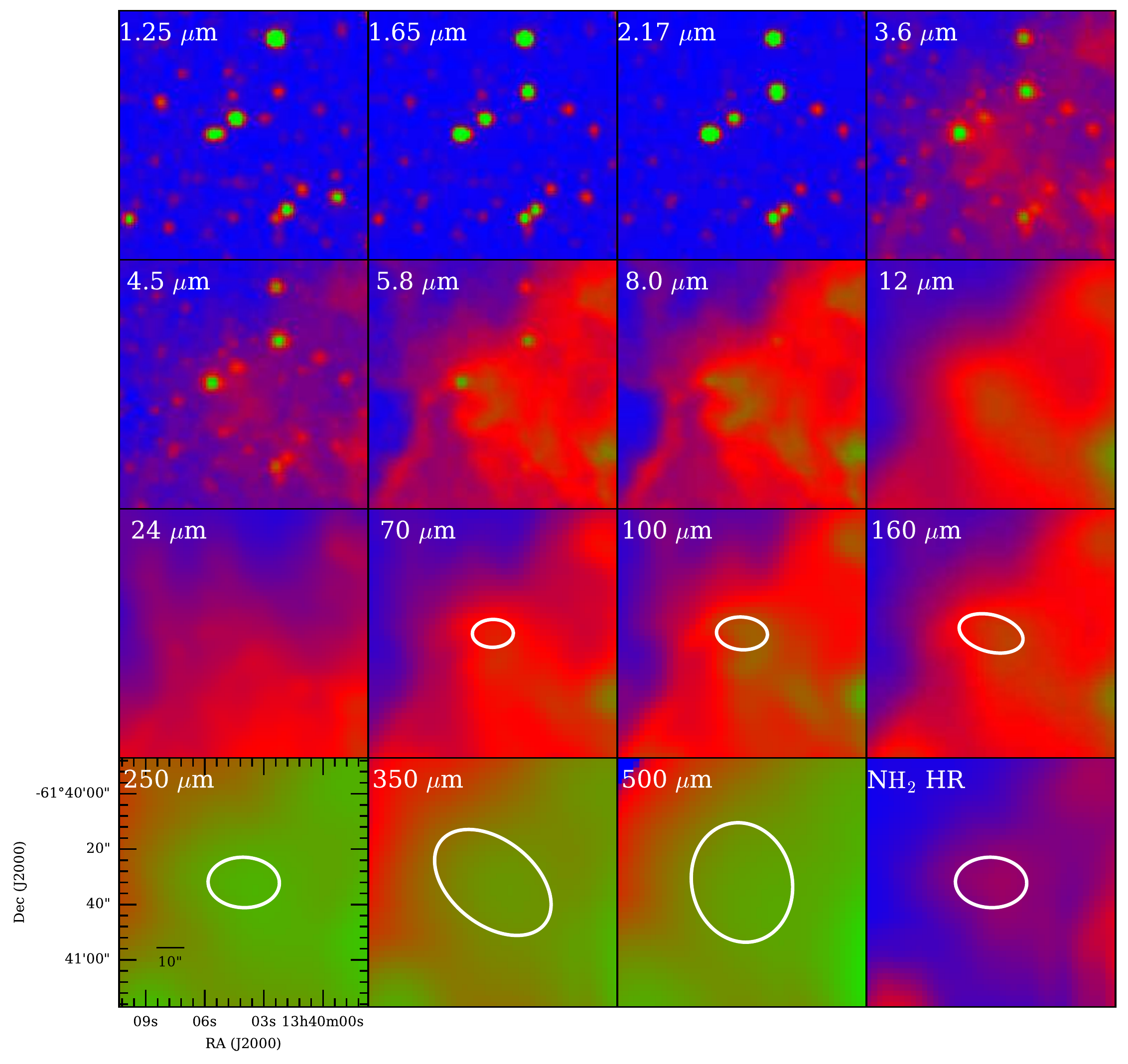}
\hskip 0.00000001cm
\includegraphics[width=3.4 in]{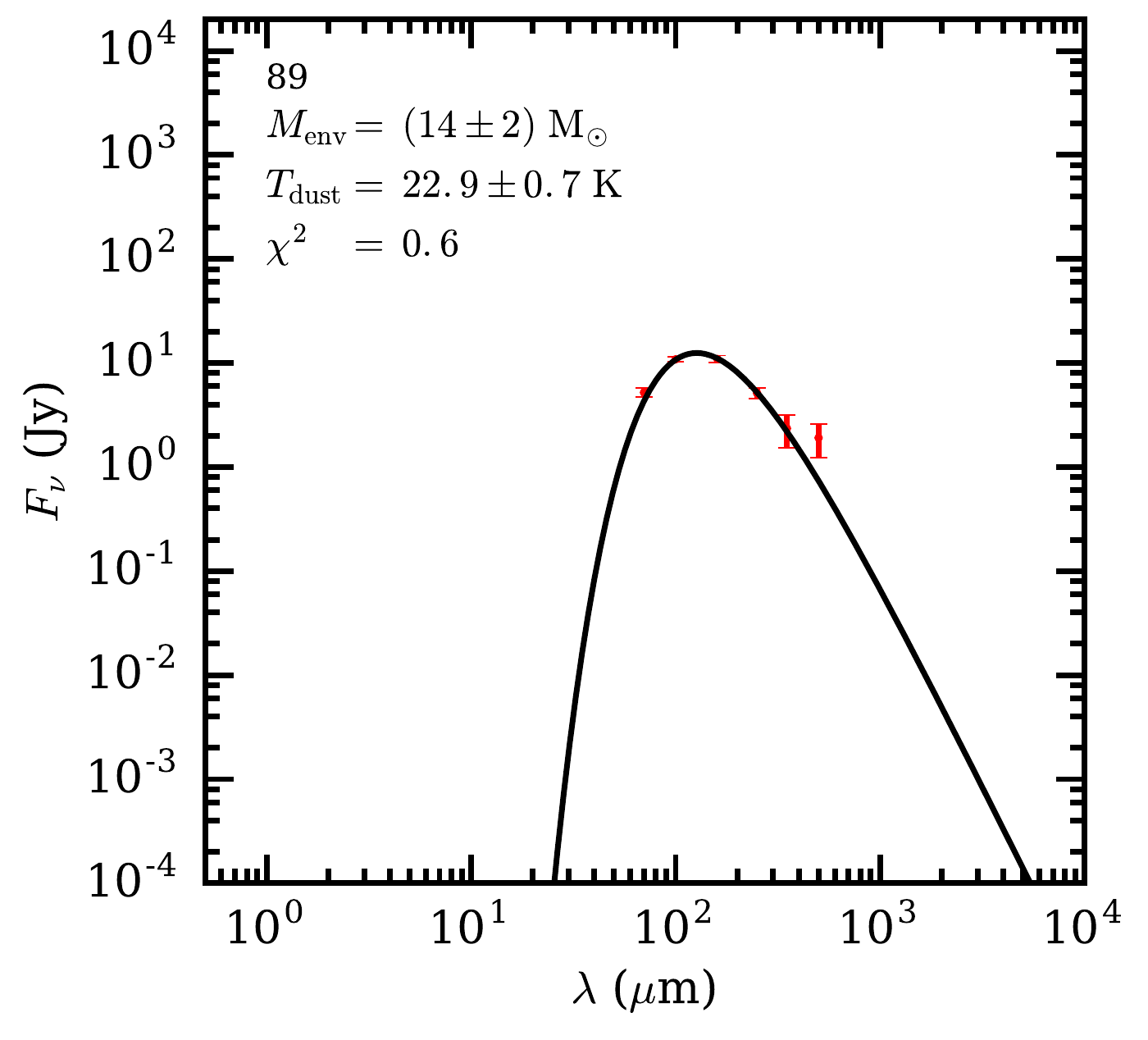}
}
\begin{flushleft}
\caption{ -- continued.}
\end{flushleft}
\end{figure*}
\begin{figure*}[!ht]
\ContinuedFloat
\centering
\subfloat{\includegraphics[width=3.4 in]{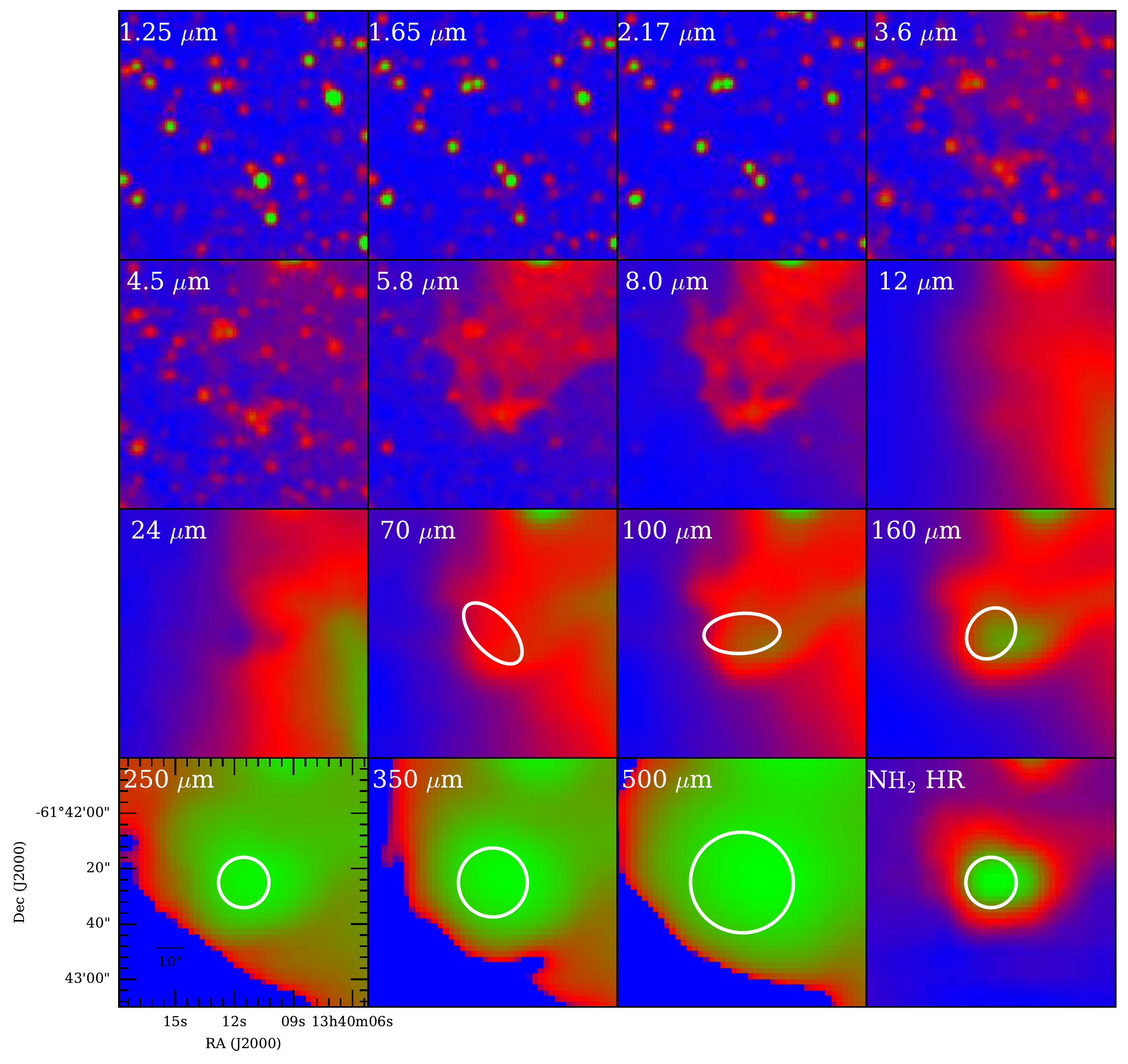}
\hskip 0.00000001cm
\includegraphics[width=3.4 in]{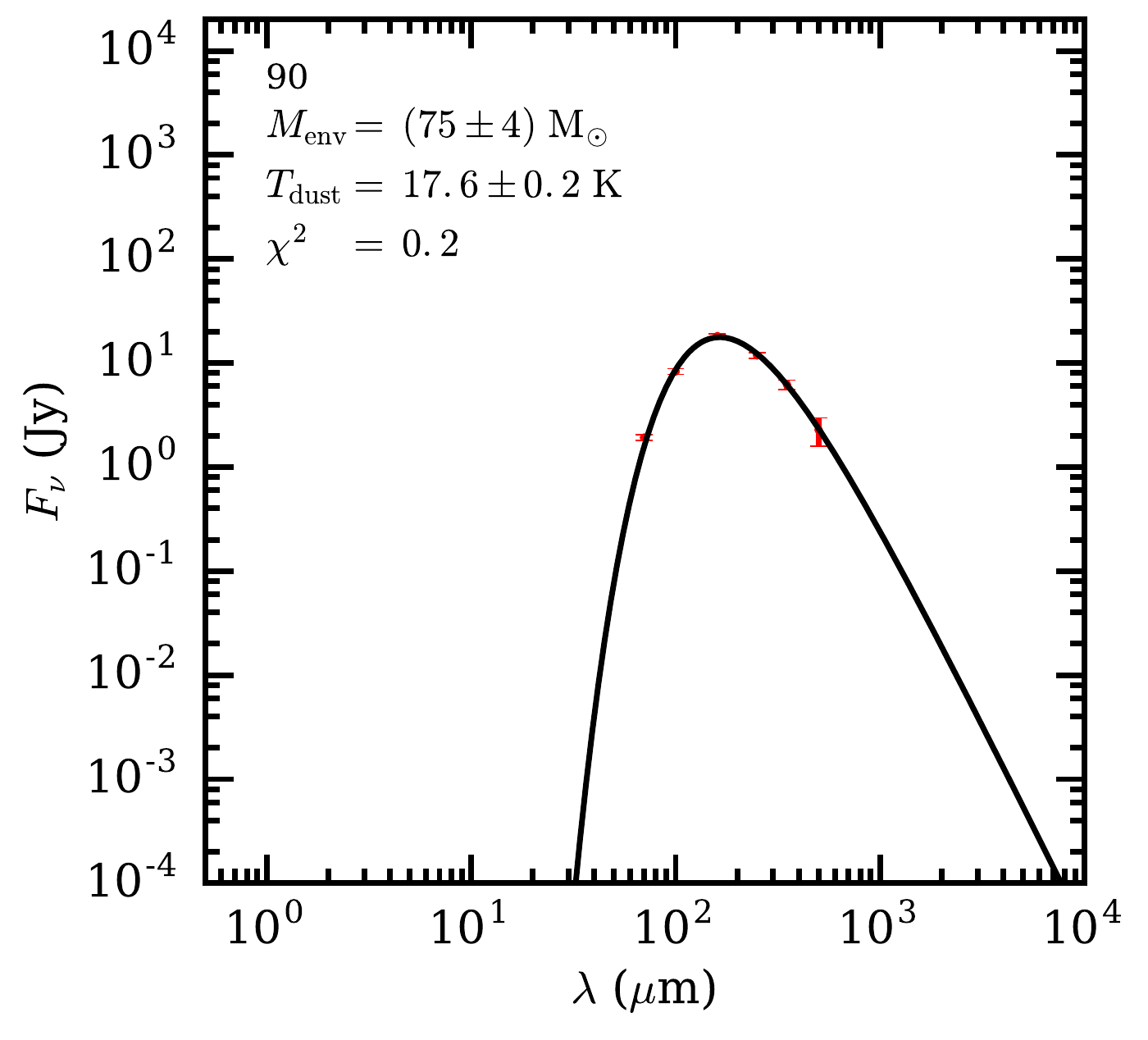}
}
\qquad
\subfloat{\includegraphics[width=3.4 in]{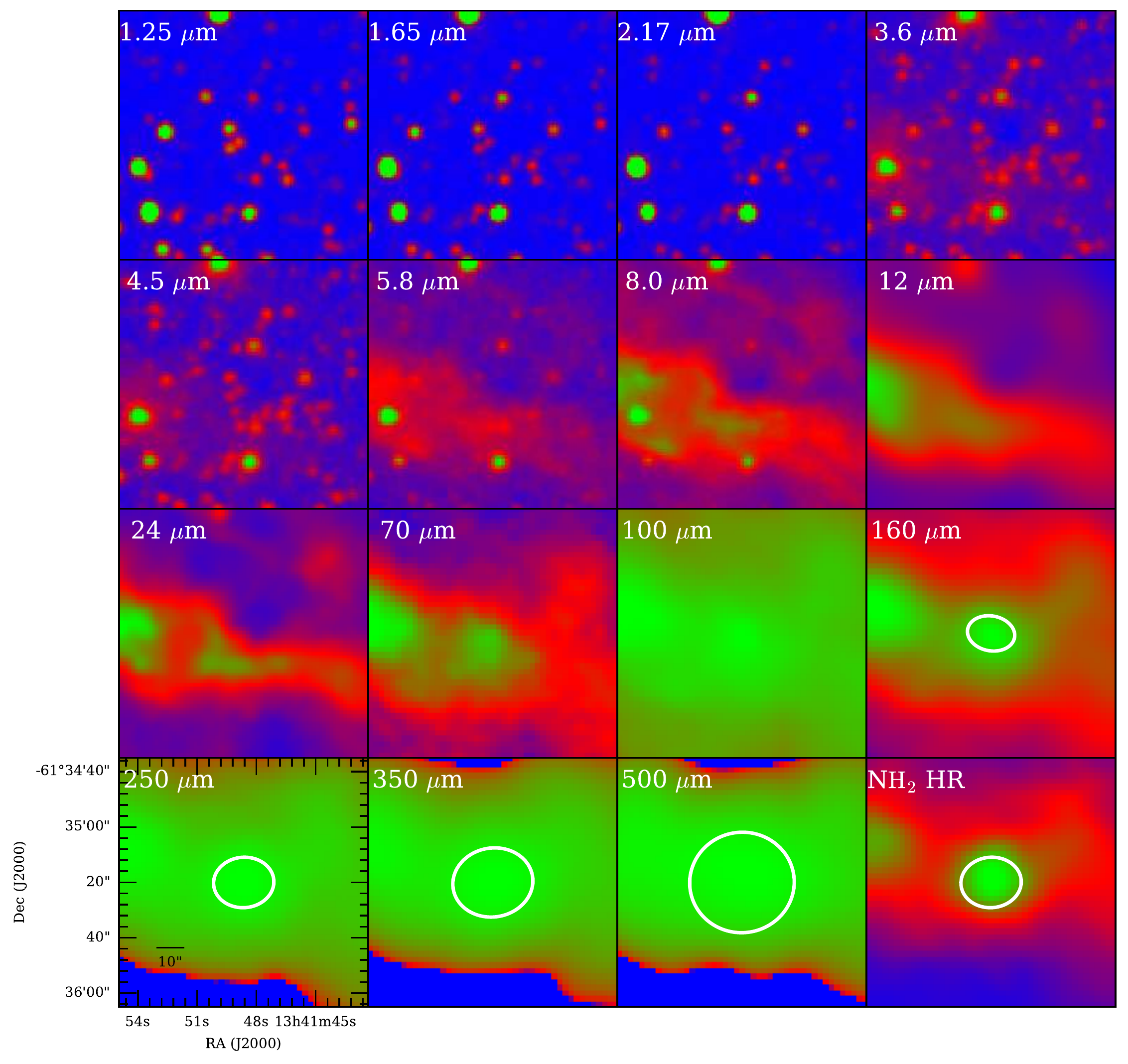}
\hskip 0.00000001cm
\includegraphics[width=3.4 in]{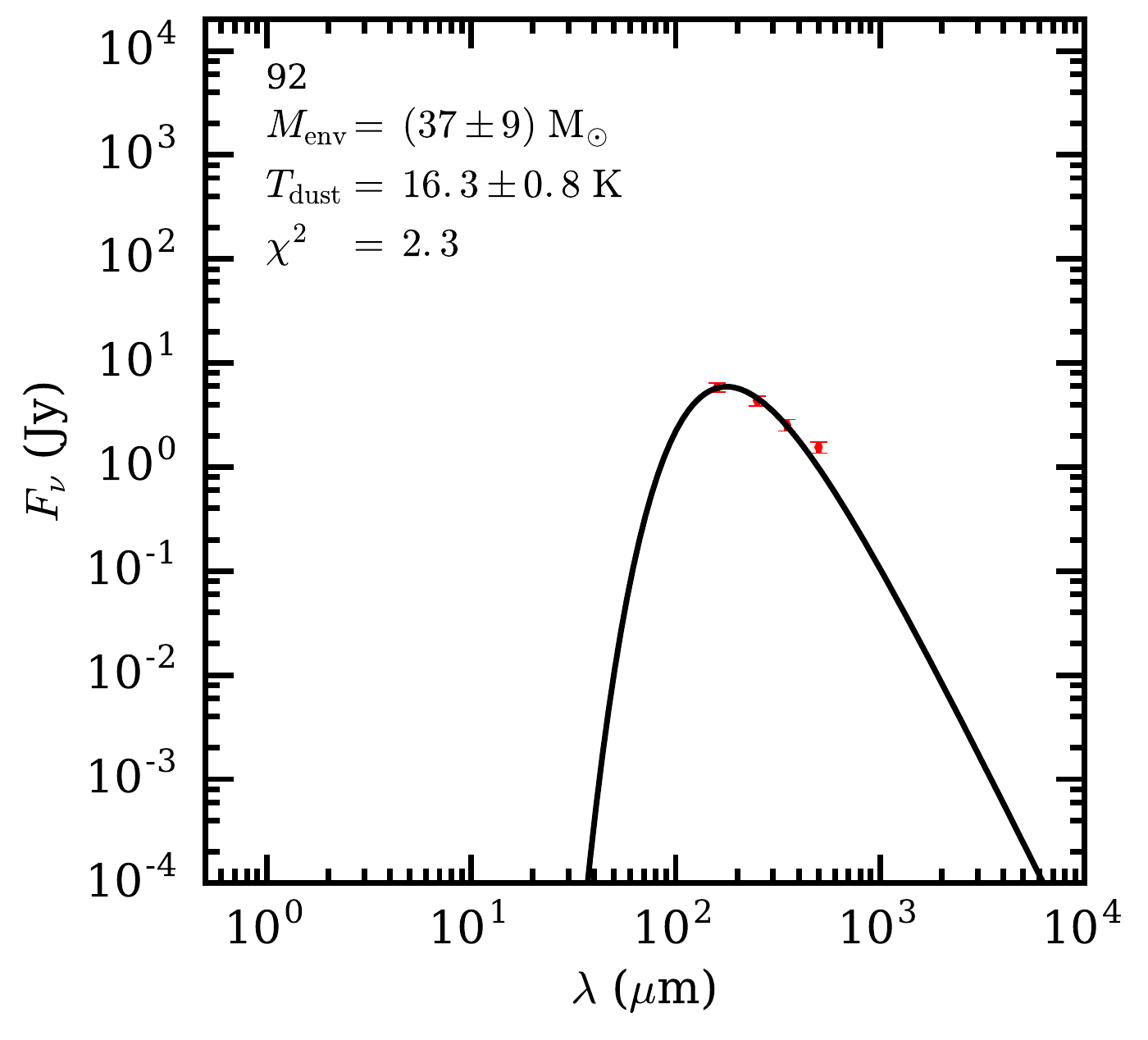}
}
\begin{flushleft}
\caption{ -- continued.}
\end{flushleft}
\end{figure*}
\clearpage
\begin{figure*}[!ht]
\ContinuedFloat
\centering
\subfloat{\includegraphics[width=3.4 in]{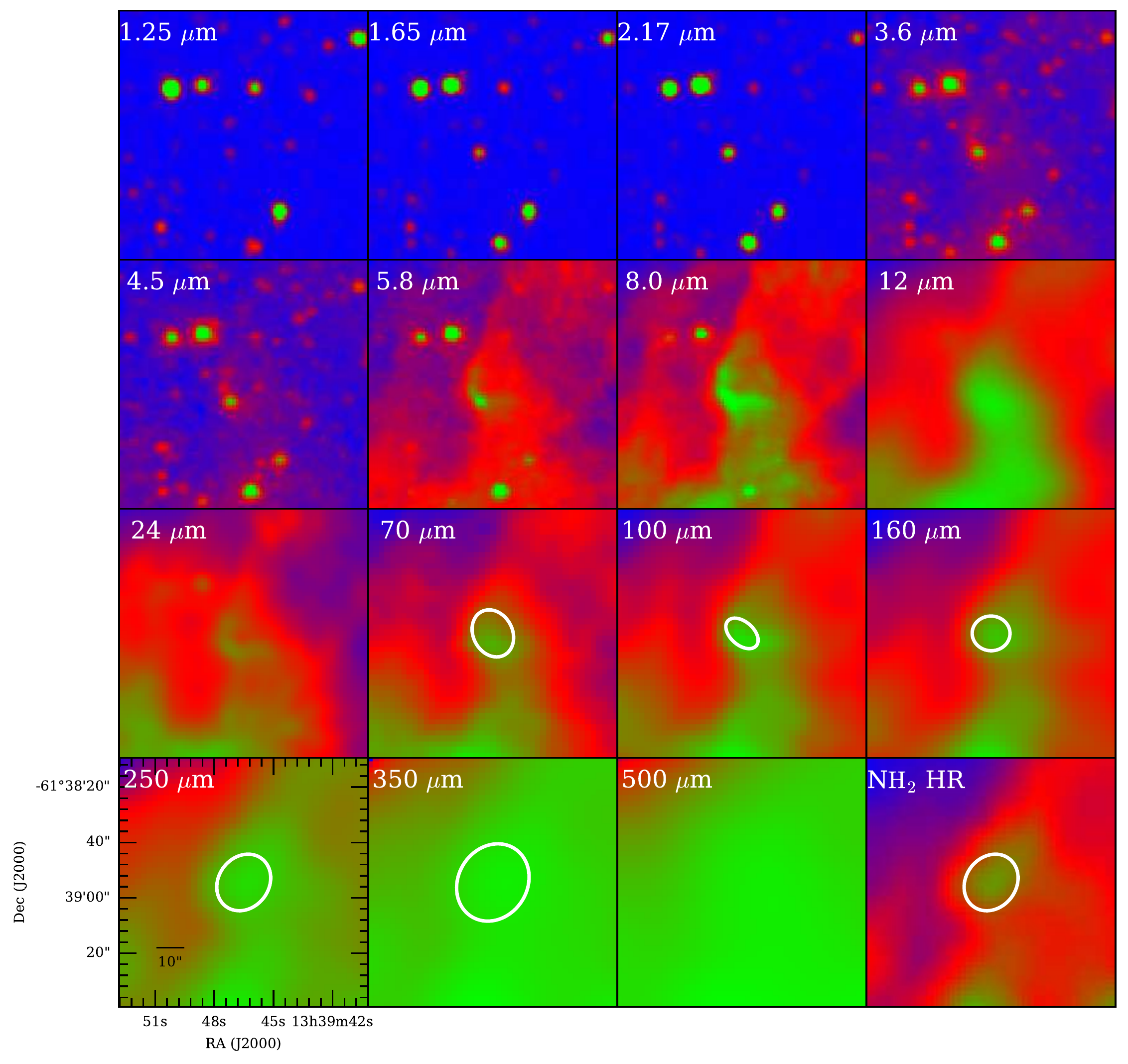}
\hskip 0.00000001cm
\includegraphics[width=3.4 in]{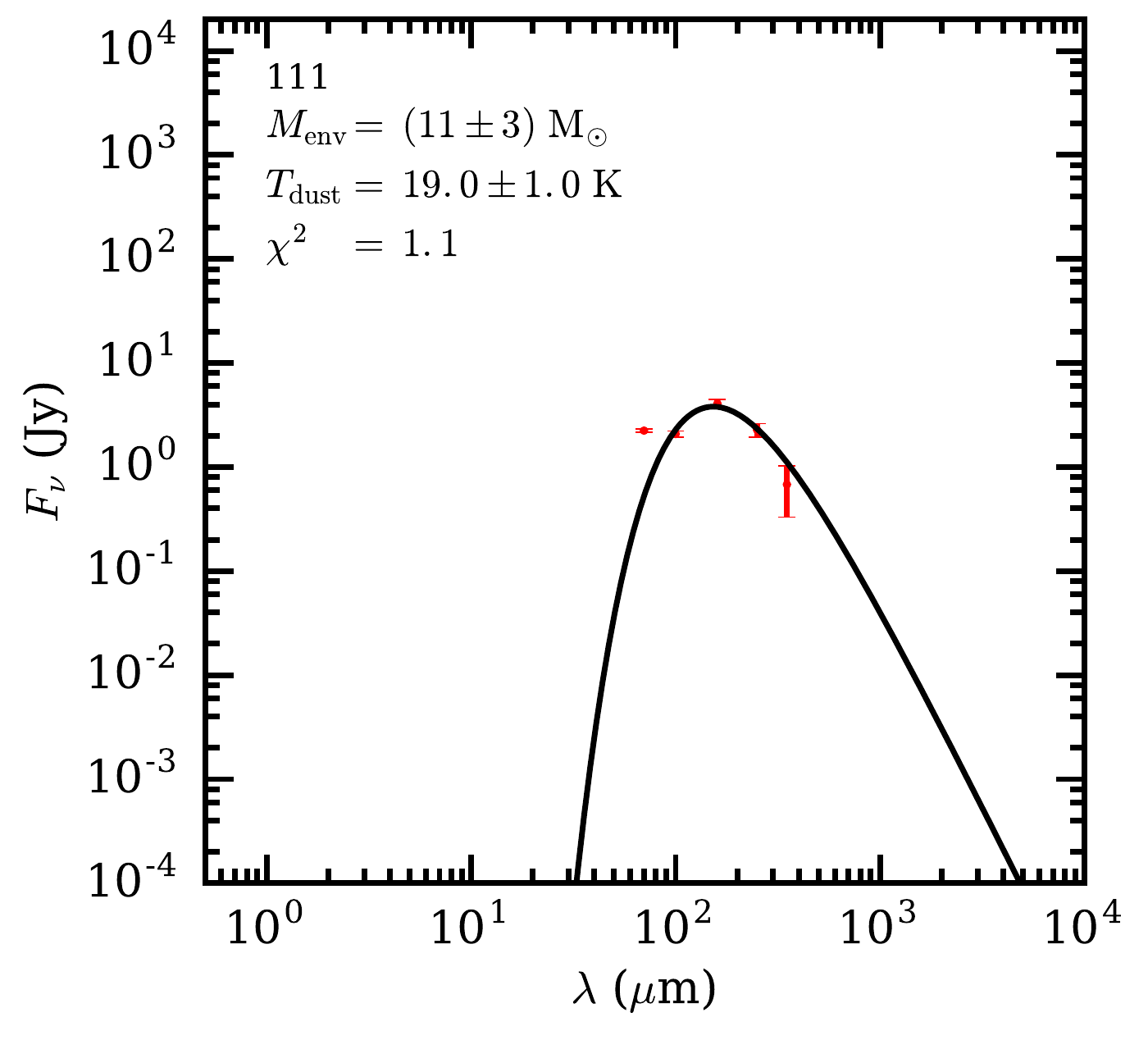}
}
\qquad
\subfloat{\includegraphics[width=3.4 in]{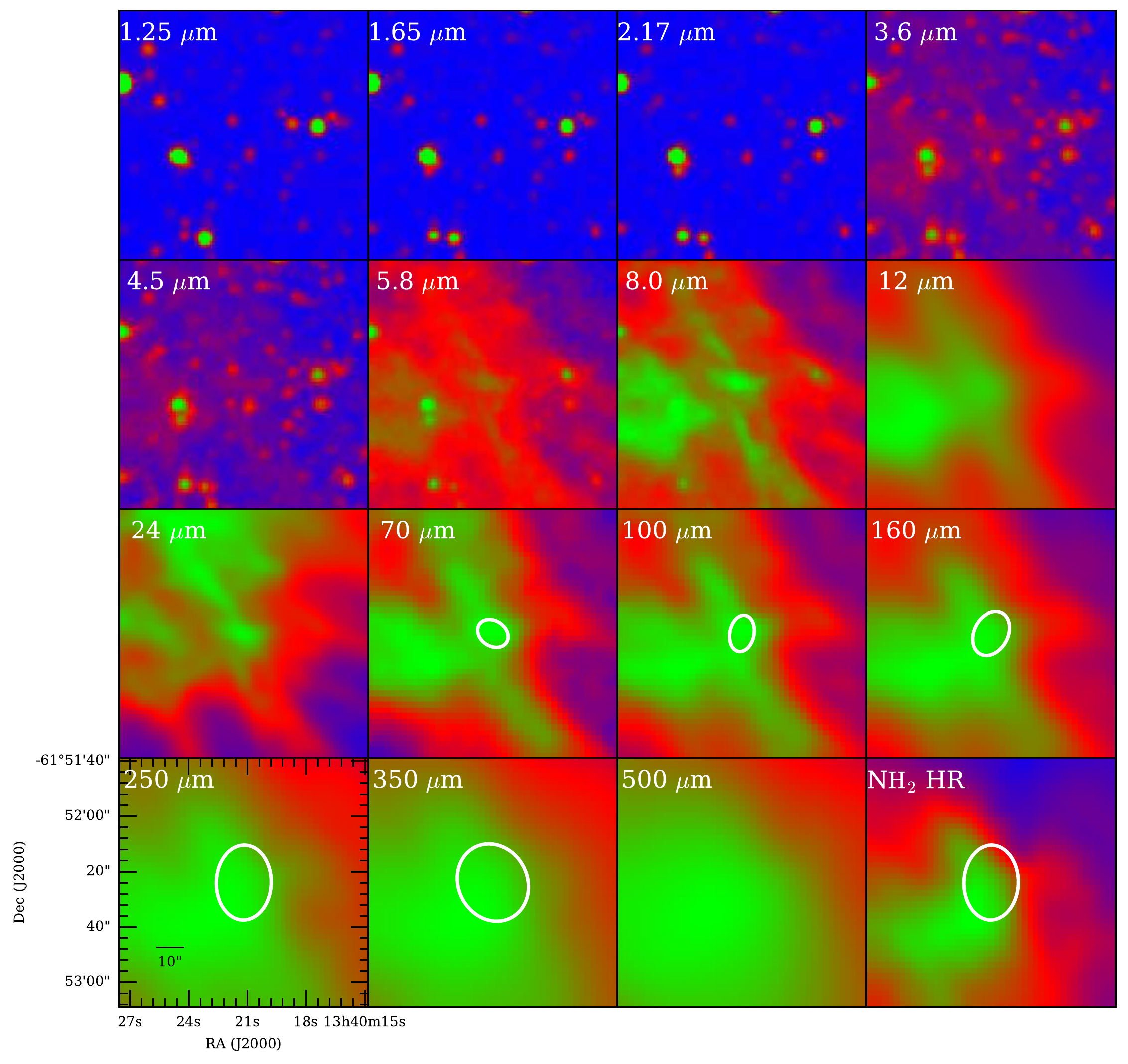}
\hskip 0.00000001cm
\includegraphics[width=3.4 in]{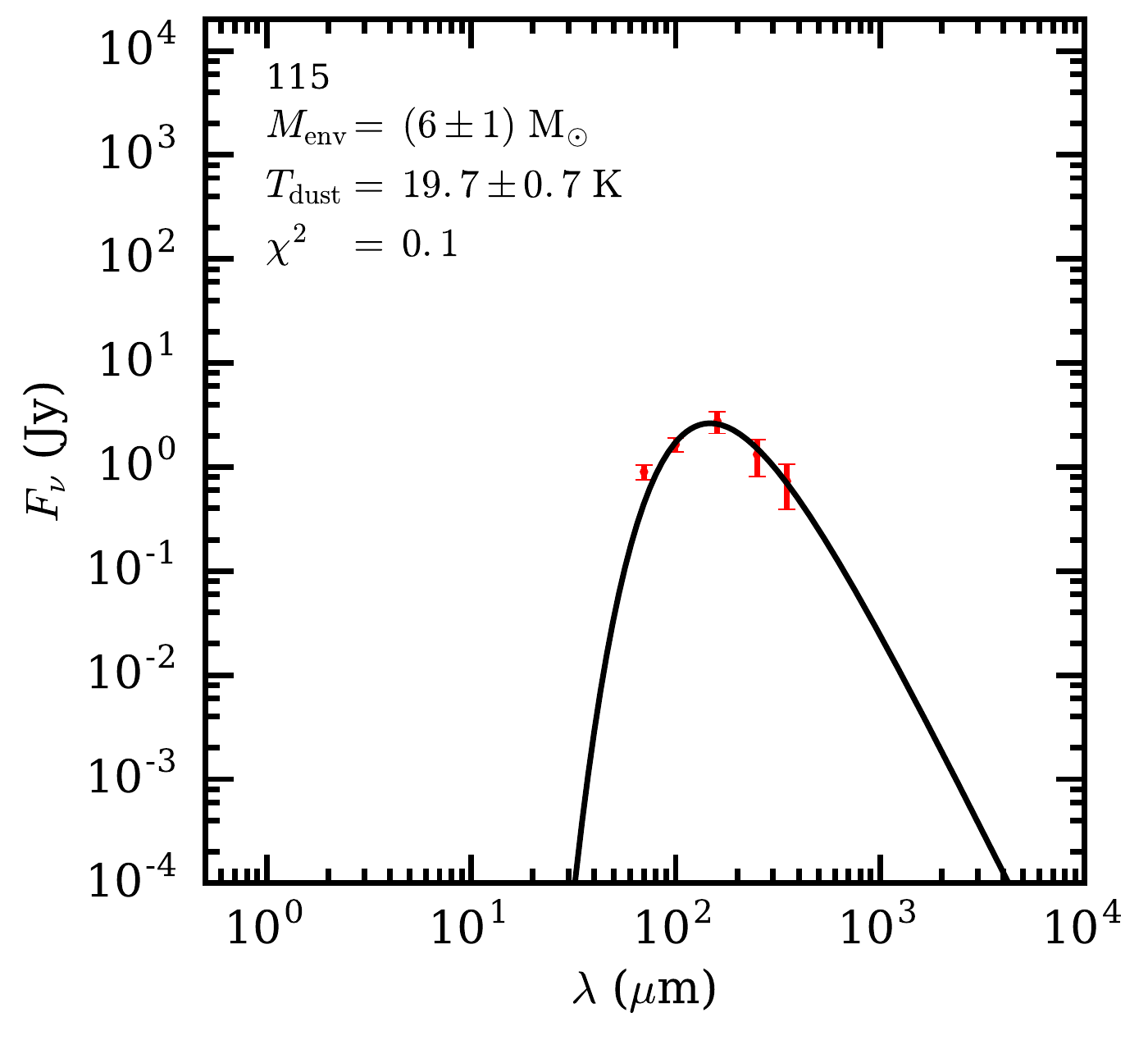}
}
\begin{flushleft}
\caption{ -- continued.}
\end{flushleft}
\end{figure*}
\begin{figure*}[!ht]
\ContinuedFloat
\centering
\subfloat{\includegraphics[width=3.4 in]{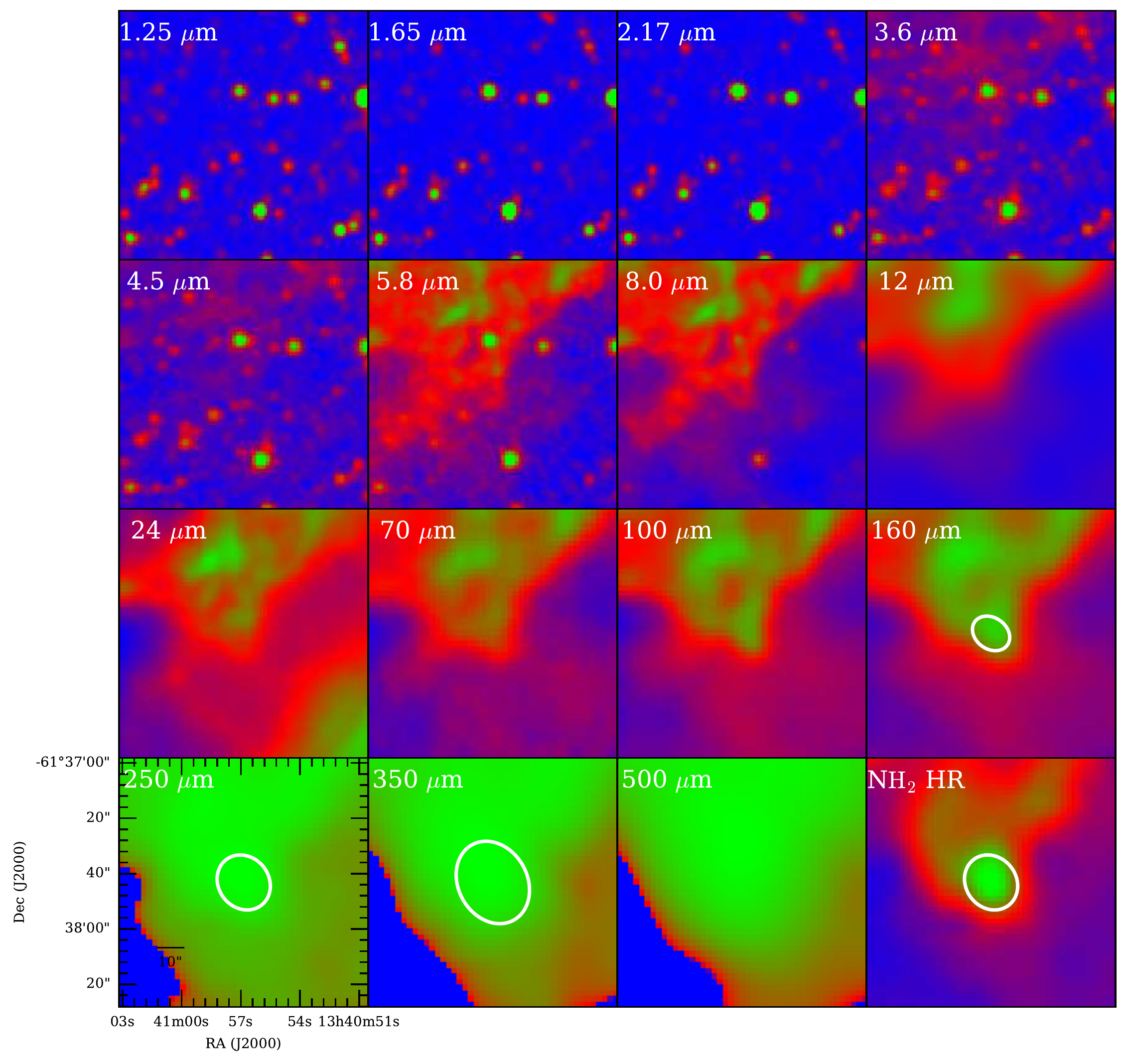}
\hskip 0.00000001cm
\includegraphics[width=3.4 in]{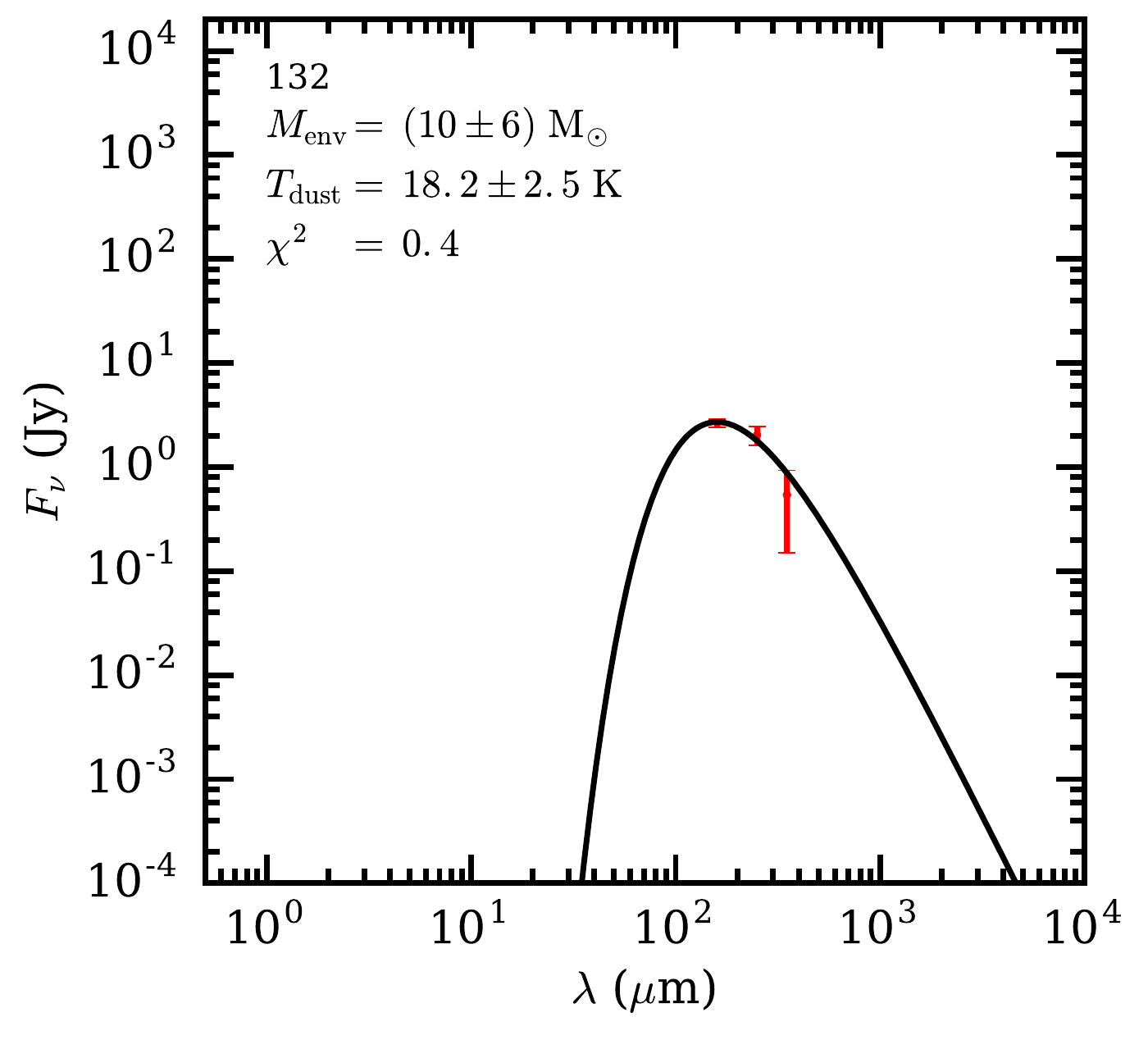}
}
\qquad
\subfloat{\includegraphics[width=3.4 in]{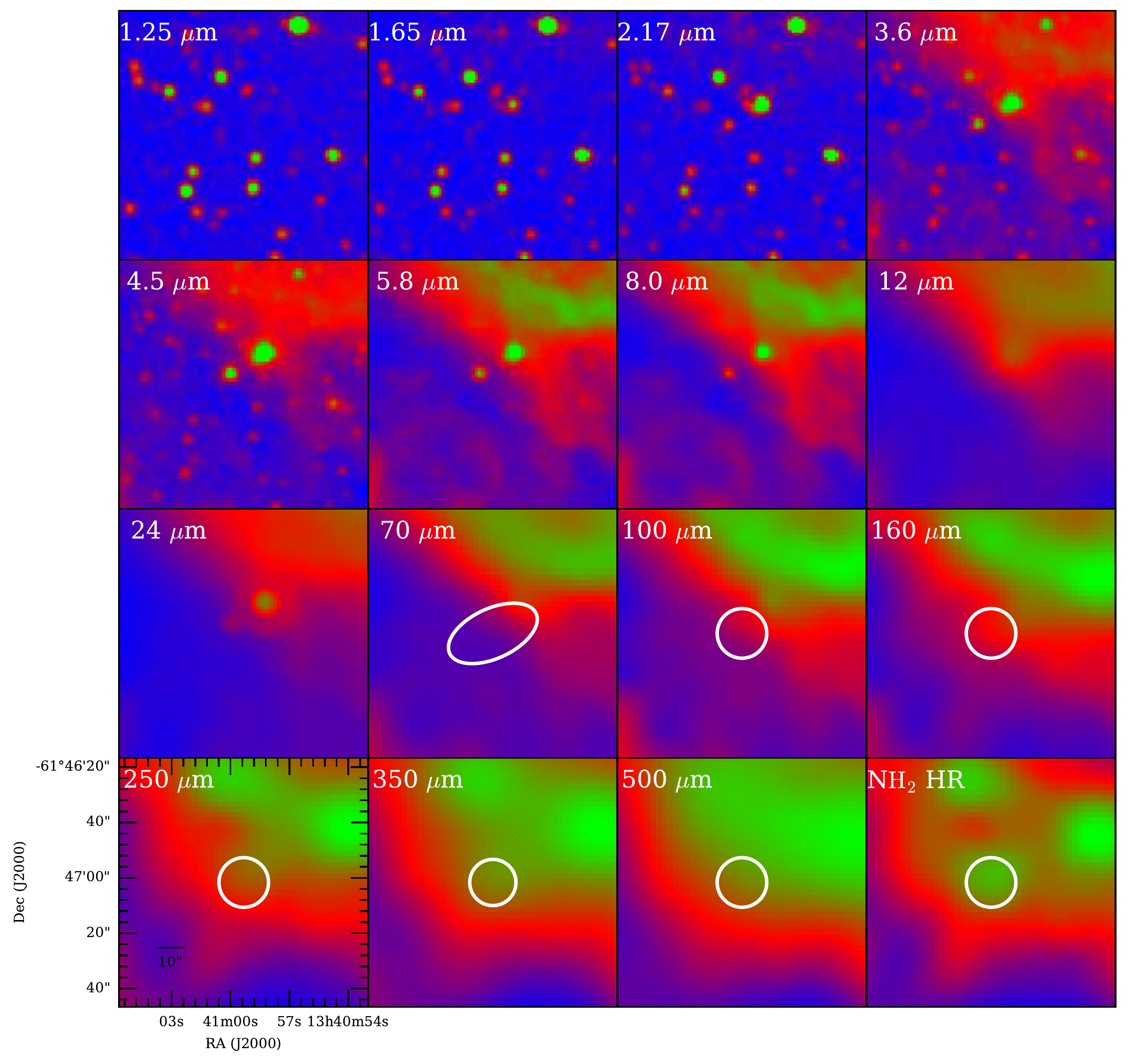}
\hskip 0.00000001cm
\includegraphics[width=3.4 in]{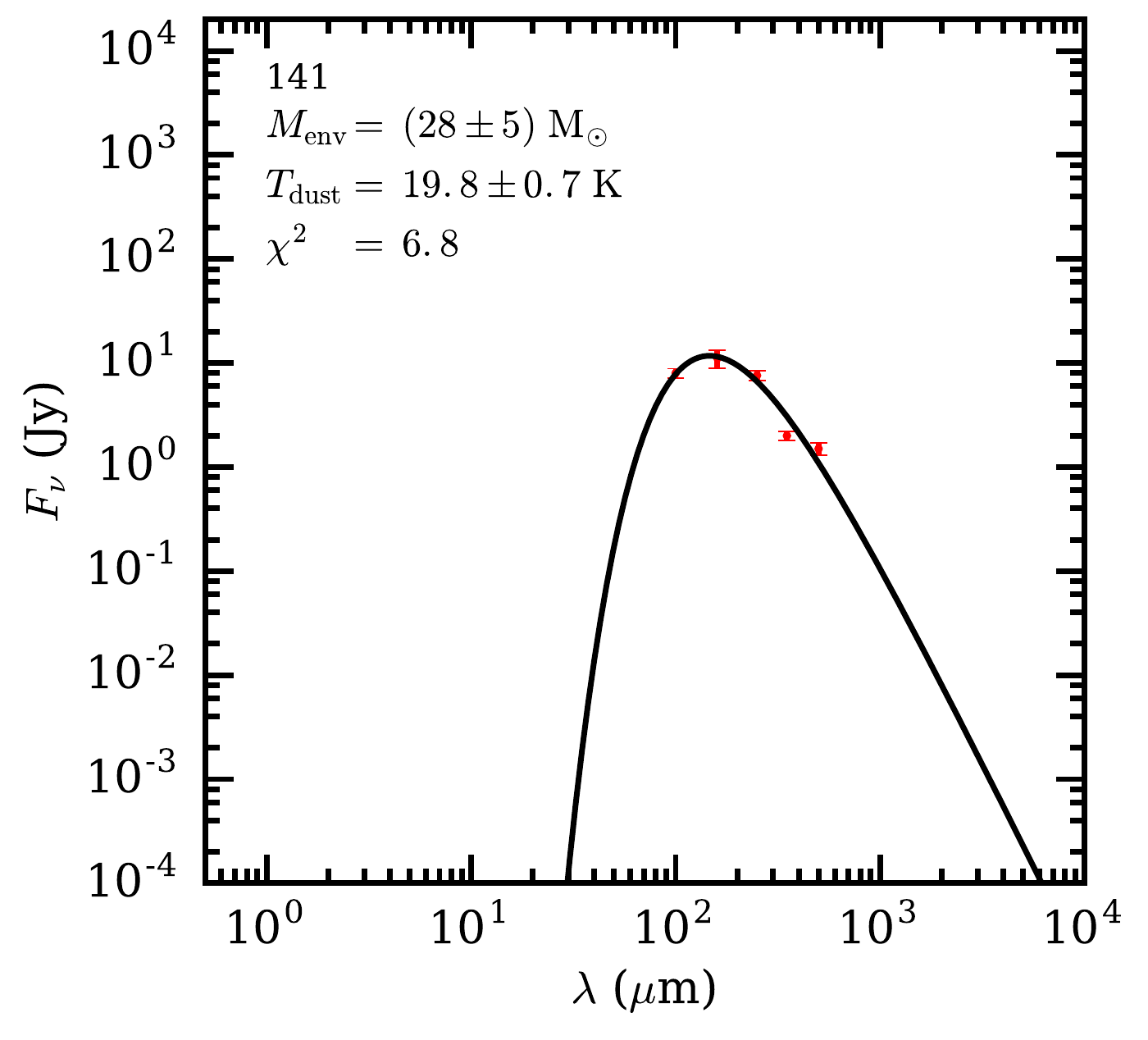}
}
\begin{flushleft}
\caption{ -- continued.}
\end{flushleft}
\end{figure*}
\begin{figure*}[!ht]
\ContinuedFloat
\centering
\subfloat{\includegraphics[width=3.4 in]{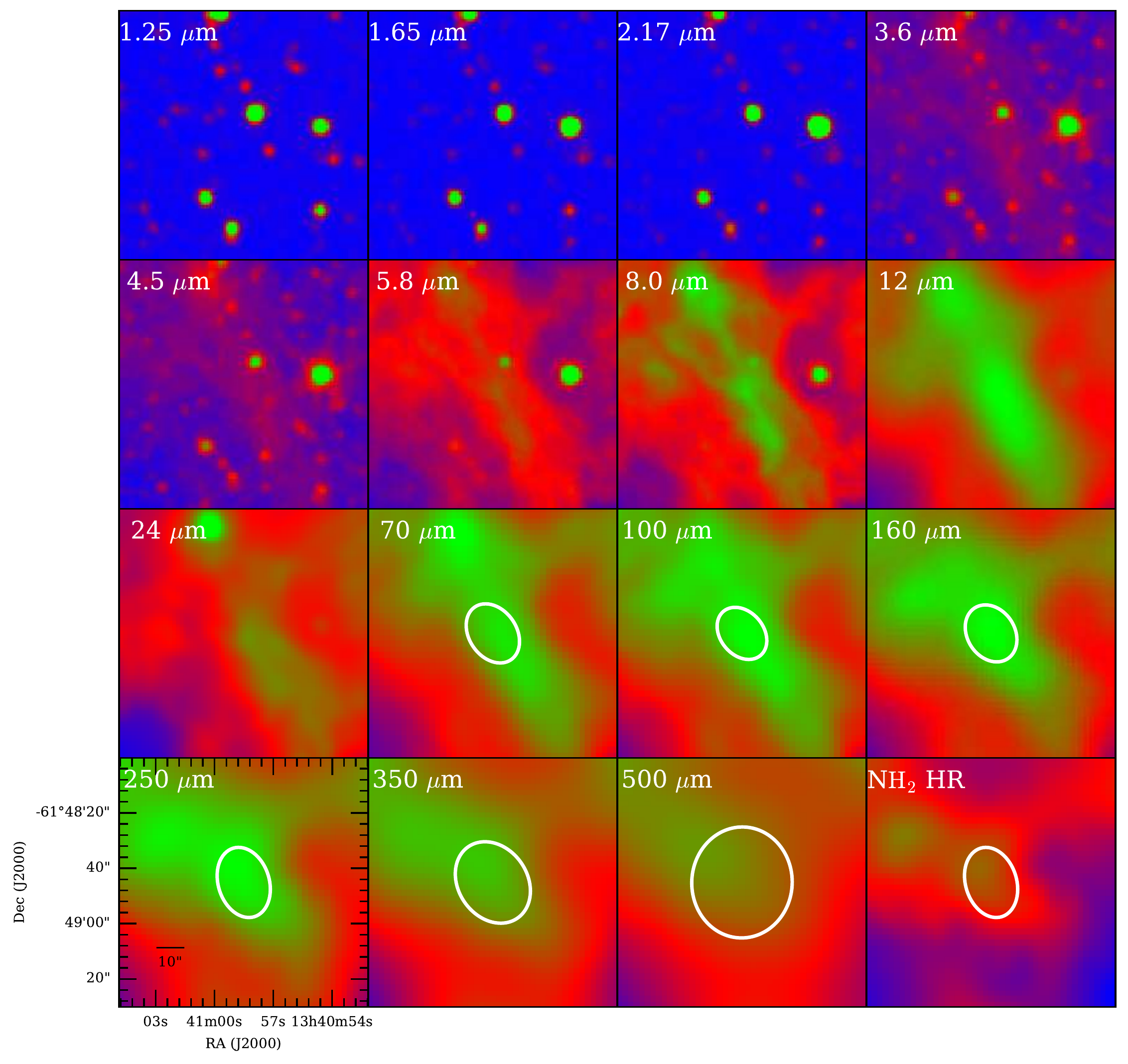}
\hskip 0.00000001cm
\includegraphics[width=3.4 in]{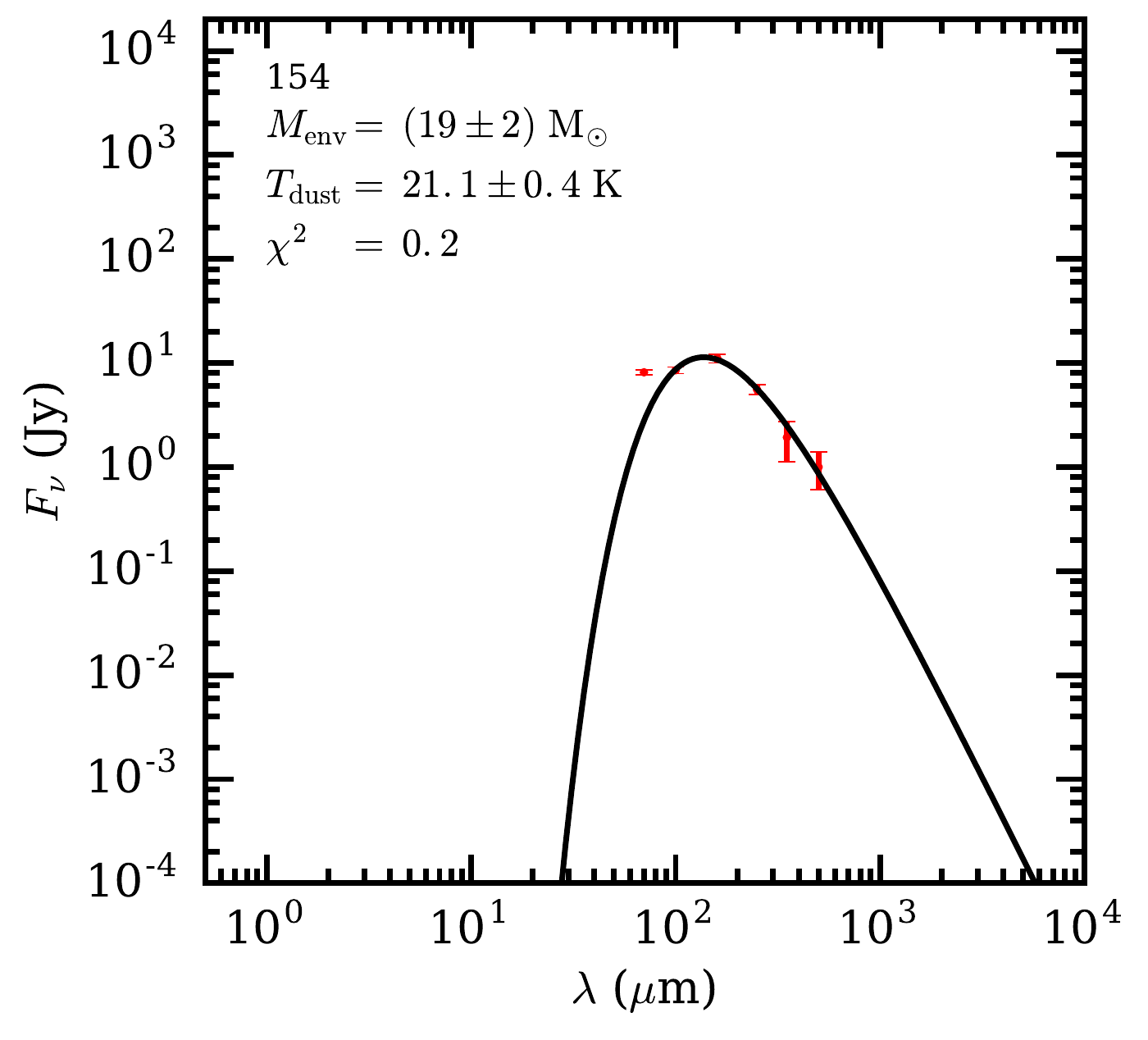}
}
\qquad
\subfloat{\includegraphics[width=3.4 in]{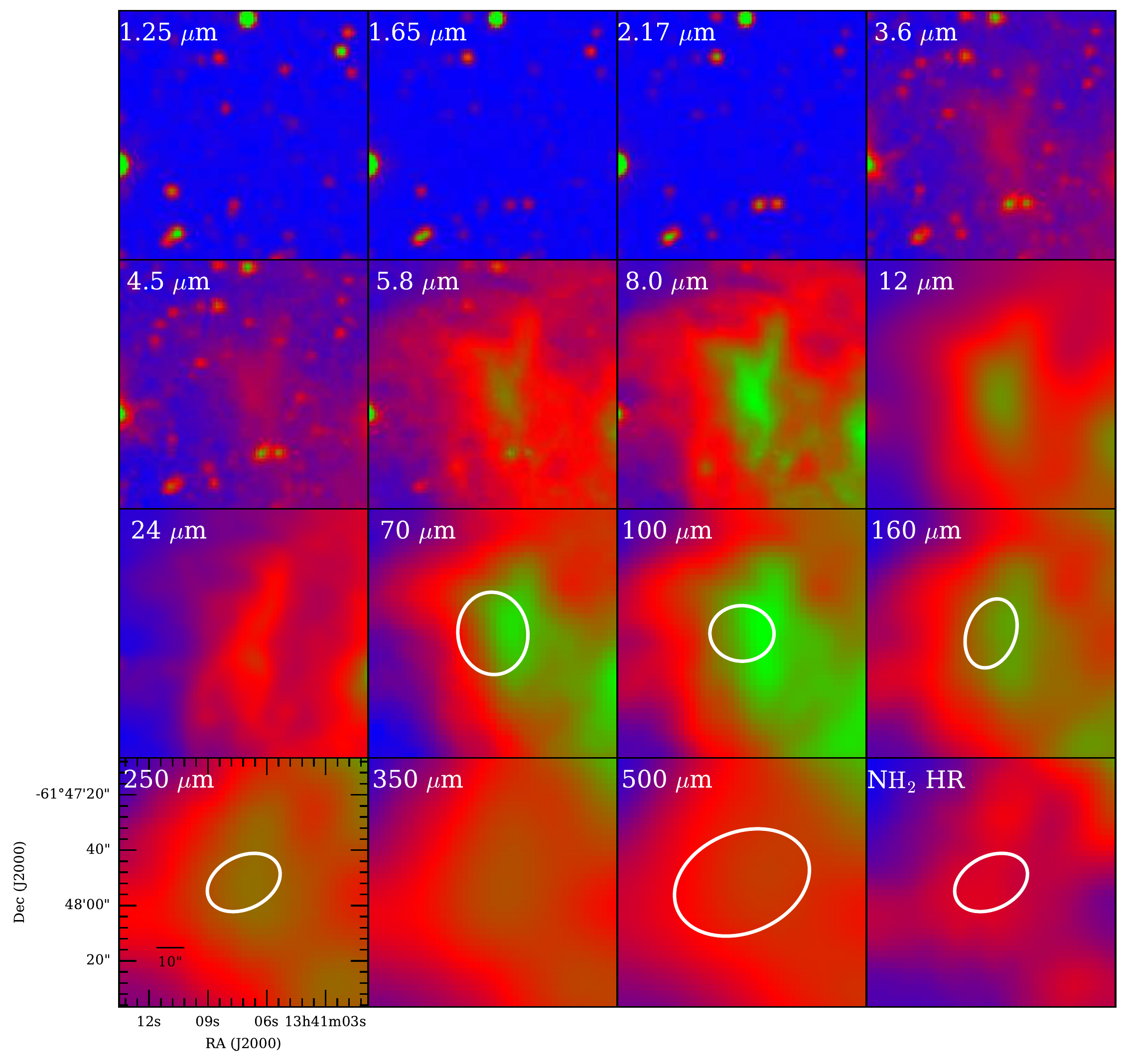}
\hskip 0.00000001cm
\includegraphics[width=3.4 in]{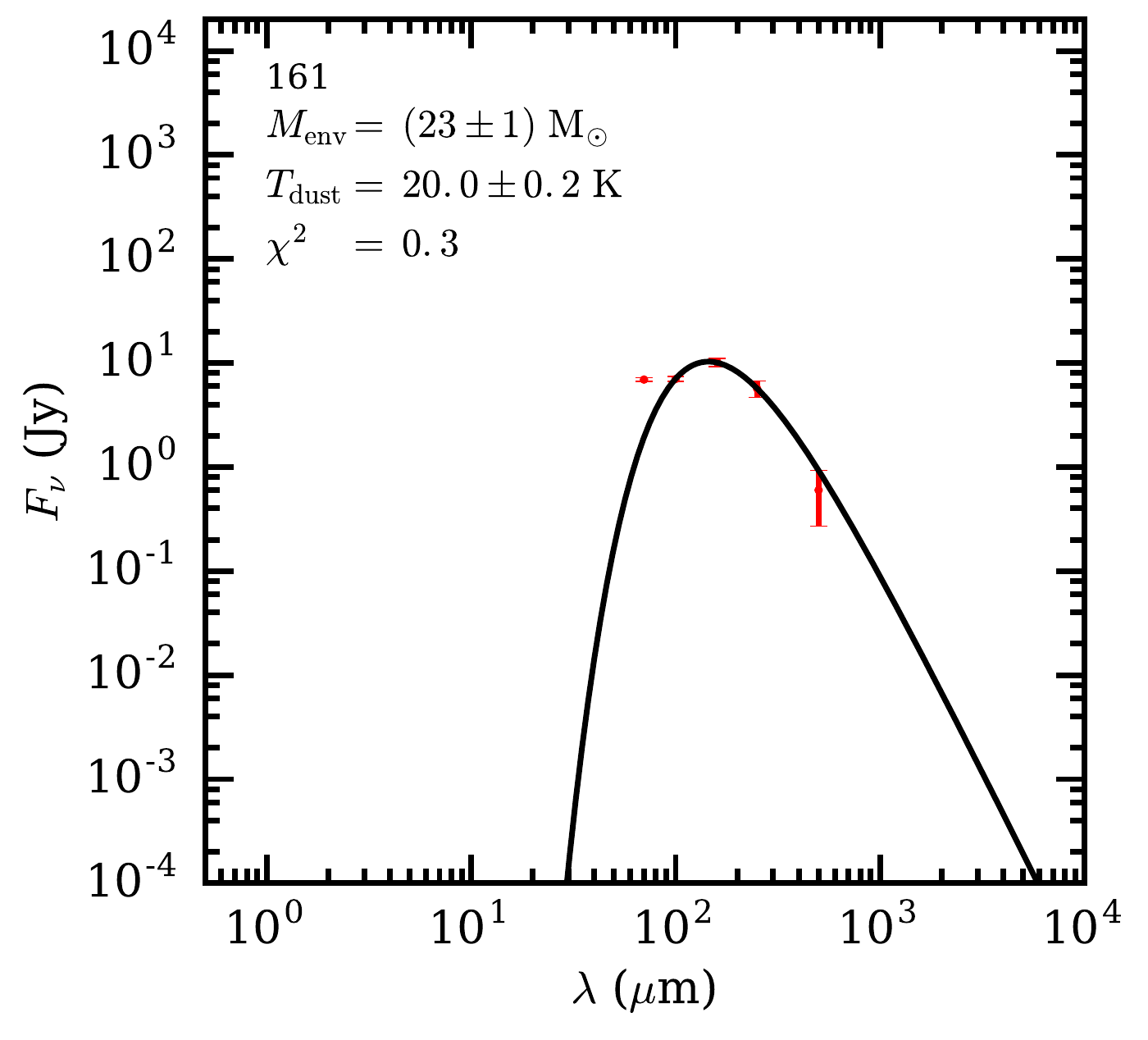}
}
\begin{flushleft}
\caption{ -- continued.}
\end{flushleft}
\end{figure*}
\begin{figure*}[!ht]
\ContinuedFloat
\centering
\subfloat{\includegraphics[width=3.4 in]{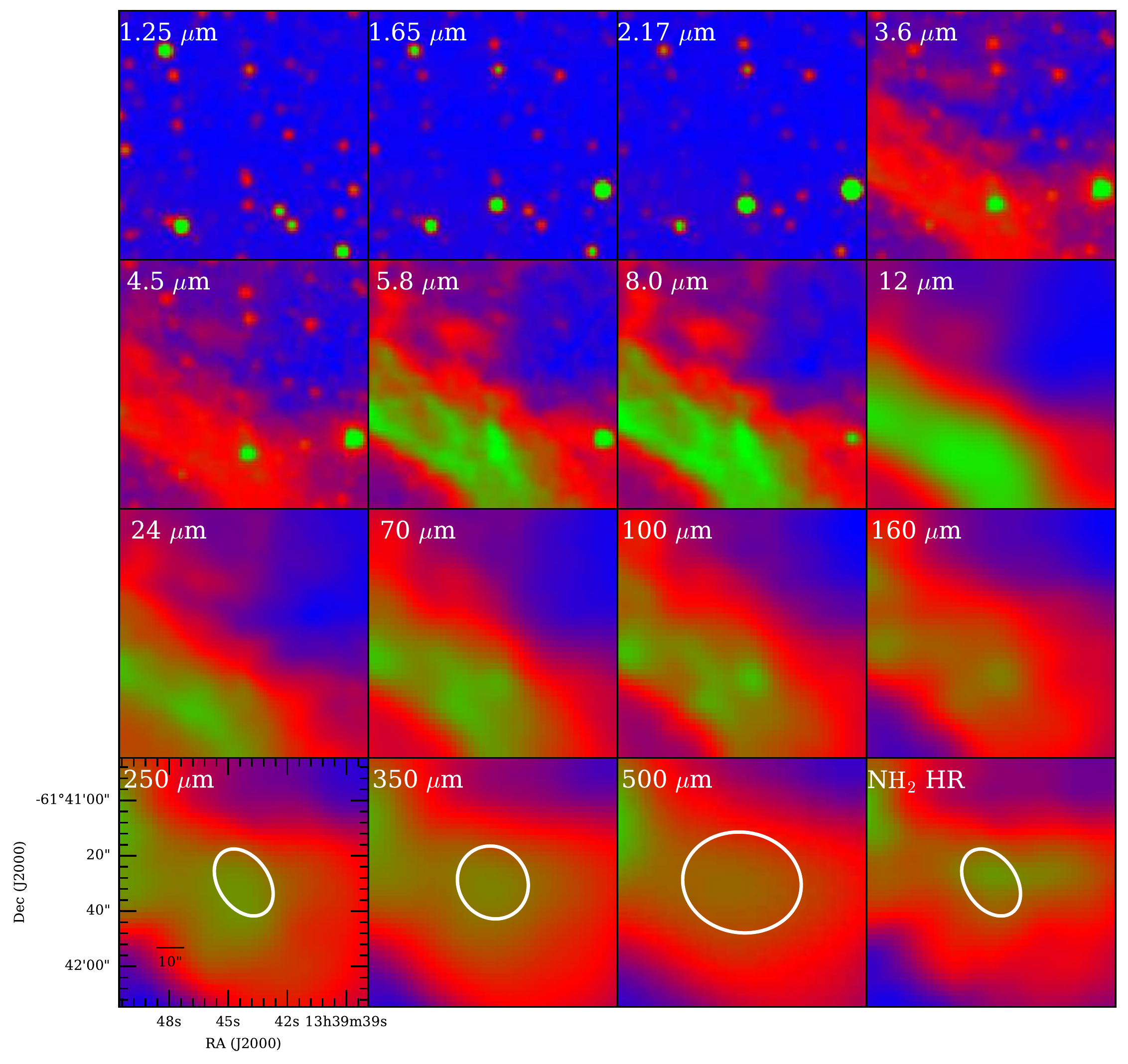}
\hskip 0.00000001cm
\includegraphics[width=3.4 in]{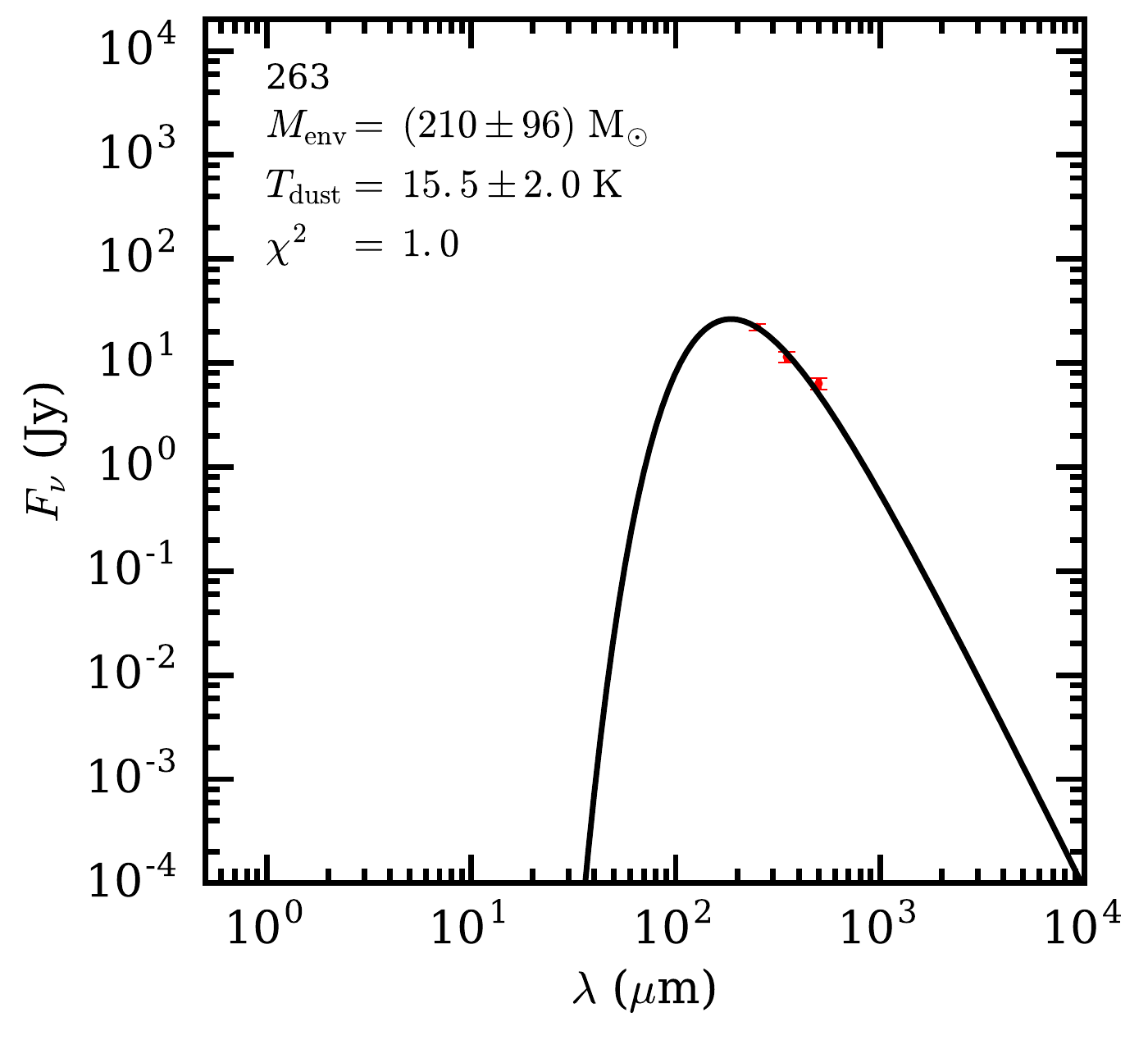}
}
\qquad
\subfloat{\includegraphics[width=3.4 in]{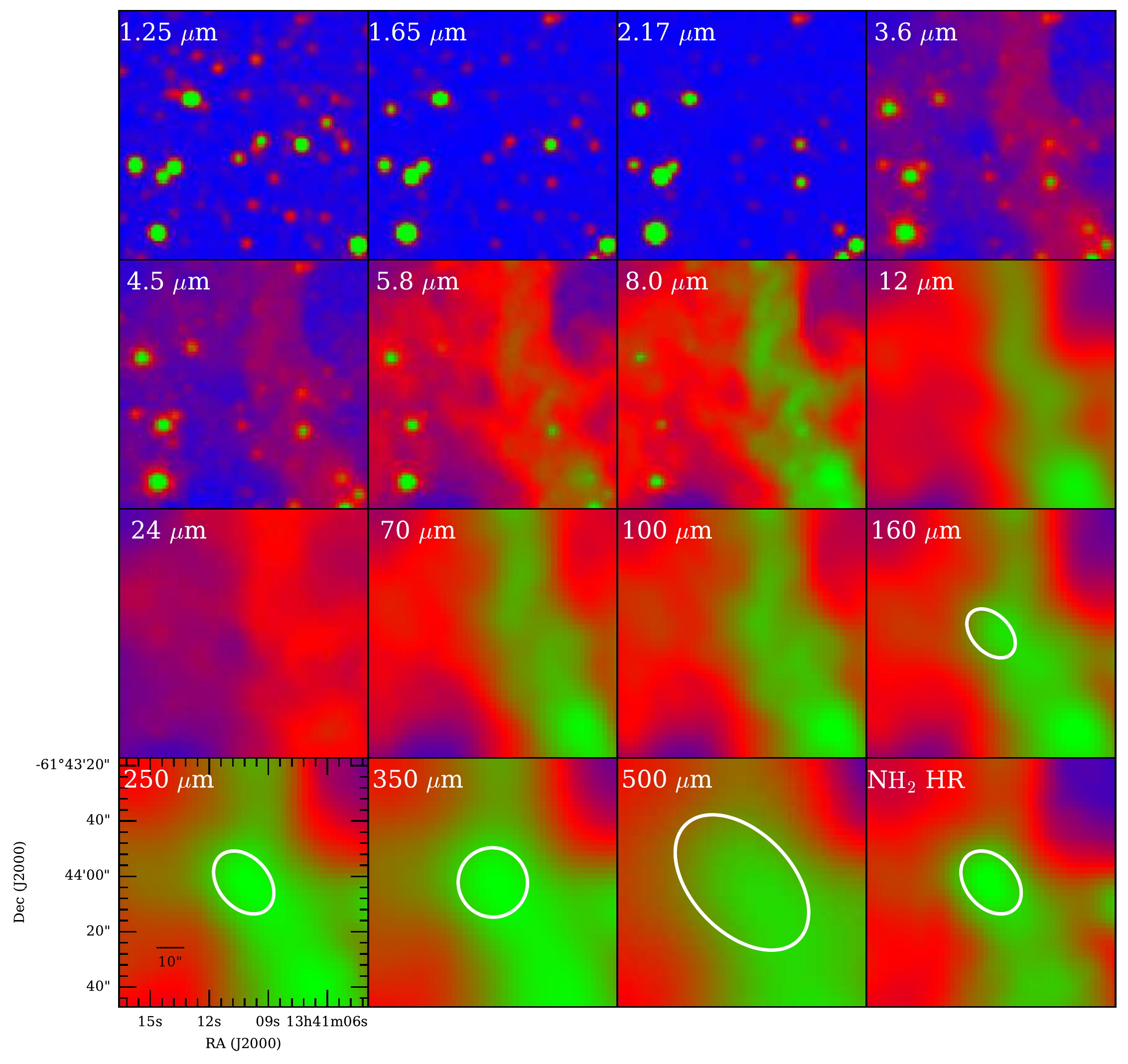}
\hskip 0.00000001cm
\includegraphics[width=3.4 in]{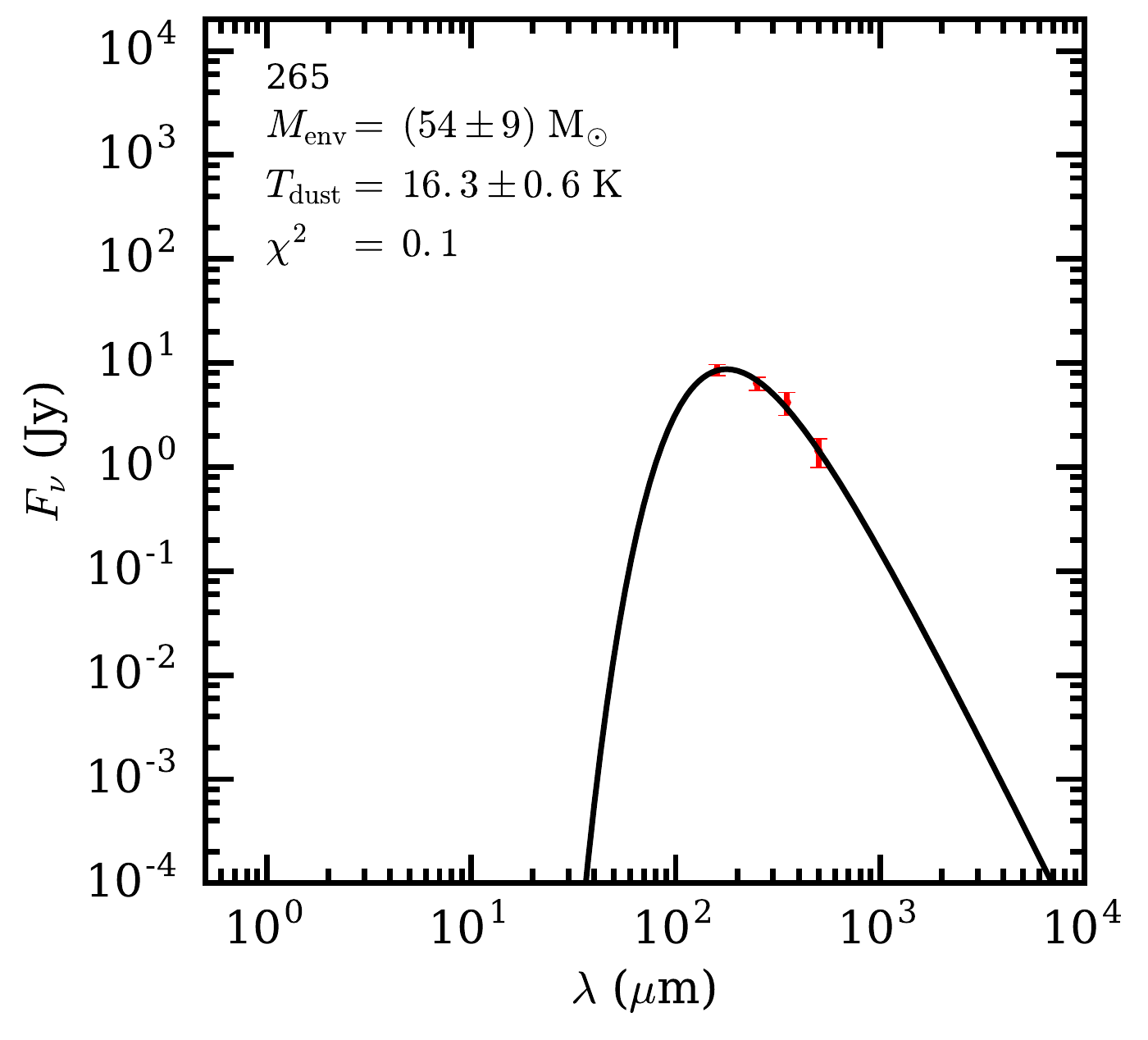}
}
\begin{flushleft}
\caption{ -- continued.}
\end{flushleft}
\end{figure*}
\begin{figure*}[!ht]
\ContinuedFloat
\centering
\subfloat{\includegraphics[width=3.4 in]{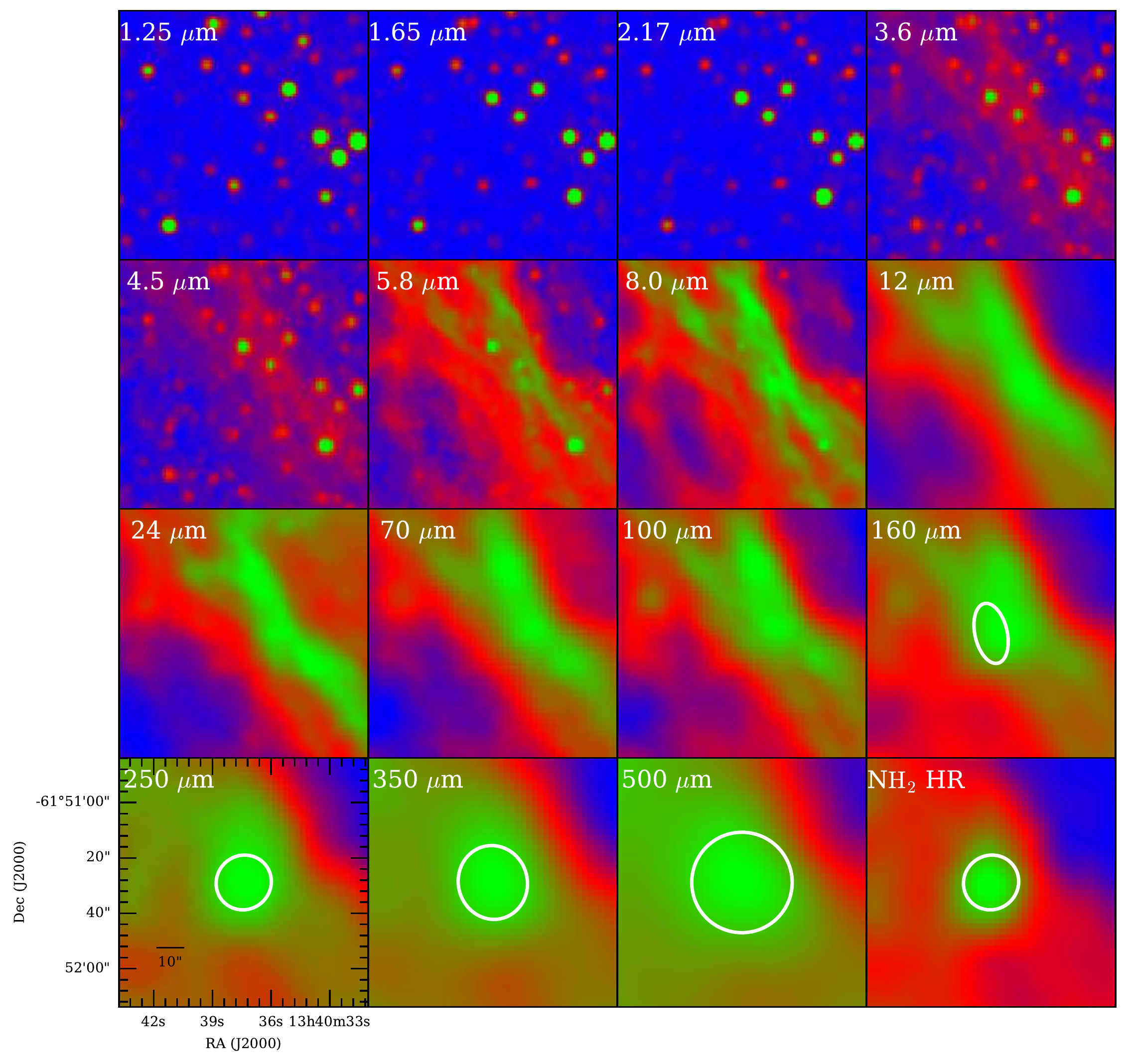}
\hskip 0.00000001cm
\includegraphics[width=3.4 in]{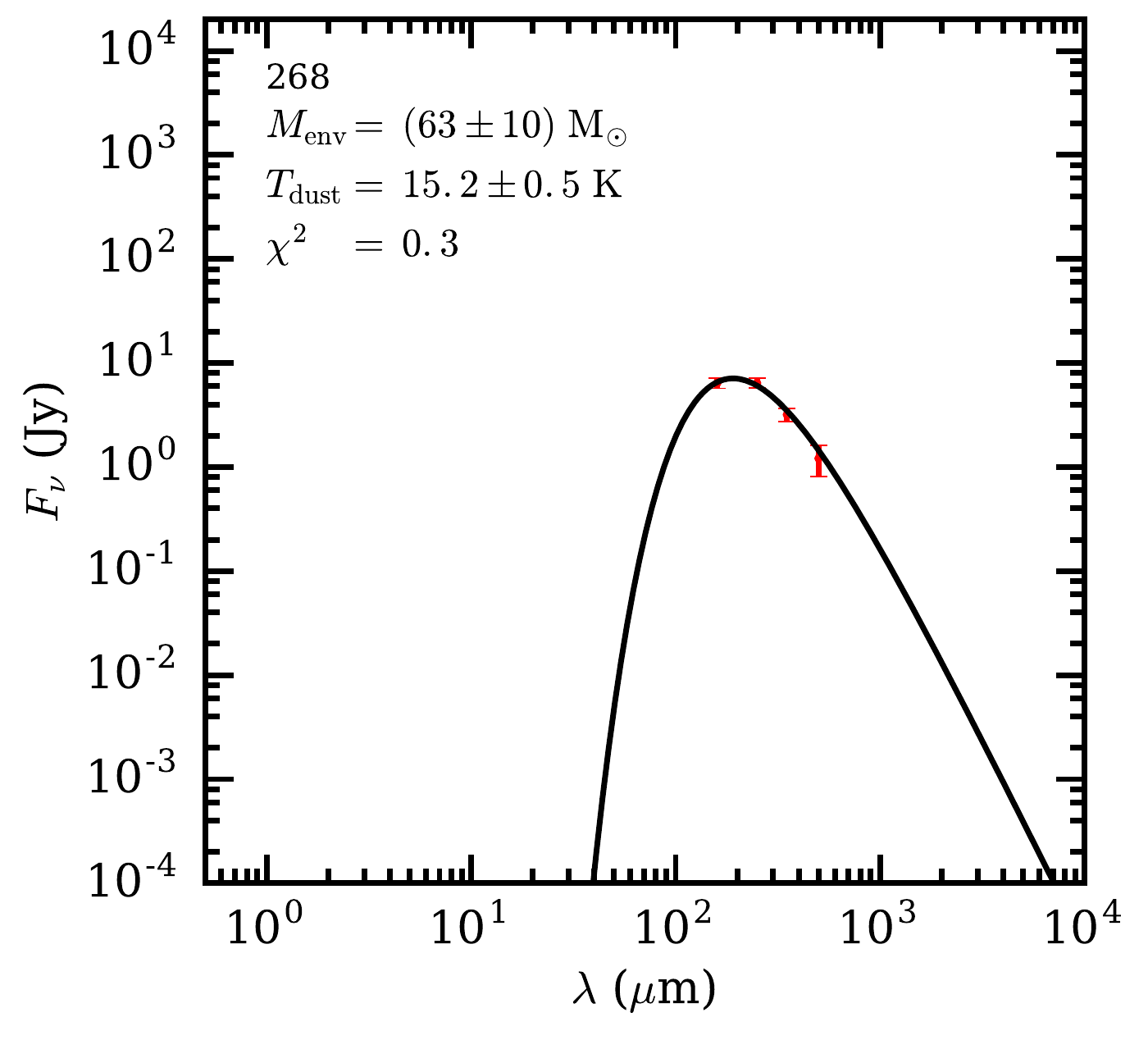}
}
\qquad
\subfloat{\includegraphics[width=3.4 in]{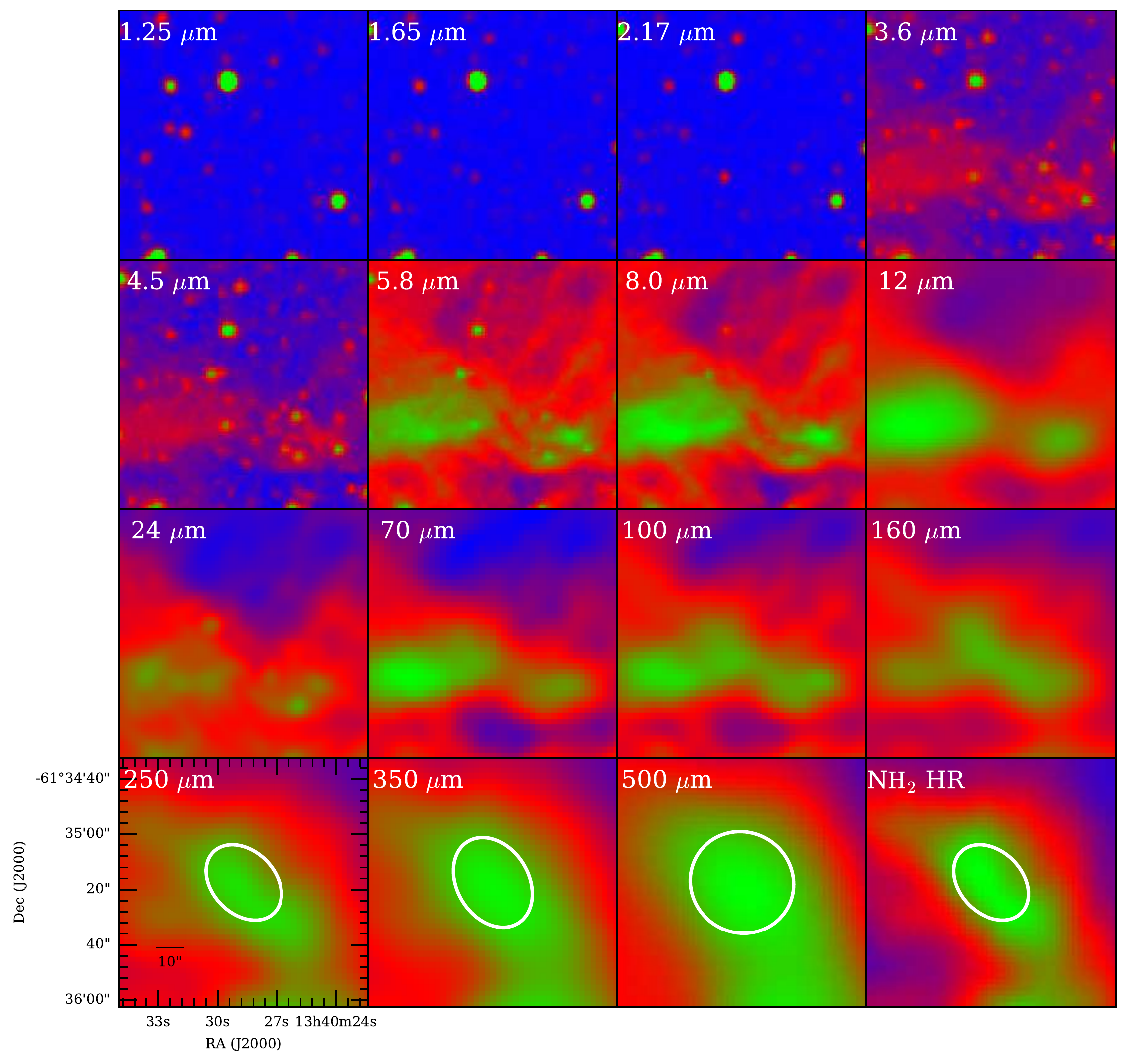}
\hskip 0.00000001cm
\includegraphics[width=3.4 in]{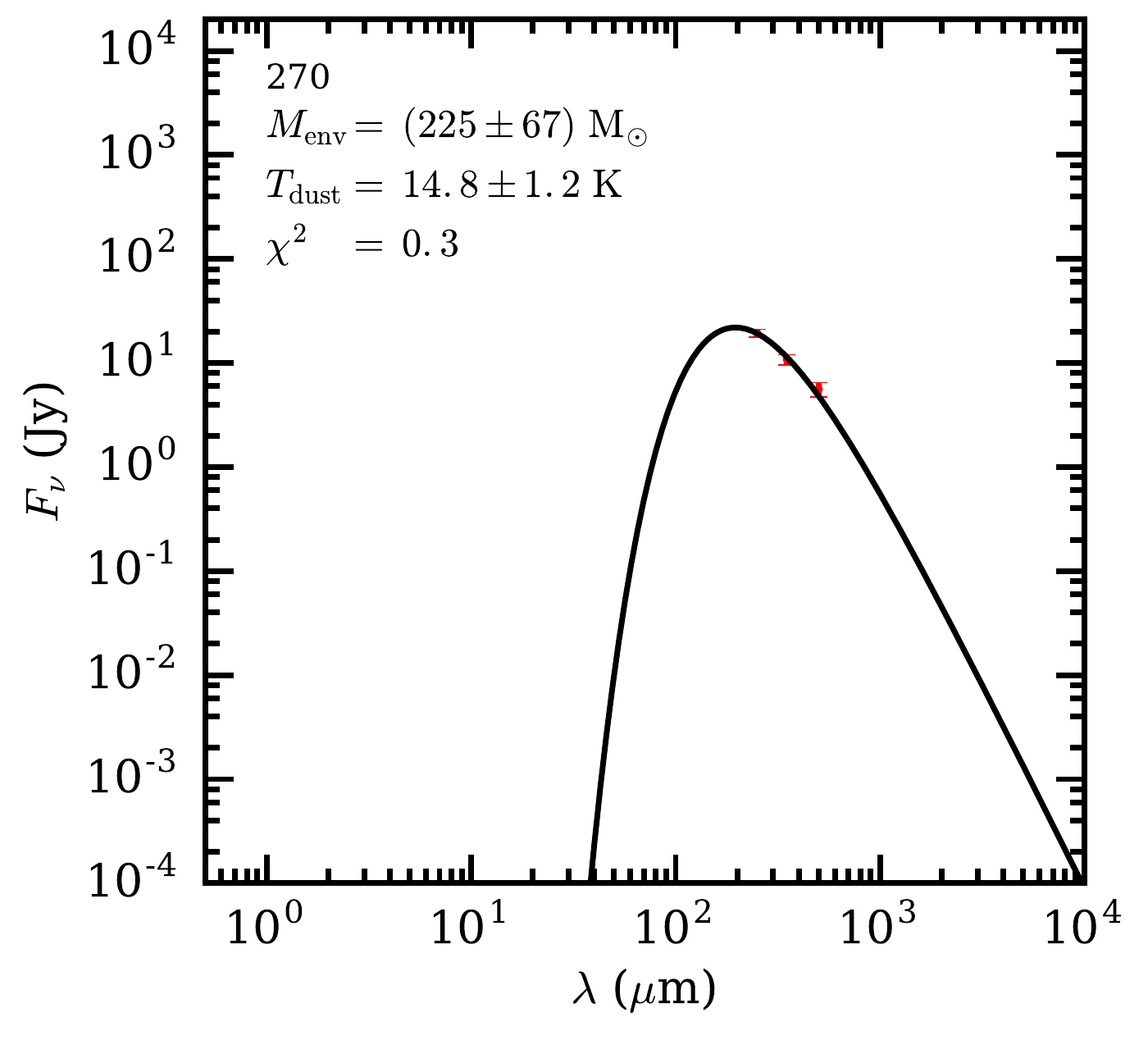}
}
\begin{flushleft}
\caption{ -- continued.}
\end{flushleft}
\end{figure*}
\begin{figure*}[!ht]
\ContinuedFloat
\centering
\subfloat{\includegraphics[width=3.4 in]{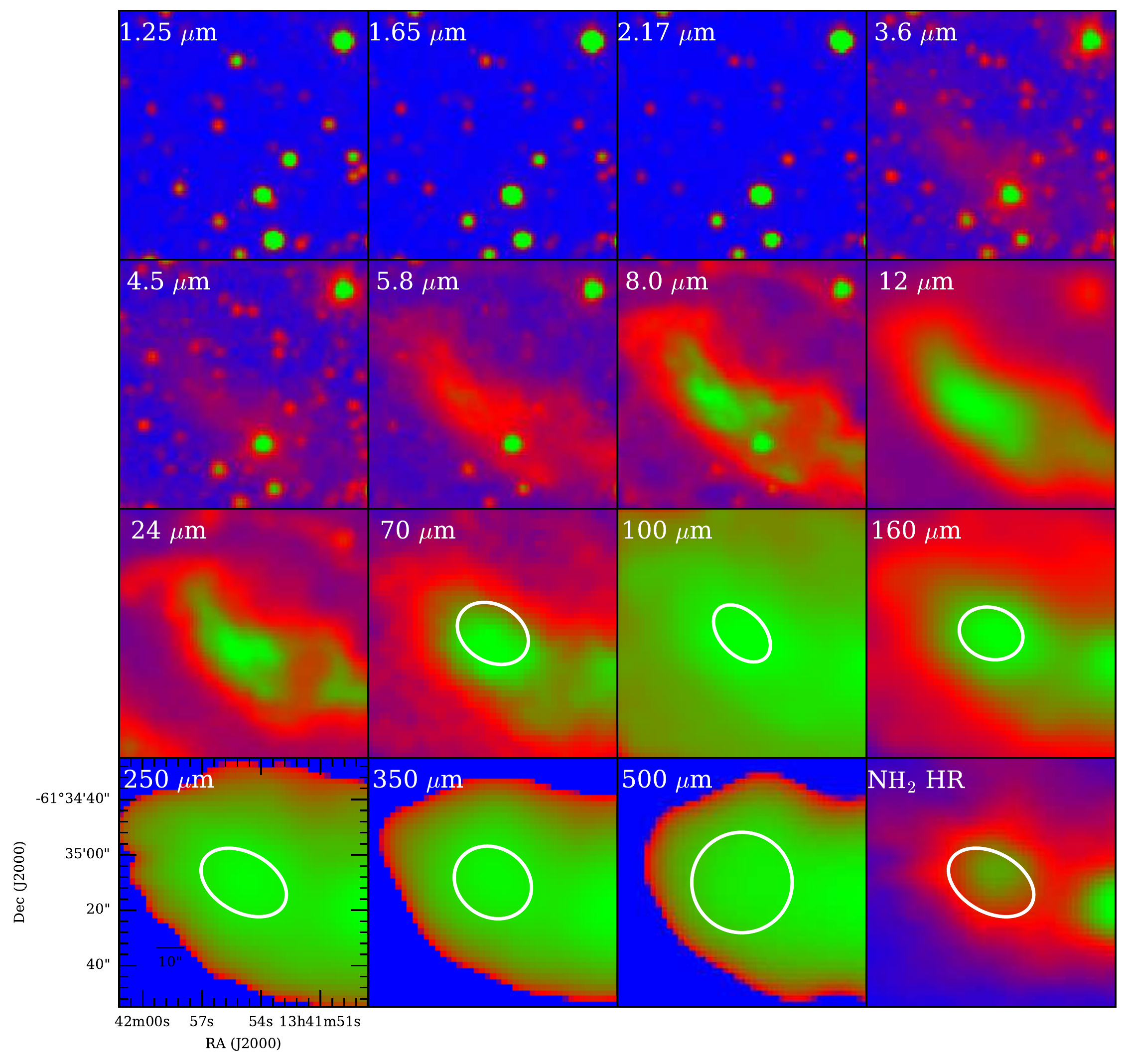}
\hskip 0.00000001cm
\includegraphics[width=3.4 in]{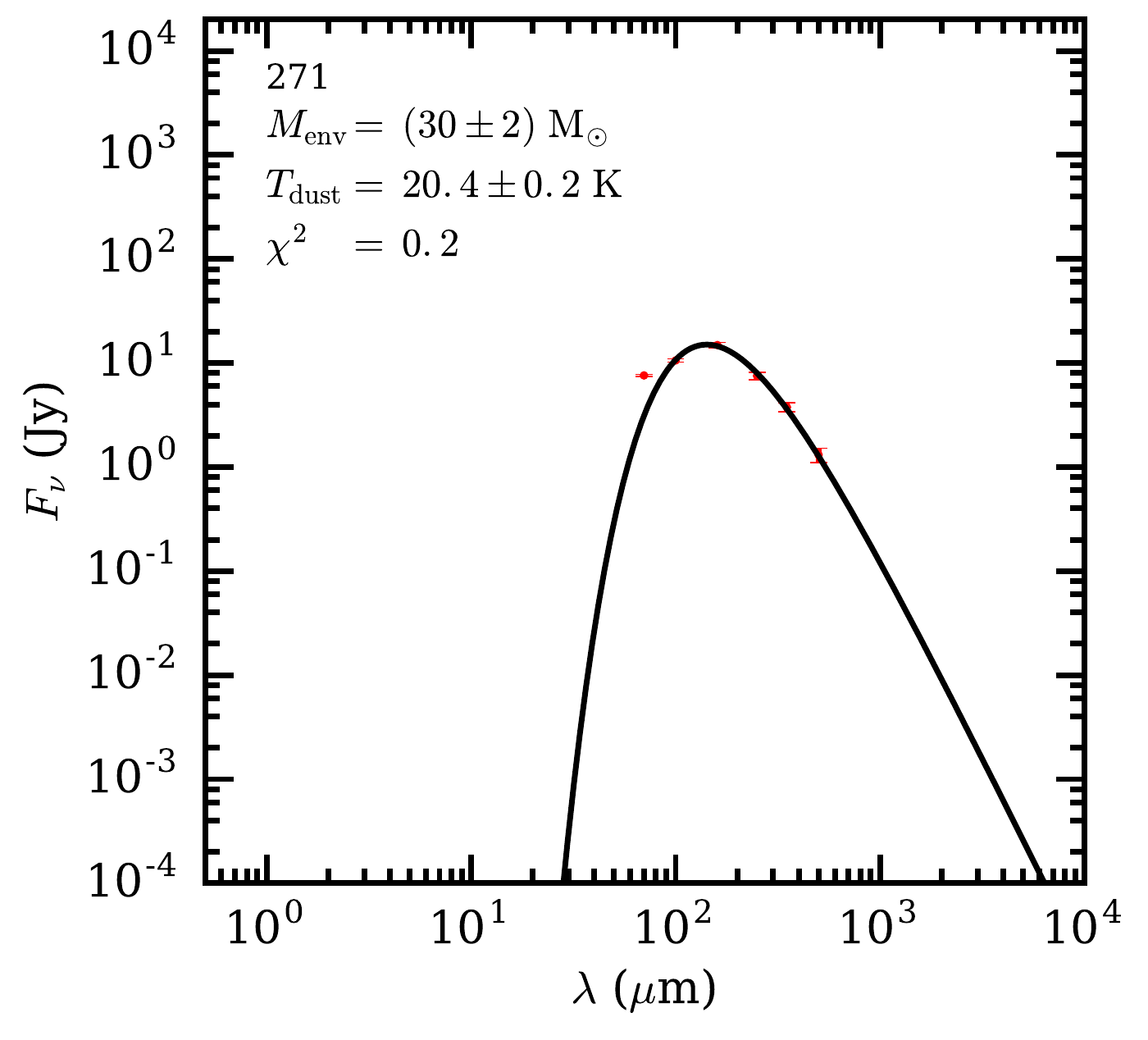}
}
\qquad
\subfloat{\includegraphics[width=3.4 in]{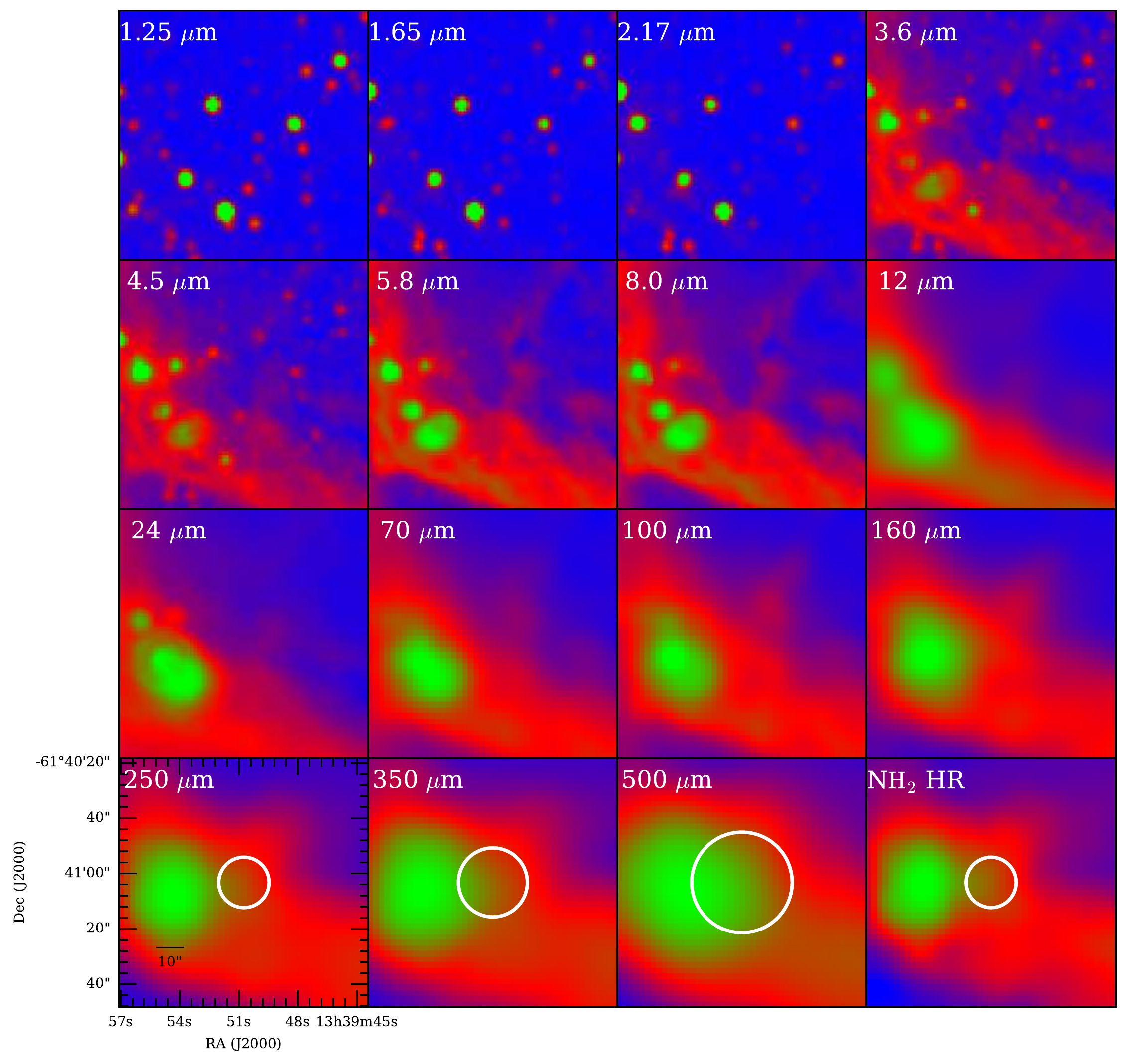}
\hskip 0.00000001cm
\includegraphics[width=3.4 in]{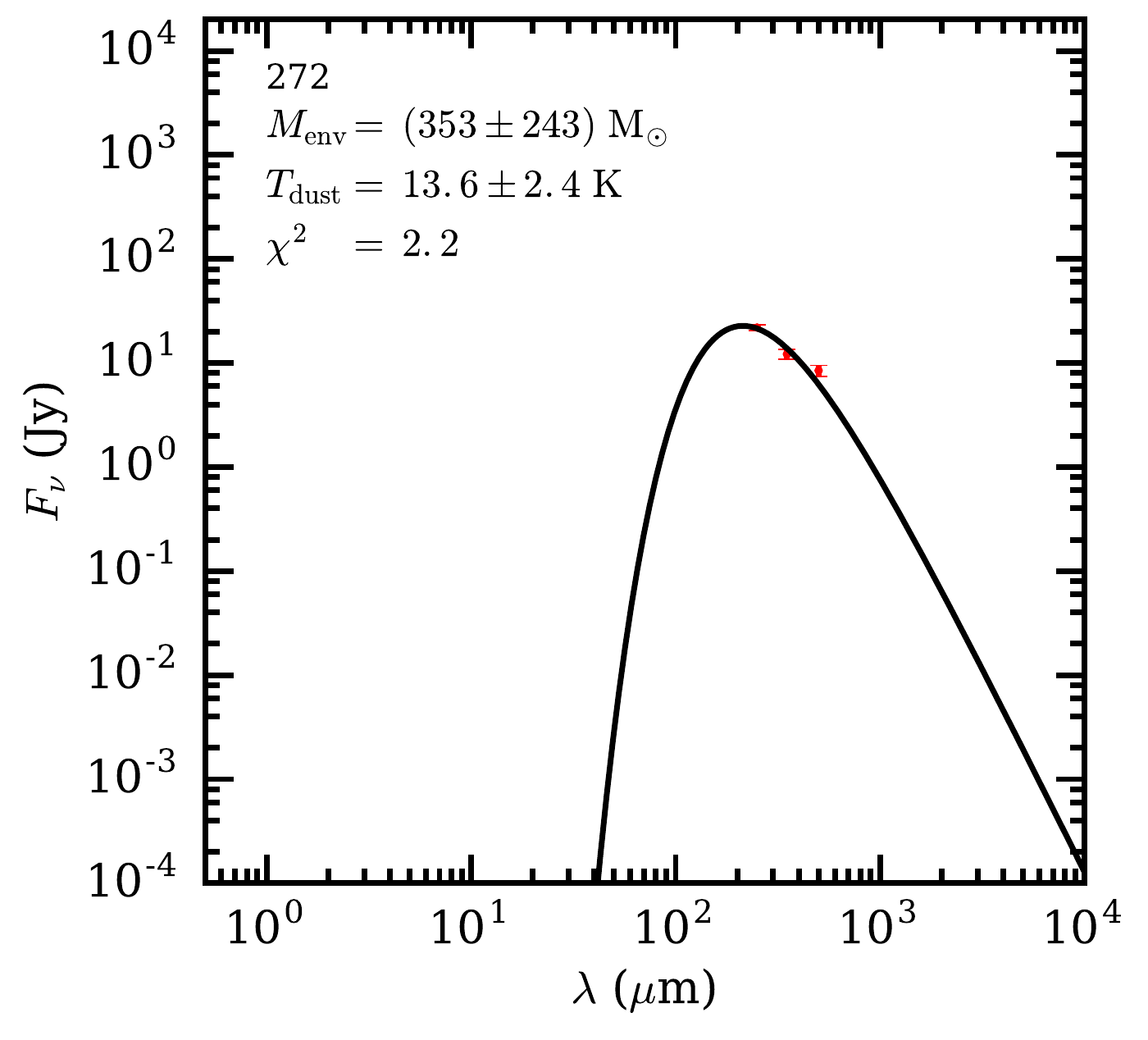}
}
\begin{flushleft}
\caption{ -- continued.}
\end{flushleft}
\end{figure*}
\begin{figure*}[!ht]
\ContinuedFloat
\centering
\subfloat{\includegraphics[width=3.4 in]{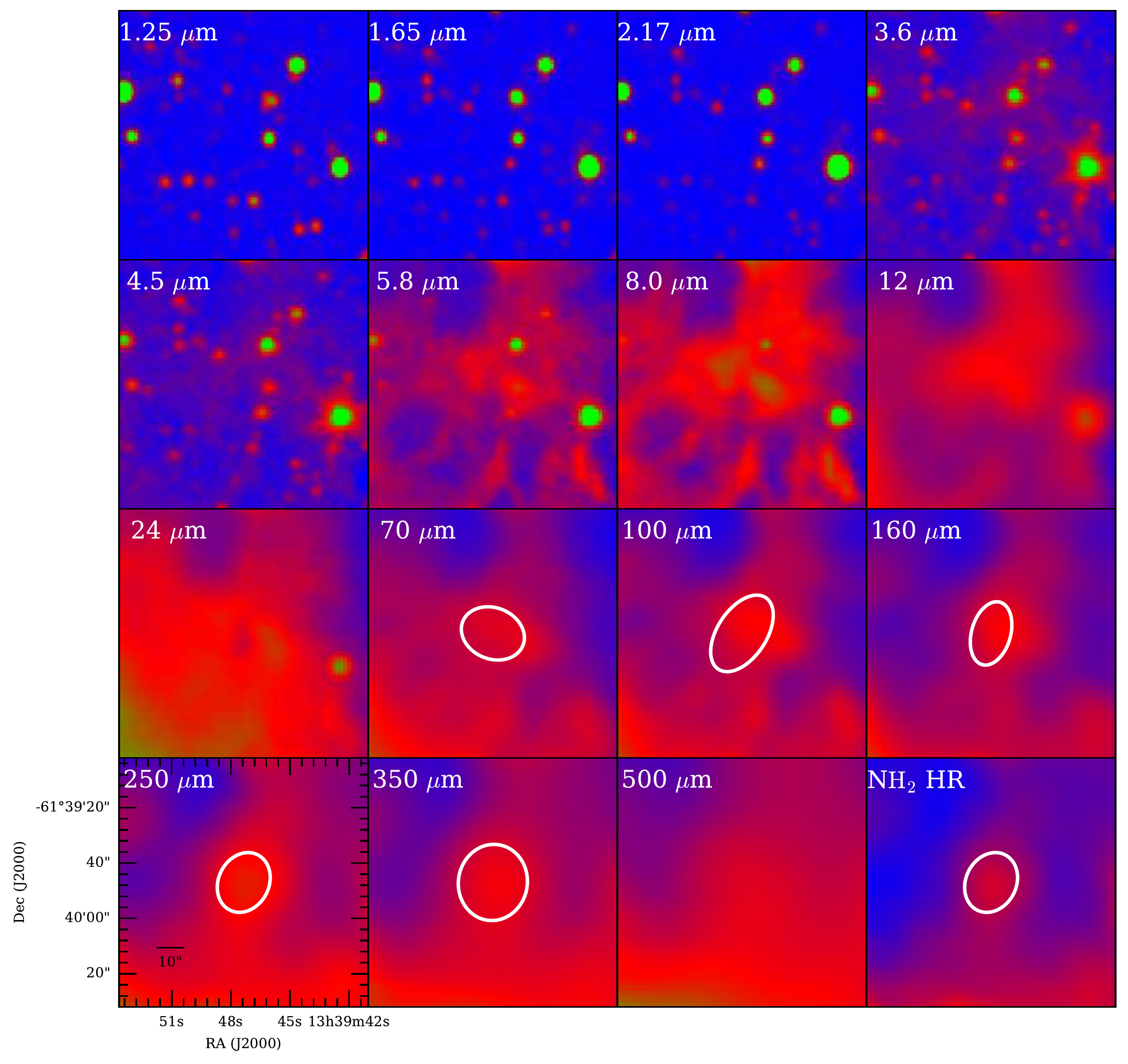}
\hskip 0.00000001cm
\includegraphics[width=3.4 in]{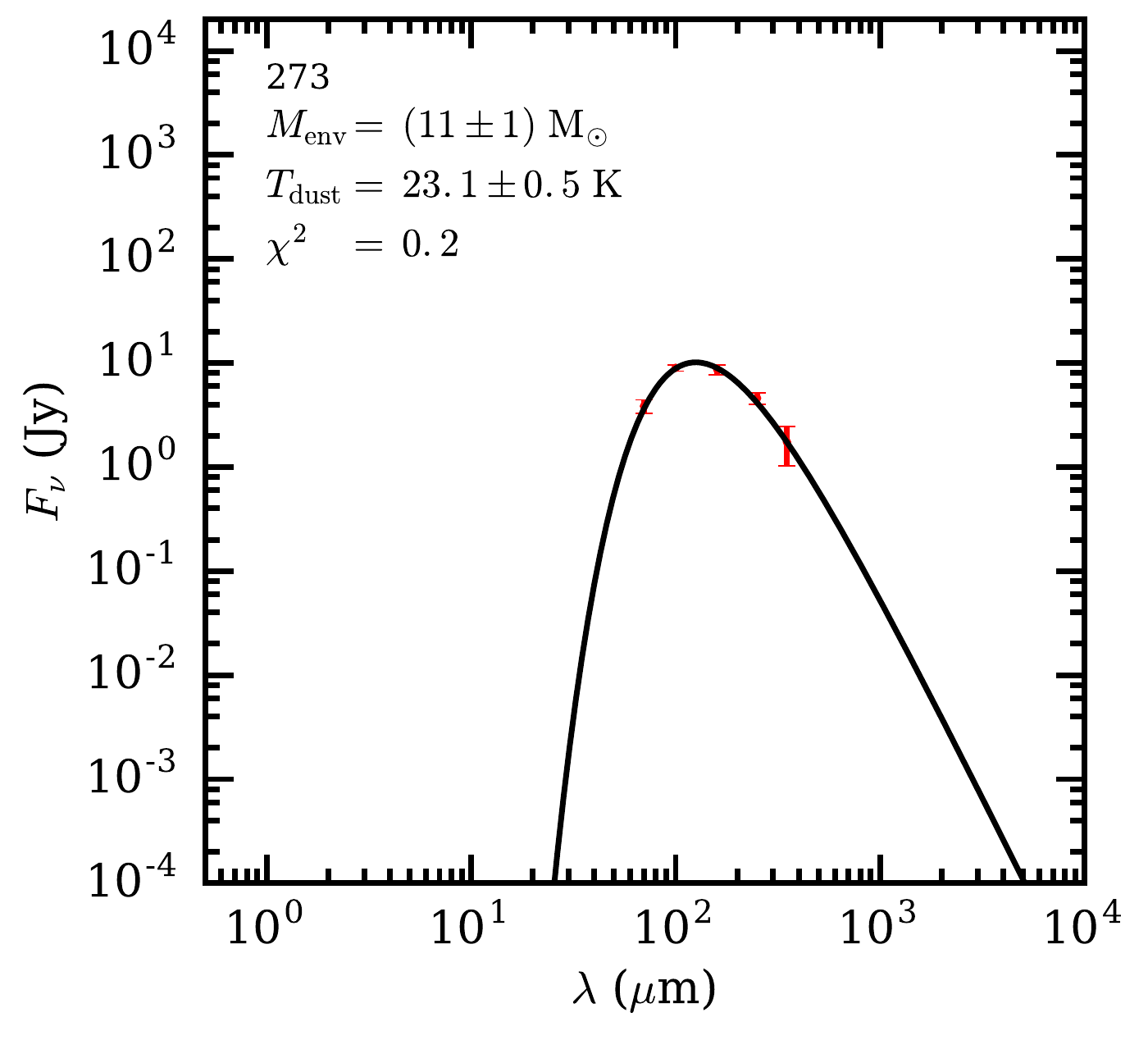}
}
\qquad
\subfloat{\includegraphics[width=3.4 in]{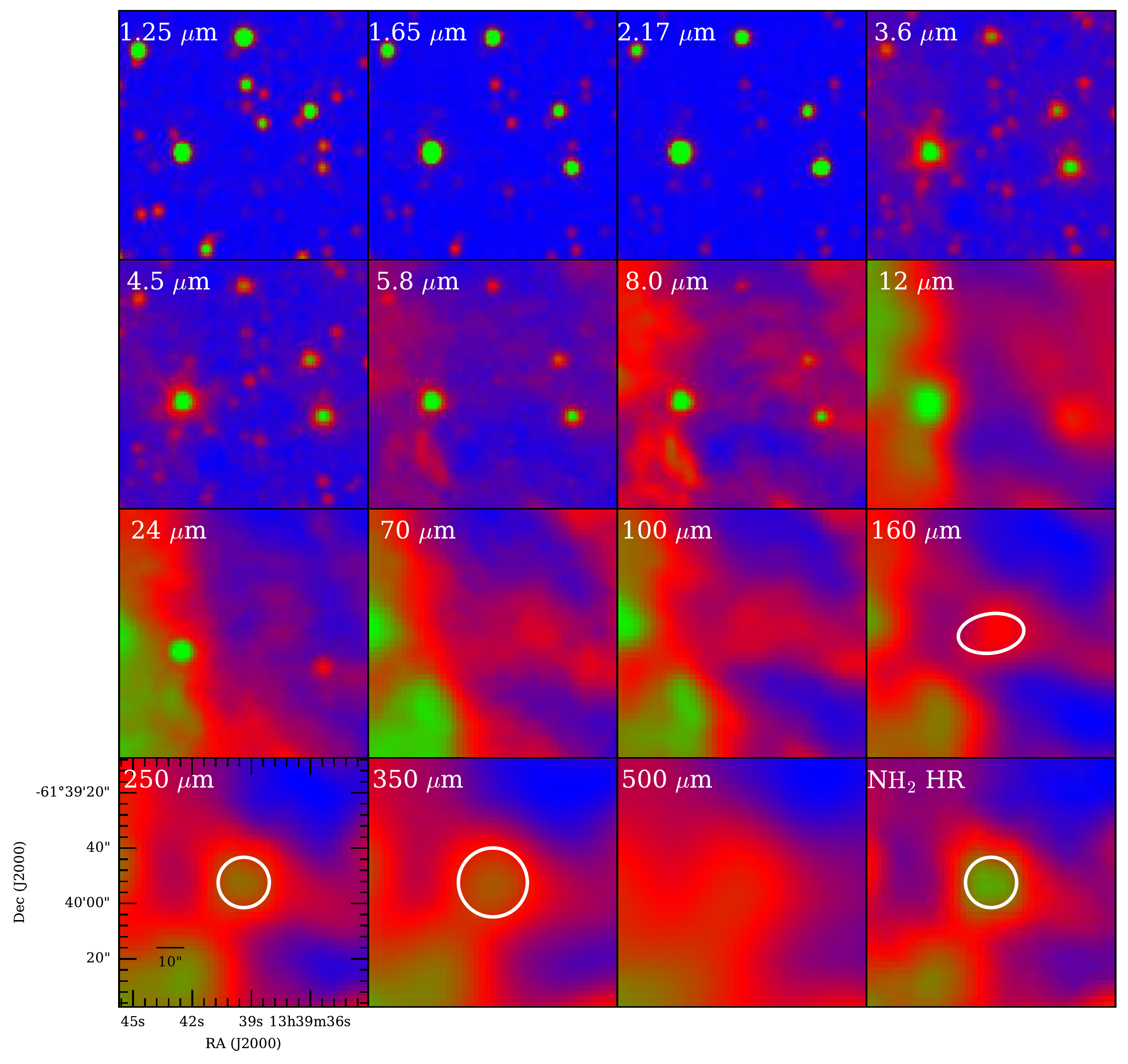}
\hskip 0.00000001cm
\includegraphics[width=3.4 in]{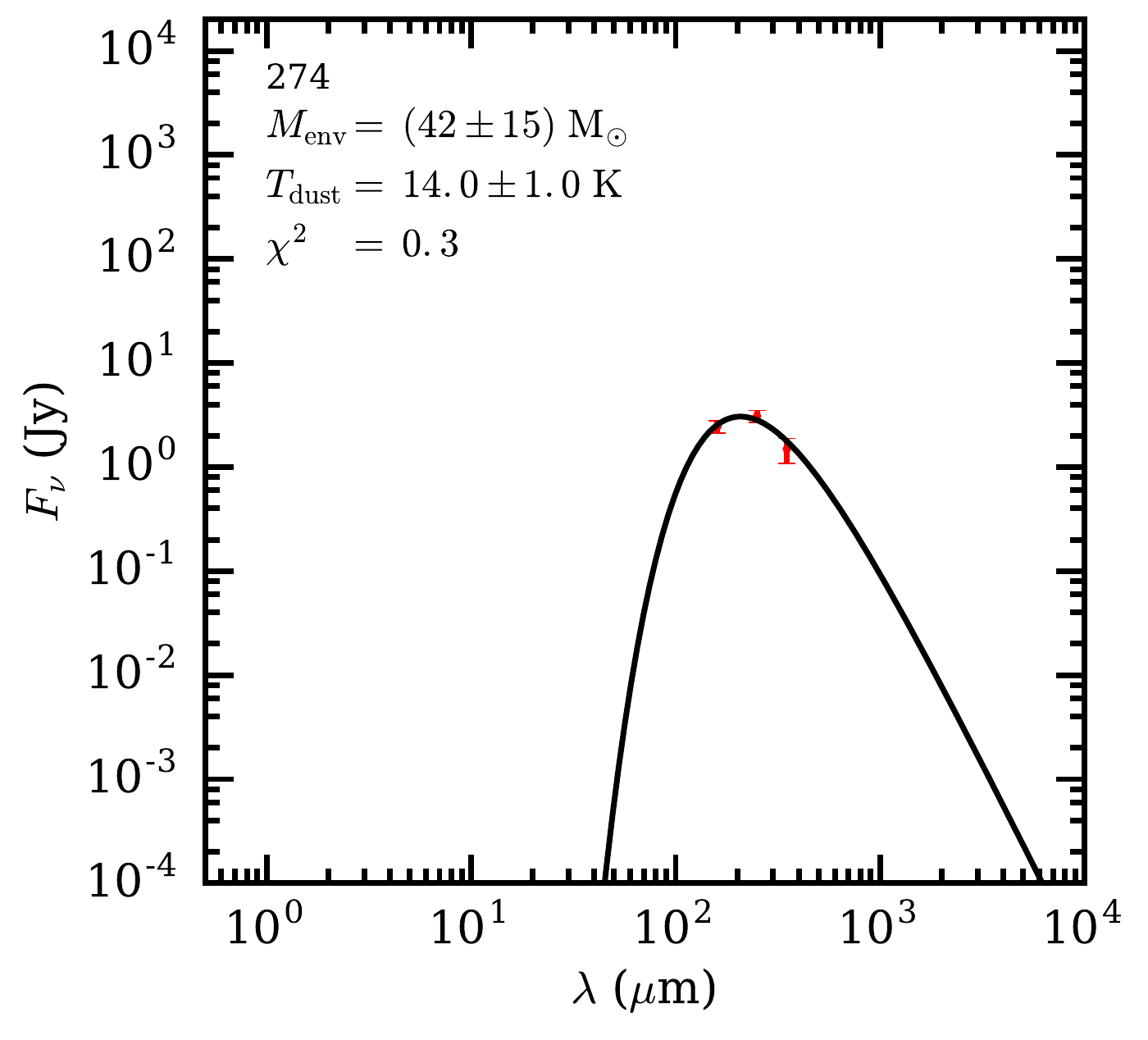}
}
\begin{flushleft}
\caption{ -- continued.}
\end{flushleft}
\end{figure*}
\begin{figure*}[!ht]
\ContinuedFloat
\centering
\subfloat{\includegraphics[width=3.4 in]{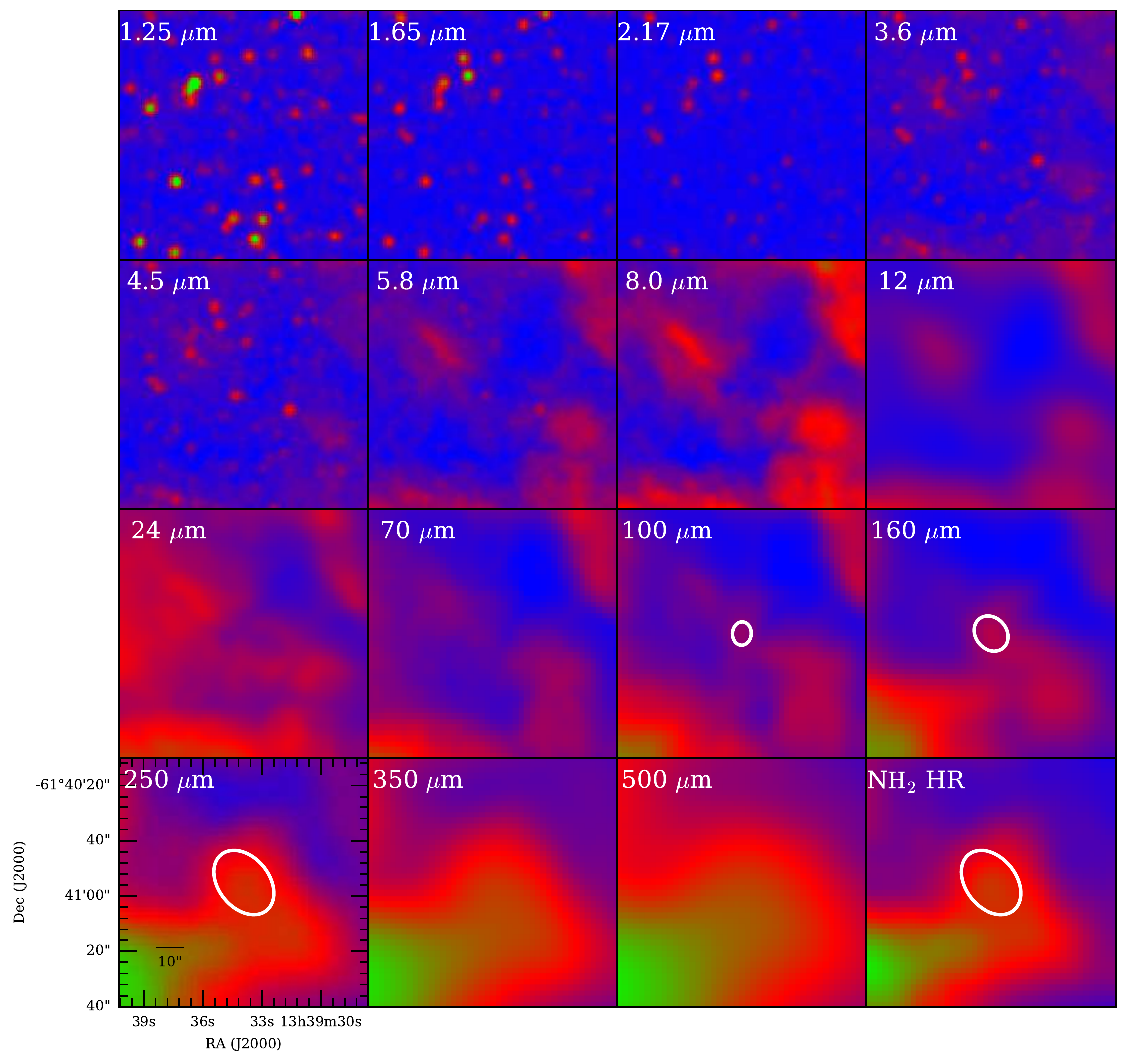}
\hskip 0.00000001cm
\includegraphics[width=3.4 in]{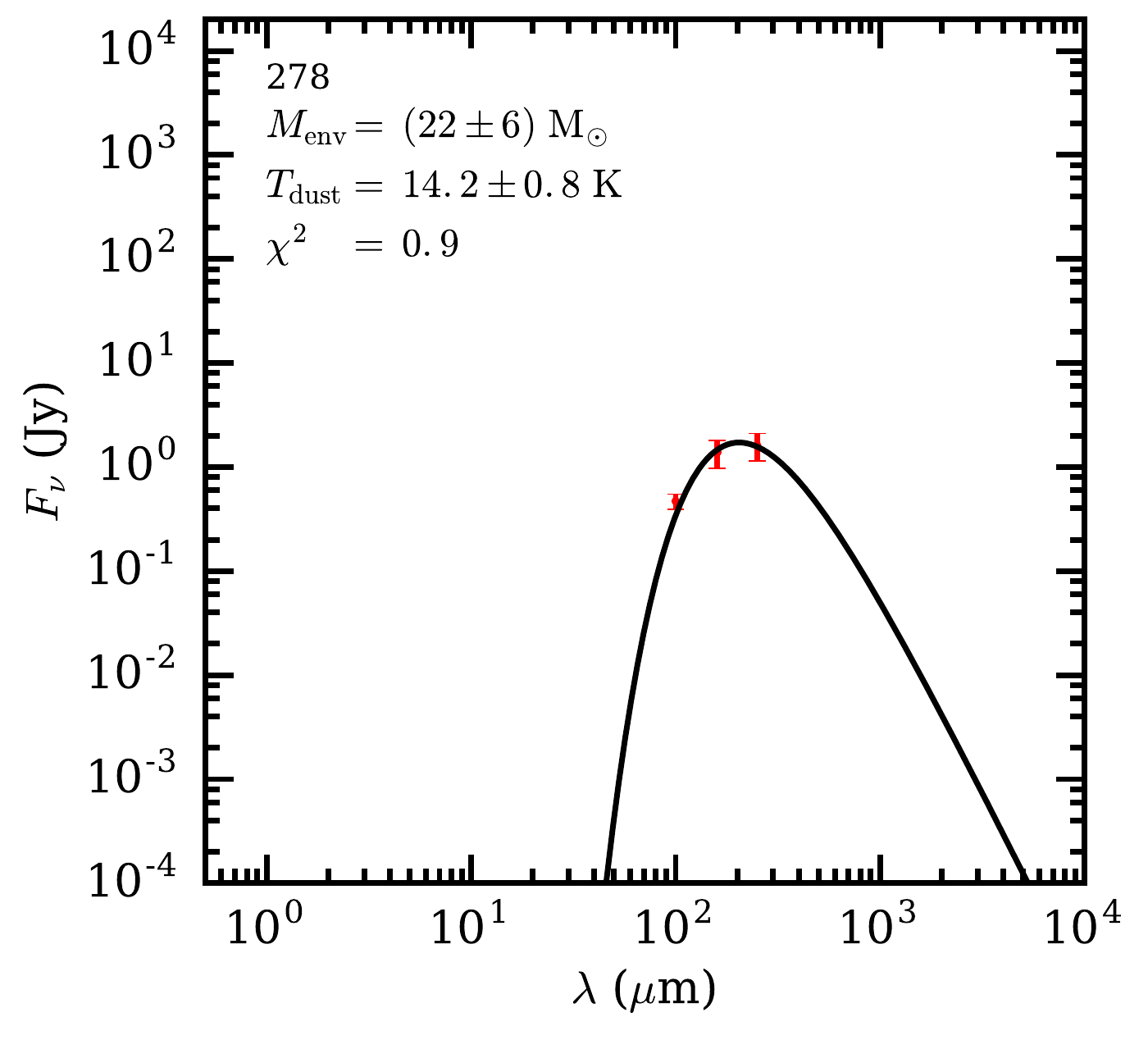}
}
\qquad
\clearpage
\subfloat{\includegraphics[width=3.4 in]{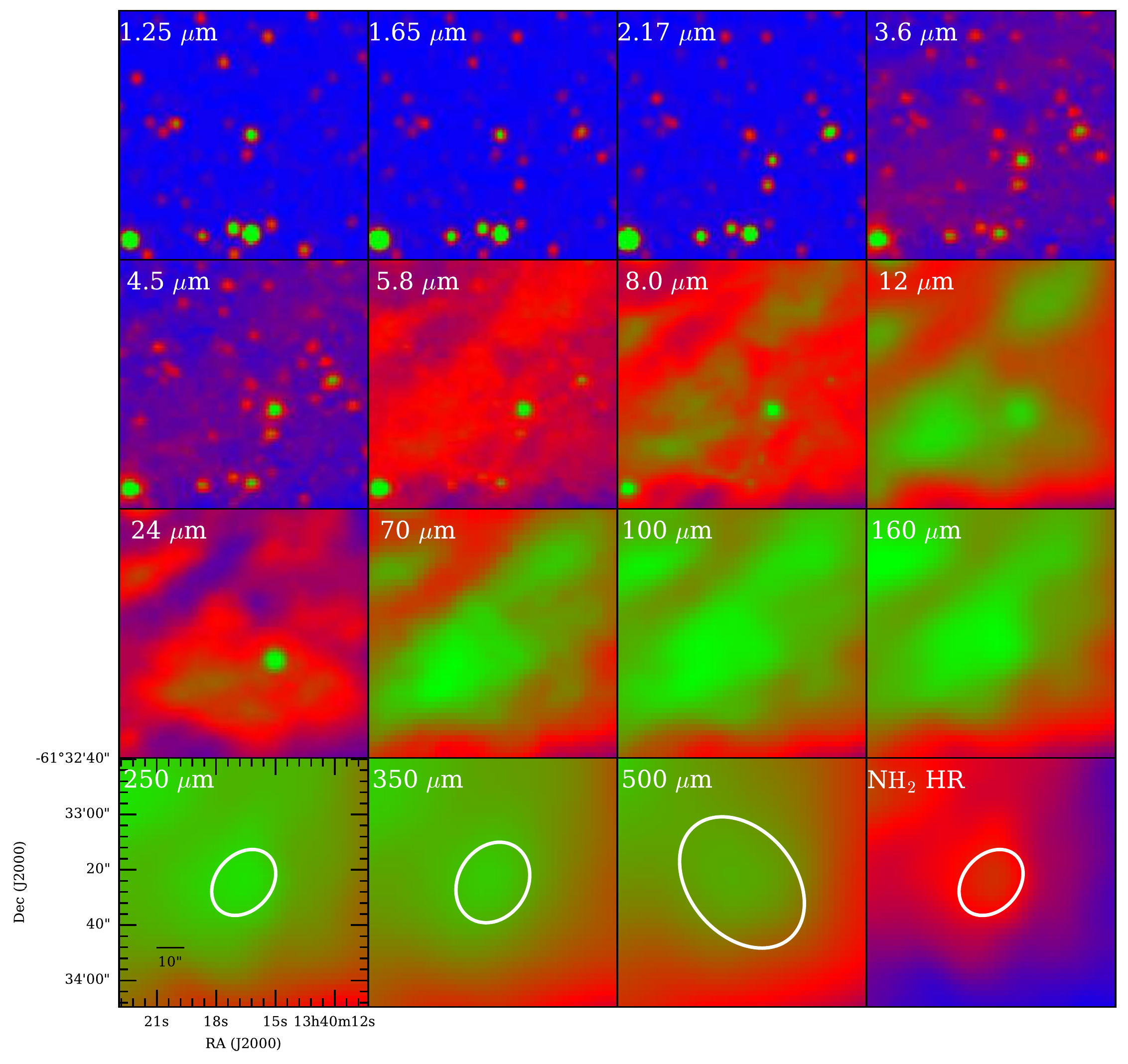}
\hskip 0.00000001cm
\includegraphics[width=3.4 in]{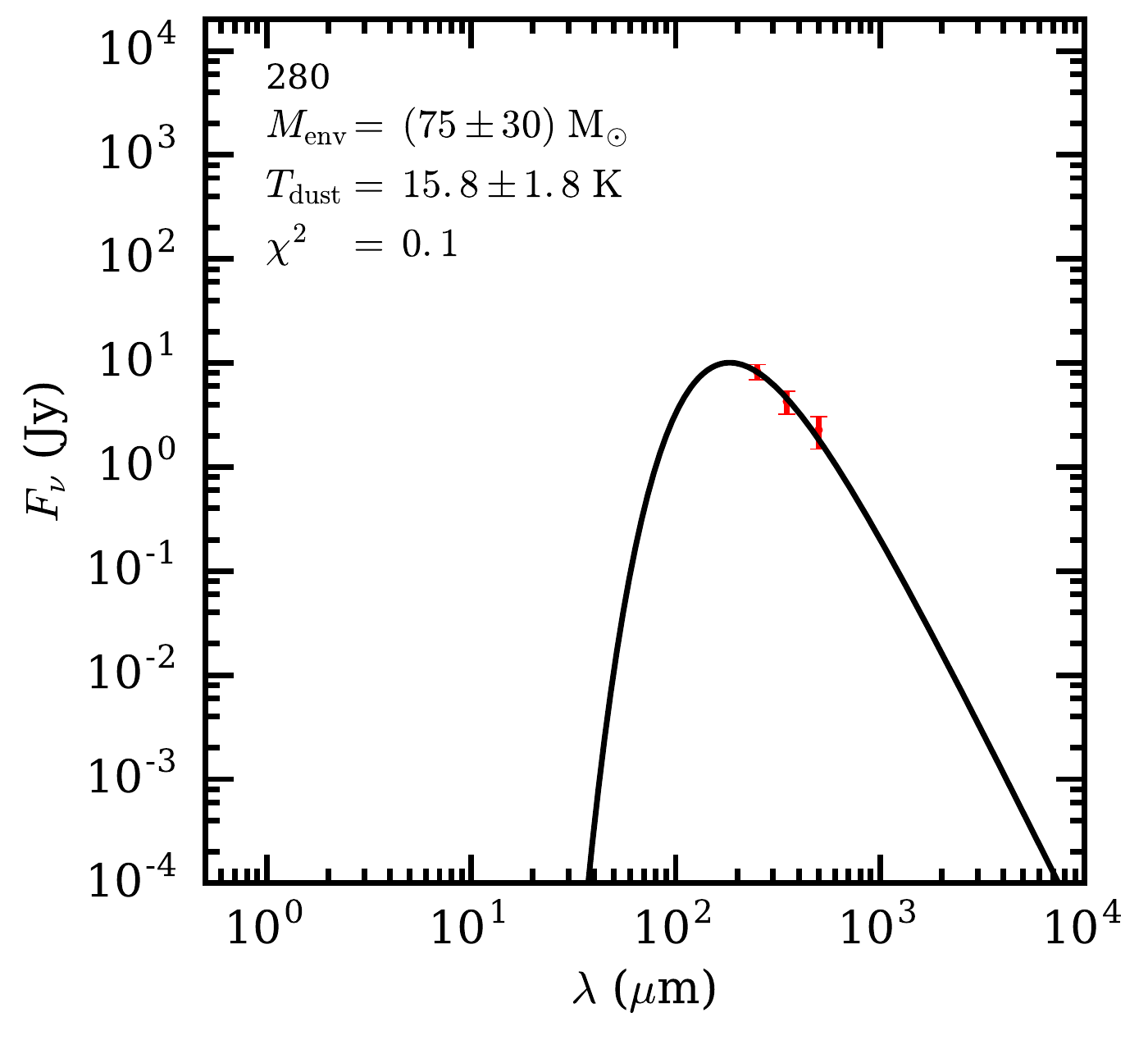}
}
\begin{flushleft}
\caption{ -- continued.}
\end{flushleft}
\end{figure*}
\begin{figure*}[!ht]
\ContinuedFloat
\centering
\subfloat{\includegraphics[width=3.4 in]{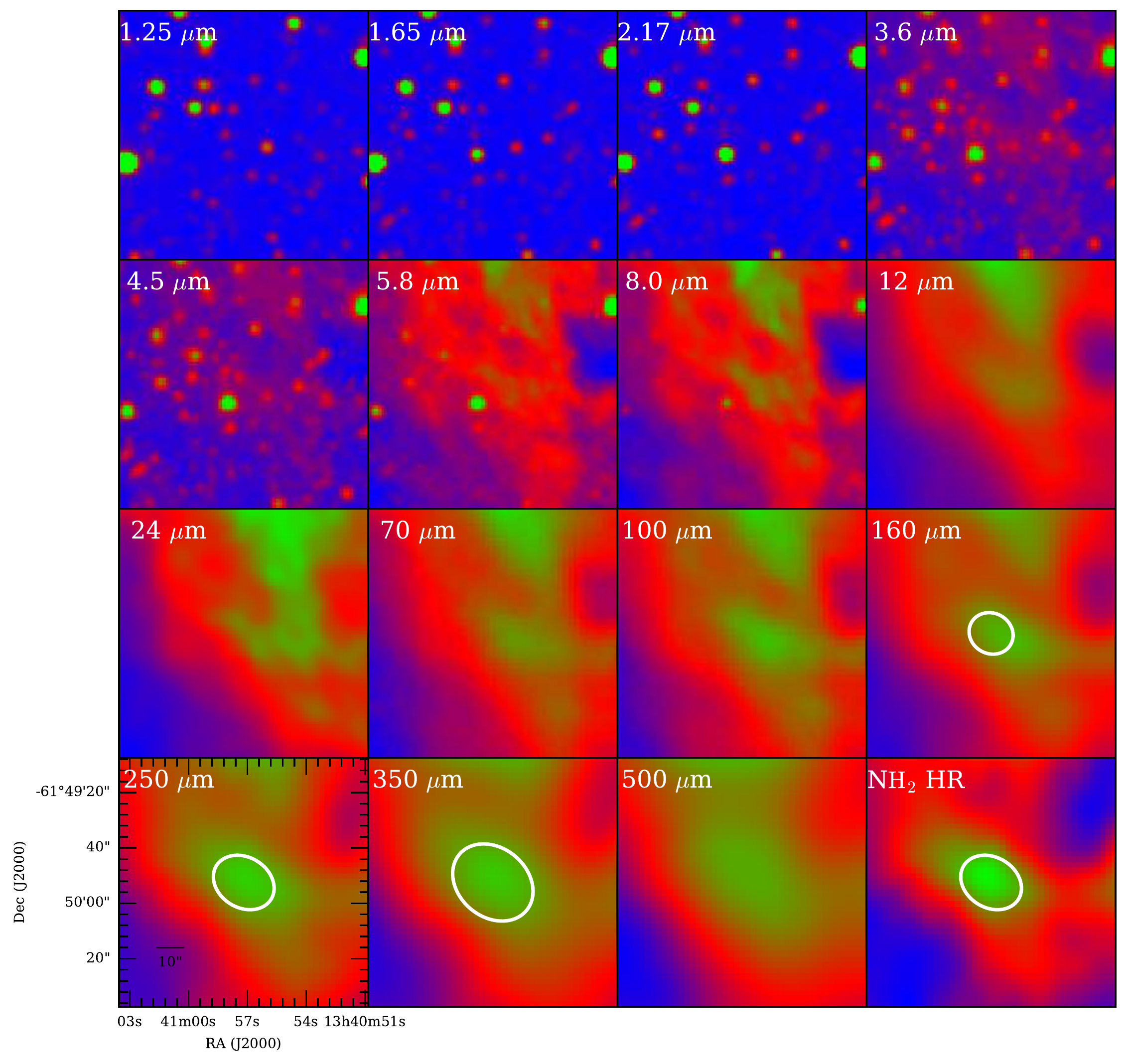}
\hskip 0.00000001cm
\includegraphics[width=3.4 in]{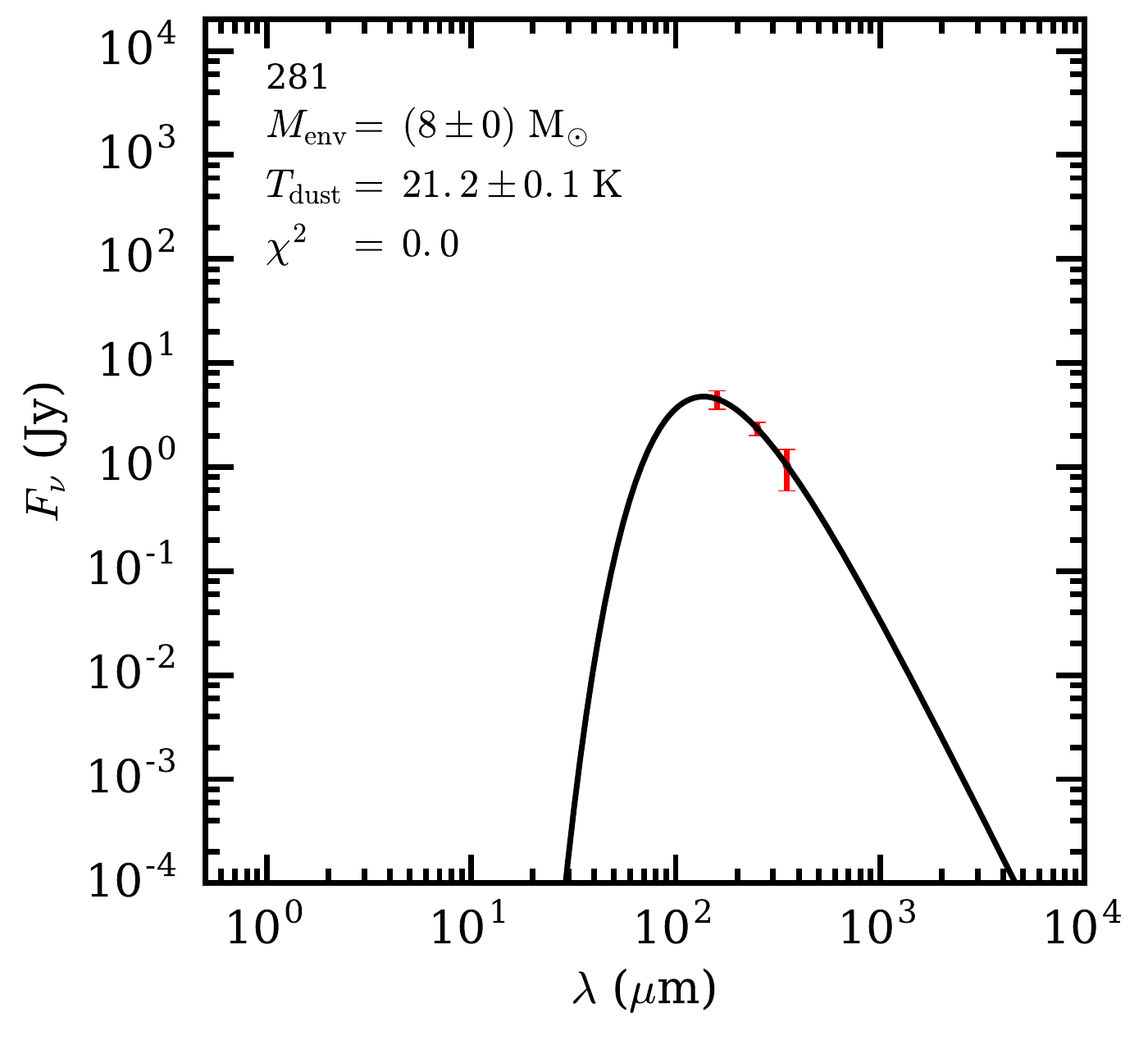}
}
\qquad
\subfloat{\includegraphics[width=3.4 in]{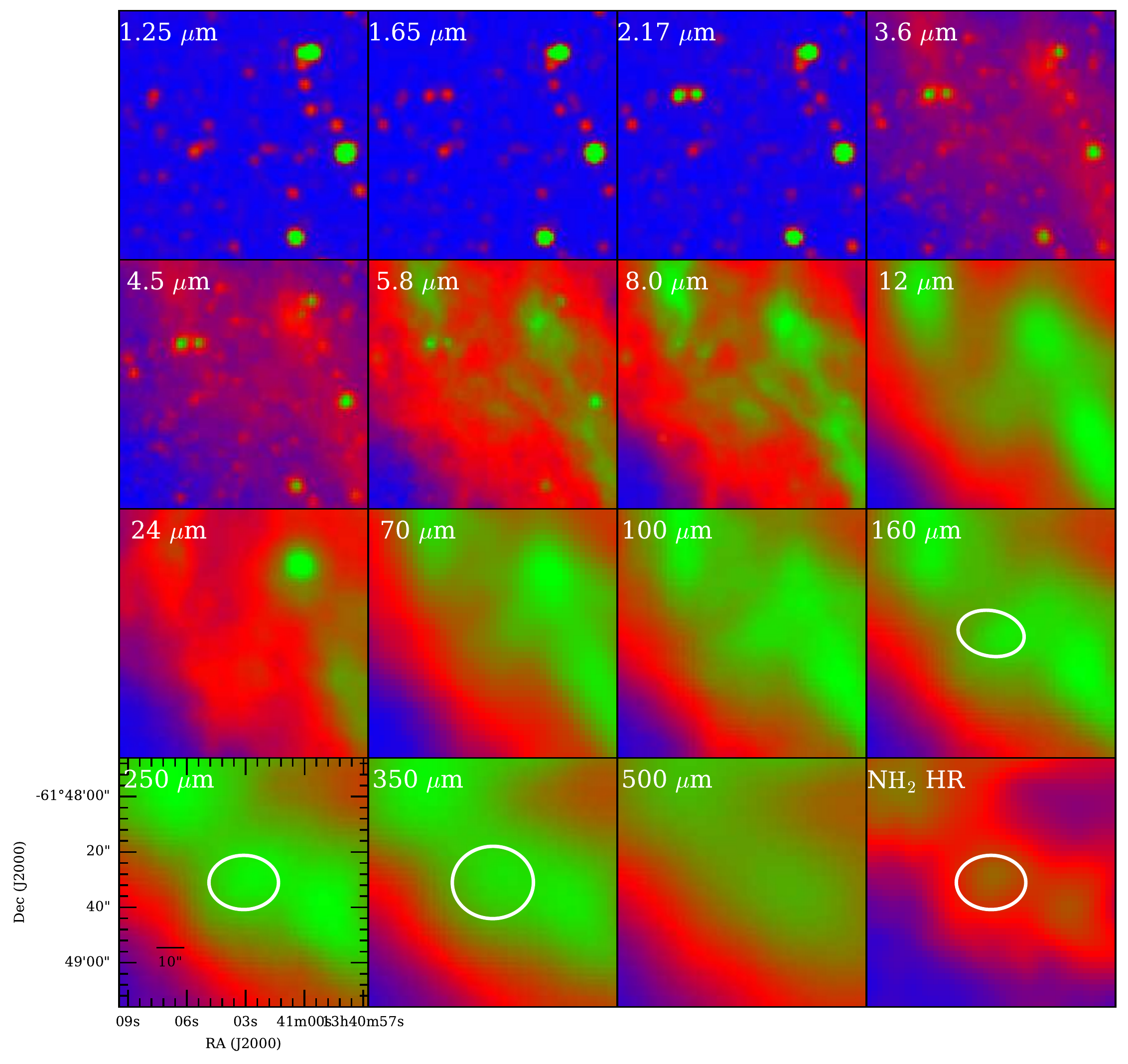}
\hskip 0.00000001cm
\includegraphics[width=3.4 in]{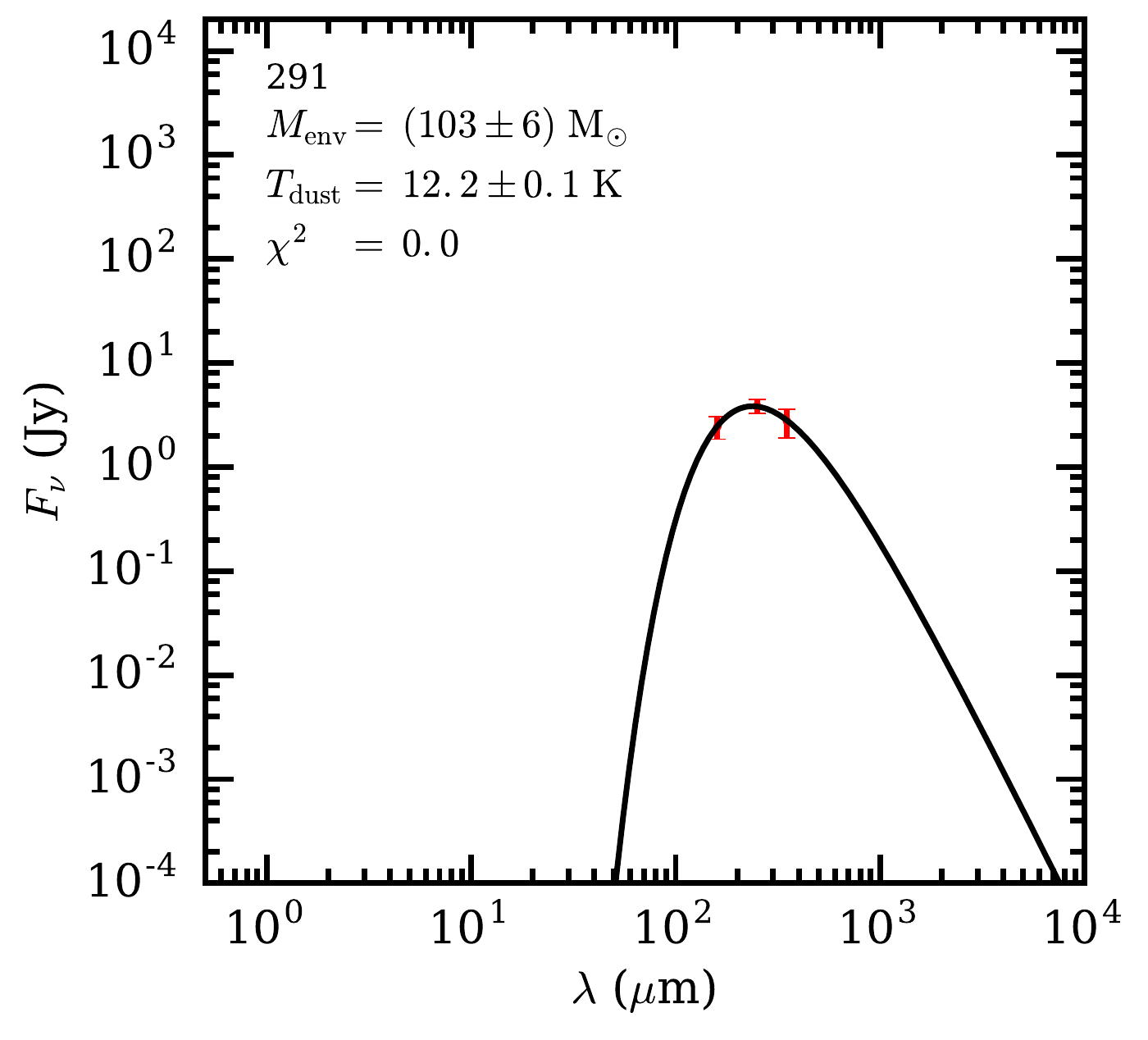}
}
\begin{flushleft}
\caption{ -- continued.}
\end{flushleft}
\end{figure*}
\begin{figure*}[!ht]
\ContinuedFloat
\centering
\subfloat{\includegraphics[width=3.4 in]{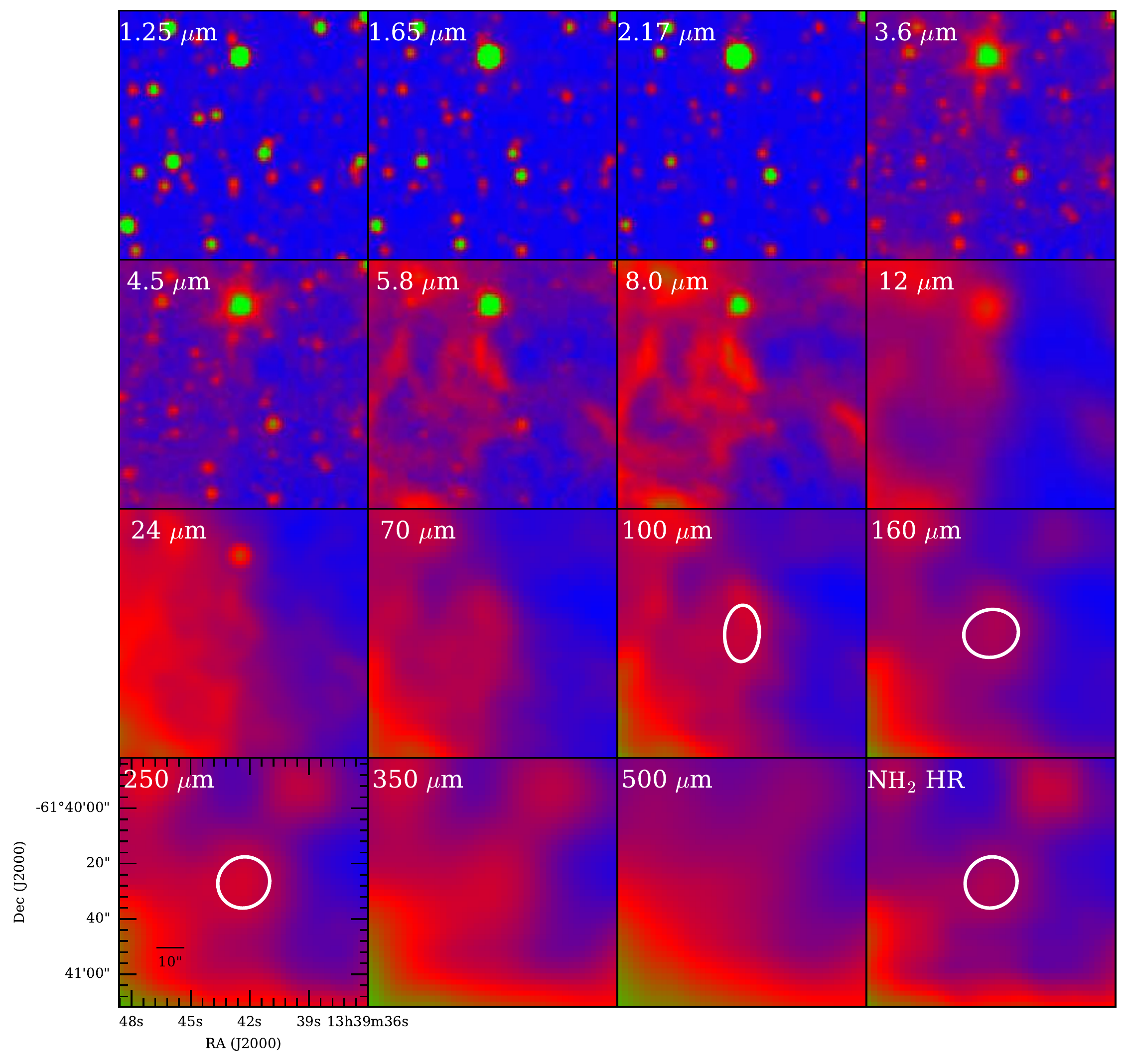}
\hskip 0.00000001cm
\includegraphics[width=3.4 in]{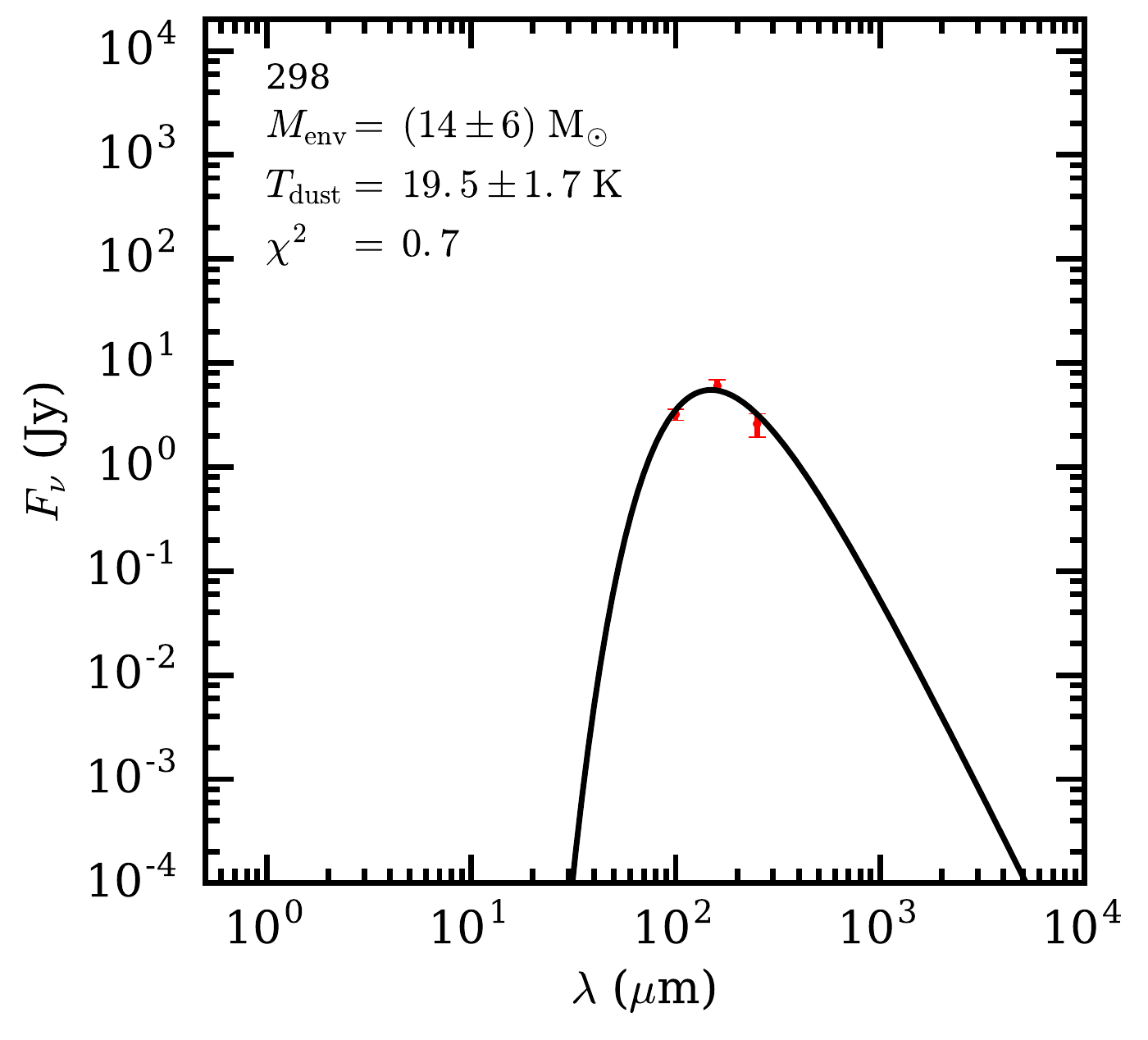}
}
\qquad
\subfloat{\includegraphics[width=3.4 in]{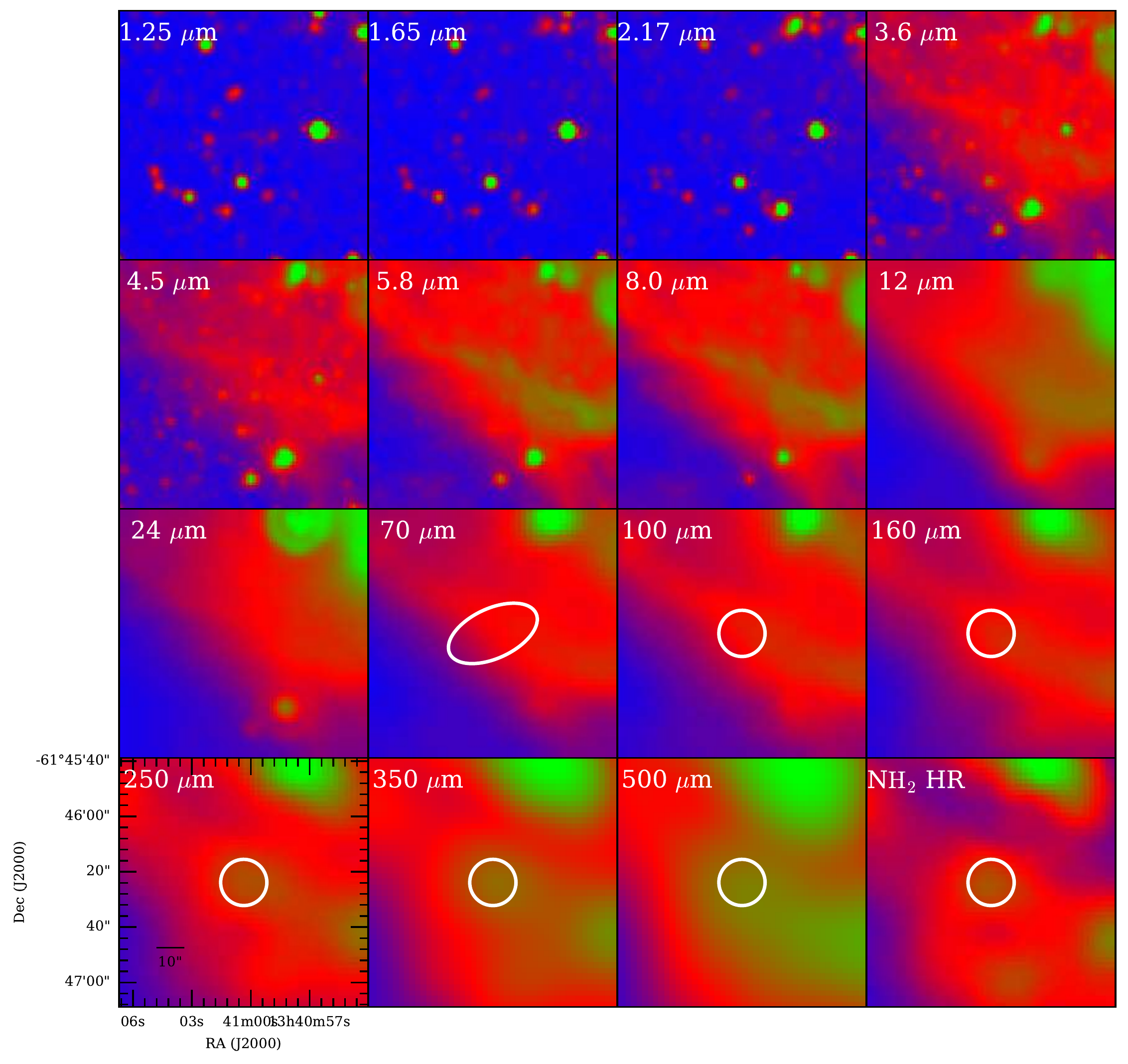}
\hskip 0.00000001cm
\includegraphics[width=3.4 in]{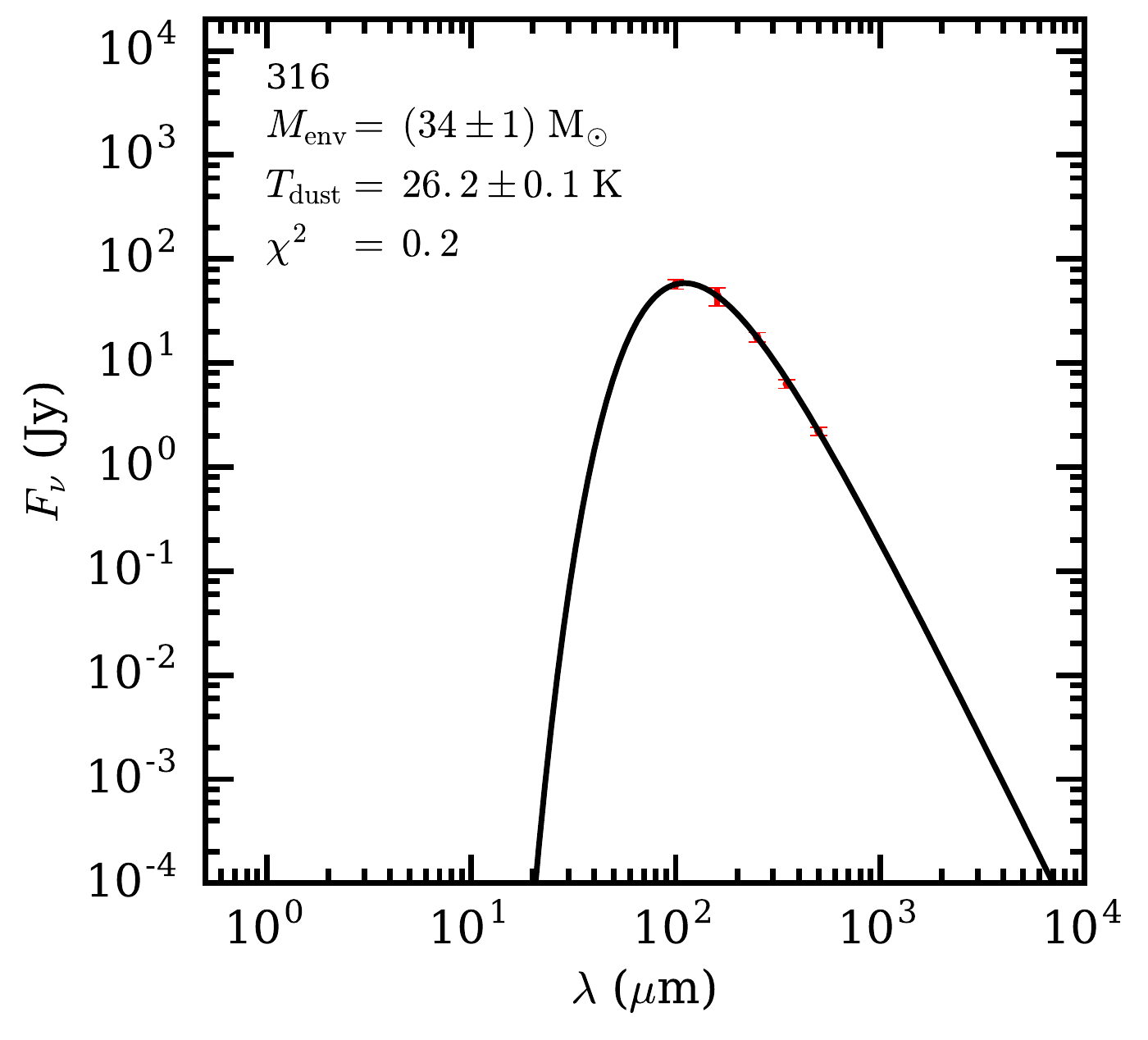}
}
\begin{flushleft}
\caption{ -- continued.}
\end{flushleft}
\end{figure*}

\end{appendix}
\end{document}

%% file: tbl/compactSour_para_proto.tbl
   1 &   13:40:57.312 &   -61:45:42.12 &   0.12 &   25.2 $\pm$    0.5 &    173 $\pm$     16 &   28.5 $\pm$    2.6 &   7513 $\pm$   1503 &  0.005 \\
   2 &   13:40:26.328 &    -61:47:54.6 &   0.12 &   20.5 $\pm$    0.5 &     92 $\pm$     11 &   15.1 $\pm$    1.7 &   1411 $\pm$    282 &  0.010 \\
   3 &   13:39:54.456 &   -61:41:09.24 &   0.16 &   24.4 $\pm$    0.2 &    204 $\pm$      9 &   13.1 $\pm$    0.6 &  12712 $\pm$   2542 &  0.003 \\
   4 &   13:40:26.448 &   -61:47:39.84 &   0.13 &   17.4 $\pm$    0.6 &    162 $\pm$     31 &   18.9 $\pm$    3.6 &    781 $\pm$    156 &  0.022 \\
   5 &   13:40:32.352 &   -61:47:20.04 &   0.16 &   21.8 $\pm$    1.5 &     42 $\pm$     15 &    3.0 $\pm$    1.1 &   1411 $\pm$    282 &  0.005 \\
   8 &   13:39:54.576 &   -61:40:59.52 &   0.12 &   18.1 $\pm$    1.4 &    270 $\pm$    106 &   44.4 $\pm$   17.4 &   1607 $\pm$    321 &  0.019 \\
  10 &   13:41:01.992 &   -61:44:14.28 &   0.12 &   23.1 $\pm$    0.8 &     12 $\pm$      2 &    2.0 $\pm$    0.4 &    399 $\pm$     80 &  0.006 \\
  11 &   13:40:09.168 &   -61:41:31.56 &   0.19 &   25.1 $\pm$    0.3 &     39 $\pm$      2 &    1.6 $\pm$    0.1 &   3938 $\pm$    788 &  0.002 \\
  15 &   13:39:49.560 &   -61:44:30.12 &   0.12 &   25.3 $\pm$    1.0 &     10 $\pm$      2 &    1.6 $\pm$    0.3 &    240 $\pm$     48 &  0.009 \\
  16 &   13:40:18.456 &   -61:47:07.44 &   0.19 &   17.6 $\pm$    1.0 &    183 $\pm$     53 &    7.4 $\pm$    2.2 &    616 $\pm$    123 &  0.032 \\
  19 &   13:40:37.704 &   -61:46:59.88 &   0.19 &   18.6 $\pm$    1.7 &     87 $\pm$     34 &    3.5 $\pm$    1.4 &    456 $\pm$     91 &  0.023 \\
  28 &   13:39:54.360 &    -61:42:59.4 &   0.19 &   20.8 $\pm$    0.5 &     25 $\pm$      3 &    1.0 $\pm$    0.1 &   1215 $\pm$    243 &  0.003 \\
  32 &   13:39:45.216 &   -61:44:41.64 &   0.16 &   25.5 $\pm$    0.7 &      9 $\pm$      1 &    0.6 $\pm$    0.1 &    225 $\pm$     45 &  0.009 \\
  33 &   13:40:36.648 &   -61:49:29.64 &   0.25 &   19.7 $\pm$    0.2 &     55 $\pm$      3 &    1.0 $\pm$    0.1 &    340 $\pm$     68 &  0.022 \\
  37 &   13:41:06.912 &   -61:45:54.72 &   0.18 &   17.1 $\pm$    0.8 &     49 $\pm$     13 &    2.2 $\pm$    0.6 &    347 $\pm$     69 &  0.014 \\
  42 &   13:40:27.240 &   -61:36:16.56 &   0.24 &   20.3 $\pm$    0.8 &     54 $\pm$     10 &    1.1 $\pm$    0.2 &    392 $\pm$     78 &  0.020 \\
  44 &   13:39:39.456 &   -61:43:31.08 &   0.19 &   23.4 $\pm$    1.7 &     23 $\pm$      8 &    0.9 $\pm$    0.3 &    360 $\pm$     72 &  0.012 \\
  45 &   13:40:29.424 &   -61:47:28.68 &   0.18 &   14.0 $\pm$    0.6 &    360 $\pm$     80 &   17.3 $\pm$    3.8 &    290 $\pm$     58 &  0.076 \\
  47 &   13:40:54.624 &   -61:46:40.08 &   0.20 &   17.0 $\pm$    0.2 &    158 $\pm$     10 &    5.3 $\pm$    0.3 &    429 $\pm$     86 &  0.036 \\
  49 &   13:40:46.584 &   -61:46:59.16 &   0.29 &   18.3 $\pm$    6.4 &     31 $\pm$     47 &    0.3 $\pm$    0.5 &    128 $\pm$     26 &  0.028 \\
  51 &   13:40:27.384 &   -61:47:48.48 &   0.12 &   15.8 $\pm$    0.3 &    121 $\pm$     12 &   19.8 $\pm$    2.0 &    290 $\pm$     58 &  0.035 \\
  58 &   13:39:58.704 &   -61:39:56.16 &   0.19 &   22.3 $\pm$    0.6 &     14 $\pm$      2 &    0.6 $\pm$    0.1 &    167 $\pm$     33 &  0.014 \\
  61 &   13:40:25.704 &   -61:32:23.28 &   0.21 &   10.9 $\pm$    1.1 &    759 $\pm$    384 &   22.4 $\pm$   11.3 &    100 $\pm$     20 &  0.234 \\
  69 &   13:39:55.704 &   -61:46:19.56 &   0.20 &   19.9 $\pm$    0.7 &     32 $\pm$      5 &    1.1 $\pm$    0.2 &    206 $\pm$     41 &  0.021 \\
  72 &   13:39:38.280 &   -61:43:06.24 &   0.19 &   24.1 $\pm$    0.7 &     17 $\pm$      2 &    0.7 $\pm$    0.1 &    310 $\pm$     62 &  0.011 \\
  85 &   13:41:03.144 &   -61:45:22.68 &   0.12 &   14.9 $\pm$    0.9 &    150 $\pm$     47 &   24.7 $\pm$    7.7 &    188 $\pm$     38 &  0.058 \\
  86 &   13:40:32.424 &   -61:42:35.64 &   0.17 &   22.4 $\pm$    0.6 &      7 $\pm$      1 &    0.4 $\pm$    0.1 &     85 $\pm$     17 &  0.014 \\
  89 &   13:40:04.032 &   -61:40:32.16 &   0.25 &   22.9 $\pm$    0.7 &     14 $\pm$      2 &    0.3 $\pm$    0.1 &    201 $\pm$     40 &  0.013 \\
  90 &   13:40:11.520 &   -61:42:24.84 &   0.19 &   17.6 $\pm$    0.2 &     75 $\pm$      4 &    3.0 $\pm$    0.2 &    253 $\pm$     51 &  0.032 \\
  92 &   13:41:48.648 &   -61:35:20.04 &   0.19 &   16.3 $\pm$    0.8 &     37 $\pm$      9 &    1.5 $\pm$    0.4 &     80 $\pm$     16 &  0.042 \\
 111 &   13:39:46.512 &    -61:38:54.6 &   0.13 &   19.0 $\pm$    1.0 &     11 $\pm$      3 &    1.3 $\pm$    0.3 &     57 $\pm$     11 &  0.025 \\
 115 &   13:40:21.192 &   -61:52:23.88 &   0.18 &   19.7 $\pm$    0.7 &      6 $\pm$      1 &    0.3 $\pm$    0.1 &     40 $\pm$      8 &  0.022 \\
 132 &   13:40:56.856 &   -61:37:43.32 &   0.12 &   18.2 $\pm$    2.5 &     10 $\pm$      6 &    1.5 $\pm$    0.9 &     40 $\pm$      8 &  0.029 \\
 141 &   13:40:59.328 &   -61:47:01.68 &   0.37 &   19.8 $\pm$    0.7 &     28 $\pm$      5 &    0.1 $\pm$    0.1 &    178 $\pm$     36 &  0.022 \\
 154 &   13:40:58.512 &      -61:48:45 &   0.25 &   21.1 $\pm$    0.4 &     19 $\pm$      2 &    0.3 $\pm$    0.1 &    177 $\pm$     35 &  0.017 \\
 161 &   13:41:07.176 &   -61:47:51.72 &   0.29 &   20.0 $\pm$    0.2 &     23 $\pm$      1 &    0.3 $\pm$    0.1 &    157 $\pm$     31 &  0.021 \\
 263 &   13:39:44.208 &   -61:41:29.76 &   0.26 &   15.5 $\pm$    2.0 &    210 $\pm$     96 &    3.4 $\pm$    1.5 &    339 $\pm$     68 &  0.050 \\
 265 &   13:41:10.248 &    -61:44:02.4 &   0.24 &   16.3 $\pm$    0.6 &     54 $\pm$      9 &    1.0 $\pm$    0.2 &    117 $\pm$     23 &  0.042 \\
 268 &   13:40:37.392 &    -61:51:28.8 &   0.19 &   15.2 $\pm$    0.5 &     63 $\pm$     10 &    2.6 $\pm$    0.4 &     88 $\pm$     18 &  0.055 \\
 270 &   13:40:28.680 &   -61:35:17.52 &   0.40 &   14.8 $\pm$    1.2 &    225 $\pm$     67 &    0.9 $\pm$    0.3 &    263 $\pm$     53 &  0.061 \\
 271 &   13:41:54.888 &   -61:35:09.96 &   0.36 &   20.4 $\pm$    0.2 &     30 $\pm$      2 &    0.2 $\pm$    0.1 &    231 $\pm$     46 &  0.019 \\
 272 &   13:39:50.760 &   -61:41:03.12 &   0.19 &   13.6 $\pm$    2.4 &    353 $\pm$    243 &   14.3 $\pm$    9.8 &    238 $\pm$     48 &  0.085 \\
 273 &   13:39:47.352 &   -61:39:47.16 &   0.19 &   23.1 $\pm$    0.5 &     11 $\pm$      1 &    0.5 $\pm$    0.1 &    164 $\pm$     33 &  0.013 \\
 274 &   13:39:39.384 &   -61:39:52.56 &   0.19 &   14.0 $\pm$    1.0 &     42 $\pm$     15 &    1.7 $\pm$    0.6 &     34 $\pm$      7 &  0.075 \\
 278 &   13:39:33.936 &    -61:40:55.2 &   0.12 &   14.2 $\pm$    0.8 &     22 $\pm$      6 &    3.6 $\pm$    1.1 &     19 $\pm$      4 &  0.072 \\
 280 &   13:40:16.608 &   -61:33:24.48 &   0.29 &   15.8 $\pm$    1.8 &     75 $\pm$     30 &    0.8 $\pm$    0.3 &    131 $\pm$     26 &  0.047 \\
 281 &   13:40:57.192 &   -61:49:52.32 &   0.20 &   21.2 $\pm$    0.1 &      8 $\pm$      0 &    0.3 $\pm$    0.1 &    178 $\pm$     36 &  0.007 \\
 291 &   13:41:03.096 &   -61:48:30.96 &   0.26 &   12.2 $\pm$    0.1 &    103 $\pm$      6 &    1.5 $\pm$    0.1 &     32 $\pm$      6 &  0.139 \\
 298 &   13:39:42.312 &   -61:40:26.76 &   0.19 &   19.5 $\pm$    1.7 &     14 $\pm$      6 &    0.6 $\pm$    0.3 &     83 $\pm$     17 &  0.023 \\
 316 &   13:41:00.360 &   -61:46:23.88 &   0.35 &   26.2 $\pm$    0.1 &     34 $\pm$      1 &    0.2 $\pm$    0.1 &   1024 $\pm$    205 &  0.008 \\

%% file: tbl/Condpara_obs.tbl
     cond1 & 13:41:08.560 & -61:44:17.43 &   25.9 &   38.2 &    514 &   1974 &    224 &     22 &    2.0 &    0.6  \\
     cond2 & 13:40:56.013 & -61:46:21.92 &  105.2 &   72.1 &   8828 &  23240 &   4200 &     22 &    2.9 &    0.6  \\
     cond3 & 13:40:26.940 & -61:47:27.88 &  122.6 &   64.5 &   7567 &  17833 &   2350 &     21 &    2.3 &    0.5  \\
     cond4 & 13:39:47.435 & -61:41:17.14 &   90.2 &   47.6 &   6198 &  12585 &   3000 &     22 &    2.9 &    0.5  \\
     cond5 & 13:40:27.041 & -61:35:42.84 &   26.8 &   58.8 &   1179 &   3088 &    795 &     20 &    2.1 &    0.7  \\
     cond6 & 13:40:08.717 & -61:41:29.01 &   15.4 &   19.1 &    272 &    574 &     81 &     23 &    1.7 &    0.6  \\
     cond7 & 13:40:10.802 & -61:42:24.84 &   20.5 &   17.1 &    281 &    665 &    144 &     21 &    2.0 &    0.5  \\
     cond8 & 13:40:25.868 & -61:32:38.16 &   26.3 &   59.3 &   1088 &   3062 &    483 &     18 &    2.8 &    0.7  \\

%% file: tbl/Condpara_der.tbl
     cond1 &    0.8$\times$   0.8 &     71 &    272 &    100 &    424 &   0.13 $\pm$   0.02 &    6.3 \\
     cond2 &    2.2$\times$   2.2 &   1219 &   3208 &   1874 &   5337 &   0.11 $\pm$   0.02 &    3.7 \\
     cond3 &    2.6$\times$   2.6 &   1094 &   2578 &   1094 &   4355 &   0.24 $\pm$   0.06 &    3.2 \\
     cond4 &    1.9$\times$   1.9 &    872 &   1772 &   1363 &   3153 &   0.33 $\pm$   0.07 &    5.9 \\
     cond5 &    1.2$\times$   1.2 &    181 &    475 &    392 &    675 &   0.39 $\pm$   0.12 &    6.1 \\
     cond6 &    0.4$\times$   0.4 &     35 &     74 &     34 &     99 &   0.39 $\pm$   0.02 &    8.3 \\
     cond7 &    0.4$\times$   0.4 &     41 &     97 &     67 &    161 &   0.47 $\pm$   0.03 &   10.2 \\
     cond8 &    1.2$\times$   1.2 &    199 &    559 &    277 &   1002 &   0.76 $\pm$   0.38 &    9.3 \\

%% file: tbl/compact_source_statistic_tex.tbl
Absorption at 8 \um\ (14) [$Group\  0$]          &     13     &    1      &     0 \\
Extended IR emission (25) [$Group\ IM$]          &     3      &    18     &     4 \\
Point-like IR objects (11) [$Group\  I$]         &     0      &    0      &     11 \\
Total (50) [$Groups\  0,IM,I$]                   &     16     &    19     &     15  \\

%% file: tbl/MDC_statitical.tbl
Cygnus\,X   &   1.7            &  40   &      0.1 &  $1.2\times 10^5$   &  $3.3\times 10^5$   &  $1.9\times 10^6$   &  Motte+2007  \\
NGC\,6334   &   1.75           &  75   &      0.1 &  $1.0\times 10^5$   &  $6.0\times 10^6$   &  $>7.0\times 10^7$   &  Tig\'{e}+2016  \\
RCW\,79     &   4.3            & 140   &      0.2 &  $9.4\times 10^5$   &  $1.6\times 10^6$   &  $4.4\times 10^6$   &  This paper \\

%% file: tbl/Herschel_flux_ori.tbl
   1 &   79.7 $\pm$    1.6 &  149.6 $\pm$    2.2 &   98.4 $\pm$    2.2 &  227.8 $\pm$    5.5 &  118.7 $\pm$    3.0 &  188.1 $\pm$    4.0 &   94.5 $\pm$    4.4 &   92.3 $\pm$    3.9 &   39.6 $\pm$    2.4 &   38.3 $\pm$    2.2 &   15.8 $\pm$    1.0 &   19.4 $\pm$    1.6  \\
   2 &   13.0 $\pm$    0.3 &   23.2 $\pm$    0.4 &   24.2 $\pm$    0.3 &   32.7 $\pm$    0.5 &   28.8 $\pm$    1.2 &   42.0 $\pm$    2.0 &   20.9 $\pm$    0.6 &   25.2 $\pm$    0.6 &   14.3 $\pm$    1.0 &   14.6 $\pm$    0.9 &    5.2 $\pm$    0.6 &    5.8 $\pm$    0.5  \\
   3 &   26.4 $\pm$    0.9 &   95.8 $\pm$    1.5 &   50.7 $\pm$    0.3 &  303.2 $\pm$    1.3 &   87.3 $\pm$    2.6 &  199.9 $\pm$    4.0 &   89.8 $\pm$    1.1 &   96.6 $\pm$    0.9 &   39.6 $\pm$    1.4 &   40.0 $\pm$    1.2 &    8.5 $\pm$    1.0 &    8.2 $\pm$    1.0  \\
   4 &    7.1 $\pm$    0.6 &   10.4 $\pm$    0.6 &   15.0 $\pm$    0.5 &   17.9 $\pm$    0.6 &   22.3 $\pm$    1.3 &   32.6 $\pm$    1.3 &   24.8 $\pm$    0.6 &   28.3 $\pm$    0.5 &   11.9 $\pm$    1.0 &   12.0 $\pm$    1.0 &    8.9 $\pm$    0.6 &    9.3 $\pm$    0.5  \\
   5 &   11.2 $\pm$    0.5 &   20.6 $\pm$    0.7 &   12.6 $\pm$    0.6 &   20.7 $\pm$    0.7 &   14.9 $\pm$    1.1 &   29.8 $\pm$    2.2 &   10.7 $\pm$    0.9 &   10.4 $\pm$    0.8 &    4.7 $\pm$    1.2 &    6.1 $\pm$    1.1 &    3.6 $\pm$    0.8 &    3.6 $\pm$    0.7  \\
   8 &    0.0 $\pm$    1.2 &    0.4 $\pm$    1.4 &   11.5 $\pm$    2.2 &   47.8 $\pm$    3.2 &   43.6 $\pm$    2.9 &   58.1 $\pm$    2.9 &   50.0 $\pm$    0.9 &   55.1 $\pm$    0.8 &   33.9 $\pm$    1.3 &   35.4 $\pm$    1.2 &   17.9 $\pm$    1.1 &   18.3 $\pm$    1.0  \\
  10 &    3.2 $\pm$    0.2 &    5.6 $\pm$    0.2 &    5.5 $\pm$    0.3 &    9.3 $\pm$    0.4 &    8.6 $\pm$    0.9 &   10.1 $\pm$    0.9 &    5.8 $\pm$    1.2 &    4.2 $\pm$    1.4 &    2.4 $\pm$    1.2 &    2.1 $\pm$    1.2 &    0.0 $\pm$    0.4 &                \\
  11 &    5.0 $\pm$    0.1 &   32.3 $\pm$    0.2 &    7.0 $\pm$    0.2 &   50.2 $\pm$    0.5 &   17.1 $\pm$    0.4 &   42.8 $\pm$    0.6 &   16.5 $\pm$    0.6 &   19.3 $\pm$    0.6 &    8.9 $\pm$    1.1 &    9.3 $\pm$    1.2 &    3.1 $\pm$    0.5 &    3.3 $\pm$    1.3  \\
  15 &    3.4 $\pm$    0.2 &    7.9 $\pm$    0.2 &    5.8 $\pm$    0.2 &   13.3 $\pm$    0.3 &    8.6 $\pm$    0.4 &   10.4 $\pm$    0.4 &    6.8 $\pm$    0.3 &    5.8 $\pm$    0.3 &    2.9 $\pm$    0.3 &    2.2 $\pm$    0.3 &    1.4 $\pm$    0.5 &    1.6 $\pm$    0.4  \\
  16 &    3.4 $\pm$    0.4 &   16.4 $\pm$    0.7 &    4.3 $\pm$    0.4 &   24.8 $\pm$    0.8 &   16.3 $\pm$    1.0 &   37.5 $\pm$    1.2 &   25.6 $\pm$    0.5 &   31.4 $\pm$    0.4 &   14.7 $\pm$    0.6 &   21.8 $\pm$    1.3 &    7.5 $\pm$    0.5 &   10.4 $\pm$    0.7  \\
  19 &    2.7 $\pm$    0.3 &   10.7 $\pm$    0.4 &    4.3 $\pm$    0.3 &   17.6 $\pm$    0.5 &   12.2 $\pm$    1.5 &   26.1 $\pm$    1.9 &   14.6 $\pm$    1.8 &   15.7 $\pm$    1.6 &   10.1 $\pm$    1.2 &   11.0 $\pm$    1.1 &    5.2 $\pm$    0.9 &    5.1 $\pm$    0.8  \\
  28 &    1.1 $\pm$    0.1 &    5.7 $\pm$    0.1 &    2.1 $\pm$    0.1 &    9.9 $\pm$    0.2 &    5.6 $\pm$    0.9 &   12.9 $\pm$    1.1 &    6.8 $\pm$    0.6 &    7.5 $\pm$    0.6 &    2.9 $\pm$    0.8 &    2.9 $\pm$    0.7 &    1.9 $\pm$    0.7 &    2.1 $\pm$    0.6  \\
  32 &    2.1 $\pm$    0.2 &    8.0 $\pm$    0.3 &    3.4 $\pm$    0.2 &   16.6 $\pm$    0.4 &    5.5 $\pm$    0.4 &   10.9 $\pm$    0.4 &    3.8 $\pm$    0.5 &    3.9 $\pm$    0.4 &    1.8 $\pm$    0.3 &    2.4 $\pm$    0.4 &    1.7 $\pm$    0.5 &    3.2 $\pm$    0.6  \\
  33 &    0.8 $\pm$    0.1 &   17.5 $\pm$    0.4 &    1.5 $\pm$    0.1 &   28.0 $\pm$    0.5 &    6.2 $\pm$    0.5 &   32.9 $\pm$    1.1 &    7.9 $\pm$    0.7 &   13.1 $\pm$    0.8 &    5.2 $\pm$    0.8 &    7.1 $\pm$    0.8 &    2.6 $\pm$    0.7 &    2.9 $\pm$    0.6  \\
  37 &    0.6 $\pm$    0.3 &    1.8 $\pm$    0.4 &    1.3 $\pm$    0.4 &    7.7 $\pm$    0.9 &    5.5 $\pm$    2.0 &   10.1 $\pm$    2.9 &    8.3 $\pm$    2.8 &    7.2 $\pm$    2.4 &    3.0 $\pm$    1.5 &    2.9 $\pm$    1.6 &    1.0 $\pm$    0.4 &    0.8 $\pm$    0.3  \\
  42 &    1.6 $\pm$    0.1 &   14.1 $\pm$    0.2 &    3.1 $\pm$    0.2 &   26.2 $\pm$    0.6 &   10.1 $\pm$    0.4 &   30.0 $\pm$    1.1 &    9.6 $\pm$    0.8 &   12.3 $\pm$    0.8 &    7.3 $\pm$    0.7 &    7.9 $\pm$    0.6 &    3.2 $\pm$    0.6 &    4.2 $\pm$    1.0  \\
  44 &    2.0 $\pm$    0.2 &   10.8 $\pm$    0.4 &    2.9 $\pm$    0.2 &   18.2 $\pm$    0.7 &    6.6 $\pm$    0.6 &   21.7 $\pm$    1.0 &    4.8 $\pm$    0.7 &    5.3 $\pm$    0.7 &    3.3 $\pm$    0.9 &    3.4 $\pm$    0.9 &    1.5 $\pm$    0.7 &    2.0 $\pm$    0.6  \\
  45 &    0.6 $\pm$    0.5 &    1.3 $\pm$    0.6 &    1.6 $\pm$    0.4 &    2.8 $\pm$    0.5 &   11.3 $\pm$    1.3 &   21.5 $\pm$    1.3 &   22.4 $\pm$    0.7 &   27.6 $\pm$    0.6 &   15.4 $\pm$    1.2 &   19.5 $\pm$    1.1 &    9.2 $\pm$    0.6 &    9.2 $\pm$    0.6  \\
  47 &    2.6 $\pm$    1.7 &   11.4 $\pm$    2.8 &    5.4 $\pm$    2.0 &   34.2 $\pm$    3.8 &   12.7 $\pm$    3.4 &   39.3 $\pm$    4.7 &   18.6 $\pm$    3.7 &   23.9 $\pm$    3.4 &   11.1 $\pm$    2.2 &   12.5 $\pm$    2.0 &    9.6 $\pm$    1.0 &   14.6 $\pm$    0.9  \\
  49 &    0.0 $\pm$    0.2 &    0.8 $\pm$    0.3 &    0.5 $\pm$    0.3 &    3.3 $\pm$    0.5 &    5.2 $\pm$    1.5 &    8.8 $\pm$    1.8 &    7.1 $\pm$    2.8 &    5.1 $\pm$    2.5 &    4.8 $\pm$    1.4 &    5.1 $\pm$    1.3 &    4.7 $\pm$    1.0 &    6.0 $\pm$    0.9  \\
  51 &    0.8 $\pm$    0.2 &    0.4 $\pm$    0.2 &    5.0 $\pm$    0.4 &    5.5 $\pm$    0.4 &   11.8 $\pm$    1.2 &   15.2 $\pm$    1.2 &   14.1 $\pm$    0.6 &   14.8 $\pm$    0.6 &    7.3 $\pm$    1.1 &    7.0 $\pm$    1.0 &    3.9 $\pm$    0.6 &    3.9 $\pm$    0.5  \\
  58 &    1.4 $\pm$    0.1 &    8.0 $\pm$    0.1 &    2.1 $\pm$    0.2 &    8.4 $\pm$    0.4 &    4.0 $\pm$    0.4 &    9.6 $\pm$    0.6 &    3.9 $\pm$    0.5 &    5.1 $\pm$    0.4 &    1.4 $\pm$    0.4 &    1.4 $\pm$    0.7 &    1.5 $\pm$    0.8 &    2.3 $\pm$    0.7  \\
  61 &    0.1 $\pm$    0.0 &    0.3 $\pm$    0.1 &    0.3 $\pm$    0.1 &    0.9 $\pm$    0.1 &    2.9 $\pm$    0.5 &    5.7 $\pm$    0.6 &   10.3 $\pm$    0.8 &   21.0 $\pm$    1.6 &    9.3 $\pm$    0.7 &   11.9 $\pm$    0.7 &    6.9 $\pm$    0.7 &    8.1 $\pm$    0.7  \\
  69 &    1.4 $\pm$    0.4 &    9.9 $\pm$    0.9 &    1.9 $\pm$    0.5 &   17.4 $\pm$    1.1 &    5.3 $\pm$    0.6 &   20.6 $\pm$    0.8 &    6.3 $\pm$    0.5 &    9.4 $\pm$    0.5 &    3.8 $\pm$    0.4 &    5.3 $\pm$    0.4 &    1.9 $\pm$    0.3 &    2.7 $\pm$    0.3  \\
  72 &    1.7 $\pm$    0.2 &    9.3 $\pm$    0.3 &    2.0 $\pm$    0.2 &   16.6 $\pm$    0.5 &    5.5 $\pm$    0.8 &   16.6 $\pm$    1.1 &    4.5 $\pm$    1.0 &    6.3 $\pm$    0.9 &    2.5 $\pm$    0.9 &    3.1 $\pm$    0.8 &    0.8 $\pm$    0.7 &    1.0 $\pm$    0.6  \\
  85 &    0.1 $\pm$    0.3 &    0.6 $\pm$    0.4 &    0.9 $\pm$    0.4 &    5.8 $\pm$    0.9 &    7.3 $\pm$    1.9 &   12.4 $\pm$    2.7 &   14.2 $\pm$    3.2 &   14.1 $\pm$    2.8 &    8.5 $\pm$    1.8 &    8.0 $\pm$    1.6 &    5.0 $\pm$    0.9 &    6.5 $\pm$    1.0  \\
  86 &    0.4 $\pm$    0.0 &    1.5 $\pm$    0.1 &    0.6 $\pm$    0.1 &    2.4 $\pm$    0.1 &    2.7 $\pm$    0.1 &    4.8 $\pm$    0.1 &    2.5 $\pm$    0.2 &    2.4 $\pm$    0.1 &    1.3 $\pm$    0.1 &    1.1 $\pm$    0.1 &    0.5 $\pm$    0.1 &    0.4 $\pm$    0.1  \\
  89 &    1.3 $\pm$    0.2 &    5.2 $\pm$    0.5 &    1.7 $\pm$    0.2 &   11.8 $\pm$    0.5 &    4.1 $\pm$    0.4 &   12.5 $\pm$    0.6 &    3.6 $\pm$    0.6 &    5.5 $\pm$    0.6 &    2.5 $\pm$    0.5 &    6.3 $\pm$    0.9 &    2.5 $\pm$    0.8 &    3.1 $\pm$    0.7  \\
  90 &    0.2 $\pm$    0.1 &    1.9 $\pm$    0.1 &    1.0 $\pm$    0.2 &    8.2 $\pm$    0.5 &    5.7 $\pm$    0.3 &   17.5 $\pm$    0.5 &   10.9 $\pm$    0.5 &   12.6 $\pm$    0.6 &    6.6 $\pm$    0.5 &    6.6 $\pm$    0.5 &    2.9 $\pm$    0.4 &    2.4 $\pm$    0.7  \\
  92 &    0.1 $\pm$    0.1 &    0.3 $\pm$    0.1 &    0.2 $\pm$    0.1 &    1.3 $\pm$    0.2 &    2.1 $\pm$    0.3 &    5.5 $\pm$    0.5 &    3.6 $\pm$    0.5 &    4.6 $\pm$    0.4 &    2.2 $\pm$    0.3 &    2.7 $\pm$    0.3 &    1.5 $\pm$    0.2 &    1.6 $\pm$    0.2  \\
 111 &    0.4 $\pm$    0.0 &    2.2 $\pm$    0.1 &    0.8 $\pm$    0.1 &    2.7 $\pm$    0.1 &    2.3 $\pm$    0.2 &    3.9 $\pm$    0.3 &    2.2 $\pm$    0.4 &    2.4 $\pm$    0.3 &    1.4 $\pm$    0.4 &    1.4 $\pm$    0.4 &    0.6 $\pm$    0.5 &    0.6 $\pm$    0.4  \\
 115 &    0.3 $\pm$    0.1 &    0.9 $\pm$    0.1 &    0.6 $\pm$    0.2 &    1.5 $\pm$    0.3 &    1.6 $\pm$    0.6 &    2.7 $\pm$    0.6 &    1.9 $\pm$    0.6 &    2.3 $\pm$    0.5 &    0.8 $\pm$    0.4 &    0.8 $\pm$    0.4 &    0.3 $\pm$    0.3 &    0.0 $\pm$    0.3  \\
 132 &    0.1 $\pm$    0.0 &    0.4 $\pm$    0.0 &    0.6 $\pm$    0.1 &    1.8 $\pm$    0.1 &    1.8 $\pm$    0.2 &    2.6 $\pm$    0.2 &    2.1 $\pm$    0.4 &    2.2 $\pm$    0.4 &    1.3 $\pm$    0.5 &    1.4 $\pm$    0.4 &    0.5 $\pm$    0.3 &    0.6 $\pm$    0.2  \\
 141 &    0.3 $\pm$    0.1 &    3.5 $\pm$    0.2 &               &    8.0 $\pm$    0.8 &               &   11.1 $\pm$    2.2 &    0.2 $\pm$    0.3 &    7.6 $\pm$    0.8 &    0.2 $\pm$    0.2 &    2.0 $\pm$    0.2 &    0.0 $\pm$    0.3 &    1.5 $\pm$    0.2  \\
 154 &    0.9 $\pm$    0.2 &    8.1 $\pm$    0.4 &    1.3 $\pm$    0.2 &   11.5 $\pm$    0.5 &    3.5 $\pm$    0.5 &   14.1 $\pm$    0.9 &    3.9 $\pm$    0.6 &    5.9 $\pm$    0.6 &    2.1 $\pm$    0.9 &    2.4 $\pm$    0.8 &    1.0 $\pm$    0.4 &    1.1 $\pm$    0.4  \\
 161 &    0.5 $\pm$    0.1 &    6.9 $\pm$    0.3 &    0.8 $\pm$    0.1 &    9.9 $\pm$    0.3 &    2.6 $\pm$    0.4 &   12.6 $\pm$    0.8 &    3.2 $\pm$    1.0 &    6.1 $\pm$    1.0 &    1.6 $\pm$    1.0 &    2.8 $\pm$    0.9 &    0.8 $\pm$    0.4 &    1.1 $\pm$    0.3  \\
 263 &    0.3 $\pm$    0.2 &    3.9 $\pm$    0.7 &    0.5 $\pm$    0.2 &   13.9 $\pm$    1.0 &    3.3 $\pm$    1.6 &    9.7 $\pm$    2.8 &   15.6 $\pm$    1.0 &   23.6 $\pm$    1.4 &    9.9 $\pm$    0.9 &   12.1 $\pm$    1.1 &    7.7 $\pm$    0.8 &    9.7 $\pm$    0.7  \\
 265 &    0.0 $\pm$    0.1 &    0.4 $\pm$    0.1 &    0.1 $\pm$    0.1 &    1.7 $\pm$    0.4 &    3.5 $\pm$    0.6 &    8.7 $\pm$    0.9 &    5.9 $\pm$    1.0 &    6.8 $\pm$    0.9 &    4.5 $\pm$    1.2 &    4.4 $\pm$    1.1 &    2.5 $\pm$    0.4 &    3.8 $\pm$    0.4  \\
 268 &               &    2.0 $\pm$    0.1 &    0.2 $\pm$    0.1 &    3.0 $\pm$    0.2 &    2.3 $\pm$    0.4 &    6.1 $\pm$    0.6 &    4.6 $\pm$    0.5 &    6.8 $\pm$    0.7 &    3.1 $\pm$    0.5 &    3.4 $\pm$    0.4 &    1.6 $\pm$    0.4 &    1.3 $\pm$    0.4  \\
 270 &    0.1 $\pm$    0.0 &               &    0.3 $\pm$    0.1 &    5.4 $\pm$    0.3 &    2.9 $\pm$    0.2 &   27.4 $\pm$    0.6 &    7.6 $\pm$    0.9 &   20.6 $\pm$    1.5 &    6.0 $\pm$    0.8 &   10.6 $\pm$    1.0 &    4.5 $\pm$    0.8 &    5.9 $\pm$    0.8  \\
 271 &    0.6 $\pm$    0.0 &    7.6 $\pm$    0.2 &    0.9 $\pm$    0.1 &   10.9 $\pm$    0.2 &    2.8 $\pm$    0.2 &   14.5 $\pm$    0.5 &    3.2 $\pm$    0.4 &    9.0 $\pm$    0.5 &    2.0 $\pm$    0.3 &    2.8 $\pm$    0.3 &    1.2 $\pm$    0.2 &    1.4 $\pm$    0.2  \\
 272 &               &               &               &    4.4 $\pm$    0.5 &    0.0 $\pm$    1.2 &    0.0 $\pm$    1.2 &   22.2 $\pm$    1.2 &   23.3 $\pm$    1.0 &   12.3 $\pm$    1.2 &   12.9 $\pm$    1.1 &    9.2 $\pm$    0.9 &    8.9 $\pm$    0.9  \\
 273 &    0.3 $\pm$    0.2 &    3.9 $\pm$    0.6 &    0.6 $\pm$    0.2 &    8.8 $\pm$    0.5 &    2.3 $\pm$    0.4 &    8.6 $\pm$    0.8 &    3.3 $\pm$    0.4 &    4.9 $\pm$    0.6 &    1.7 $\pm$    0.7 &    1.8 $\pm$    0.7 &    0.7 $\pm$    0.4 &    0.7 $\pm$    0.4  \\
 274 &    0.0 $\pm$    0.0 &    1.4 $\pm$    0.1 &    0.1 $\pm$    0.0 &    1.3 $\pm$    0.1 &    0.9 $\pm$    0.2 &    2.3 $\pm$    0.3 &    2.5 $\pm$    0.4 &    3.3 $\pm$    0.4 &    1.7 $\pm$    0.4 &    1.6 $\pm$    0.4 &    0.6 $\pm$    0.4 &    0.2 $\pm$    0.3  \\
 278 &    0.0 $\pm$    0.1 &    0.0 $\pm$    0.1 &    0.3 $\pm$    0.1 &    0.5 $\pm$    0.1 &    1.2 $\pm$    0.4 &    1.3 $\pm$    0.4 &    1.7 $\pm$    0.6 &    1.7 $\pm$    0.5 &    1.2 $\pm$    0.6 &    2.1 $\pm$    0.5 &    1.1 $\pm$    0.4 &    1.1 $\pm$    0.3  \\
 280 &    0.1 $\pm$    0.0 &    4.3 $\pm$    0.1 &    0.3 $\pm$    0.1 &    6.7 $\pm$    0.2 &    1.8 $\pm$    0.6 &   17.4 $\pm$    1.6 &    3.4 $\pm$    0.7 &    8.8 $\pm$    1.4 &    2.6 $\pm$    0.7 &    4.4 $\pm$    1.1 &    2.4 $\pm$    0.6 &    4.7 $\pm$    0.8  \\
 281 &    0.2 $\pm$    0.1 &    0.4 $\pm$    0.1 &    0.3 $\pm$    0.1 &    0.7 $\pm$    0.1 &    1.8 $\pm$    0.5 &    4.8 $\pm$    0.9 &    2.1 $\pm$    0.4 &    2.5 $\pm$    0.3 &    1.5 $\pm$    0.5 &    1.7 $\pm$    0.5 &    1.0 $\pm$    0.3 &    1.0 $\pm$    0.3  \\
 291 &    0.0 $\pm$    0.1 &    0.1 $\pm$    0.1 &    0.3 $\pm$    0.0 &    2.5 $\pm$    0.1 &    1.3 $\pm$    0.5 &    3.1 $\pm$    0.6 &    2.7 $\pm$    0.7 &    4.1 $\pm$    0.6 &    2.2 $\pm$    1.0 &    3.2 $\pm$    0.9 &    1.2 $\pm$    0.4 &    1.2 $\pm$    0.4  \\
 298 &    0.3 $\pm$    0.1 &    1.9 $\pm$    0.3 &    0.4 $\pm$    0.1 &    3.1 $\pm$    0.4 &    1.4 $\pm$    0.4 &    5.8 $\pm$    0.7 &    1.9 $\pm$    0.6 &    2.8 $\pm$    0.7 &    1.0 $\pm$    0.6 &    1.0 $\pm$    0.5 &    0.2 $\pm$    0.3 &    0.1 $\pm$    0.3  \\
 316 &    0.3 $\pm$    0.1 &    3.5 $\pm$    0.2 &               &   57.1 $\pm$    5.7 &               &   44.1 $\pm$    8.8 &    0.2 $\pm$    0.3 &   17.8 $\pm$    1.8 &    0.2 $\pm$    0.2 &    6.3 $\pm$    0.6 &    0.0 $\pm$    0.3 &    2.2 $\pm$    0.2  \\

%% file: tbl/Herschel_flux_corr.tbl
   1 &  233.9 $\pm$    9.0 &  187.5 $\pm$   10.2 &   86.8 $\pm$    5.7 &   36.1 $\pm$    3.2 &   18.5 $\pm$    2.0  \\
   2 &   33.5 $\pm$    1.1 &   43.1 $\pm$    3.0 &   23.7 $\pm$    1.3 &   13.7 $\pm$    1.3 &    5.5 $\pm$    0.6  \\
   3 &  227.0 $\pm$    6.9 &  199.3 $\pm$   10.7 &   90.8 $\pm$    4.6 &   37.7 $\pm$    2.9 &    7.8 $\pm$    1.1  \\
   4 &   17.9 $\pm$    0.8 &   34.0 $\pm$    2.2 &   26.6 $\pm$    1.4 &   11.3 $\pm$    1.2 &    8.9 $\pm$    0.8  \\
   5 &   21.2 $\pm$    1.0 &   30.0 $\pm$    2.7 &    9.8 $\pm$    0.9 &    2.4 $\pm$    1.1 &    3.5 $\pm$    0.7  \\
   8 &   47.7 $\pm$    3.5 &   60.5 $\pm$    4.3 &   51.8 $\pm$    2.7 &   33.4 $\pm$    2.6 &   17.4 $\pm$    1.5  \\
  10 &    9.5 $\pm$    0.5 &   10.1 $\pm$    1.0 &    4.0 $\pm$    1.3 &               &                \\
  11 &   51.5 $\pm$    1.6 &   42.5 $\pm$    2.2 &   18.1 $\pm$    1.1 &    8.7 $\pm$    1.2 &    3.1 $\pm$    1.2  \\
  15 &   13.7 $\pm$    0.5 &   10.4 $\pm$    0.6 &    5.5 $\pm$    0.4 &    2.1 $\pm$    0.3 &    1.6 $\pm$    0.4  \\
  16 &   25.0 $\pm$    1.1 &   38.9 $\pm$    2.3 &   29.5 $\pm$    1.5 &   20.6 $\pm$    1.9 &    9.9 $\pm$    1.0  \\
  19 &               &   27.2 $\pm$    2.4 &   14.8 $\pm$    1.7 &   10.4 $\pm$    1.3 &    4.9 $\pm$    0.9  \\
  28 &   10.1 $\pm$    0.4 &   13.1 $\pm$    1.3 &    7.1 $\pm$    0.7 &    2.7 $\pm$    0.7 &    2.0 $\pm$    0.6  \\
  32 &   12.5 $\pm$    0.6 &   10.7 $\pm$    0.7 &    3.7 $\pm$    0.4 &    1.2 $\pm$    0.3 &    0.8 $\pm$    0.6  \\
  33 &   14.6 $\pm$    0.7 &   22.4 $\pm$    1.6 &   12.3 $\pm$    1.0 &    6.2 $\pm$    0.9 &    2.8 $\pm$    0.6  \\
  37 &    4.0 $\pm$    0.9 &   10.5 $\pm$    3.1 &    6.8 $\pm$    2.3 &               &                \\
  42 &   17.6 $\pm$    0.8 &   26.0 $\pm$    1.7 &   11.5 $\pm$    0.9 &    7.5 $\pm$    0.8 &    4.0 $\pm$    1.0  \\
  44 &   18.7 $\pm$    0.9 &   21.8 $\pm$    1.5 &    5.0 $\pm$    0.7 &    3.2 $\pm$    0.8 &    1.9 $\pm$    0.6  \\
  45 &    3.4 $\pm$    0.5 &   22.4 $\pm$    1.7 &   25.9 $\pm$    1.4 &   12.2 $\pm$    1.3 &    8.8 $\pm$    0.8  \\
  47 &               &   30.6 $\pm$    5.1 &   22.5 $\pm$    3.4 &   11.8 $\pm$    2.1 &    5.3 $\pm$    0.9  \\
  49 &               &    9.1 $\pm$    0.9 &    4.8 $\pm$    0.5 &    2.0 $\pm$    0.2 &    5.7 $\pm$    0.6  \\
  51 &    5.5 $\pm$    0.4 &   16.0 $\pm$    1.5 &   13.9 $\pm$    0.9 &    6.6 $\pm$    1.1 &    3.8 $\pm$    0.6  \\
  58 &    8.7 $\pm$    0.5 &    9.7 $\pm$    0.8 &    4.8 $\pm$    0.5 &    1.4 $\pm$    0.7 &                \\
  61 &               &    5.6 $\pm$    0.7 &   19.8 $\pm$    1.8 &   11.3 $\pm$    1.0 &    7.7 $\pm$    0.9  \\
  69 &    9.6 $\pm$    1.1 &   12.4 $\pm$    1.1 &    8.8 $\pm$    0.6 &    3.2 $\pm$    0.4 &    1.1 $\pm$    0.3  \\
  72 &   17.0 $\pm$    0.8 &   16.5 $\pm$    1.4 &    6.0 $\pm$    0.9 &    2.9 $\pm$    0.8 &                \\
  85 &    5.7 $\pm$    0.9 &   13.1 $\pm$    2.9 &   13.3 $\pm$    2.7 &    7.5 $\pm$    1.6 &    6.2 $\pm$    1.0  \\
  86 &               &    4.8 $\pm$    0.3 &    2.3 $\pm$    0.2 &    1.1 $\pm$    0.1 &                \\
  89 &   10.8 $\pm$    0.6 &   10.9 $\pm$    0.8 &    5.1 $\pm$    0.6 &    2.4 $\pm$    0.8 &    1.9 $\pm$    0.7  \\
  90 &    8.3 $\pm$    0.5 &   18.1 $\pm$    1.0 &   11.8 $\pm$    0.8 &    6.2 $\pm$    0.6 &    2.3 $\pm$    0.7  \\
  92 &               &    5.8 $\pm$    0.6 &    4.3 $\pm$    0.5 &    2.5 $\pm$    0.3 &    1.6 $\pm$    0.2  \\
 111 &    2.1 $\pm$    0.1 &    4.1 $\pm$    0.4 &    2.3 $\pm$    0.3 &    0.7 $\pm$    0.3 &                \\
 115 &    1.6 $\pm$    0.3 &    2.8 $\pm$    0.7 &    1.3 $\pm$    0.5 &    0.7 $\pm$    0.3 &                \\
 132 &               &    2.7 $\pm$    0.2 &    2.0 $\pm$    0.4 &    0.5 $\pm$    0.4 &                \\
 141 &    8.0 $\pm$    0.8 &   11.1 $\pm$    2.2 &    7.6 $\pm$    0.8 &    2.0 $\pm$    0.2 &    1.5 $\pm$    0.2  \\
 154 &    8.5 $\pm$    0.6 &   11.1 $\pm$    1.1 &    5.6 $\pm$    0.6 &    1.9 $\pm$    0.8 &    1.0 $\pm$    0.4  \\
 161 &    7.0 $\pm$    0.4 &   10.2 $\pm$    0.9 &    5.7 $\pm$    1.0 &               &    0.6 $\pm$    0.3  \\
 263 &               &               &   22.2 $\pm$    1.7 &   11.4 $\pm$    1.3 &    6.3 $\pm$    0.8  \\
 265 &               &    8.6 $\pm$    1.1 &    6.4 $\pm$    0.9 &    4.2 $\pm$    1.0 &    1.4 $\pm$    0.4  \\
 268 &               &    6.5 $\pm$    0.7 &    6.4 $\pm$    0.7 &    3.2 $\pm$    0.5 &    1.2 $\pm$    0.4  \\
 270 &               &               &   19.4 $\pm$    1.7 &   10.8 $\pm$    1.2 &    5.6 $\pm$    0.9  \\
 271 &   10.6 $\pm$    0.4 &   14.9 $\pm$    0.9 &    7.5 $\pm$    0.6 &    3.8 $\pm$    0.4 &    1.3 $\pm$    0.2  \\
 272 &               &               &   21.9 $\pm$    1.5 &   12.2 $\pm$    1.3 &    8.5 $\pm$    1.0  \\
 273 &    9.0 $\pm$    0.6 &    8.6 $\pm$    0.9 &    4.6 $\pm$    0.6 &    1.7 $\pm$    0.7 &                \\
 274 &               &    2.5 $\pm$    0.3 &    3.1 $\pm$    0.4 &    1.5 $\pm$    0.4 &                \\
 278 &    0.5 $\pm$    0.1 &    1.4 $\pm$    0.4 &    1.6 $\pm$    0.5 &               &                \\
 280 &               &               &    8.3 $\pm$    1.4 &    4.3 $\pm$    1.1 &    2.3 $\pm$    0.8  \\
 281 &               &    4.5 $\pm$    0.9 &    2.4 $\pm$    0.3 &    1.0 $\pm$    0.4 &                \\
 291 &               &    2.5 $\pm$    0.6 &    3.9 $\pm$    0.6 &    2.8 $\pm$    0.8 &                \\
 298 &    3.2 $\pm$    0.4 &    6.0 $\pm$    0.8 &    2.6 $\pm$    0.7 &               &                \\
 316 &   57.1 $\pm$    5.7 &   44.1 $\pm$    8.8 &   17.8 $\pm$    1.8 &    6.3 $\pm$    0.6 &    2.2 $\pm$    0.2  \\

%% file: tbl/compact_source_comments_v3_tex.tbl
1   & (100,160,250,350,500) \um\   &  (3.6,4.5,5.8,8.0,12,24,70) \um\                   &      no          &     cond-2       &      I    &   91C2, class\,I, high-mass YSO  \\
2   & (100,160,250,350,500) \um\   &  (2.17,3.6,4.5,5.8,8.0,12,24,70) \um\              &      51          &     cond-3       &      IM   &   3C3, class\,I                  \\
3   & (100,160,250,350,500) \um\   &  (3.6,4.5,5.8,8.0,24,70) \um\                      &      8           &     cond-4       &      I    &   2C4, class\,I                  \\
4   & (100,160,250,350,500) \um\   &  (1.25,1.65,2.17,3.6,4.5,5.8,8.0,24,70) \um\       &      no          &     cond-3       &      IM   &   class\,I                       \\
5   & (100,160,250,350,500) \um\   &  (1.25,1.65,2.17,3.6,4.5,5.8,8.0,12,24,70) \um\    &      2,51        &     cond-3       &      I    &   2C3, class\,I                  \\
8   & (100,160,250,350,500) \um\   &  (3.6,4.5,5.8,8.0) \um\                            &      3           &     cond-4       &      IM   &   3C4, class\,I                  \\
10  & (100,160,250) \um\           &  (3.6,4.5,5.8,8.0,12,24,70) \um\                   &      no          &     cond-1       &      I    &   237C2, class\,I, high-mass YSO \\
11  & (100,160,250,350,500) \um\   &  (1.25,1.65,2.17,3.6,4.5,5.8,8.0,70) \um\          &      no          &     cond-9       &      I    &   class\,I                       \\
15  & (100,160,250,350,500) \um\   & Filaments at [3.6,4.5,5.8,8.0,12,24,70] \um\       &      no          &     SW           &      I    &   I                              \\
16  & [100](160,250,350,500) \um\  & Filaments at [3.6,4.5,5.8,8.0,12,24,70] \um\       &      no          &     cond-3       &      0    &   class\,0                       \\
19  & (160,250,350,500) \um\       &  (3.6,4.5,5.8,8.0,24) \um\                         &      no          &     cond-3       &      IM   &   1C3, class\,I                  \\
28  & (100,160,250,350,500) \um\   &  (1.65,2.17,3.6,4.5) \um\                          &      no          &     cond4below   &      I    &   class\,I                       \\
32  & (100,160,250,350,500) \um\   &  Filaments at [3.6,4.5,5.8,8.0,12,24,70] \um\      &      no          &     SW           &      I    &   class\,I                       \\
33  & [100](160,250,350,500) \um\  &             [8.0,12,24,70] \um\                    &      no          &     cond2below   &      IM   &   IM                             \\
37  & (100,160,250) \um\           &  [5.8,8.0,12,24,70] \um\                           &      no          &     cond-2       &      IM    &  IM                             \\
42  & (100,160,250,350,500) \um\   &  [5.8,8.0,12,24,70] \um\                           &      no          &     cond-5       &      IM   &   IM                             \\
44  & (100,160,250,350,500) \um\   &  Filaments at [3.6,4.5,5.8,8.0,12,24,70] \um\      &      no          &     SW           &      IM   &   IM                             \\
45  & (100,160,250,350,500) \um\   &  [5.8,8.0] \um\                                    &      no          &     cond-3       &      0    &   class\,0                       \\
47  & (160,250,350,500) \um\       &  [5.8,8.0,12,24,70] \um\                           &      no          &     cond-2       &      0    &   class\,0                       \\
49  & (160,250,350,500) \um\       &  [5.8,8.0,12,24,70] \um\                           &      no          &     cond-2       &      IM    &   IM                            \\
51  & (100,160,250,350,500) \um\   &  (2.17,3.6,4.5,5.8,8.0) \um\                       &      2,4         &    cond-3        &     0     &   5C3, class\,I                  \\
58  & (100,160,250,350) \um\       &  [5.8,8.0,12,24,70] \um\                           &      no          &    cond-4        &     IM    &   IM                             \\
61  & [160](250,350,500) \um\      &  Absorption at [5.8,8.0,12] \um\                   &      no          &    cond-8        &     0     &   class\,0                       \\
69  & (100,160,250,350,500) \um\   &  Filaments at [3.6,4.5,5.8,8.0,12,24,70] \um\      &      no          &    SW            &     IM    &   IM                             \\
72  & (100,160,250,350) \um\       &  Filaments at [3.6,4.5,5.8,8.0,12,24,70] \um\      &      no          &    SW            &     IM    &   IM                             \\
85  & (100,160,250,350,500) \um\   &  [3.6,4.5,5.8,8.0,12,24] \um\                      &      no          &    cond-2        &     0     &   class\,0                       \\
86  & (100,160,250,350,500) \um\   &  [3.6,4.5,5.8,8.0,70] \um\                         &      no          &    Center        &     IM    &   IM                             \\
89  & (100,160,250,350,500) \um\   &  [5.8,8.0,12,24,70] \um\                           &      95          &    cond-4        &     IM    &   IM                             \\
90  & [100](160,250,350,500) \um\  &  Absorption at [8,12,24] \um\                      &      no          &    cond-7        &     0     &   class\,0                       \\
92  & (160,250,350,500) \um\       &  Absorption at [8,12,24] \um\                      &      no          &    NE            &     0     &   class\,0                       \\
111 & (100,160,250,350,500) \um\   &  [5.8,8.0,12,24,70] \um\                           &      no          &    cond-4        &     IM     &  IM                             \\
115 & (100,160,250,350) \um\       &  [5.8,8.0,12,24,70] \um\                           &      no          &    cond3below    &     IM     &  IM                             \\
132 & (160,250,350) \um\           &  [12,24,70] \um\                                   &      no          &    NE            &     IM     &  IM                             \\
141 & (160,250,350,500) \um\       &  Absorption at [8.0] \um                           &      no          &    cond-2        &     IM     &   AC2, class\,I                  \\
154 & (100,160,250,350,500) \um\   &  Filaments at [5.8,8.0,12,24,70] \um\              &      no          &    cond2below    &     IM    &   IM                             \\
161 & [100](160,250,500) \um\      &  Filaments at [5.8,8.0,12,24,70] \um\              &      no          &    cond2below    &     IM    &   IM                             \\
263 & (250,350,500) \um\           &  Absorption at [8.0,12,24] \um\                    &      no          &    cond-4        &     0     &   class\,0                       \\
265 & (160,250,350,500) \um\       &  Absorption at [8.0,12,24] \um\                    &      no          &    cond-1        &     0     &   class\,0                       \\
268 & (160,250,350,500) \um\       &  Absorption at [8.0,12,24] \um\                    &      no          &    cond3below    &     0     &   class\,0                       \\
270 & (250,350,500) \um\           &  Absorption at [8.0,12,24,70] \um\                 &      no          &    cond-5        &     0     &   class\,0                       \\
271 & (100,160,250,350,500) \um\   &  [8.0,12,24,70] \um\                               &      no          &    NE            &     IM    &   IM                             \\
272 & (250,350,500) \um\           &  Absorption at [8.0,12,24,70] \um\                 &      no          &    cond-4        &     0    &    class\,0                       \\
273 & (100,160,250,350) \um\       &   [8.0,12,24,70] \um\                              &      no          &    cond-4        &     IM    &   IM                             \\
274 & (160,250,350) \um\           &  Absorption at [8.0,12,24,70] \um\                 &      no          &    cond-4        &     0     &   class\,0                       \\
278 & (100,160,250) \um\           &  Absorption at [8.0,12,24,70] \um\                 &      no          &    cond-4        &     0     &   class\,0                       \\
280 & (250,350,500) \um\           &  Absorption at [8.0,12,24,70] \um\                 &      no          &    cond-9        &     0     &   class\,0                       \\
281 & (160,250,350) \um\           &  [8.0,12,24,70] \um\                               &      no          &    cond2below    &     I     &   class\,I                       \\
291 & (160,250,350) \um\           &  [5.8,8.0,12,24,70] \um\                           &      no          &    cond-2        &     0     &   class\,0                       \\
298 & (100,160,250) \um\           &  [5.8,8.0,12,24,70] \um\                           &      no          &    cond-4        &     IM    &   IM                             \\
316 & [100](160,250,350,500) \um\  &  [5.8,8.0,12,24,70] \um\                           &      no          &    cond-2        &     I     &   class\,I                       \\

%% file: tbl/NIR_MIR_phot_final.tbl
   1 & G308.7543+00.5486  &              &               &               &   20.4 $\pm$    0.8 &  107.7 $\pm$   13.8 &  156.7 $\pm$    8.6 &  210.0 $\pm$    9.3 &  353.5 $\pm$    7.4 & 6959.4 $\pm$   57.3 \\
   2 & G308.6874+00.5240  &             &               &    6.2 $\pm$    0.3 &   57.9 $\pm$   10.0 &  278.6 $\pm$   17.7 &  532.4 $\pm$   17.1 &  487.6 $\pm$   16.6 &   74.5 $\pm$    2.5 & 1272.5 $\pm$   26.4 \\
   3 & G308.6467+00.6458  &              &               &               &    9.7 $\pm$    3.0 &   23.2 $\pm$    3.8 &  178.6 $\pm$   19.6 &  456.8 $\pm$   67.7 &               &  946.0 $\pm$   23.6 \\
   4 & G308.6884+00.5279  &   1.0 $\pm$    0.5 &    6.1 $\pm$    1.9 &    3.8 $\pm$    0.3 &    5.0 $\pm$    0.7 &   12.6 $\pm$    1.6 &   16.8 $\pm$    1.0 &   27.2 $\pm$    2.6 &               &  242.2 $\pm$   10.0 \\
   5 & G308.7009+00.5315  &   3.7 $\pm$    0.1 &    7.8 $\pm$    0.3 &   12.5 $\pm$    0.4 &   43.0 $\pm$   13.8 &   43.5 $\pm$    4.6 &  259.0 $\pm$   12.8 &  668.2 $\pm$   36.3 &  680.0 $\pm$    8.0 & 6633.2 $\pm$   56.8 \\
   8 & G308.6465+00.6503  &              &               &               &   20.1 $\pm$    1.6 &   49.1 $\pm$    3.6 &   95.4 $\pm$    3.9 &  129.5 $\pm$    7.5 &               &               \\
  10 & G308.7681+00.5706  &              &               &               &   11.0 $\pm$    0.7 &   45.0 $\pm$    2.3 &  112.7 $\pm$    4.1 &  162.3 $\pm$    9.0 &   42.2 $\pm$    1.8 &  759.4 $\pm$   14.4 \\
  11 & G308.6729+00.6335  &   6.5 $\pm$    0.8 &   16.7 $\pm$    0.8 &   51.0 $\pm$    1.6 &   86.5 $\pm$    9.2 &   95.3 $\pm$    8.1 &  128.8 $\pm$   11.8 &  178.8 $\pm$   25.8 &               &               \\
  19 & G308.7129+00.5349  &              &               &               &   34.5 $\pm$    2.9 &   65.5 $\pm$    3.4 &  114.2 $\pm$    7.5 &  122.0 $\pm$   27.8 &               &  168.9 $\pm$    6.1 \\
  28 &                    &              &    1.3 $\pm$    1.8 &    3.5 $\pm$    3.3 &    1.6 $\pm$    0.1 &    1.3 $\pm$    0.1 &               &               &               &               \\
  51 & G308.6895+00.5253  &              &               &    3.7 $\pm$    0.3 &   31.4 $\pm$    3.5 &   78.0 $\pm$    6.4 &  134.9 $\pm$    9.0 &  140.6 $\pm$    7.2 &               &                \\